\documentclass[english,twoside,a4paper,10pt]{report}

\usepackage[latin1]{inputenc}
\usepackage[T1]{fontenc}
\usepackage[english]{babel}
\usepackage{amsmath}
\usepackage{amsfonts}
\usepackage{graphicx}
\usepackage{subfigure}
\usepackage{a4wide}
\usepackage{amssymb}
\usepackage{fancyhdr}
\usepackage{mathrsfs}
\usepackage[toc,page]{appendix}

\begin{document}


\begin{titlepage}

\Huge
\center{ \textsc{Some aspects of generalized modified gravity models} } \\
\large
\vspace{2cm}
R. Myrzakulov$^+$\footnote{Email: rmyrzakulov@gmail.com; rmyrzakulov@csufresno.edu}\,,
L. Sebastiani$^{+}$\footnote{E-mail address:l.sebastiani@science.unitn.it
}\, and S. Zerbini$^{++}$\footnote{E-mail address:zerbini@science.unitn.it}\\
\vspace{0.5cm}
\begin{small}
$^+$ Eurasian International Center for Theoretical Physics and  Department of General
\end{small}\\
\begin{small} 
Theoretical Physics, Eurasian National University, Astana 010008, Kazakhstan
\end{small}\\
\begin{small}
$^{++}$ Dipartimento di Fisica, Universit\`a di Trento, Italy and 
\end{small}\\
\begin{small}
 Gruppo Collegato di Trento, Istituto Nazionale di Fisica Nucleare, Sezione di Padova, Italy
\end{small}\\
\vspace{2cm}

In the present work, we review some general aspects of modified gravity theories, investigating mathematical and physical properties and, more specifically, the feature of viable and realistic models able to reproduce the dark energy epoch and the early-time inflation. We will discuss the black hole solutions in generalized theories of gravity: it is of fundamental interest to understand  
how properties and laws of black holes in General Relativity can be addressed in the framework of modified theories. In particular, we will discuss the energy issue and the possibility to derive the First Law of thermodynamics from the field equations. Then, in the analysis of cosmological solutions, we will pay a particular attention to the occurrence of finite-time future singularities and to the possibility to avoid them in $\mathcal F(R,G)$-gravity. Furthermore, realistic models of $F(R)$-gravity will be analyzed in the detail. A general feature occurring in matter era will be shown, namely the high derivatives of Hubble parameter may be influenced by the high frequency oscillation of the dark energy and some correction term is required in order to stabilize the theory at high redshift. The inflationary scenario is also carefully analyzed and an unified description of the universe is risen. In the final part of the work, we will have a look at the last developments in modified gravity, namely we will investigate cosmological and black hole solutions in a covariant field theory of gravity and we will introduce the extended ``teleparallel'' $F(T)$-gravity theories. A nice application to the dark matter problem will be presented.\\
\vspace{0.5cm}
\vspace{0.5cm}


\end{titlepage}





\addcontentsline{toc}{chapter}{Contents}
\tableofcontents


\chapter{Introduction}

\paragraph*{} Recent observational data imply -against any previous belief- that the current expansion of the universe
is accelerating~\cite{SN11}-\cite{WMAP}. Since this discovery, the so called Dark Energy issue has become the ``Mystery
of the Millennium'' \cite{Padmanabhan:2006ag}. Today, dark
energy is probably the most ambitious and tantalizing field of
research because of its implications in fundamental physics. There exist several descriptions of the acceleration of the universe. Among them, the simplest one is the introduction of small positive 
Cosmological Constant in the framework of General Relativity, the so called $\Lambda$CDM model, where the dark energy, whose energy density is given by Cosmological Constant, drives the accelerated expansion of the universe. Alternatively, accelerating Friedmann-Robertson-Walker universe may be  described by quintessence/phantom-fluid or other kind of
inhomogenous fluid, satisfying suitable Equation of State.  That the dark fluid has an Equation of State parameter $\omega_{\mathrm{DE}}$
very close to minus one represents an important point in favour
of a Cosmological Constant-like representation of the dark energy, but in principle quintessence/phantom-fluid is not excluded.
However, the estimated extremely small value of Cosmological Constant leads to several well-know problems. The first one, is the so called `cosmological constant problem'. In quantum field theory, the Cosmological Constant appears as the vacuum energy density, which has to be included in gravity theory, as the vacuum effect~\cite{Casimir} may suggest. On the other hand, the expected value of vacuum energy density results to be of 122 orders (!) of magnitude larger than the observed value. Supersymmetry and strings theories aim to solve this problem by different ways, but up to now a successful answer seems to be far away~\cite{Maggiore}. 

Other questions arise from standard cosmology (the so-called `coincidence problem', linked with the same order of magnitude of matter and dark energy density in the universe today, the origin of dark matter, the absence of a consistent quantum theory of gravity and so on), and, despite the successful results obtained by General Relativity in describing the universe and the Solar System, it is well accepted the idea according to
which General Relativity plus Cosmological Constant is not the ultimate theory of gravity,
but an extremely good approximation valid in the present day
range of detection.

The existence of an early accelerated epoch in our universe, namely the `hot universe' scenario or inflation, adds a new problem to the standard cosmology, and various proposals to construct acceptable inflationary model exist (scalar, spinor, (non-)abelian vector theory and so on). Otherwise, the scenarios to describe the early-time and the late-time accelerations are usually very similar and is quite natural to expect that same theory lies behind both they. Since General Relativity with matter and radiation correctly describes the intermediate (decelerated) expansion of the universe, it is reasonable to expect that a different gravitational theory dictates the (Friedman-Robertson-Walker) background evolution at high and small energy (curvatures) without the introduction of any other dark components. 

The modified theories of gravity represent a generalization of Einstein's gravity, where some combination of curvature invariants (The Riemann tensor, the Weyl tensor, the Ricci tensor and so on) replaces or is added into the classical Hilbert-Einstein action formed by the Ricci scalar term $R$. Thus, in this framework, the early-time and the late-time acceleration may be caused by the fact that some (sub)-dominant terms of gravitational action become essential at high or small curvatures. Moreover, some other related problem of Cosmological Constant could be solved in this way. Of course, the complete understanding of gravity and the fundamental theory remains to be an open problem of modern physics. 

The original idea of taking the gravitational action to be a more general invariant of the Riemann tensor has been contemplated long time ago by Buchdahl~\cite{B}.
In 1980 Starobinsky proposed the introduction of a correction to 
Hilbert-Einstein action in the form of $R+R^2$
in order to solve many of the problems left open by the
inflation~\cite{Starobinsky:1980te},  so that the Starobinsky model can be
considered as the first modified gravity inflationary model.
Finally, after the discover of cosmic acceleration,
the
interest in models of modified gravity grew up in the last then years. 
In Refs.~\cite{Capozziello001}-\cite{turner} is possible to find some examples. Here, the first candidate proposed to explain the current acceleration was the model $R-\mu_0^4/R$, with $\mu_0$ on the same order of Hubble parameter today, but this theory is subject to cosmological instabilities. In Ref.~\cite{lnR}, one of the first consistent models which passes the cosmological test was investigated, and the first work of a viable unification of the early- and late-time acceleration was studied by Nojiri \& Odintsov
in Ref.~\cite{inlation+DE}. Moreover, in Ref.~\cite{dark matter+dark energy}, Capozziello \emph{et al.} suggested that both, dark matter and dark energy, are curvature effects of some modification to standard gravity. The origin of F(R)-gravity from string was proposed in Ref.~\cite{Mtheory1, Mtheory2},
where new gravitational physics comes from $M$-theory.  
For a general review of dark energy see Ref.~\cite{DEreview}.

The mathematical structure of modified theories of gravity
and their physical properties are an exciting field of
research.
Furthermore, despite the arena of modified gravity-models is in principle infinite, 
the very accurate data arisen from observation of our universe, restrict the field of viable models.  

The aim of this work is to present the both, some mathematical and physical general aspects of modified gravity, and, more specifically, the proprieties of viable, realistic models of modified gravity which can be used to reproduce the inflation and the dark energy epoch of universe today. 

The work is organized as the following. In Chapter {\bf 2}, the formalism of $\mathcal{F}(R,G)$-modified gravity is presented. In this kind of theories, the modification to the Hilbert-Einstein action is given by the function $\mathcal{F}(R,G)$ of the Ricci scalar $R$ and the Gauss-Bonnet invariant $G$. 
A remark is in order. As a rule, modification of gravity may contain a huge list of invariants. Otherwise, we often work with the above specific class of   
models. 
The popularity of $F(R)$-gravity, where the modification is a function of the Ricci scalar only, is clearly motivated by the easier formalism and by the prospect to find a final theory of gravity in its simplest form. Furthermore, the Gauss-Bonnet modified gravity is a string-inspired theory which also has been proposed as a good candidate for the inflation and the dark energy scenario. In Chapter {\bf 3}, we will discuss the black hole solutions in modified gravity. Our aim is to address in the framework of generalized theories of gravity proprieties and laws of the black holes of General Relativity.
The black hole solutions in modified gravity are not expected to share the same
laws of their Einsteinian counterparts. Some of the physical quantities one
would like to define for modified gravity are their mass, their
horizon entropy, their temperature and so on. 
In particular, the mass problem results to be an interesting issue and
here we propose and identify the mass with
a quantity proportional to the constant of integration, which appears 
in the explicit black hole solutions, making use of 
derivation of the First Law of black hole thermodynamics from the equations of motion (where it is possible), and
evaluating independently the entropy via Wald method and the 
Hawking temperature via quantum mechanical methods in curved space-times. The results are
investigated in $F(R)$-, $\mathcal F(R,G)$-gravity and therefore an attempt to extend them to general classes of modified gravity theories is done.
In Chapter {\bf 4}, we consider the (Friedman-Robertson-Walker) cosmological solutions of $\mathcal{F}(R,G)$-gravity and we study the occurence of the finite-time future singularities. It is explicitly demonstrated that the Gauss-Bonnet modified gravity as well as the $F(R)$-gravity may show singularities during cosmological evolution. However, the introduction of specific form of curvature terms may naturally solve the problem of singularities in generalized theories of gravity: moreover, we will see that this kind of terms may also induce the early-time acceleration.  In other words, the non-singualar dark energy scenario is strictly related with the inflation and an unified description is suggested. In Chapter {\bf 5}, as a prosecution of Chapter {\bf 4}, we study inhomogeneous viscous fluids, especially related with modified gravity and singularities. In Chapters {\bf 6}, {\bf 7}, {\bf 8}, we restrict our analysis to realistic models of $F(R)$-modified gravity producing (unstable) de Sitter inflation and (stable) de Sitter solution of dark energy epoch, the so called `one step'- and `two steps'-models, paying particular attention to exponential gravity corrections of General Relativity.
 It is quite interesting to note how, despite this models mimic with high precision the $\Lambda$CDM Model, the dynamical behaviour of Equation of State and the introduction of new degree of freedom in the equations of motion, involve a very accurate analysis in order to reach the feasibility of the models and in order to fit the all the most recent and accurate
observational data. 
In Chapter {\bf 7}, 
we show 
that 
the behavior of higher derivatives of the Hubble parameter in $F(R)$-gravity may be 
influenced by large frequency oscillations of effective dark energy during matter era, 
which makes solutions singular and unphysical at a high redshift. 
As a consequence, 
we examine an additional correction term to the models 
in order to remove any instability with keeping the viability properties. 
This analysis is very interesting, since points out a general feauture of realistic $F(R)$-gravity,
as it is carefully demonstrated in analytical and numerical way for power-law and exponential gravity.
Some comments about future universe evolution and the growth of matter perturbations are also given.
In Chapter {\bf 8}, the numerical analysis of viable inflation is carried out.
To conclude, in the last Chapters we will have a look at the last stadies of modified gravity, by considering very recent proposals. 
In Chapter {\bf 9}, the black hole and de Sitter solutions are considered in a covariant-renormalizable field theory of gravity. The popularity of this kind of theories is related with the possibility to reach a quantum theory of gravity. 
Finally, in Chapter {\bf 10}, an introduction to $F(T)$-models is given. These models are based on the ``teleparallel'' equivalent of General
Relativity and instead of using
the curvature defined via the Levi-Civita connection, we use the
Weitzenb$\ddot{o}$ck connection, 
and work with the torsion $T$ instead the curvature $R$: an interesting application to dark matter problem is presented. 

\paragraph*{} The present work is based on the Refs. \cite{ProceedingUno, SSSsolutions, Weyl, GBSSS, SSSEnergy, Entropy, GBSingularities, Viscousfluidsingularities, Viscousfluids, Twostepmodels0, Twostepmodels, malloppone, omegaDE, covarianttheory, davoodDMmodel}, published in the referred journals and conference proceedings. A part of this work was carried out also in Ref.~\cite{tesi}.\\

\paragraph*{} {\bf Units:}
We use units of $k_{\mathrm{B}} = c = \hbar = 1$ and denote the
gravitational constant $G_N$ by $\kappa^2\equiv 8 \pi G_{N}$, such that
$G_{N}^{-1/2} =M_{\mathrm{Pl}}$, being  
$M_{\mathrm{Pl}} =1.2 \times 10^{19}$GeV the Planck mass.



\chapter{The formalism of $\mathcal{F}(R,G)$-theories of gravity.}

\paragraph*{} In modified theories of gravity the Hilbert-Einstein action of General Relativity (GR), that is the Ricci scalar $R$, is substituted by a more general combination of curvature invariants like the Riemann and Ricci tensors or tensors formed from these by the operations of taking duals, contractions or covariant differentiations.
The simplest class of modified gravitational
theories is the $F(R)$-gravity, where the Lagrangian is a function $F(R)$ of the Ricci scalar only~\cite{Review1}-\cite{Review4}. 
Among them, the most reasonable choice is 
$F(R)=R+f(R)$, where the modification is given by the function $f(R)$ of the Ricci scalar itself. In this way, one can 
describe the universe where we live
by treating the modification of gravity like an effective fluid producing acceleration of dark energy (DE) epoch.    
An other interesting class of modified theories is string-inspired Gauss-Bonnet
gravity, the so-called
$F(G)$-gravity~\cite{F(G)review1}-\cite{F(G)review21},
where $F(G)$ is a function of the Gauss-Bonnet four dimensional topological invariant $G$.
In this Chapter, 
we explore the formalism of $\mathcal{F}(R,G)$-gravity models, where the modification to GR is given by a general combination of the both, the Ricci scalar and Gauss-Bonnet invariant, 
and we briefly derive the gravitational field equations for  the Friedman Robertson Walker (FRW) and spherically static symmetric (SSS) metrics. In the end of the Chapter, we will give some elments of scalar tensor theories.  

\section{The action and FRW equations of motion}

\paragraph*{} The action of $\mathcal{F}(R,G)$-gravity in four dimension is given by
\begin{eqnarray}
I = \int_\mathcal{M} d^4 x \sqrt{-g} \left[ \frac{\mathcal{F}(R,G)}{2\kappa^2}
+{\mathcal{L}}^{\mathrm{(matter)}} \right]\,,
\label{azione}
\end{eqnarray}
where $g$ is the determinant of the metric tensor $g_{\mu\nu}$,
${\mathcal{L}}^{\mathrm{(matter)}}$ is the matter Lagrangian
and $\mathcal{M}$ is the space-time manifold. 
$\mathcal{F}(R,G)$ is a generic function of the Ricci scalar $R$ and the Gauss Bonnet four dimensional topological invariant\footnote{
In four dimension, any linear combination of the Gauss-Bonnet invariant does not contribute to the effective Lagrangian.
} $G$:
\begin{equation}
G=R^{2}-4R_{\mu\nu}R^{\mu\nu}+R_{\mu\nu\xi\sigma}R^{\mu\nu\xi\sigma}\,.\label{GaussBonnet}
\end{equation}
In fact, the Gauss-Bonnet is a combination of the Riemann Tensor $R_{\mu\nu\xi\sigma}$, the Ricci Tensor $R_{\mu\nu}=R^{\rho}_{\mu\rho\nu}$ and its trace $R=g^{\alpha\beta}R_{\alpha\beta}$.\\ 
The gravitational field equations are
derived from the action (\ref{azione}) and read
\begin{eqnarray}
{\mathcal{F}}'_{R}\left( R_{\mu\nu}-\frac{1}{2}R g_{\mu\nu}\right)
&=& \kappa^2 T^{(\mathrm{matter})}_{\mu \nu}
+\frac{1}{2}g_{\mu\nu} \left(\mathcal{F}-{\mathcal{F}}'_{R}R\right)
+{\nabla}_{\mu}{\nabla}_{\nu}
{\mathcal{F}}'_{R} -g_{\mu\nu} \Box {\mathcal{F}}'_{R}
\nonumber \\ \nonumber \\
&& \hspace{-30mm}
+\bigl(-2RR_{\mu\nu} +4R_{\mu\rho}R_{\nu}{}^{\rho}
-2R_{\mu}{}^{\rho\sigma\tau}R_{\nu\rho\sigma\tau}
+4g^{\alpha\rho}g^{\beta\sigma}R_{\mu\alpha\nu\beta}R_{\rho\sigma}
\bigr){\mathcal{F}}'_{G}
\nonumber \\ \nonumber \\
&& \hspace{-30mm}
+2\left({\nabla}_{\mu}{\nabla}_{\nu} {\mathcal{F}}'_{G} \right)R
-2g_{\mu \nu}\left(\Box {\mathcal{F}}'_{G} \right)R
+4\left(\Box {\mathcal{F}}'_{G} \right)R_{\mu \nu}
-4\left({\nabla}_{\rho}{\nabla}_{\mu} {\mathcal{F}}'_{G} \right)
R_{\nu}{}^{\rho}
\nonumber \\ \nonumber \\
&& \hspace{-30mm}
-4\left({\nabla}_{\rho}{\nabla}_{\nu} {\mathcal{F}}'_{G} \right)
R_{\mu}{}^{\rho}
+4g_{\mu \nu}\left({\nabla}_{\rho}{\nabla}_{\sigma}
{\mathcal{F}}'_{G} \right)R^{\rho\sigma}
-4\left({\nabla}_{\rho}{\nabla}_{\sigma} {\mathcal{F}}'_{G} \right)
g^{\alpha\rho}g^{\beta\sigma}R_{\mu\alpha\nu\beta}\,.
\label{Field equation}
\end{eqnarray}
In the above equation, $\mathcal{F}(R,G)$ has been replaced with $\mathcal{F}$ and we have used the following expressions:
\begin{eqnarray}
{\mathcal{F}}'_{R} \equiv
\frac{\partial \mathcal{F}}{\partial R}\,, \quad
{\mathcal{F}}'_{G} \equiv
\frac{\partial \mathcal{F}}{\partial G}\,.
\label{convention}
\end{eqnarray}
Furthermore, ${\nabla}_{\mu}$ is the covariant derivative
operator associated with $g_{\mu \nu}$,
$\Box\phi\equiv g^{\mu\nu}\nabla_{\mu}\nabla_{\nu}\phi$ is the
covariant d'Alembertian for a scalar field $\phi$, and
$T^{(\mathrm{matter})}_{\mu \nu} = \mathrm{diag} \left(\rho_{\mathrm{m}}, p_{\mathrm{m}}, p_{\mathrm{m}}, p_{\mathrm{m}}
\right)$
is the contribution to the stress energy-momentum tensor from
all ordinary matters\footnote{In general, it includes matter and radiation.},
with
$\rho_{\mathrm{m}}$ and $p_{\mathrm{m}}$ being, respectively, the matter energy-density and
pressure.
By putting $\mathcal{F}(R,G)=R$, we recover the Einstein's Equation.

The most general flat  FRW space-time is described by the metric
\begin{equation}
ds^{2}=-N^2(t)dt^{2}+a(t)^2\left(dx^2+dy^2+dz^2\right)\,,
\label{metric}
\end{equation}
where $a(t)$ is the scale factor of the universe and $N(t)$ is an arbitrary function of the cosmic time $t$.
In what follows, we take the gauge $N(t)=1$.

In the FRW background, 
from $(\mu,\nu)=(0,0)$ component 
and the trace part of $(\mu,\nu)=(i,j)$, with $i,j=1,2,3$, in
Eq.~(\ref{Field equation}), we obtain the equations of motion (EOMs):
\begin{eqnarray}
\left(\frac{3}{\kappa^{2}}H^{2}\right)\mathcal{F}'_R &=&
\rho_{\mathrm{m}} +
\frac{1}{2\kappa^{2}}
\left[ \left( {\mathcal{F}}'_{R}R+G{\mathcal{F}}'_{G}-\mathcal{F} \right)
-6H{\dot{\mathcal{F}}}'_{R}
-24H^3{\dot{\mathcal{F}}}'_{G}
\right]\,,\label{EOM1}
\\ \nonumber\\
-\frac{1}{\kappa^{2}} \left( 2\dot H+3H^{2} \right)\mathcal{F}_R' &=&
p_{\mathrm{m}} +\frac{1}{2\kappa^{2}} \Bigl[
-\left( {\mathcal{F}}'_{R}R+G{\mathcal{F}}'_{G}-\mathcal{F} \right)
+4H{\dot{\mathcal{F}}}'_{R}+2{\ddot{\mathcal{F}}}'_{R}
\nonumber \\ \nonumber\\
&&+16H\left(\dot{H} +H^2 \right){\dot{\mathcal{F}}}'_{G}+8H^2 {\ddot{\mathcal{F}}}'_{G}
\Bigr]\,. \label{EOM2}
\end{eqnarray}
Here, $H=\dot{a}(t)/a(t)$ is the Hubble parameter and
the dot denotes the time derivative.
Moreover, the Ricci scalar and the Gauss-Bonnet invariant read
\begin{eqnarray}
R &=& 6 \left(2H^{2}+\dot H \right)\,, \label{R}
\\
G &=& 24H^{2} \left( H^{2}+\dot H \right)\,.\label{G}
\end{eqnarray} 

In a large class of modified gravity theories which reproduce the cosmology of Standard Model 
at the level of Solar System or at high curvatures of matter era, one has
\begin{equation}
\mathcal{F}(R,G)=R+f(R,G)\,.\label{actiontwo}
\end{equation}
Thus, the modification to gravity is encoded in the function $f(R,G)$ of $R$ and $G$, which is added to the classical action of General Relativity as a suitable gravitational term yelding inflation and/or current cosmic acceleration without invoking any dark component.
In this text, we will often discuss modified gravity in this form, 
such that, in analogy with the theory of Einstein, we can rewrite the field equations as 
\begin{equation}
R_{\mu\nu}-\frac{1}{2}Rg_{\mu\nu}=\kappa^2 \left
(T^{{\mathrm{MG}}}_{\mu\nu}
+\tilde{T}^{\mathrm{(matter)}}_{\mu\nu}\right) \,.\label{fieldequation}
\end{equation}
Here, the part of modified
gravity is
formally included into the `modified gravity' stress-energy tensor
$T^{{\mathrm{MG}}}_{\mu\nu}$, 
\begin{eqnarray}
T_{\mu\nu}^{{\mathrm{MG}}}&\equiv&\frac{1}{\kappa^2
\mathcal{F}'(R)}\biggl\{
\frac{1}{2}g_{\mu\nu} \left(\mathcal{F}-{\mathcal{F}}'_{R}R\right)
+{\nabla}_{\mu}{\nabla}_{\nu}
{\mathcal{F}}'_{R} -g_{\mu\nu} \Box {\mathcal{F}}'_{R}
\nonumber \\ \nonumber \\
&& \hspace{-15mm}
+\bigl(-2RR_{\mu\nu} +4R_{\mu\rho}R_{\nu}{}^{\rho}
-2R_{\mu}{}^{\rho\sigma\tau}R_{\nu\rho\sigma\tau}
+4g^{\alpha\rho}g^{\beta\sigma}R_{\mu\alpha\nu\beta}R_{\rho\sigma}
\bigr){\mathcal{F}}'_{G}
\nonumber \\ \nonumber \\
&& \hspace{-15mm}
+2\left({\nabla}_{\mu}{\nabla}_{\nu} {\mathcal{F}}'_{G} \right)R
-2g_{\mu \nu}\left(\Box {\mathcal{F}}'_{G} \right)R
+4\left(\Box {\mathcal{F}}'_{G} \right)R_{\mu \nu}
-4\left({\nabla}_{\rho}{\nabla}_{\mu} {\mathcal{F}}'_{G} \right)
R_{\nu}{}^{\rho}
\nonumber \\ \nonumber \\
&& \hspace{-15mm}
-4\left({\nabla}_{\rho}{\nabla}_{\nu} {\mathcal{F}}'_{G} \right)
R_{\mu}{}^{\rho}
+4g_{\mu \nu}\left({\nabla}_{\rho}{\nabla}_{\sigma}
{\mathcal{F}}'_{G} \right)R^{\rho\sigma}
-4\left({\nabla}_{\rho}{\nabla}_{\sigma} {\mathcal{F}}'_{G} \right)
g^{\alpha\rho}g^{\beta\sigma}R_{\mu\alpha\nu\beta}\biggr\}\,.
\end{eqnarray}
Hence, one must not forget that gravitational terms enter in both sides of the
Eq.~(\ref{fieldequation}). Furthermore,
$\tilde{T}^{\mathrm{(matter)}}_{\mu\nu}$ is given by the non-minimal
coupling
of the ordinary matter stress-energy tensor
$T^{\mathrm{(matter)}}_{\mu\nu}$
with geometry, namely,
\begin{equation}
\tilde{T}^{\mathrm{(matter)}}_{\mu\nu}=\frac{1}{\mathcal{F}'_R}T^{\mathrm{(matter)}}_
{\mu\nu}\,.
\end{equation}
It should be noted that only 
$T^{\mathrm{(matter)}}_{\mu\nu}$ is covariant conserved, and formally $\kappa^2/\mathcal{F}'_R$ may be interpreted
as an effective gravitational constant.
Now equations~(\ref{EOM1})--(\ref{EOM2}) read
\begin{equation}
\rho_{\mathrm{eff}}=\frac{3}{\kappa^{2}}H^{2}\label{EOM1bis}\,,
\end{equation}
\begin{equation}
p_{\mathrm{eff}}=-\frac{1}{\kappa^{2}} \left( 2\dot H+3H^{2} \right)\label{EOM2bis}\,,
\end{equation}
where $\rho_{\mathrm{eff}}$ and $p_{\mathrm{eff}}$ are
the effective energy density and pressure of the universe, respectively, and
these are defined as
\begin{eqnarray}
\rho_{\mathrm{eff}} &\equiv&
\frac{1}{{\mathcal{F}}'_{R}} \left\{ \rho_{\mathrm{m}} +
\frac{1}{2\kappa^{2}}
\left[ \left( {\mathcal{F}}'_{R}R+G{\mathcal{F}}'_{G}-\mathcal{F} \right)
-6H{\dot{\mathcal{F}}}'_{R}
-24H^3{\dot{\mathcal{F}}}'_{G}
\right] \right\}\,,
\label{rhoeffRG0} \\ \nonumber\\
p_{\mathrm{eff}} &\equiv&
\frac{1}{{\mathcal{F}}'_{R}} \biggl\{ p_{\mathrm{m}} +
\frac{1}{2\kappa^{2}} \Bigl[
-\left( {\mathcal{F}}'_{R}R+G{\mathcal{F}}'_{G}-\mathcal{F} \right)
+4H{\dot{\mathcal{F}}}'_{R}+2{\ddot{\mathcal{F}}}'_{R}
+16H\left(\dot{H} +H^2 \right){\dot{\mathcal{F}}}'_{G}
\nonumber \\ \nonumber\\
&& \hspace{0mm}
+8H^2 {\ddot{\mathcal{F}}}'_{G}
\Bigr]
\biggr\}\,.
\label{peffRG0}
\end{eqnarray}
On shell, namely by using Eqs. (\ref{EOM1bis})--(\ref{EOM2bis}) and by writing $\mathcal F'_R$ as $1+(\mathcal F'_R-1)$, one has
\begin{eqnarray}
\rho_{\mathrm{eff}} &\equiv&
\rho_{\mathrm{m}} +
\frac{1}{2\kappa^{2}}
\left[ \left( {\mathcal{F}}'_{R}R+G{\mathcal{F}}'_{G}-\mathcal{F} \right)-6H^2(\mathcal{F}'_R-1)
-6H{\dot{\mathcal{F}}}'_{R}
-24H^3{\dot{\mathcal{F}}}'_{G}
\right]\,,
\label{rhoeffRG} \\ \nonumber\\
p_{\mathrm{eff}} &\equiv&
p_{\mathrm{m}} +
\frac{1}{2\kappa^{2}} \Bigl[
-\left( {\mathcal{F}}'_{R}R+G{\mathcal{F}}'_{G}-\mathcal{F} \right)+(4\dot{H}+6H^2)(\mathcal{F}'_R-1)
+4H{\dot{\mathcal{F}}}'_{R}+2{\ddot{\mathcal{F}}}'_{R}
\nonumber \\ \nonumber\\
&& \hspace{0mm}
+16H\left(\dot{H} +H^2 \right){\dot{\mathcal{F}}}'_{G}
+8H^2 {\ddot{\mathcal{F}}}'_{G}
\Bigr]\,.
\label{peffRG}
\end{eqnarray}
For General Relativity with $\mathcal{F}(R,G)=R$, we get
$\rho_{\mathrm{eff}} = \rho_{\mathrm{m}}$ and $p_{\mathrm{eff}} = p_{\mathrm{m}}$ and
therefore Eqs.~(\ref{EOM1bis})--(\ref{EOM2bis}) are the usual Friedman equations.

The following matter conservation law can be derived,
\begin{equation}
\dot \rho_{\mathrm{m}}+3H(\rho_{\mathrm{m}}+p_{\mathrm{m}})=0\,.\label{conservationlaw}
\end{equation}
For a perfect fluid, it gives the Equation of State (EoS)
\begin{equation}
p_{\mathrm{m}}=\omega\rho_{\mathrm{m}}\,,
\end{equation}
$\omega$ being the thermodynamical EoS-parameter of matter. For standard matter, $\omega=0$ and $\rho_{\mathrm{m}}= \rho_{\mathrm{m}(0)} a(t)^{-3}$, and for radiation, $\omega=1/3$ and $\rho_{\mathrm{r}}=\rho_{\mathrm{r}(0)} a(t)^{-4}$, $\rho_{\mathrm{r}}$ being the radiation energy density and $\rho_{\mathrm{m}(0)}$, $\rho_{\mathrm{r}(0)}$ generic constants given by boundary conditions.

We also can introduce the effective EoS by using the corresponding parameter $\omega_{\mathrm{eff}}$,
\begin{equation}
\omega_{\mathrm{eff}}\equiv\frac{p_\mathrm{eff}}{\rho_{\mathrm{eff}}}\,,\label{omegaeffdef}
\end{equation}
such that (on shell)
\begin{equation}
\omega_{\mathrm{eff}}=-1-\frac{2\dot{H}}{3H^2}\,.\label{omegaeff}
\end{equation}
Thus, if the strong energy condition (SEC) is violated
($\omega_{\mathrm{eff}}<-1/3$),
the universe expands in an accelerating way, and vice-versa.

\section{$F(R)$-gravity: critical points and stability of cosmological perturbations}

\paragraph*{} This Section is devoted to the specific study of $F(R)$-gravity and in the next Section we will generalize some results to a more general class of modified gravity theories.

The action of $F(R)$-gravity is given by
\begin{equation}
I=\int_{\mathcal{M}} d^4x\sqrt{-g}\left[
\frac{F(R)}{2\kappa^2}+\mathcal{L}^{\mathrm{(matter)}}\right]\,.\label{actionF(R)}
\end{equation}
Now, $F(R)$ is a function of the Ricci scalar $R$ only.
Eq.~(\ref{Field equation}) simply reads\\
\phantom{line}
\begin{equation}
F'(R)\left(R_{\mu\nu}-\frac{1}{2}Rg_{\mu\nu}\right)=\kappa^2T^{{\mathrm{(matter)}}}_{\mu\nu}+\left\{\frac{1}{2}g_{\mu\nu}[F(R)-RF'(R)]
+(\nabla_{\mu}\nabla_{\nu}-g_{\mu\nu}\Box)F'(R)\right\}\,.
\label{fieldequationF(R)}
\end{equation}
\phantom{line}\\
The prime denotes derivative with respect to the curvature $R$. 
The starting point is the trace of field equations, which is 
trivial in Einstein gravity, $R=-\kappa^2 \mathrm{T}^{\mathrm{(matter)}}$, with $\mathrm{T}^{\mathrm{(matter)}}=g^{\mu\nu}T^{\mathrm{(matter)}}_{\mu\nu}$ the trace of the matter stress energy tensor, but now it is
\begin{equation}
3\Box F'(R)+RF'(R)-2F(R)=\kappa^2 \mathrm{T}^{\mathrm{(matter)}}\,.\label{scalaroneq}
\end{equation}
We can rewrite this equation as
\begin{equation}
\Box F'(R)=\frac{\partial V_{\mathrm{eff}}}{\partial F'(R)}\,,\label{scalaroneeqbis}
\end{equation}
where
\begin{equation}
\frac{\partial V_{\mathrm{eff}}}{\partial
F'(R)}=\frac{1}{3}\left[2F(R)-RF'(R)+\kappa^2 \mathrm{T}^{\mathrm{(matter)}}\right]\,,\label{Veff}
\end{equation}
$F'(R)$ being the so-called `scalaron' or the effective scalar degree
of freedom.
On the critical points of the
theory, the effective potential $V_{\mathrm{eff}}$ has a maximum (or
minimum), such that
\begin{equation}
\Box F'(R)=0\,,\label{criticalpoint}
\end{equation}
and
\begin{equation}
2F(R)-R F'(R)=-\kappa^2
\mathrm{T}^{\mathrm{(matter)}}
\,.\label{criticalpointbis}
\end{equation}
For example, in the absence of matter, i.e. $\mathrm{T}^{\mathrm{(matter)}}=0$, 
one finds that the de Sitter/Anti-de Sitter (dS/AdS) critical point 
associated with a constant scalar curvature $R_{\mathrm{dS}} $ must satisfy the relation
\begin{equation}
2F(R_{\mathrm{dS}})-R_{\mathrm{dS}}F'(R_{\mathrm{dS}})=0\,.\label{dScondition}
\end{equation}
We have derived the de Sitter condition without using a specific metric Ansatz. It is valid in FRW space-time, as well as in the SSS one, like for the well-known case of Scwarzshild-dS/AdS solution.

Performing the variation of Eq.~(\ref{scalaroneq}) with respect to $R=R^{(0)}+\delta R$,
where $R^{(0)}$ is not necessarly a constant,
if we
remember that 
$\Box F'(R)=F''(R)\Box R+F'''\nabla^{\mu} R\nabla_{\nu} R$,
we get, to first order in $\delta R$,\\
\phantom{line}
\begin{eqnarray}
&& \Box
R^{(0)}+\frac{F'''(R^{(0)})}{F''(R^{(0)})}g^{\mu\nu}\nabla_{\mu}R^{(0)}\nabla_{\nu}R^{(0)}-\frac{1}{3F''(R^{(0)}
)}
\left[2F(R^{(0)})-R^{(0)}F'(R^{(0)})+\kappa^2 \mathrm{T}^{\mathrm{matter}}\right] \nonumber \\ \nonumber \\
&&\hspace{10mm} +\Box \delta
R+\left\{\left[\frac{F''''(R^{(0)})}{F''(R^{(0)})}-\left(\frac{F'''(R^{(0)})}{F''(R^{(0)})}\right)^2\right]
g^{\mu\nu}\nabla_{\mu}R^{(0)}\nabla_{\nu}R^{(0)}
+\frac{R^{(0)}}{3}-\frac{F'(R^{(0)})}{3F''(R^{(0)})}
\right. \nonumber \\ \nonumber\\
&& \hspace{10mm}\left. +\frac{F'''(R^{(0)})}{3(F''(R^{(0)}))^2}\left[ 2F(R^{(0)})-R^{(0)}F'(R^{(0)})+\kappa^2 \mathrm{T}^{\mathrm{(matter)}}
\right]\right\}\delta R \nonumber\\ \nonumber\\
&&\hspace{10mm}+2\frac{F'''(R^{(0)})}{F''(R^{(0)})}g^{\mu\nu}\nabla_{\mu}R^{(0)}\nabla_{\nu}\delta R
+\mathcal{O}(\delta R^2) -\frac{\kappa^2}{3F''(R^{(0)})}\delta\mathrm{T}^{\mathrm{(matter)}}\simeq 0\,.\label{completeperturbationEq}
\end{eqnarray}
\phantom{line}\\
Here, $\delta\mathrm{T}^{\mathrm{(matter)}}$ is the variation of the trace of matter stress energy tensor. 
The above equation can be used to study perturbations around critical
points. 
The simplest case is given by a constant value of $R^{(0)}$. 
For the dS/AdS solution ($R^{(0)}=R_{\mathrm{dS}}$), by neglecting the contribute of matter and   
by using the relation (\ref{dScondition}), we obtain
\begin{equation}
\left(\Box-m^2\right)\delta R\simeq 0\,,
\label{boxdeltaR}
\end{equation}
where
\begin{equation}
m^2=\frac{1}{3}\left(\frac{F'(R_{\mathrm{dS}})}{F''(R_{\mathrm{dS}})}-R_{\mathrm{dS}}
\right)\,.\label{scalaronmass}
\end{equation}
Note that
\begin{equation}
m^2=\frac{\partial^2 V_{\mathrm{eff}}}{\partial
F'(R_{\mathrm{dS}})^2}\,.
\end{equation}
The second derivative of the effective potential represents the
effective
mass of the scalaron. Thus, if $m^2>0$ (in the sense of the quantum
theory,
the scalaron, which is a new scalar degree of freedom, is not a
tachyon),
one gets a stable solution. In this case the stability consition reads
\begin{equation}
\frac{F'(R_{\mathrm{dS}})}{R_{\mathrm{dS}}
F''(R_{\mathrm{dS}})}>1\,.\label{dSstability}
\end{equation}
Again, this result is independent on the metric background. We can take the FRW metric for which Eq.~(\ref{boxdeltaR}) corresponds to\\
\phantom{line}\\
\begin{equation}
-\left(\ddot{\delta R}\pm 3H_{dS}\dot{\delta R}+m^2\delta R\right)\simeq 0\,, 
\end{equation}
\phantom{line}\\
where $H_{\mathrm{dS}}=\pm\sqrt{R_{\mathrm{dS}}/12}$ (the signs minus are for the AdS solution), and cleary we see that perturbation $\delta R$ decreases or oscillates with time if $m^2$ is
positive, and exponentially diverges if $m^2$ is negative. 
As an example, let us consider modified gravity in the form
\begin{equation}
F(R)=R+\alpha R^n\,,
\end{equation}
where $\alpha$ is a constant dimensional parameter and $n$ is a positive number. In vacuum, this model leads to the de Sitter solution
\begin{equation}
 R_{\mathrm{dS}}=\left(\frac{1}{\alpha(n-2)}\right)^{\frac{1}{n-1}}\,,\quad n\neq 2\,,\label{RdeSitter}
\end{equation}
as a consequence of Eq.~(\ref{dScondition}). We assume $\alpha>0$ if $n>2$ and $\alpha<0$ if $0<n<2$. The stability condition (\ref{dSstability}) reads 
\begin{equation}
\frac{1}{n}>1\,. 
\end{equation}
It means that if $0<n<1$ the de Sitter point is stable and vice versa. We also observe that the term $R^2$ is trivial on the the Sitter solution, since in this case Eq.~(\ref{dScondition}) is identically zero. 

\section{De Sitter solution and stability in $\mathcal{F}(R,P,Q)$-modified gravity}

\paragraph*{} In this Section, we deal with modified generalized models described by the Lagragian density $\mathcal{F}(R,P,Q)$~\cite{ProceedingUno},\\
\phantom{line}\\
\begin{equation}
I=\int_{\mathcal{M}} d^4x\sqrt{-g}\left[
\frac{\mathcal{F}(R,P,Q)}{2\kappa^2}+\mathcal{L}^{\mathrm{(matter)}}\right]\,,\label{actionF(RPQ)} 
\end{equation}
\phantom{line}\\
where $\mathcal{F}(R,P,Q)$ is a function of the Ricci scalar and the quadratic curvature invariants
\begin{equation}
P=R_{\mu \nu}R^{\mu \nu}\,,
\quad  
Q=R_{\mu \nu \xi \sigma}R^{\mu \nu \xi \sigma}\,.\label{PQinvariants}
\end{equation}
The Gauss-Bonnet corresponds to $G=R^2-4P+Q$, according with Eq.~(\ref{GaussBonnet}). 

The field equations for such class of models read \cite{easson}:\\
\phantom{line}
\begin{eqnarray}
\mathcal{F}'_R\left(R_{\mu\nu}-\frac{1}{2}g_{\mu\nu}R\right)&+&\frac{1}{2}g_{\mu\nu}(\mathcal{F}'_R R-\mathcal{F})
+2\mathcal{F}'_P\,R^\alpha{}_\mu\,R_{\alpha\nu}
+2\mathcal{F}'_Q\,R_{\alpha\beta\gamma\mu}\,R^{\alpha\beta\gamma}{}_\nu\nonumber\\ \nonumber\\
&+&g_{\mu\nu}\,\Box \mathcal{F}_R -\nabla_\mu\nabla_\nu \mathcal{F}'_R
-2\nabla_\alpha\nabla_\beta[\mathcal{F}'_P\,R^\alpha{}_{\mu}{}\delta^\beta{}_{\nu}]
+\Box(\mathcal{F}'_P\,R_{\mu\nu})\nonumber\\ \nonumber\\
&+&g_{\mu\nu}\,\nabla_\alpha\nabla_\beta(\mathcal{F}'_P\,R^{\alpha\beta})
-4\nabla_\alpha\nabla_\beta[\mathcal{F}'_Q\,R^\alpha{}_{\mu\nu}{}^\beta]
=8\pi G\,T_{\mu\nu}\ .\label{fieldequationRPQ}
\end{eqnarray}
\phantom{line}\\
For simplicity we have putted $\mathcal{F}(R,P,Q)\equiv\mathcal{F}$ and
\begin{eqnarray}
{\mathcal{F}}'_{R} \equiv
\frac{\partial \mathcal{F}}{\partial R}\,, \quad
{\mathcal{F}}'_{P} \equiv
\frac{\partial \mathcal{F}}{\partial P}\,, \quad
{\mathcal{F}}'_{Q} \equiv
\frac{\partial \mathcal{F}}{\partial Q}\,.
\label{convention2}
\end{eqnarray}
The trace of Eq.~(\ref{fieldequationRPQ}) is given by\\
\phantom{line}
\begin{equation}
\nabla^2 \left(3\mathcal{F}'_R+R\mathcal{F}'_P \right)+2\nabla_\mu \nabla_\nu \left[ \left(\mathcal{F}'_P+2\mathcal{F}'_Q\right)R^{\mu \nu}\right]-2\mathcal{F}+R\mathcal{F}'_R
+2\left(\mathcal{F}'_P+\mathcal{F}'_Q\right)=\kappa^2 \mathrm{T}^{\mathrm{(matter)}}\,.
\end{equation}  
\phantom{line}\\
For $R=R_{0} $, $P=P_{0} $, and $Q=Q_{0}$, where $R_{0}$, $P_{0}$, $Q_{0}$ are constants, one has the dS/AdS condition in vacuum  
\begin{equation}
2\mathcal{F}_{(0)}-R_0\mathcal{F}'_{R(0)}-2P_0\mathcal{F}'_{P(0)}-2Q_0\mathcal{F}'_{Q(0)}=0 \, .
\end{equation}
The adding subscript `0' indicates that the functions are evaluated at the de Sitter point.
Perturbing around  the de Sitter-space, namely   $R=R_0+\delta R $,  $P=P_0+\delta P $, and $Q=Q_0+\delta Q $, 
observing that\footnote{In the specific, the following relations hold true:
\begin{equation*}
\delta P=2R_{\mu\nu}\delta R^{\mu\nu}=\frac{2}{4}R_{\mu\nu}g^{\mu\nu}\delta R=\frac{R_0}{2}\delta R\,,           
\end{equation*}
\begin{equation*}
\delta Q=2R_{\mu\nu\xi\sigma}\delta R^{\mu\nu\xi\sigma}=\frac{2}{6}R_{\mu\nu\xi\sigma}\frac{(g^{\mu\nu}g^{\xi\sigma}-g^{\nu\sigma}g^{\xi\mu})}{2}\delta R=\frac{R_0}{3}\delta R\,. 
\end{equation*}
}
$P_0=R_0^2/4$ and $Q_0=R_0^2/6$, and
$\delta P= (R_0/2)\,\delta R $, and $\delta Q= (R_0/3)\,\delta R $, 
one obtains to first order in $\delta R$
\begin{equation}
\left(\Box-M^2\right)\delta R\simeq 0\,,
\end{equation}
in which the scalaron effective mass reads
\begin{equation}
M^2= \frac{R_0}{3}\left(\frac{\mathcal{F}'_{R(0)}+\frac{2R_0}{3}
\left(\mathcal{F}'_{P(0)}+\mathcal{F}'_{Q(0)} \right)}{R_0 \left[A_{R(0)}+A_{P(0)}+A_{Q(0)}+\frac{2}{3}\left(\mathcal{F}'_{P(0)}+\mathcal{F}'_{Q(0)}\right)\right]}-1\right)\,,
\end{equation}
with\\
\phantom{line}
\begin{eqnarray}
\left\{\begin{array}{l}
A_{R(0)} =\phantom{\frac{R}{1}}\left(\mathcal{F}''_{R R}+\frac{R}{2}\mathcal{F}''_{R P}+\frac{R}{3}\mathcal{F}''_{R Q}\right)\Big\vert_{R_0,P_0,Q_0}\,,\\ \\
A_{P(0)}=\frac{R}{3}\left( \mathcal{F}''_{R Q}+\frac{R}{2}\mathcal{F}''_{Q P}+\frac{R}{3}\mathcal{F}''_{Q Q}\right)\Big\vert_{R_0,P_0,Q_0}\,,\\ \\
A_{Q(0)} =\frac{R}{2}\left(\mathcal{F}''_{R P}+\frac{R}{2}\mathcal{F}''_{P P}+\frac{R}{3}\mathcal{F}''_{P Q}\right)\Big\vert_{R_0,P_0,Q_0}\,.
\end{array}\right.
\end{eqnarray}
Thus, if  $M^2>0$, one has  stability of the de Sitter solution. In the particular case $\mathcal{F}(R,P,Q)=\mathcal{F}(R,G)$, we get \cite{Monica, Cognola} 
\begin{equation}
\frac{9\mathcal{F}'_{R}}{R [9\mathcal{F}''_{R R}+6 R \mathcal{F}''_{R G}+R^2\mathcal{F}''_{G G}] }\Big\vert_{R_0,\,G_0}> 1\,.
\end{equation}
\phantom{line}\\
We note that $G_0=R_0^2/6$. For $F(R)$-gravity, one also recovers the condition (\ref{dSstability}).

\section{Topological static spherically symmetric metric: Lagrangian derivation\label{1.4}} 

\paragraph*{} 
In this Section we derive the EOMs for topological SSS vacuum solutions in $\mathcal{F(R,G)}$-gravity. A convenient Lagrangian derivation is presented. 

We shall look for static, (pseudo-)spherically symmetric solutions
with various topologies, and write the metric element as
\begin{equation}
ds^2=-e^{2\alpha(r)}B(r)dt^2+\frac{d r^2}{B(r)}+r^2\,\left(\frac{d\rho^2}{1-k\rho^2}+\rho^2 d\phi^2\right)\label{SSS}\,,
\end{equation}
where $\alpha(r)$ and $B(r)$ are functions of the radius $r$, and the manifold will be either a sphere $S_2$, a torus $T_2$ or a compact hyperbolic manifold $Y_2$, according to whether $k=1,0,-1$, respectively. 

With this metric Ansatz, the scalar curvature and the Gauss-Bonnet read
\begin{eqnarray}
R  &=&
-\frac{1}{r^2}\left[3r^2\,\left({\frac{d B(r)}{dr}}\right)\left({\frac{d\alpha(r)}{dr}}\right)
+2r^2\,B\left(r\right)\left({\frac{d\alpha(r)}{dr}}
\right)^{2}+r^2\left({\frac{d^{2}B(r)}{d{r}^{2}}}\right)
\right.\nonumber\\ \nonumber\\
&&+2r^2\,B\left(r\right)\left({\frac{d^{2}\alpha(r)}{d{r}^{2}}}\right)\left.+4r\,\left({\frac{d B(r)}{dr}}\right)
+4 r B(r)\,\left({\frac{d\alpha(r)}{dr}}\right)+2B(r)
-2k\right]\,,\label{Rspheric}\\ \nonumber\\
G &=&\frac{4}{r^2}\left[\left(\frac{d\alpha(r)}{dr}\right)\left(\frac{d B(r)}{dr}\right)(5B(r)-3)+\left(\frac{d B(r)}{dr}\right)^2+\left(\frac{d^2 B(r)}{dr^2}\right)(B(r)-k)\right.\nonumber\\ \nonumber\\
&&\left.+2(B(r)-k)B(r)\left(\left(\frac{d\alpha(r)}{dr}\right)^2 + \frac{d^2\alpha(r)}{dr^2}\right)\right]\,.\label{Gspheric} 
\end{eqnarray}
By plugging this expression into the action (\ref{azione}), and by neglecting the contribute of matter, one obtains a higher derivative Lagrangian theory. 
In order to work with a first derivatives Lagrangian system, we may use the  method of Lagrangian multipliers, adopted 
for the FRW space-time in  Refs.~\cite{Monica,Vilenkin,Capozziello}. In the spherically static case we are dealing with, the method  
permits to consider as independent Lagrangian coordinates 
the scalar curvature $R=R(r)$, the Gauss Bonnet invariant $G=G(r)$ and the quantities $\alpha(r)$ and $B(r)$, appearing in the metric. 
The main difference with respect to the other general approaches is that we do not directly use  field equations (\ref{Field equation}).   
By introducing the Lagrangian multipliers $\lambda$ and $\mu$ and making use of Eq.~(\ref{Rspheric})--(\ref{Gspheric}), the (effective) action  may be written as 
\begin{eqnarray}
\hat I &\equiv& \frac{1}{2\kappa^2}\int dt\int d{ r}\left(\mathrm{e}^{\alpha(r)}r^2 \right)\left\{ \mathcal{F}(R,G)-\lambda \left[R+3\,\left({\frac{d}{dr}}B\left(r\right)\right){\frac{d}{dr}}
\alpha\left(r\right)\right.\right.\nonumber\\ \nonumber\\
&& +2\,B\left(r\right)\left({\frac{d}{dr}}
\alpha\left(r\right)\right)^{2}+{\frac{d^{2}}{d{r}^{2}}}
B\left(r\right)+2\,B\left(r\right){\frac{d^{2}}{d{r}^{2}}}\alpha\left(r\right)+\frac{4}{r}\frac{d}{dr}B(r)\nonumber\\ \nonumber\\ 
&&+4\,{\frac{B\left(r\right)}{r}}{\frac{d}{dr}}\alpha\left(r\right)+2\,{\frac{B\left(r\right)}{{r}^{2}}}
\left.-\frac{2k}{{r}^{2}}\right]\nonumber\\ \nonumber\\
&&\hspace{-15mm}-\mu\left[G- 
\frac{4}{r^2}\left[\left(\frac{d\alpha(r)}{dr}\right)\left(\frac{d B(r)}{dr}\right)(5B(r)-3k)+\left(\frac{d B(r)}{dr}\right)^2+\left(\frac{d^2 B(r)}{dr^2}\right)(B(r)-k)\right.\right.\nonumber\\ \nonumber\\
&&\left.\left.+2(B(r)-k)B(r)\left(\left(\frac{d\alpha(r)}{dr}\right)^2 + \frac{d^2\alpha(r)}{dr^2}\right)\right]
\right\}\,.
\end{eqnarray}
The variations with respect to $R$ and $G$ lead to
\begin{equation}
\lambda=\mathcal{F}'_R(R,G)\,,
\quad
\mu=\mathcal{F}'_G(R,G)\,.
\end{equation}
Thus, by substituting this values and after a partial integration, the total Lagrangian $\mathcal L$ of the system assumes the form (we use convention (\ref{convention}))
\begin{eqnarray}
&&\hspace{-15mm}\mathcal L(\alpha, d\alpha/dr, B, d B/dr, R, d R/dr, G, d G/dr)=\mathrm{e}^{\alpha(r)}\left\{r^2\left(\mathcal{F}-\mathcal{F}'_R R-\mathcal{F}'_G G\right)\phantom{\frac{1}{1}}\right.
\nonumber\\ \nonumber\\
&&\hspace{15mm}+2\mathcal{F}'_R\left(k-r\frac{d B(r)}{dr}- B(r)\right)
+\mathcal{F}''_{RR}\frac{d R}{d r}r^2\left(\frac{d B(r)}{d r}+2B(r)\frac{d \alpha(r)}{dr}\right)\nonumber\\ \nonumber\\
&&\left.\hspace{15mm}-\mathcal{F}_{GG}''\frac{d G}{dr}\left(4\frac{d B(r)}{dr}+8B(r)\frac{d\alpha(r)}{de}\right)(B(r)-k)
\right\}\,.
\end{eqnarray}
This is a standard Lagrangian and depends on the first derivatives of the coordinates.
It is also easy to see that, if $\mathcal{F}'_G=\mathrm{const}$, namely $G$ simply is a linear term, the contribute of Gauss-Bonnet disappears.
Making the variation with respect to $\alpha(r)$ and with respect to $B(r)$, and by dropping $1/(2\kappa^2)$, one finally gets the EOMs:\\ 
\phantom{line}
\begin{eqnarray}
&&\hspace{-5mm}\mathrm{e}^{\alpha(r)}\left\{r^2(\mathcal{F}-\mathcal{F}'_R R-\mathcal{F}'_G G)+2\mathcal{F}'_R\left[k-r\left(\frac{d B(r)}{dr}\right)-B(r)\right]-\frac{d\mathcal{F}'_R}{dr}\left[r^2 \left(\frac{d B(r)}{dr}\right)+4rB(r)\right]\right. \nonumber\\ \nonumber\\
&&\hspace{-8mm}\left.-2r^2B(r)\frac{d^2\mathcal{F}'_R}{dr^2}+4(3B(r)-k)\left(\frac{d B(r)}{dr}\right)\frac{d\mathcal{F}'_G}{dr}
+8B(r)(B(r)-k)\frac{d^2\mathcal{F}'_G}{dr^2}\right\}=0\,,\label{EOM1SphRG}\\ \nonumber\\
&&\hspace{-8mm}\mathrm{e}^{\alpha(r)}\left\{\frac{d\alpha(r)}{dr}\left(2r\mathcal{F}'_R+r^2\frac{d\mathcal{F}'_R}{dr}-4(3B(r)-k)\frac{d\mathcal{F}'_G}{dr}\right)-r^2\frac{d^2\mathcal{F}'_R}{dr^2}+4(B(r)-k)\frac{d^2\mathcal{F}'_G}{dr^2}\right\}=0\label{EOM2SphRG}. 
\end{eqnarray}
\phantom{line}\\
The above equations with Eqs.~(\ref{Rspheric})--(\ref{Gspheric}) form a system of four ordinary differential equations in the four unknown quantities $\alpha(r)$, $B(r)$, $R(r)$ and $G(r)$. By explicitly written $R$ and $G$ in Eqs.~(\ref{EOM1SphRG})--(\ref{EOM2SphRG}) as functions of $B(r)$ and $\alpha(r)$, we reduce the system to the two fourth order (independent) differential equations of (\ref{Field equation}), but we will see in Chapter {\bf 3} that the present form results to be much more useful in  finding explicit solutions.\\ 

Let us see for some well-known case in order to check the formalism.
When $\mathcal{F}(R,G)=R$, the equations lead to the topological Schwarzschild solution, 
\begin{equation}
\left\{\begin{array}{l}
\alpha(r)=\mathrm{const}\,,\\\\
B(r)=\left(k-\frac{c_0}{r}\right)\,.\label{S2}
\end{array}\right.
\end{equation}
Here, $c_0$ is an integration constant. Furthermore, the Ricci scalar results to be $R=0$.

Another well known solution when $\mathcal{F}(R,G)$ is a function of $R$ only, i.e. $\mathcal{F}(R,G)=F(R)$, is the one associated with $R$ constant. As a result, with  $\alpha(r)=\mathrm{const}$, 
Eq.~(\ref{EOM2SphRG}) is trivially satisfied, and Eq.~(\ref{Rspheric}) with Eq.~(\ref{EOM1SphRG}) leads to the topological Schwarschild-dS/AdS solution
\begin{equation}
B(r)=\left(k-\frac{c_0}{r}-\frac{\Lambda
r^2}{3}\right)\,,\label{S2bis}
\end{equation}
when the de Sitter condition (\ref{dScondition}) is verified, namely $2F(R)-RF'(R)=0$.
Here, $\Lambda=\mathrm{const}$ such that 
$R=4\Lambda$.

In the same way, it is also possible to verify that in general $\mathcal F(R,G)$-gravity the pure topological dS/AdS solution,
\begin{equation}
B(r)=\left(k-\frac{\Lambda r^2}{3}\right)\,,\label{puredS} 
\end{equation}
with $\alpha(r)=\mathrm{const}$, exists
provided by the condition $\mathcal F-R \mathcal F'_R-G\mathcal F'_G+2\Lambda\mathcal F'_R=0$, where $R=4\Lambda$ and $G=8\Lambda^2/3$ are constants. For example, the model $\mathcal F(R,G)=R+\gamma G^2$, with $\gamma$ a dimensional parameter, exhibits the de Sitter solution with $\Lambda=[9/(32\gamma)]^{1/3}$.

\section{Conformal transformations in $F(R)$-gravity\label{conformal}}

\paragraph*{} In (non-minimally) scalar-tensor theories of gravity, a scalar field strongly coupled to the metric field through the Ricci scalar in the action is used. The first model of scalar-tensor theory was proposed by Brans \& Dicke in 1961~\cite{BransDicke}, trying to incorporate Mach's principle into the theory of gravity. In Brans-Dicke theory a scalar field $\phi$, whose kinetic term is proportional to $1/\phi$, is coupled with the Ricci scalar. Furthermore, in scalar tensor theories, a potential $V(\phi)$ of scalar field may appear. The success of this kind of theories principally is related with the possibility to reproduce the primordial acceleration of the universe, namely the inflation.      

A modified gravity theory may be rewritten in scalar-tensor or Einstein frame
form.
We analyze the case of $F(R)$-gravity.
One can
introduce
in the so called Jordan
frame action
of Eq.~(\ref{actionF(R)}) a scalar
field
which couples to the curvature.
Of course, this is not exactly physically-equivalent formulation, as
it is
explained in Ref.~\cite{Troisi}.
However, Einstein frame formulation may be used for getting some of
intermediate results in simpler form (especially, when the matter is
not
accounted for).

Let us introduce the field $A$ into
Eq.~(\ref{actionF(R)}):\\
\phantom{line}
\begin{equation}
I_{JF}=\frac{1}{2\kappa^{2}}\int_{\mathcal{M}}\sqrt{-g}\left[
F'(A) \, (R-A)+F(A)\right] d^{4}x\label{JordanFrame}\,.
\end{equation}
\phantom{line}\\
Here `$JF$' means `Jordan frame' and we neglect the contribute of matter. By making the variation of the
action
with respect to $A$, we obtain $A=R$. The scalar field $\sigma$ is
defined
as
\begin{equation}
\sigma = -\ln [F'(A)]\label{sigma}\,.
\end{equation}
Consider now the following conformal transformation of the metric,
\begin{equation}
\tilde g_{\mu\nu}=\mathrm{e}^{-\sigma}g_{\mu\nu}\label{conforme}\,,
\end{equation}
for which we get the `Einstein frame' action of the scalar field $\sigma$\\
\phantom{line}
\begin{eqnarray}
I_{EF} &=& \frac{1}{2\kappa^2}\int_{\mathcal{M}} d^4 x \sqrt{-\tilde{g}} \left\{ \tilde{R} -
\frac{3}{2}\left(\frac{F''(A)}{F'(A)}\right)^2
\tilde{g}^{\mu\nu}\partial_\mu A \partial_\nu A - \frac{A}{F'(A)}
+ \frac{F(A)}{F'(A)^2}\right\} \nonumber\\ \nonumber\\
&=&\frac{1}{2\kappa^2}\int_{\mathcal{M}} d^4 x \sqrt{-\tilde{g}} \left( \tilde{R} -
\frac{3}{2}\tilde{g}^{\mu\nu}
\partial_\mu \sigma \partial_\nu \sigma - V(\sigma)\right)\,,\label{EinsteinFrame}
\end{eqnarray}
\phantom{line}\\
where
\begin{equation}
V(\sigma)\equiv\frac{A}{F'(A)} - \frac{F(A)}{F'(A)^2}=\mathrm{e}^{\sigma}R(\mathrm{e}^{-\sigma})
-\mathrm{e}^{2\sigma}F[R(\mathrm{e}^{-\sigma})]\label{V(sigma)}\,.
\end{equation}
Here, $R(\mathrm{e}^{-\sigma})$ is the solution of
Eq.~(\ref{sigma}) with $A=R$, becoming $R$ a function of $\mathrm{e}^{-\sigma}$, and $\tilde{R}$ denotes the Ricci scalar evaluated in the conformal metric $\tilde{g}_{\mu\nu}$. Furthermore, $\tilde{g}$ is the determinant of conformal metric, namely $\tilde{g}=\mathrm{e}^{-4\sigma}g$.

At this point, it results to be interesting to make a comparison with conformal transformation of matter Lagrangian. After the scale transformation $g_{\mu\nu}\to\mathrm{e}^{-\sigma} g_{\mu\nu}$ is done,
there appears a coupling of the scalar field $\sigma$
with matter. For example, if matter is a scalar field $\Phi$, with mass
$M_{\Phi}$, whose action is given by\\
\phantom{line} 
\begin{equation}
I_{JF(\Phi)}=\frac{1}{2}\int_{\mathcal{M}} d^4x\sqrt{-g}\left(-g^{\mu\nu}\partial_\mu\Phi
\partial_\nu\Phi - M_{\Phi}^2 \Phi^2\right)\, ,
\end{equation}
\phantom{line}\\
then there appears a coupling with $\sigma$ in Einstein frame,\\
\phantom{line}
\begin{equation}
I_{EF(\Phi)}=\frac{1}{2}\int d^4x\sqrt{-\tilde{g}} \left(-\mathrm{e}^{\sigma}
\tilde{g}^{\mu\nu}\partial_\mu\Phi \partial_\nu\Phi
 - M_{\Phi}^2 \mathrm{e}^{2\sigma}\Phi^2\right)\, .
\end{equation}
\phantom{line}\\
The strength of the coupling is of the same order as that of the gravitational coupling
$\kappa^2$ in Eq.~(\ref{EinsteinFrame}). Unless the mass corresponding to $\sigma$, which is defined by $m^2_\sigma$\\
\phantom{line}
\begin{equation}
m_\sigma^2 \equiv \frac{3}{2}\frac{d^2 V(\sigma)}{d\sigma^2}
=\frac{3}{2}\left\{\frac{A}{F'(A)} - \frac{4F(A)}{\left(F'(A)\right)^2} +
\frac{1}{F''(A)}\right\}\, ,
\end{equation}
\phantom{line}\\
is big, the system is unstable. Sometimes in modified gravity 
it is necessary to check stability of the solutions with a detailed investigation on the mass of $\sigma$ in conformal transformation framework.  

\subsubsection{Conformal FRW metric}

\paragraph*{} By using a Lagrangian derivation similar to the one presented in \S~\ref{1.4}, we complete this Section by giving the FRW and the SSS-conformal equations of motion.
 
Let us consider the conformal transformation (\ref{conforme}) of FRW metric (\ref{metric}), namely
\begin{equation}
d\tilde s^2=-N(t)^2\mathrm{e}^{-\sigma(t)}dt^2+a(t)^2 \mathrm{e}^{-\sigma(t)}\left(dx^2+dy^2+dz^2\right)\,.\label{SSSconformalFRW}
\end{equation}
Here, $\sigma(t)$ is a function of $t$. The scalar curvature $\tilde R$ reads\\
\phantom{line}
\begin{eqnarray}
\tilde{R}&=&6e^{\sigma(t)}\left(\frac{\ddot{a}(t)}{a(t)N(t)^2}+\frac{\dot{a}(t)^2}{a(t)^2N(t)^2}-\frac{\dot{a}(t)\dot{N}(t)}{a(t)N(t)^3}\right)\nonumber\\\nonumber\\
&&+3e^{\sigma(t)}\left(\frac{\dot{\sigma}(t)^2}{2N(t)^2}+\frac{\dot{N}(t)\dot{\sigma}(t)}{N(t)^3}-\frac{3\dot{\sigma}(t)\dot{a}(t)}{a(t)N(t)^2}-\frac{\ddot{\sigma}(t)}{N(t)^2}\right)\,. 
\end{eqnarray}
\phantom{line}\\
If we put $\sigma(t)=0$ and $N(t)=1$, we obtain Eq.~(\ref{R}). By plugging this expression into the Einstein frame action (\ref{EinsteinFrame}) with conformal metric (\ref{SSSconformalFRW}) and by making an integration by part, one arrives at the Lagrangian,\\
\phantom{line}
\begin{eqnarray}
\mathcal L\left(a(t),\dot{a}(t),N(t),\dot{N}(t),\sigma(t),\dot{\sigma(t)}\right)&=&\nonumber\\\nonumber\\
&&\hspace{-30mm}\frac{6e^{-\sigma(t)}}{N(t)}\left[\dot{a}(t)a(t)^2\dot{\sigma}(t)-\dot{a}(t)^2
a(t)\right]-V(\sigma(t))a(t)^3N(t)e^{-2\sigma(t)}\,,
\end{eqnarray}
\phantom{line}\\
and we deal with a first derivative Lagrangian system.
The Hamilton-Jacobi equations for $N(t)$, $a(t)$ and $\sigma(t)$ give the following EOM in the gauge\footnote{We stress that, following this Lagrangian derivation, we must assume the most general form for the metric, and only once the EOMs have been derived the gauge $N(t)=1$ can be choosen. On the contrary, in the derivation of the EOMs through the direct differentiation of the action with respect to the metric tensor, we can fix the gauge from the beginning.}
 $N(t)=1$:
\phantom{line}\\
\begin{eqnarray}
6H\left(H-\dot{\sigma}(t)\right)&=&V(\sigma(t))e^{-\sigma(t)}\,, 
\\\nonumber\\
2\left(3H^2+2\dot{H}\right)&=&V(\sigma(t))e^{-\sigma(t)}\,, 
\\\nonumber\\
3\left(2H^2+\dot{H}\right)&=&e^{-\sigma(t)}\left(V(\sigma(t))-\frac{1}{2}\frac{d V(\sigma(t))}{d\sigma}\right)\,. 
\end{eqnarray}
\phantom{line}\\
Note that, due to the presence of scalar field $\sigma(r)$, in conformal theories we work with an additional equation of motion. If $V(\sigma(t))=0$, the last equation is redundant. 

\subsubsection{Conformal topological SSS metric}

\paragraph*{} Let us consider the conformal transformation (\ref{conforme})  of topological SSS metric (\ref{SSS}), namely
\begin{equation}
d\tilde s^2=-B(r)\mathrm{e}^{2\alpha(r)-\sigma(r)}dt^2+\frac{dr^2}{B(r)\mathrm{e}^{\sigma(r)}}+r^2\mathrm{e}^{-\sigma(r)}
\left(\frac{d\rho^2}{1-k\rho^2}+\rho^2 d\phi^2\right)\,.\label{SSSconformal}
\end{equation}
Here, the field $\sigma(r)$ is a function of $r$. The scalar curvature $\tilde R$ reads\\
\phantom{line}
\begin{eqnarray}
\tilde R &=& -2 \mathrm{e}^{\sigma\left(r\right)}B\left(r\right)\left(\frac{d\alpha(r)}{d r}\right)^2-\frac{4 \mathrm{e}^{\sigma\left(r\right)}B\left(r\right)}{r}\left(\frac{d\alpha(r)}{d r}\right)-3 \mathrm{e}^{\sigma\left(r\right)}B_r\left(r\right)\left(\frac{d\alpha(r)}{d r}\right)\nonumber\\\nonumber\\&&
-2 \mathrm{e}^{\sigma\left(r\right)}B\left(r\right)\left(\frac{d^2\alpha(r)}{d r^2}\right)+3 \mathrm{e}^{\sigma\left(r\right)}B\left(r\right)\left(\frac{d\alpha(r)}{d r}\right)\left(\frac{d\sigma(r)}{dr}\right)+\frac{6 \mathrm{e}^{\sigma\left(r\right)}B\left(r\right)}{r}\left(\frac{d\sigma(r)}{dr}\right)\nonumber\\\nonumber\\&&
+3 \mathrm{e}^{\sigma\left(r\right)}\left(\frac{d B(r)}{dr}\right)\left(\frac{d\sigma(r)}{dr}\right)-\frac{3}{2}\mathrm{e}^{\sigma\left(r\right)}B\left(r\right)\left(\frac{d\sigma(r)}{dr}\right)^2+3 \mathrm{e}^{\sigma\left(r\right)}B\left(r\right)\left(\frac{d^2\sigma(r)}{dr^2}\right)\nonumber\\\nonumber\\&&
-\frac{4 \mathrm{e}^{\sigma\left(r\right)}}{r}\left(\frac{d B(r)}{dr}\right)
-\mathrm{e}^{\sigma\left(r\right)}\left(\frac{d^2 B(r)}{dr^2}\right)
-\frac{2 \mathrm{e}^{\sigma\left(r\right)}B\left(r\right)}{r^2}+\frac{2k \mathrm{e}^{\sigma\left(r\right)}}{r^2}\,.
\end{eqnarray}
\phantom{line}\\
If we put $\sigma(r)=0$, we obtain Eq.~(\ref{Rspheric}). 
By plugging this expression into the Einstein frame action (\ref{EinsteinFrame}) with conformal metric (\ref{SSSconformal}), and by making an integration by part, one arrives at the Lagrangian,\\
\phantom{line}
\begin{eqnarray}
\mathcal{L}(\alpha(r), d\alpha(r)/d r, B(r), d B(r)/d r, \sigma(r), d\sigma(r)/d r)&=&\nonumber\\\nonumber\\
&&\hspace{-95mm}e^{\alpha(r)-\sigma(r)}\left(2k-2B(r)-2\frac{B(r)}{d r}r-\frac{\sigma(r)}{d r}\frac{B(r)}{d r}r^2-2\frac{\alpha(r)}{d r}\frac{\sigma(r)}{d r}B(r)r^2-e^{-\sigma(r)}V(\sigma(r))r^2\right)\,.
\end{eqnarray}
The Hamilton-Jacobi equations for $\alpha(r)$, $B(r)$ and $\sigma(r)$ give the following EOM:\\
\phantom{line}
\begin{eqnarray}
& &\hspace{-10mm} 2k-2\left(\frac{d B(r)}{dr}\right)r-2B(r)+\left(\frac{d B(r)}{dr}\right)\left(\frac{d \sigma(r)}{dr}\right)r^2+2B(r)\left(\frac{d^2 \sigma(r)}{dr^2}\right)r^2\nonumber\\\nonumber\\&&
\hspace{+0mm}+4r\left(\frac{d \sigma(r)}{dr}\right)B(r)
-2r^2\left(\frac{d \sigma}{dr}\right)^2B(r)=r^2\mathrm{e}^{-\sigma(r)}V(\sigma(r))\,,
\end{eqnarray}
\begin{equation}
\label{three}2r\left(\frac{d \alpha(r)}{dr}\right)-\left(\frac{d \alpha(r)}{dr}\right)\left(\frac{d \sigma(r)}{dr}\right)r^2+\left(\frac{d^2 \sigma(r)}{dr^2}\right)r^2-\left(\frac{d \sigma(r)}{dr}\right)^2r^2=0\,,
\end{equation}
\begin{eqnarray}
&&\hspace{-5mm}-\left(\frac{d^2 B(r)}{dr^2}\right)r^2-4\left(\frac{d B(r)}{dr}\right)r-3\left(\frac{d B(r)}{dr}\right)\left(\frac{d \alpha(r)}{dr}\right)r^2-2\left(\frac{d \alpha(r)}{dr}\right)^2B(r)r^2\nonumber\\\nonumber\\&&\hspace{-10mm}
-2\left(\frac{d^2 \alpha(r)}{dr^2}\right)B(r)r^2-4\left(\frac{d \alpha(r)}{dr}\right)B(r)r-2B(r)+2k
=r^2\mathrm{e}^{-\sigma(r)}\left(2V(\sigma(r))-\frac{d V(\sigma(r))}{d \sigma}\right)\,.\nonumber\\
\end{eqnarray}
\phantom{line}\\
Again, the last equation is redundant if $V(\sigma(r))=0$.\\

The scalar field formulation has many application also in Gauss-Bonnet modified gravity. For example, in Ref.~\cite{nonlocalgravity} a non-local model of modified gravity, which depends on Gauss Bonnet and other higher-derivative invariants (like $\Box^{-1}G$), is presented. By introducing a scalar field coupled with the metric through the Gauss-Bonnet invariant, it is shown that a local form can be obtained, and the analysis of the model results considerably simplified. Furthermore, it is interesting to note that results and methods discussed in this 
Section can be extended for ghost free massive $F(R)$-bigravity, recently proposed 
in Ref.~\cite{bigravity1, bigravity2}.     


\chapter{Black hole solutions and energy issue in modified gravity}

\def\hs{\qquad} 
\def\beq{\begin{eqnarray}} 
\def\eeq{\end{eqnarray}} 
\def\ap{\left.} 
\def\at{\left(} 
\def\aq{\left[} 
\def\ag{\left\{} 
\def\cp{\right.} 
\def\ct{\right)} 
\def\cq{\right]} 
\def\cg{\right\}} 
\def\Det{\mathop{\rm Det}\nolimits} 
\def\tr{\mathop{\rm tr}\nolimits} 
\def\Tr{\mathop{\rm Tr}\nolimits} 
\def\lap{\Delta\,} 
\def\ii{\infty}
\def\segue{\qquad\Longrightarrow\qquad} 
\def\prece{\qquad\Longleftarrow\qquad} 
\def\al{\alpha}
\def\be{\beta}
\def\ga{\gamma}
\def\de{\delta}
\def\ep{\varepsilon}
\def\ze{\zeta}
\def\io{\iota}
\def\ka{\kappa}
\def\la{\lambda}
\def\ro{\varrho}
\def\si{\sigma}
\def\om{\omega}
\def\ph{\varphi}
\def\th{\theta}
\def\te{\vartheta}
\def\up{\upsilon}
\def\Ga{\Gamma}
\def\De{\Delta}
\def\La{\Lambda}
\def\Si{\Sigma}
\def\Om{\Omega}
\def\Te{\Theta}
\def\Th{\Theta}
\def\Up{\Upsilon}
\newcommand{\bea}{\begin{eqnarray}}
\newcommand{\eea}{\end{eqnarray}}
\newcommand{\beaa}{\begin{eqnarray*}}
\newcommand{\eeaa}{\end{eqnarray*}}
\newcommand{\Lhat}{\widehat{\mathcal{L}}}
\def\nn{\nonumber}
\newcommand{\e}{{\rm e}}
\def\h{\tilde h}
\def\txi{\tilde\xi}
\def\B{\tilde B}
\def\C{\tilde C}
\def\R{R_0}
\def\g{\hat g}
\def\V{\tilde V}

\paragraph*{} 
Static, spherically symmetric solutions have been investigated in several papers and in the Chapter {\bf 2} we have derived the corresponding field equations for $\mathcal F(R,G)$-gravity. 
Typically, modified models
admit the de Sitter space as a solution, but the issue to find exact (non trivial) SSS metrics appears a formidable task, since also for a reasonable model, the equations of motion are much more complicated with respect to the ones of General Relativity.
One should note that
in order to verify the consistence of a gravity theory in the weak-field limits with the standard solar-system tests of GR, it is extremely important to know all the possible generalizations of the analogue of the Schwarzschild metric in the generalized theory.
Furthermore, the interest in SSS solutions is related with the possibility in describing black holes (BHs).
If a modified theory lies behind our universe, it is crucial to (re)write proprieties and laws of GR in the framework of extended gravity,
where the black holes are not expected to share the same laws of their Einsteinian counterparts: for this reason, following Visser in Ref.~\cite{Visser:1993nu} , we shall refer to them as 
``dirty black holes''. 
Some of the physical quantities one would like to address to dirty black holes are their mass, the horizon entropy, their temperature and so on. Thanks to the large amount of work carried over in the last decade, we can say that the issues of entropy and temperature  represent a well posed problem within the class of higher order gravitational models, but 
the issue associated with the energy (mass) of these black holes is still a debated question, despite tha fact that
several attempts in order to find a satisfactory answer to the problem 
have been investigated (see for example Refs.~\cite{D1,D2,Visser,Cai} and references therein). 
In this Chapter, we will exhibit some non trivial topological SSS vacuum solutions in $F(R)$ 
and $\mathcal F(R,G)$-gravity. 
Thus, an attempt to give an expression for the energy of the associated BHs is proposed and identified with
a quantity proportional to the constant of integration, which appears 
in the explicit solutions. 
The identification is achieved making use of 
derivation of the First Law of black hole thermodynamics from the equations of motion, 
evaluating independently the entropy via Wald method~\cite{Wald} and the 
Hawking temperature~\cite{HT} via quantum mechanical methods in curved space-times. 
An attempt to extend the results to
general classes of modified gravity theories is done and several non trivial examples are discussed. A particular attention is posed in Weyl conformal gravity. 
In this Chapter, even if not specified,
we denote with $c_i$, where $i=0,1,2$, generic dimensional or adimensional constants appearing in the solutions.


\section{$F(R)$-static spherically symmetric topological vacuum solutions\label{2.1}}

\paragraph*{} The number of exact non-trivial static black hole solutions so far known in modified theories of gravity is small:
for the specific choice $R^{1+\delta}$, with $\delta$ a real parameter, a class of 
exact SSS solutions has been presented by Barrow \& Clifton (2005) in Ref.~\cite{CB} and in
a simple class of Weyl gravity,
namely by adding to Hilbert-Einstein Lagrangian a
non-polynomial contribution of the type $\sqrt{C^2}$, with $C^2$ the square of the Weyl tensor,
a SSS solution has been proposed by Deser, Sarioglu \&
Tekin (2008) in Ref.~\cite{Deser}.
Furthermore, in pure Weyl conformal gravity whose action is given by $C^2$ only, an interesting solution has been derived by Riegert (1984) in Ref.~\cite{Riegert}.
General discussions on SSS solutions can be also found in 
Refs.~\cite{Multamaki1}-\cite{Saffari2}, where one can check for further references. The extremal limit of Schwarzshild-de Sitter solution in $F(R)$-gravity has been
recently studied in Ref.~\cite{extremal}.

In this Section,
we would like to investigate topological SSS solutions in the simple class of $F(R)$-gravity models,
by starting from the EOMs derived in  \S \ref{1.4}.
We will see how our Lagrangian derivation allows us to deal with a simple first order differential equation system which generates a large number of exact solutions \cite{SSSsolutions, Weyl}. 

By putting $\mathcal{F}(R,G)=F(R)$, Eqs.~(\ref{EOM1SphRG})-(\ref{EOM2SphRG}) read:\\
\phantom{line}
\begin{eqnarray}
\label{one}&&\hspace{-5mm}\mathrm{e}^{\alpha(r)}\left\{\left[F(R)-R\,F'(R)\right]+\frac{2F'(R)}{r^2}\left[k-B(r)-r\frac{d B(r)}{d r}\right]
-\frac{d F'(R)}{d r}\left[\frac{d B(r)}{d r}+\frac{4B(r)}{r}\right]
\right.\nonumber\\ \nonumber\\
& & \left.
-2 B(r)\frac{d^2 F'(R)}{d r^2}\right\}=0
\,,\\ \nonumber\\
&&\hspace{-5mm}\mathrm{e}^{\alpha(r)}\left[\frac{d\alpha(r)}{dr}\left(\frac{2}{r}+\frac{d F'(R)}{d r}\right)-\frac{d^2 F'(R)}{d r^2}\right]=0\,.\label{two}
\end{eqnarray}
\phantom{line}\\
Once $F(R)$ is given, together with 
equation (\ref{Rspheric}), the above equations form  a system of three (second order) differential equations for the three unknown 
quantities $\alpha(r), B(r)$ and $R(r)$. 
On the other hand, it is also possible to try to reconstruct the models by starting from the solutions. 
In fact, by taking the derivative with respect to $r$ of Eq. (\ref{one}), it is easy to see that all the system only depends on $F'(R)$.
As a consequence,
fixing the form of $\alpha(r)$, one may reconstruct the model by using Eq. (\ref{two}) and therefore $B(r)$ from Eq. (\ref{one}). 
In general the Lagrangian one eventually finds is not unique since we have to infer its form starting from the value it assumes on the solution. For example, all Lagrangians of the form $F(R) = R\,g(R)$ with $g(R)$ a generic function of $R$ such that $\lim_{R\rightarrow 0} \,g(R) = 0$, have the Schwarzschild solution.
We note that one advantage of our system is that $B(r)$ does not explicitly appear in Eq.~(\ref{two}) and vice versa for $\alpha(r)$.
In what follows, we shall consider two important cases, that is, $\alpha(r) =\text{Const}$ and $\alpha(r)=\log[(r/r_0)^z]$. The second choice corresponds to the Lifshitz solutions with redshift parameter $z$. Among this class, we will also present the topological version of SSS Clifton-Barrow solution.

\subsection{Solutions with constant $\alpha(r)$}

\paragraph*{} Now, let us consider the case of $\alpha(r)=\textrm{Const}$. We can directly put $\alpha(r)=0$ without loss of generality. The metric (\ref{SSS}) reads
\begin{equation}
ds^2=-B(r)dt^2+\frac{d r^2}{B(r)}+r^2\,\left(\frac{d\rho^2}{1-k\rho^2}+\rho^2 d\phi^2\right)\,.
\label{metric00}
\end{equation}
As a consequence, the Ricci scalar (\ref{Rspheric}) becomes\\
\phantom{line}
\begin{equation}
R(r)=-{\frac{d^{2}B(r)}{d{r}^{2}}}
-\frac{4}{r}\frac{d B\left(r\right)}{dr}
-2\,{\frac{B\left(r\right)}{{r}^{2}}}
+\frac{2k}{{r}^{2}}\,,\label{Ralphaconst}
\end{equation}
\phantom{line}\\
and in general is not a constant.
To see how the  reconstruction procedure works, we must 
take the derivative of equation (\ref{one}) with respect to $r$, such that the two equations (\ref{one})--(\ref{two}) give, after some algebra,
\phantom{line}\\
\begin{equation}
\left[\frac{2(B(r)-k)}{r^3}-\frac{1}{2}\frac{d^2 B(r)}{d r^2}\right]F'(r)+\left(\frac{2B(r)}{r^2}-\frac{1}{r}\frac{d B(r)}{d r}\right)\frac{d F'(r)}{d r}=0\label{genericequation} \,,
\end{equation}
\begin{equation}
\frac{d^2}{dr^2}F'(r)=0\,. 
\end{equation}
Thus, from the second equation, we directly obtain
\begin{equation}
F'(R)=ar+b\,. \label{Fprime}
\end{equation}
Here, $a$ and $b$ are two constants, $b$ being adimensional. 
Putting this result into Eq.~(\ref{genericequation}) and integrating, one gets\\
\phantom{line}
\begin{equation}
\begin{array}{lll}
1)&a=0,b\neq0\,,&\quad
 B(r)=k+\frac{c_1}{r}+c_2r^2\,;\\
2)&a\neq0,b=0\,,&\quad
 B(r)=\frac{k}2+\frac{c_1}{r^2}+c_2r^2\,;\\
3)&a,b\neq0\,,&\quad
 B(r)=\frac{k}{2}+\frac{b}{3ar}+c_2\,r^2
       +c_0\,\left[\frac{b^2}{2a^2}-\frac{br}{a}-\frac{b^3}{3a^3r}
+r^2\,\,\log\left[\frac{ar+b}{ar}\right]\right]\,.
\end{array}
\label{Bab}\end{equation}
\phantom{line}\\
The last expression is the more general form which the $B(r)$ function can assume
if the metric (\ref{metric00}) is a solution of a generic $F(R)$-theory. We see that it depends, at most, on
four independent parameters ($a,b,c_0,c_2$), being implicitly the EOMs fourth order differential equations.

The corresponding scalar curvatures are given by\\
\phantom{line}
\begin{equation}
\begin{array}{lll}
1)&a=0,b\neq0\,,&\quad R=-12c_2\,;\\
2)&a\neq0,b=0\,,&\quad R=-12c_2+\frac{k}{r^2}\,;\\
3)&a,b\neq0\,,&\quad
R=-12c_2+\frac{k}{r^2}+c_0\,\left[-12\,\log\left[\frac{ar+b}{ar}\right]
        +\frac{b(12a^3r^3+4ab^2r+18a^2br^2-b^3)}{a^2r^2(b+ar)^2}\right]\,.
\end{array}
\label{Rab}
\end{equation}
\phantom{line}\\
Now, in some cases we are able to ``infer'' the function $F(R)$ which generates
the solutions in (\ref{Bab}).
In the first case, namely $a=0\,,b\neq0$,
using expressions (\ref{Bab})--(\ref{Rab}) in Eq. (\ref{one}) one may  recover  the result of GR,\\
\phantom{line} 
\begin{equation}
a=0,b\neq0\,,\quad
\left\{\begin{array}{l}
   B(r)=k-\frac{c_1}{r}-\frac{\Lambda r^2}3\,,\\
   R=4\Lambda\,,\\
   F(R)=(R-2\Lambda)=2\Lambda\,.
\end{array}
\right.
\label{deS1}
\end{equation}
\phantom{line}\\
Here, we have set $b=1$, $c_2=-\Lambda/3$ and putted $c_1\rightarrow -c_1$. This is the topological Schwarzshild-dS/AdS solution. We stress that the latter equation represents the value of $F(R)$ evaluated on the
solution. Of course, there are infinitely many functions $F(R)$ which reduce (on shell) to
$F(R)=2\Lambda$ for $R=4\Lambda$ and, in general, they do not give the correct solution\footnote{This is a consequence of the fact that $R$ does not depend on $r$ and the reconstruction of the model can not be well defined. In particular, we must also verify that $F'(R)=b$, where in this case $b=1$.}, but we have also other possibilities (in particular, we will discuss in \S\ref{R^2} the case of $R^2$ model).

In the second case here considered, that is $a\neq0\,,b=0$, we have
\begin{equation}
a\neq0,b=0\,,\quad
\left\{\begin{array}{l}
 B(r)=\frac{k}2+\frac{c_1}{r^2}+c_2r^2\,,\\
R=-12c_2+\frac{k}{r^2}\,,\\
F(R)=\frac{2a\,k}r\,.
\end{array}
\right.\label{mod2}
\end{equation}
Thus, we easily get
\begin{equation}
r=\sqrt{\frac{k}{R+12c_2}}\,,\quad F(R)=2a\,\sqrt{k(R+12c_2)}\,,\quad k\neq 0\,.
\label{deS2}
\end{equation}
One can directly verify, using Eqs. (\ref{one})--(\ref{two}), that this Lagrangian effectively admits the vacuum solution in (\ref{mod2}). 

In the third case here considered, that is $a\neq0\,,b\neq0$, the method does not work
owing to the presence of logarithmic terms which  prevent the explicit reconstruction
of the Lagrangian, which formally reads
\begin{equation}
F(R)=\frac12\,bR+\frac{k(b+4ar)}{2r^2}-\frac{c_0b^5}{2a^2r^2(ar+b)^2}\,.\label{GGG}
\end{equation}
However, if we choose $c_0=0$, then
the logarithm disappears from all equations and we get
\begin{equation}
a\neq0,b\neq0\,,\quad c_0=0
\,,\quad
\left\{\begin{array}{l}
B(r)=\frac{k}2+\frac{b}{3ar}+c_2\,r^2\,,\\
R=-12c_2+\frac{k}{r^2}\,,\\
F(R)=\frac12\,bR+\frac{k(b+4ar)}{2r^2}\,.\label{modellonesol}
\end{array}
\right.
\end{equation}
These equations can be easily solved and one has
\begin{equation}
r=\sqrt{\frac{k}{R+12c_2}}\,,\quad F(R)=b(R+6c_2)+2a\,\sqrt{k(R+12c_2)}\,.\label{modellone}
\end{equation}
Also in this case one can directly verify that the field equations are
satisfied. For $k=0$, the model reduces to the corresponding topological case of (\ref{deS1}).

\subsection{Lifshitz like solutions\label{Lif}}

\paragraph{}  Now we are going to consider the second case, that is $\alpha(r)=\log[r/r_0]^z$, where
$z$ is an (in principle) arbitrary parameter and $r_0$ a dimensional constant.
With this special but important choice, Eq.~(\ref{two}) can be explicitly
solved, to get
\begin{equation}
F'(r)=a r^{\frac12\left(1+z+p(z)\right)}+b r^{\frac12\left(1+z-p(z)\right)}\,,\quad
p(z)=\sqrt{z^2+10z+1}\,.
\label{F1sol}
\end{equation}
Here, $a,b$ are arbitrary (dimensional) constants.
By choosing $b=0$ (or $a=0$)
also the derivative of Eq. (\ref{one}), namely\footnote{In this equation, we give the general formalism valid for any choice of $\alpha(r)$.}\\
\phantom{line}
\begin{eqnarray}
&&\frac{2F'(r)}{r^3}\left(2k-2B(r)+r^2\frac{d^2 B(r)}{d r^2}\right)-\frac{d F'(r)}{d r}\left[
2B(r)\frac{d^2\alpha(r)}{d r^2}+\frac{d\alpha(r)}{d r}\left(3\frac{d B(r)}{d r}+\frac{4 B(r)}{r}\right)
\right.\nonumber\\
&&\hspace{-10mm}\left.
+2B(r)\left(\frac{d\alpha(r)}{d r}\right)^2+\frac{4 B(r)}{r^2}-\frac{2}{r}\frac{d B(r)}{d r}\right]
+\frac{d^2F'(r)}{d r^2}\left(\frac{4B(r)}{r}+3\frac{d B(r)}{d r}\right)+2B(r)\frac{d^3 F'(r)}{d r^3}=0\,,\nonumber\\&&
\end{eqnarray}
\phantom{line}\\
 can be solved, obtaining in this way
a complicated  expression for $B(r)$, valid for any $z$. It depends on two
arbitrary integration constants, say $c_{1,2}$,  and for $z\neq1$ it assumes the form
\begin{equation}
B(r)=\gamma_0\,k+\frac{c_1}{r^{\gamma_+}}+\frac{c_2}{r^{\gamma_-}}\,,
\label{BGen}
\end{equation}
$\gamma_0,\gamma_\pm$ being given function of $z$, which read
\begin{eqnarray}
a\neq0,b=0\,,\quad
\left\{\begin{array}{lll}
\gamma_0&=& \frac{2}{(z-1)[p(z)-3(z+1)]}\,,\\
\gamma_\pm&=& r^{\frac{7z-p(z)-1}{4}
        \pm\frac{i}2\,\sqrt{\frac{z^3+[p(z)-3]\,z^2-[8p(z)-69]\,z-23p(z)+41}{p(z)-3(z+1)}}}\,;
\end{array}\right.\end{eqnarray}
\begin{eqnarray}
a=0,b\neq0\,,\quad
\left\{\begin{array}{lll}
\gamma_0&=& -\frac{2}{(z-1)[p(z)+3(z+1)}\,,\\
\gamma_\pm&=& r^{\frac{7z+p(z)-1}{4}
    \pm\frac{1}2\,\sqrt{\frac{z^3-[p(z)+3]\,z^2+[8p(z)+69]\,z-23p(z)+41}
                         {p(z)+3(z+1)}}}\,.
\end{array}\right.\end{eqnarray}
The corresponding metric can be considered as a generalization of the one of
Clifton-Barrow (see below).
In the special case $z=1$, a logarithmic term is also present.

For example, for the following choices of the parameters, we get\\
\phantom{line}
\begin{eqnarray}
1)&\:\:\:a\neq0\,,b=0\,,z=1/2\,,\quad
&\left\{\begin{array}{lll}
B(r)&=&\frac{4k}{7}+\frac{c_1}{r^{7/2}}+c_2r\,,\\
R&=&-\frac{9c_2}{r}\,,\\
F(R)&=& a\left( 6k-\frac{81c_2^2}{R}\right)\,;
\end{array}\right.\\&&\nonumber\\
2)&\:\:\:a\neq0\,,b=0\,,z=2\,,\quad
&\left\{\begin{array}{lll}
B(r)&=&-\frac{k}{7}+\frac{c_1}{r^7}+\frac{c_2}{r^2}\,,\\
R&=& \frac{4k}{r^2}\,,\\
F(R)&=& a\left(30c_2-\frac{16k^2}{R}\right)\,;
\end{array}\right.\\&&\nonumber\\
3)&\:\:\:a\neq0\,,b=0\,,z=1\,,\quad
&\left\{\begin{array}{lll}
B(r)&=&-\frac{c_1r^{-(3+\sqrt{3})}}{3+\sqrt{3}}-\frac{2 k\log\left(r\right)}{3+\sqrt{3}}+c_2\,,\\
R&=& \frac{12k\log r+2(9+\sqrt{3})k-6(3+\sqrt3)c_2}{(3+\sqrt3)\,r^2}\,,\\
F(R)&=&a\left[\frac{R}{r}-\frac{4k}{r^3}
     +\frac{(54+30\sqrt3)c_2-12k\log r^{2+\sqrt3}}{(3+\sqrt{3})\,r^3}\right]\,r^{2+\sqrt{3}}\,;
\end{array}\right.\\&&\nonumber\\
4)&\:\:\:a=0\,,b\neq0\,,z=1/2\,,\quad
&\left\{\begin{array}{lll}
\quad\quad B(r)&=&2k+\frac{c_1}{r}+rc_2\,,\\
\quad\quad R&=& -\frac{5 k}{r^2}-\frac{9 c_2}{r}\,,\\
4a)\,F(R)&=& b\,R^{5/4}\,,\quad\quad\mbox{ if }c_2=0\,,\\
4b)\,F(R)&=& b\,R^{3/2}\,,\quad\quad\mbox{ if }k\,\,=0\,;\\
\end{array}\right.\\&&\nonumber\\
5)&\:\:\:a=0\,,b\neq0\,,z=2\,,\quad
&\left\{\begin{array}{lll}
B(r)&=&-\frac{k}{2}+\frac{c_1}{r^2}\,,\\
R&=&\frac{9 k}{r^2}\,,\\
F(R)&=&b\,R^{3/2}\,;
\end{array}\right.\\&&\nonumber\\
6)&\:\:\:a=0\,,b\neq0\,,z=1\,,\quad
&\left\{\begin{array}{lll}
B(r)&=&\frac{c_1 r^{-3+\sqrt{3}}}{-3+\sqrt{3}}+\frac{2 k\log\left(r\right)}{-3+\sqrt{3}}+c_2\,,\\
R&=&\frac{12k\log r+2(9-\sqrt3)k
     -6\left(3-\sqrt3\right)c_2}{(3-\sqrt3)\,r^2}\,,\\
F(R)&=&b\left[\frac{R}{r}-\frac{4k}{r^3}
     +\frac{(54-30\sqrt3)c_2-12k\log r^{2-\sqrt3}}{(3-\sqrt{3})\,r^3}\right]\,r^{2-\sqrt{3}}\,.
\end{array}\right.
\end{eqnarray}
\phantom{line}\\
The function $F(R)$ for the two examples with $z=1$ has not been explicitly
written in terms of the Ricci scalar, because it is a quite complicated expression
which we do not will use in what follows, while in the fourth example
we have also set $c_2=0$ (or $k=0$), in order to explicitly  build up the action.
We observe that the first part of the fourth example and the fifth one 
are particular cases of the Clifton-Barrow metric that we will analyze in the next Subsection
(in the specific, they correspond to the cases $\delta=1/2$ and  $\delta=1/4$, respectively).
A last remark is in order.
In the previous Subsection we have observed that a given metric can be the solution of different
actions. In the same way, a given action can give rise to different SSS-metrics.
For example, $F(R)=b\,R^{3/2}$ has the Schwarzschild solution, but also the solutions
in the fourth and fifth examples above.

\subsection{Topological Clifton-Barrow solutions}

\paragraph{}  Here we present  the topological generalization of the Clifton-Barrow solution~\cite{CB},
which is a SSS metric of the form considered in the previous Subsection, with\\
\phantom{line}
\begin{eqnarray}
\left\{\begin{array}{l}
\alpha(r)=\log\left[\left(\frac{r}{r_0}\right)^{\delta(1+2\delta)/(1-\delta)}\left(\frac{(1-2\delta+4\delta^2)(1-2\delta-2\delta^2)}{(1-\delta)^2}\right)^{1/2}\right]\,,\\\\
B(r)=\frac{(1-\delta)^2}{(1-2\delta+4\delta^2)(1-2\delta-2\delta^2)}\left(k-\frac{c_0}{r^{(1-2\delta+4\delta^2)/(1-\delta)}}\right)\,.
\label{CBB}
\end{array}\right.
\end{eqnarray}
\phantom{line}\\
Here, $\delta\neq1$ is an arbitrary parameter of the model.

The Ricci scalar reads\\
\phantom{line}
\begin{equation}
R=\frac{6\delta(1+\delta)}{(2\delta^2+2\delta-1)}\left(\frac{1}{r^2}\right)\,,
\label{CBR}
\end{equation}
\phantom{line}\\
and on the solution we obtain\\
\phantom{line}
\begin{equation}
F(R)=(\kappa^2)^\delta\left[ R^{\delta+1}\right]\,,\quad\delta\neq 1\,.\label{CB}
\end{equation}
\phantom{line}\\
Motivated by dimensional reasons, we have introduced the Newton Constant in $\kappa^2$.
This is a non trivial modification to GR, whose Hilbert-Einstein action corresponds to the particular choice $\delta=0$, for which we recover the Schwarzshild metric.
One can directly verify that (\ref{CBB}) are solutions of the field equations
of the Clifton-Barrow Lagrangian (\ref{CB}).
The Clifton-Barrow metric is a particular case of someone considered in the previous
Subsection, and, in fact,
it can be obtained using the method described above, as it is  shown in Ref. \cite{SSSsolutions}.

\section{$\mathcal F(R,G)$-static spherically symmetric topological vacuum solutions}

\paragraph*{} By using the reconstruction methods adopted for $F(R)$-gravity, we can try to find solutions in the more general framework of $\mathcal F(R,G)$-theories \cite{GBSSS}. We start from the EOMs~(\ref{EOM1SphRG})--(\ref{EOM2SphRG}). 
Since now in the second equation $B(r)$ explicitly appears, the reconstruction of the solutions results to be more complicated and requires some additional assumption. In what follows, we will give some examples.

\subsubsection*{Case $\alpha(r)=\text{Const}$}

\paragraph{}  Let us start by the important case of $\alpha(r)=\text{Const}$ (or, equivalently, $\alpha(r)=0$) into the metric (\ref{SSS}). 
From the derivative with respect to $r$ of the first EOM~(\ref{EOM1SphRG}) and from the second EOM~(\ref{EOM2SphRG}) we get\\
\phantom{line}
\begin{eqnarray}
\label{unounouno}
&&\left[\frac{2(B(r)-k)}{r^3}-\frac{1}{2}\frac{d^2 B(r)}{d r^2}\right]F_R'(r)+\left(\frac{2B(r)}{r^2}-\frac{1}{r}\frac{d B(r)}{d r}\right)\frac{d F_R'(r)}{d r}
\nonumber\\
&&+\frac{d \mathcal F'_G}{d r}\left[\frac{4}{r^2}\left(\frac{d B(r)}{d r}\right)^2
+\frac{4 B(r)}{r^2}\frac{d^2 B(r)}{d r^2}-\frac{4(3B(r)-k)}{r^3}\frac{d B(r)}{d r}
\right]=0\,,
\end{eqnarray}
\begin{equation}
\frac{d^2}{dr^2}\mathcal F_R'-\frac{4}{r^2}(B(r)-k)\frac{d^2}{d r^2}\mathcal F_G'=0\,.\label{dueduedue}
\end{equation}
\phantom{line}\\
Here, we have used Eqs. (\ref{Rspheric})--(\ref{Gspheric}) also.
A simple choice is to investigate the models with $d^2 \mathcal F_R/d r^2=0$ and $d^2\mathcal F'_G/d r^2=0$ which easily satisfy Eq. (\ref{dueduedue}), namely
\begin{equation}
\mathcal F'_R=ar+b\,,\quad  \mathcal F'_G= c r\,,\label{FprimeG}
\end{equation}
where $a$, $b$ and $c$ are integration constants and we avoid the linear contribution of Gauss-Bonnet invariant.
Unfortunately, equation (\ref{unounouno}) cannot be solved for general values of the parameters. Otherwise, it may be used to find some specific metric.
For example, if $a=0$ and $b=1$, by replacing $c\rightarrow c_0/4$, a possible solution is given by
\begin{equation}
B(r)=-k+\frac{r}{c_0}\,,\quad R=\frac{4 k-6 r/c_0}{r^2}\,,\quad G=\frac{4 }{(c_0 r)^2}\label{BGB}\,.
\end{equation}
This topological SSS vacuum solution is generated by the Lagrangian
\begin{equation}
\mathcal F(R,G)=R+\sqrt{G}\,,\label{GBmodel}
\end{equation}
and satisfies the field equations.

\subsubsection*{Case $G(r)=\text{Const}$}

\paragraph{}  The other class of solutions we will consider leads to a constant value $G_0$ for the Gauss-Bonnet invariant.
We start by considering $\alpha(r)=0$ again.
By putting $G=G_0$ in Eq. (\ref{Gspheric}), one derives the form of $B(r)$,
\begin{equation} 
B(r)=k\pm\frac{\sqrt{G_0 r^4+c_1+c_2\,r}}{2\sqrt{6}}\,.\label{Gconst}
\end{equation}
Now, we have two very simple cases. The first one corresponds to $\mathcal F'_R=0$, it means that the model $\mathcal F(R,G)=F(G)$ depends on $G$ only and we deal with a pure Gauss-Bonnet theory. Thus,  Eq. (\ref{EOM2SphRG}) trivially is satisfied and Eq. (\ref{EOM1SphRG}) reads
\begin{equation}
F(G_0)-G_0 F'_G(G_0)=0\,.\label{GconstMod}
\end{equation}
Of course, there are infinite functions which reduce (on shell) to this expression for $G=G_0$.
For example, one finds that the model
\begin{equation}
F(G)=\gamma\mathrm{e}^{G/G_0}\,,\label{GconstMod1}
\end{equation}
where $\gamma$ is a dimensional constant,
exibits the topological SSS solution with $\alpha(r)=0$ and $B(r)$ given by Eq. (\ref{Gconst}), but
also the class of models
\begin{equation}
F(G)=\gamma G^n+\Lambda\,,\label{GconstMod2}
\end{equation}
where $\Lambda=\gamma(n-1)G_0^n$, $\gamma$ dimensional constant and $n$ generic parameter, exhibits the same kind of solution.

A second possibility is to take constant the Ricci scalar also, 
so that Eq. (\ref{Gconst}) reduces to the de Sitter solution (\ref{puredS}) already discussed in $\S~\ref{1.4}$.\\

To conclude, we finally observe that, in analogy with the case of Schawarschild solution for $F(R)$-gravity, all the $F(G)$-models in the form of $F(G) = G\,g(G)$ with $g(G)$ a generic function of $G$ such that $\lim_{G\rightarrow 0} \,g(G)=0$, trivially satisfy the EOMs when $G=0$.
For example, by equaling $G_0$ to zero in 
Eq. (\ref{Gconst}), we can say that the model $F(G)=\gamma G^n$, where $\gamma$ and $n$ are generic parameters with $n\geq 2$, exhibits the topological SSS vacuum solution with 
\begin{equation}
\alpha(r)=0\,,\quad B(r)=k\pm c_1\sqrt{1+c_2r}\,,\quad G=0\,.
\end{equation}
We also can verify that the Gauss-Bonnet is null provided by
\begin{equation}
\alpha(r)\neq 0\,,\quad B(r)=k\,,\quad k\neq 0\,,\quad G=0\,.
\end{equation}
Here, $\alpha(r)$ assumes an arbitrary form. However, we see that the only acceptable topology is the one of the sphere, namely $k=1$, otherwise the metric signature changes and the solution results to be patologic.
\\

In Refs. \cite{Weyl, GBSSS} electro-vacuum solutions of $F(R)$ and $\mathcal F(R,G)$-gravity in the presence of Maxwell field are also considered and some explicit metrics are therefore derived.

\section{Black holes in modified gravity\label{F(R)BH}\label{2.3}}

\paragraph*{} The SSS metric may actually describe a black hole. We recall that an event horizon exists as soon as there exists a positive solution $r_H$ of
\begin{eqnarray}
B(r_H)=0\,, \quad
\frac{d B(r)}{d r}\Big\vert_{r_H}\gneq 0\,.\label{BHconditions} 
\end{eqnarray}
The second restriction guarantees that the (Killing) surface gravity at the horizon is positive and the metric segnature is preserved out of the horizon. Furthermore,
we also require $d B(r_H)/dr\neq 0$ in order to avoid extremal BHs. 
For example, in the well known case of constant curvature, namely (\ref{deS1})
with $k=0$ or $k=-1$,
the conditions (\ref{BHconditions}) are satisfied if 
$c_1$ is positive and the cosmological constant $\Lambda$ is negative (AdS solution);
on the contrary, if $k=1$, one has a black hole solution
whatever the sign of $\Lambda$ if  $c_1>0$ again\footnote{Note that in this case $B(r)=0$ has two roots, namely $r_{\pm}$, the first one corresponding to the event horizon of the BH ($d B(r_+)/d r>0$) and the second one corresponding to the horizon of the cosmological dS/AdS background where the black hole is immersed ($d B(r_{-})/d r<0$).}. 

For the solution ($\ref{mod2}$) we obtain
\begin{equation}
r_{H}=\sqrt{\frac{-k\pm\sqrt{1-16c_1 c_2}}{4c_2}}\,,\quad k\neq 0\,,
\end{equation}
and one has several possibilities. In particular, in order to satisfy the both conditions (\ref{BHconditions}), we observe that if $k=1$ we get $c_1<0$ whatever the choice of $c_2$, and if $k=-1$ we get $c_2>0$ and $c_1<0$ again.

For the topological Clifton-Barrow solution (\ref{CBB}), one explicitly finds
\begin{equation}
r_{H}=(c_0/k)^{(1-\delta)/(1-2\delta+4\delta^2)}\,,\label{CBr}
\end{equation}
which is positive if $c_0/k>0$. In order to have a black hole, one has also to impose
$(1-2\delta+4\delta)/(1-\delta)>0$.

Then for the metric in Eq. (\ref{BGB}), it turns out that only in the topological case with $k=1$ (sphere) we can describe a black hole as soon as $r_H=c_0$, $c_0>0$. 

Since the solutions found in the previous Sections can describe the BHs  only for some values of the parameters, in the following it is always understood that all free parameters will be restricted to values which give rise to physical BH solutions.

\section{The First Law of BH-thermodynamics}

\paragraph*{} In order to study the issue associated with the energy of black hole solutions in modified gravity, let us remind the case of GR, in which several notions of quasi-local 
energies may be introduced. 
In particular we mention the so called Misner-Sharp mass, 
which has the important property to be  defined for dynamical, 
spherically symmetric space-time~\cite{Hayward}, where 
the use of invariant quantities play a crucial role~\cite{Roberto11}-\cite{Roberto2}.  
For the sake of completeness, we recall that in four dimensions, 
any spherically symmetric (dynamical) metric can locally be expressed in the form   
\begin{equation}
\label{metric1}
ds^2 =\gamma_{ij}(x^i)dx^idx^j+ {\mathcal R}^2(x^i) d\Omega_2^2\,,\qquad i,j \in \{0,1\}\;,
\end{equation}
where $ d\Omega_2^2 $ here is the usual metric on the two sphere $S_2$, but it could be
the metric of a generic two-dimensional maximally symmetric space. 
Of course, in such cases the black hole will have different topologies as the ones we have considered in the previous Sections. 
The two-dimensional metric
\begin{equation} d\gamma^2=\gamma_{ij}(x^i)dx^idx^j\,,
\label{nm}
\end{equation}
is referred to as the normal one. The related coordinates are $\{x^i\}$, while
$ {\mathcal R}(x^i)$ is the areal radius, considered as a scalar field in the two dimensional
normal space. 
A relevant scalar quantity in the reduced normal space is 
\begin{equation}
\chi(x^i)=\gamma^{ij}(x^i)\partial_i  {\mathcal R}(x^i)\partial_j  {\mathcal R}(x^i)\,, \label{sh} 
\end{equation} 
since the dynamical trapping horizon, if it exists, is located in
correspondence of 
\begin{equation} 
\chi(x^i)\Big\vert_H = 0\,,\quad \partial_i\chi(x^i)\vert_H \gneq 0\,.\label{ho} 
\end{equation}
Here and in the following, we use the suffix `$H$' for all quantities evaluated on the horizon and
where is not necessary also the argument of such quantities will be dropped. 
In the static case, the above conditions are equivalent to (\ref{BHconditions}).
The quasi-local Misner-Sharp gravitational energy is defined by~\cite{MS}
\begin{equation}
E_{MS}(x^i):=\frac{1}{2G_N}{\mathcal R}(x^i)\left[1-\chi(x^i) \right]\,,\quad E:=E_{MS}(x^i)\Big\vert_H=\frac{1}{2G_{N}}{\mathcal R(x^i)}\vert_H\,.\label{MS}
\end{equation}
The Misner Sharp mass is an invariant quantity on the normal space and at
the horizon it reduces to the mass $E$ of the black hole. 

By means the Killing vector fields $\xi^\mu$ such that
\begin{equation}
\nabla_{\mu}\xi^{\nu}(x^\nu)+\nabla^{\nu}\xi_{\mu}(x^\nu)=0\,, 
\end{equation}
which are the generators of the metric isometries,
we recall that in particular, in a non dynamical space-time (static or stationary), one has the time-like Killing vector field 
\begin{equation}
K^\mu=\left(1,0,0,0\right)\,,\label{Killingvector}
\end{equation}
with the associated Killing surface gravity $\kappa_K$ given by the relation
\begin{equation}
\kappa_K K^{\mu}(x^\nu)=K^{\nu}\nabla_\nu K^\mu(x^\nu)\,.\label{KSG}
\end{equation}
In addition, the Misner Sharp mass corresponds to the charge of the conserved current $J_{\mu}=G_{\mu\nu}K^\nu$, where $G_{\mu\nu}$ is the Einstein tensor of GR. 

In the spherical symmetric, dynamical case, the real geometric object which generalizes the Killing vector field is 
the Kodama vector field $\mathcal K(x^i)$~\cite{Kodama}. 
Given the metric (\ref{metric1}), it is defined by
\begin{equation} 
\mathcal K^i(x^i):=\frac{1}{ \sqrt{-\gamma}}\,\varepsilon^{ij}\partial_j{\mathcal R}(x^i)\,,\,\,i=0,1\,;
\qquad \mathcal K^i:=0\,,\,\,i\neq 0,1\,.
\end{equation} 
Here, $\varepsilon^{ij}$ is the completely antisymmetric Levi-Civita tensor on the normal 
space and $\gamma$ the determinant of $\gamma_{ij}$ metric tensor.
The Kodama/Hayward surface gravity associated with dynamical
horizon is given by the normal-space scalar 
\begin{equation}
\kappa_H:=\frac{1}{2}\Box_{\gamma} {\mathcal R(x^i)}\Big\vert_H\,, \label{H} 
\end{equation} 
where $\Box_\gamma$ is the Laplacian corresponding to the $\gamma_{ij}$ metric.\\

Assuming Einstein equations of GR, in a generic four-dimensional 
spherically symmetric space-time, a geometric dynamical identity holds true. 
This can be derived as follows. 
Let us introduce the normal space invariant 
\begin{equation}
\mathrm{T}^{(2)}(x^i)=\gamma^{ij}T^{\mathrm{(matter)}}_{ij}(x^i)\,,
\end{equation} 
which is the reduced trace of the matter stress energy tensor. 
Then, making use of Einstein equations and Hayward surface gravity (\ref{H}), it is possible to show that, 
on the dynamical trapping horizon~\cite{Hayward},
\begin{equation}
\kappa_H=\frac{1}{2 {\mathcal R}_H G_N}+2\pi  {\mathcal R}_H \frac{\mathrm{T}^{(2)}_H}{G_N}\,.
\label{vanzo}
\end{equation}
Introducing the horizon area $\mathcal A_H$ 
and the (formal) three-volume $\mathcal V_H$ enclosed by the horizon, 
with their respective `thermodynamical' differentials 
$d\mathcal A_H=8\pi  {\mathcal R}_H d {\mathcal R}_H\,$
and $d \mathcal V_H=4\pi  {\mathcal R}_H^2 d {\mathcal R}_H$
(we are assuming a horizon with the topology of a sphere), we get
\begin{equation}
\frac{\kappa_{H}}{8 \pi G_N}d \mathcal A_H =d\left(\frac{ {\mathcal R}_H}{2G_N}\right) 
+\frac{ \mathrm{T}_H^{(2)}}{2 G_N} d\mathcal V_H\,. 
\end{equation}
This equation can be recast in the form of a geometrical identity, 
once the Misner-Sharp energy at the horizon (\ref{MS}) has been introduced. It reads 
\begin{equation}
\Delta E=\frac{\kappa_{H}}{2 \pi} d\left( \frac{\mathcal A_H}{4G_N}\right)
-\frac{\mathrm{T}_H^{(2)}}{2G_N} d\mathcal V_H\,. 
\label{fl}
\end{equation}
We will return below on this equation and we will 
study its thermodynamical implications.\\

Let us restrict the discussion to the static case
where the metric in Eq.~(\ref{metric1}) can be written in the simpler form of SSS metric of Eq.~(\ref{SSS}).
Of course the general formalism is also valid in the static case, and  leads 
to the horizon conditions (\ref{BHconditions}).
 
It is well known \cite{HT} that all black holes have a characteristic
temperature related to the existence of an  event horizon and, with the associated entropy,  they have  thermodynamical
properties. For the static black hole,
the so called Killing/Hawking temperature $T_K$ is proportional to the Killing surface gravity
\begin{equation}
\kappa_K:=\frac{\mathrm{e}^{\alpha(r)}}{2}\frac{d B(r)}{d r}\Big\vert_{r_H}\,,
\end{equation}
and assumes the explicit form
\begin{equation}  
T_K:=\frac{\kappa_K}{2\pi}=\frac{\mathrm{e}^{\alpha(r)}}{4\pi}\frac{d B(r)}{dr}\Big\vert_{r_H}\,.\label{HawkingTemperature}
\end{equation}
This result can be justified in several ways, 
for example  making use of standard derivations of Hawking radiation~\cite{Visser2},
or by eliminating the conical singularity in the corresponding Euclidean metric, 
or making use of  the tunneling method, 
recently introduced in Refs.~\cite{PW,Nadalini2}, and discussed in details in several papers.
All derivations of Hawking radiation (see Appendix A for a brief review) 
lead to a 
semi-classical expression for the black hole radiation rate $\Gamma$,
\begin{equation}
\Gamma\equiv \mathrm{e}^{-\frac{\Delta E_K}{T_K}}\,,
\label{ratekill}
\end{equation}
in terms of the change $\Delta E_K$ of the Killing energy $E_K$ \cite{Nadalini1,Nadalini2}. 
This fact may be interpreted as the First Law of black hole thermodynamics
as soon as  
\begin{equation}
\Gamma \equiv \mathrm{e}^{-\Delta S}\,,\quad \Delta E_K=T_{K} \Delta S\,.\label{FirstLaw} 
\end{equation}
Here, $\Delta S$ is the change
of the entropy $S$ of the irradiating black hole itself and from Hawking radiation it is possible to confirm the so called Area Law, 
\begin{equation}  
S=\frac{\mathcal{A}_H}{4G_N}\,,\label{AreaLaw}
\end{equation} 
where, as stated before, $\mathcal{A}_H$ is the area of event horizon, namely $\mathcal{A}_H=4\pi r_H^2$ for the case of the sphere. For example, in vacuum the black hole is described by Schwarzshild metric and the energy corresponds to the integration constant of the solution, so that Eq. (\ref{AreaLaw}) can be easily verified from (\ref{FirstLaw}).

\paragraph*{} In analogy with the First Law for static BHs, we can also try to write a Gibbs relation for the dynamical case, where the Killing surface gravity is replaced by Kodama/Hayward surface gravity. As a consequence, we get  the Kodama/Hayward temperature $T_H$,
\begin{equation}
T_H:=\frac{\kappa_H}{2\pi}\,,
\end{equation}
and Eq. (\ref{fl}) leads to
\begin{equation}
T_H \Delta S=\Delta E+pd\mathcal V_H\,,
\end{equation}
where we have introduced the BH entropy and the reduced trace of matter stress energy tensor gives the working term $p=\mathrm{T}_H^{(2)}/2G_N$.

Now it is interesting to note that in the static case the 
Kodama vector simply reads
\begin{equation}
\mathcal K^\mu=\left(\mathrm{e}^{-\alpha(r)}, \vec 0\right)\,,\label{Kodamavector}
\end{equation}
and in general does not coincide with the Killing vector in Eq. (\ref{Killingvector}).
As a consequence, also the Kodama surface gravity and thus the Hayward temperaure in principle are different from the corresponding Killing quantities,
\begin{equation}
T_H:=\frac{\kappa_H}{2\pi}=\frac{1}{4\pi}\frac{d B(r)}{d r}\Big\vert_{r_H}\,.\label{HaywardTemperature}
\end{equation}
This result is related with the fact that the Killing vector cannot be defined unambiguously when the space-time is not asymptotically flat. 

In GR this is not a big problem. 
The expression of the black hole radiation rate (\ref{ratekill}) can be rewritten as
\begin{equation}
\Gamma\equiv \mathrm{e}^{-\frac{\Delta E_H}{T_H}}\,,
\label{ratekoda}
\end{equation}
in terms of the change of the Kodama energy $E_H$ for the emitted particle, due to the relationship $\Delta E_H=\mathrm{e}^{-\alpha(r)}\Delta E_K$. As a consequence,
from the Eqs. (\ref{ratekill}) and (\ref{ratekoda}) one arrives at the identity
\begin{equation}
\label{kk}
\frac{\Delta E_H}{T_H}=\frac{\Delta E_K}{T_K}\,,
\end{equation}
so that the tunneling probability is 
invariant scalar whatever the choice for the temperature or energy.
Moreover, in the vacuum case, since $\alpha(r)=0$, the two temperature coincide, and the identification of the Killing energy of Schwarshild BH with the integration constant of the solution is a robust result.

A detailed discussion about this issue in GR can be 
found in Ref.~\cite{Roberto222,Roberto2}, in which also the dynamical case is analyzed.\\

Now we come to the key point of our proposal. For a generic modified gravity theory, it seams very difficult to define in a reasonable way the analogue of the local 
Misner-Sharp mass, since a conserved current cannot be found from fourth order differential field equations. 
As we will see, an exception is the higher-dimensional Lovelock gravity.
For this reason, an attempt is made for obtaining an  expression 
of energy associated with  black holes 
solutions in modified theories of gravity. 
The proposal~\cite{SSSEnergy} consists in the  identification of the black hole energy with a quantity proportional to the constant of integration, which appears in the explicit solutions, and positive defined.
The identification is 
achieved making use of derivation (where it is possible) of the First Law of black hole thermodynamics 
from the equations of motion, evaluating in an 
independent way the related black hole entropy via Wald method and the Killing/Hawking temperature via the 
quantum mechanics in curved space-time, 
as in the case of General Relativity.
Within  modified gravity
it happens to deal with (non asymptotically flat) vacuum black hole solutions where $\alpha(r)\neq 0$ and $T_K\neq T_H$, so that
in the present proposal the Killing energy seems to be preferable with respect to the Hayward energy.

At first, we will revisit the Lovelock theories, and then we will analyze the $F(R)$-gravity, where the derivation of the First Law from the equations of motion and an explicit expression for the Killing energy will be given. After that, an attempt to generalize the formalism to other classes of modified gravity models is done.  

This approach is also supported by the results obtained in Refs. \cite{ram,eli}, where, on quite general grounds, generalizing the Jacoboson results on GR~\cite{jacob}, the equations of a modified gravitational theories are shown to be equivalent to the First Law of black hole thermodynamics. As it is well known, this issue may be of high relevance in substantiating the idea that gravitation might be a manifestation of thermodynamics of quantum vacuum~\cite{pad1}-\cite{pad3}. 

\section{Lovelock black hole solutions}

\paragraph*{} In this Section, as  warm up, we review Lovelock theory~\cite{lovelock} with the related static 
and spherically symmetric black hole solutions. 
This theory is  a very interesting higher dimensional generalization of Einstein gravity introduced by Lovelock in 1971. It is the most general theory of gravity which conserves second order equations of motion in arbitrary dimensions. 
In general, by making use of higher order geometrical invariants in the action,
in the metric formalism for the field equations one obtains 
fourth order partial differential equations. 
However, as Lovelock had shown, one can derive second order differential equations by
making use of higher dimensional extended Euler densities, 
the so called $m$-th order Lovelock terms $\mathcal{L}_m$ defined by
\begin{eqnarray}
{\cal L}_m = \frac{1}{2^m} 
\delta^{\lambda_1 \sigma_1 \cdots 
\lambda_m \sigma_m}_{\rho_1 \kappa_1 \cdots \rho_m \kappa_m}
R_{\lambda_1 \sigma_1}{}^{\rho_1 \kappa_1} 
\cdots  R_{\lambda_m \sigma_m}{}^{\rho_m \kappa_m}\ ,
\qquad m=1,2,3,...\,.
\end{eqnarray}
Here,  $R_{\lambda \sigma}{}^{\rho \kappa}$ is the Riemann tensor in 
arbitrary $D$-dimensions and $\delta^{\lambda_1\sigma_1
\cdots\lambda_m\sigma_m}_{\rho_1 \kappa_1 \cdots \rho_m \kappa_m}$ is the 
generalized totally antisymmetric Kronecker delta defined by\\
\phantom{line}  
\begin{eqnarray}
\delta^{\mu_1\mu_2\cdots \mu_p}_{\nu_1\nu_2\cdots\nu_p}={\rm det}
\left(
\begin{array}{cccc}
\delta^{\mu_1}_{\nu_1}&\delta^{\mu_1}_{\nu_2}&\cdots&\delta^{\mu_1}_{\nu_p}\\
\delta^{\mu_2}_{\nu_1}&\delta^{\mu_2}_{\nu_2}&\cdots&\delta^{\mu_2}_{\nu_p}\\
\vdots&\vdots&\ddots&\vdots\\
\delta^{\mu_p}_{\nu_1}&\delta^{\mu_p}_{\nu_2}&\cdots&\delta^{\mu_p}_{\nu_p}
\end{array}
\right)\ .
\nonumber
\end{eqnarray}
\phantom{line}\\
The action for Lovelock gravitational theory reads 
\begin{eqnarray}
I=\int_{\mathcal{M}} d^Dx \sqrt{-g}\left[-2\Lambda+\sum_{m=1}^s\left(\frac{a_m}{m}{\cal L}_m\right)\right]\,,
\label{total_action}
\end{eqnarray}  
where we defined the maximum order $s\equiv [(D-1)/2]$ and  $a_m$ are arbitrary constants.
Here, $[z]$ represents the maximum integer satisfying $[z] \leq z$.  
Hereafter, we set $a_1=1$.  The cosmological constant in $D$-dimension corresponds to  $\Lambda=(D-1)(D-2)/(2L^2)$, $L$ being a lenght size. 

For such a kind of theory, the equations of motion in vacuum are second order quasi-linear 
partial differential equations in the metric tensor and  read   
\begin{eqnarray}
{\cal G}_{\mu}{}^{\nu}=0\,,
\end{eqnarray}
where the Lovelock tensor ${\cal G}_{\mu}{}^{\nu}$ is given by 
\begin{eqnarray}
{\cal G}_{\mu}{}^{\nu}=\Lambda \delta_{\mu}^{\nu}
-\sum_{m=1}^{s}\frac{1}{2^{m+1}}\frac{a_m}{m} 
	 \delta^{\nu \lambda_1 \sigma_1 \cdots \lambda_m \sigma_m}_{\mu \rho_1 \kappa_1 
\cdots \rho_m \kappa_m}
       R_{\lambda_1 \sigma_1}{}^{\rho_1 \kappa_1} 
\cdots  R_{\lambda_m \sigma_m}{}^{\rho_m \kappa_m}\,.
\label{EOM}
\end{eqnarray}
As we said in previous Sections, we shall focus our attention on topological static, 
spherically symmetric metric. The vacuum solutions of Lovelock theory  assume the (Schwarzshild-dS/AdS like) form
\begin{eqnarray}
ds^2=-B(r)dt^2+\frac{dr^2}{B(r)}+r^2d\Omega^2_{(k,n)},
\label{metric_ansatz}
\end{eqnarray}
where $d\Omega^2_{(k,n)}$ is the metric of a topological $n$-surface (one has $n=D-2$), namely a sphere $S_n$ (for $k=1$), a thorus $T_n$ (for $k=0$) or a compact hyperbolic manifold $Y_n$ (for $k=-1$).
Such kind of theories become quite interesting for $D>4$, 
the four-dimensional case being equivalent to Schwarzschild-de Sitter,
since ${\cal L}_1=R$ and ${\cal L}_2$ is equal to the Gauss Bonnet quadratic term,
which in four-dimensions is a topological invariant.

In the following, we generalize the rusults of Ref.~\cite{wheeler} to the topological case~\cite{Cai3,Cai333, Davood}. 
A direct evaluation of field equations gives
\begin{eqnarray}
&&{\cal G}_t^t={\cal G}_r^r=-\frac{n}{2r^n}\frac{d\left[r^{n+1}W(r)\right]}{dr}
\ ,\label{tt}\\ \nonumber\\
&&{\cal G}_i^j=-\frac{1}{2r^{n-1}}\frac{d^2 
\left[r^{n+1}W(r)\right]}{d^2 r}\ ,
\end{eqnarray}
where $W(r)$ is\\
\phantom{line} 
\begin{eqnarray}
W(r)=\sum_{m=2}^{s}\frac{\alpha_m}{m} [k-B(r)]^m r^{-2m}+[k-B(r)]r^{-2}
-\frac{2\Lambda}{n(n+1)}\,,\label{W(r)} 
\end{eqnarray}
\phantom{line}\\
with $\alpha_m=a_m\prod_{p=1}^{2m-2}(n-p)$. 

For example, for $D=4$, one has the topological Schwarzschild-de Sitter solution, while 
for $D=5$, there is one Lovelock non trivial term 
(the Gauss-Bonnet, which in five-dimensions is not a topological invariant) 
and one has the Boulware-Deser solution \cite{BD}. 
For  higher dimensions we get an algebraic equation of increasing complexity, 
but, as we shall see in the following, 
for our purposes it will be not necessary to know explicitly the expression 
for the solution $B(r)$.    

For the static metric in Eq.~(\ref{metric_ansatz}) one can define the Killing vector 
$K^\mu=(1,\vec{ 0})$ and since 
\begin{equation}
\nabla_\nu{\cal G}_{\mu}^{\nu}=0\,,\qquad  {\cal G}_{\mu\nu}={\cal G}_{\nu\mu}\,,
\end{equation}
the vector $J_\mu={\cal G}_{\mu \nu}{}K^{\nu}$ 
is covariantly conserved and gives rise to a Killing conserved charge. 
This corresponds to the quasi-local generalized Misner-Sharp mass which reads\\
\phantom{line}
\begin{equation}
E_{MS}(r)\equiv-\frac{1}{\tilde\kappa^2}\int_\Sigma  J^\mu d \Sigma_\mu
=\frac{nV(\Omega_{(k,n)})}{2\tilde\kappa^2}\int_0^r d\rho \frac{d(\rho^{n+1}W)}{d \rho}
=\frac{nV(\Omega_{(k,n)})}{2\tilde\kappa^2}r^{n+1}W(r)\,.
\label{EMS}
\end{equation}
\phantom{line}\\
Here, $\tilde \kappa^2$ is the generalized version of $\kappa^2$ in $D$-dimension, namely $\tilde\kappa^2=8\pi \left(G_N\right)^{n/2}$. Furthermore, $\Sigma$ is a spatial volume at fixed time,  $d\Sigma_\mu=(d\Sigma, \vec{0})$, and
$V(\Omega_{(k,n)})$ is the $(n+1)$-volume depending on the topology. For example,
assuming the horizon of the sphere ($k=1$), one has 
$V(\Omega_{1,n})=2\pi^{(n+1)/2}/\Gamma((n+1)/2)$, with $\Gamma(z)$ the Euler-Gamma function.

For Lovelock solutions in the absence of matter, Eq.~(\ref{tt}) can be integrated as
\begin{eqnarray}
r^{n+1}W(r)=c_0\,,
\label{C}
\end{eqnarray}
$c_0$ being a constant of integration which we will show to be  related to the mass 
of the black hole. 
In particular, on shell, that is at the horizon $r=r_H$ such that $B(r_H)=0$,  
Eqs.~(\ref{EMS})--(\ref{C}) lead to the (Killing) energy of the black hole $E_{MS}(r_H)\equiv E_K$, 
\begin{equation}
E_K=\frac{nV(\Omega_{(k,n)})}{2\kappa^2}c_0 \,.
\label{E}
\end{equation}

\paragraph*{} Now, let us show that a First Law of black hole thermodynamics holds true, with the energy 
of the black hole solution, namely the Killing charge here 
obtained,  proportional to constant of integration $c_0$.
In the case of Lovelock gravity the validity of the First Law of black hole thermodynamics  
has been investigated 
in many places~\cite{meyer,pad2,maeda1,maeda2,Cai2}.
For the static case we present a direct and simple proof.

First of all we introduce the horizon defined by the existence 
of the largest positive root $r_H$ of $B(r)$ which satisfies conditions~(\ref{BHconditions}).
Then, from Eq. (\ref{W(r)}) and Eq.~(\ref{C}), we have the identity
\begin{equation}
c_0=r_H^{n+1}W(r_H)=\sum_{m=2}^{s}\frac{k^m\,\alpha_m}{m} r_H^{n+1-2m}+k\, r_H^{n-1}-\frac{2\Lambda r_H^{n+1}}{n(n+1)}\,.
\label{evaluation0}
\end{equation}
On the other hand, taking the derivative with respect to $r$ of Eq. (\ref{C}) and putting $r=r_H$, we obtain\\
\phantom{line}
\begin{eqnarray}
\sum_{m=2}^{s}\frac{k^m\alpha_m(n+1-2m)}{m} r_H^{n-2m}+(n-1)k\, r_H^{n-2}-\frac{2\Lambda r_H^{n}}{n}&=&\nonumber\\\nonumber\\
&&\hspace{-5cm}\frac{d B(r)}{dr}\Big\vert_{r_H}\left(\sum_{m=2}^{s}k^{m-1}\alpha_m r_H^{n+1-2m}+ r_H^{n-1}\right)\,.
\label{evaluation}
\end{eqnarray}
\phantom{line}\\
Now, let us compute the `thermodynamical' change of $c_0$ with respect 
to a small change of $r_H$.
From  Eq.~(\ref{evaluation0}) one has\\
\phantom{line}
\begin{equation}
dc_0=\left(\sum_{m=2}^{s}\frac{k\,\alpha_m(n+1-2m)}{m} r_H^{n-2m}+(n-1) k\,r_H^{n-2}-
\frac{2\Lambda r_H^{n}}{n}\right)d r_H\,.
\label{f}
\end{equation}
\phantom{line}\\
Making use of Eq.~(\ref{evaluation})  this expression may be rewritten in the form\\
\phantom{line}
\begin{equation}
dc_0=\frac{d B(r)}{d r}\Big\vert_{r_H}\left(\sum_{m=2}^{s}k^{m-1}\alpha_m r_H^{n+1-2m}+ r_H^{n-1}\right) dr_H\,.
\label{finalform}
\end{equation}
\phantom{line}\\
Let us interpret the right side of the last identity.  Here we are dealing with a 
static, spherically symmetric metric admitting a Killing vector.
If there is an event horizon located at $r_H$, then the Killing-Hawking temperature of the related 
black hole is given by  Eq.~(\ref{HawkingTemperature}). 
Now, all thermodynamical quantities associated with these black holes solutions
can be computed by standard methods. In particular,  
the entropy $S_W$ can be calculated by the Wald method~\cite{Visser:1993nu,Wald,FaraoniEntropy} 
or other methods, and one has~\cite{pad21,maeda1,maeda2,olea}:\\
\phantom{line}
\begin{equation}  
S_W=\frac{2\pi V(\Omega_{(k,n)})}{\tilde\kappa^2}r_H^n\left(1
+n\sum_{m=2}^{s}\frac{k^{m-1}\alpha_m}{n+2-2m} r_H^{2-2m}\right)\,.
\label{we}
\end{equation}
\phantom{line}\\ 
As a result, from Eqs.~(\ref{HawkingTemperature}), (\ref{E}), (\ref{finalform}) and (\ref{we}), 
one derives the First Law of black hole thermodynamics for Lovelock gravity, that is
\begin{equation}  
T_K\,\Delta S_W=\Delta E_K\,.
\label{fl1bis}
\end{equation} 
We have shown that for a generic Lovelock gravity in vacuum, 
the  First Law of black hole thermodynamics  holds and one can identify the energy of
a topological static, spherically symmetric black hole with the constant of integration 
and Killing conserved charge.

\section{$F(R)$-four dimensional modified gravity}

\paragraph*{} Now we come back to black hole solutions in four dimensional $F(R)$-gravity.
The entropy associated to these black holes 
can be calculated by the Wald method.
Following Refs.~\cite{Visser:1993nu,Wald,FaraoniEntropy}, the explicit calculation of the BH entropy is provided by the formula\\
\phantom{line}
\begin{equation}
S_W = - 2\pi \int_{\Sigma} \left(\frac{\delta \mathcal L}{\delta R_{\mu\nu\alpha \beta}}\right)\Big\vert_{H}\, 
e_{\mu \nu} e_{\alpha \beta} d \Sigma\,,
\label{Wald}
\end{equation}
\phantom{line}\\
where $\mathcal L = \mathcal L(R_{\mu\nu \alpha \beta}, R_{\mu\nu}, R,g_{\mu\nu}...) $ is the Lagrangian density of any general theory of gravity and 
$e_{\alpha\beta}=-e_{\beta\alpha}$ 
is the binormal vector to the (bifurcate) horizon. 
It is normalized so that $e_{\alpha\beta}e^{\alpha\beta}=-2$. 
For the SSS metric (\ref{SSS}), the binormal turns out to be
\begin{equation}
e_{\alpha\beta}= e^{\alpha(r)}(\delta^0_\alpha \,\delta^1_\beta - \delta^1_\alpha \,\delta^0_\beta )\,,\label{binormal}
\end{equation}
$\delta^{\alpha}_{\beta}$ being the Kronecker delta.
The induced area form, on the bifurcate surface \{$r=r_H$, $t=\mathrm{const}$\}, is represented by $d\Sigma$.
Finally, the subscript `$H$' indicates, as usually, that the partial derivative  is evaluated on the horizon, and the variation of the Lagrangian density with respect to $ R_{\mu\nu\alpha \beta}  $ is performed as if $R_{\mu\nu\alpha \beta} $ and 
the metric $g_{\alpha \beta}$ are independent. Since
\begin{equation}
\frac{\delta R}{\delta R_{\mu \nu \alpha \beta }}=
\frac{1}{2}\left(g^{\alpha \mu}g^{\nu \beta}-g^{\nu \alpha}g^{\mu \beta}  \right)\,, \label{Gaetano}
\end{equation}
for $F(R)$-theories where $\mathcal L=F(R)/(2\kappa^2)$, one obtains
\begin{equation}
S_W=\frac{{\cal A}_H\,F'(R_H)}{4 G_N}\,.\label{waldF(R)}
\end{equation}
Here, $\mathcal A_H=V_k r_H^2$ and $V_k$ is the volume of the ``horizon'' manifold, namely $V_1=4\pi$ (the sphere),
$V_0=|\Im\,\tau|$, with $\tau$ the Teichm\"{u}ller
parameter for the torus, and finally
$V_{-1}=4\pi g$, $g>2$, for the compact hyperbolic manifold with genus $g$ (see, for example Ref. \cite{Vanzo}).
If $F(R)=R$, we recover the Area Law of GR.

Taking the equation of motion (\ref{one}) evaluated on the event horizon ($B(r_H)=0$) and recalling the expression (\ref{waldF(R)}) for the Wald entropy, we easily get\\
\phantom{line}
\begin{equation}
\mathrm{e}^{\alpha(r_H)}B'(r_H)\frac{\partial S_W}{\partial r_H}=
\mathrm{e}^{\alpha(r_H)}V_k\left(\frac{k\,F'(R_H)}{2G_N}-\frac{R_HF'(R_H)-F(R_H)}{4G_N}r_H^2\right)\,.
\label{EquazionePrincipe}
\end{equation}
\phantom{line}\\
The Killing temperature appears in a natural way in  this equation.
Furthermore, if the entropy depends only on $r_H$, and not on the (free) integration constants of the solution, its partial derivative can be read as a thermodynamical variation and
Eq.~(\ref{EquazionePrincipe}) can be used to derive  the First Law of black holes thermodynamics
and employed to define a specific BH  energy~\cite{SSSEnergy}.
Thus, we may write\\
\phantom{line}
\begin{equation}
\Delta E_{K}:=T_{K}\Delta S_W=
\frac{V_k e^{\alpha(r_H)}}{4\pi }\,
\left(\frac{k\,F'(R_H)}{2G_N}-\frac{R_HF'(R_H)-F(R_H)}{4 G_N}r_H^2\right)\,dr_H\,.
\label{differentialform}
\end{equation}
\phantom{line}\\
Of course, this expression is reasonable as soon as only one parameter, which may be identified with the mass of the black hole, arises from the solution:
otherwise, one should expect that some thermodynamical potentials appear 
and the expression for the energy
can not be computed by means of the above expression.
If the differential of the Killing energy 
is interpreted  as  variation  due to an infinitesimal
change of the size of the black hole, namely if the right side of this equation depends only on $r_H$, one may try to get an explicit expression for the Killing energy,\\
\phantom{line}
\begin{equation}
 E_{K}:=\frac{V_k}{4\pi}\int\,e^{\alpha(r_H)}
\left(\frac{k\,F'(R_H)}{2G_N}-\frac{R_HF'(R_H)-F(R_H)}{4 G_N}r_H^2\right)d r_H\,.
\label{BHEnergy}
\end{equation} 
\phantom{line}\\
In fact, the derivation of the First Law of thermodynamics from the EOMs and the expression for the Killing energy in the latter equation are valid iff  the curvature
$R$ is an explicit function of $r_H$ only,
as it happens in the models of $F(R)$-gravity we have derived in \S\ref{2.1}, where one integration constant occurs in the solutions. An exception in this sense is represented by the $R^2$-model that we will treat apart.
Furthermore,
we have extended the validity of the Killing/Hawking temperature,
whose expression is derived in General Relativity by using quantum mechanics and is given by the metric,
to topological $F(R)$-gravity. Such result seems in favor of the Killing temperature with respect to the Kodama/Hayward one, at least for the vacuum case.

In what follows, 
by making use of exact solutions,  
we will provide a  support for our identification. 
We will see that Eq.~(\ref{BHEnergy})
allows to identify the energy of the black hole with the suitable (mass) constant
which appears in the explicit solution and is positive defined.

This proposal  should  be compared with a
similar one contained in  Ref. \cite{Cai4}. 
In Ref. \cite{Cai} an attempt to define a local Misner-Sharp 
mass has been presented. There, however, the proposed formula is not really satisfactory, 
because the quasi-local form is only present in some particular cases.

\subsection{BH energy in $F(R)$ gravity: Examples}

\paragraph{}  Here we explicitly compute the energy for some models we have derived in \S \ref{2.1}.
To start with, let us consider the dS/AdS-Schwarzshild solution (\ref{deS1}) for the model $F(R)=R-2\Lambda$.
Here, the only free (positive) parameter is $c_1$, and from (\ref{waldF(R)}),
(\ref{HawkingTemperature}), and (\ref{BHEnergy}), we get\\
\phantom{line}
\begin{eqnarray}
&&\hspace{-0.8cm}S_W=\frac{A_H}{4\pi G_N}\,,  \quad
T_K=\frac{k-\Lambda r_H^2}{4\pi r_H}\,,\quad
E_K=\frac{V_k}{8\pi G_N}r_H\left(k-\frac{\Lambda}{3}r_H^2\right)=\frac{V_k}{8\pi G_N}\,c_1\,.
\label{STE1}
\end{eqnarray}
\phantom{line}\\
We have used the fact that $B_H=0$. The energy is positive defined and we recover the result of Lovelock gravity in four dimension.


As a second example let us consider the non trivial model in (\ref{mod2})--(\ref{deS2}).
The energy can be computed by the method described above:
entropy, temperature
and energy, respectively, follow from (\ref{waldF(R)}), (\ref{HawkingTemperature}) and (\ref{BHEnergy}) 
and we obtain
\begin{equation}
S_W=\frac{A_H}{4 G_N}\,a\,k\,r_H\,,  \quad\quad
T_K=\frac{1}{4\pi}\,\left(\frac{k}{r_H}+4c_2 r_H\right)\,, \quad\quad
E_K=-\frac{3\,a\,k\,V_k}{16\pi G_N}\,c_1\,.
\end{equation}
We see that the entropy is positive only if the parameters $a$ and $k$ have the same sign. In addition, in \S\ref{2.3}, we have seen that $c_1$ must be negative in order to have a BH solution with positive temperature. As a consequence, the energy is positive defined.


The third example is given by the model in (\ref{modellone}) with solution (\ref{modellonesol}). In this case we obtain from (\ref{differentialform}),
\begin{equation} 
\Delta E_K=\frac{3a\,k V_k}{8\pi G_N}\frac{d}{d r_H}\left[\frac{k r_H}{2}+\frac{b}{3a}+c_2 r_H^3\right]\,dr_H\,.
\end{equation}
By using the BH condition $B(r_H)=0$, it is easy to verify that this equation leads to $\Delta E_K=0$. It means that the Killing energy (and therefore the entropy) is a constant. This fact is not surprising, since the solution appears without integration constants, so that the First Law trivially is satisfied. 


As a further example we consider the topological Clifton-Barrow solution
(\ref{CBB}) of model (\ref{CB}). Since $\alpha(r)\neq 0$, the Killing temperature differs from the Hayward temperature 
and one easily finds
\begin{eqnarray}
&&S_W=\frac{\mathcal{A}_H}{4 G_N^{1-\delta}}(1+\delta)\left[\frac{6\delta(1+\delta)}{(2\delta^2+2\delta-1)r_H^2}\right]^\delta\,,\quad
T_K=\frac{(1-\delta)c_0}{(1-2\delta-2\delta^2)}r^{\frac{(3\delta-4\delta^2-2)}{(1-\delta)}}\,,
\nonumber\\
&&E_K=\frac{\Psi_\delta}{r_0^{\delta(1+2\delta)/(1-\delta)}}\,\left(\frac{c_0}{k}\right)\,,
\label{CBEn}
\end{eqnarray}
where we have introduced the dimensionless constant\\
\phantom{line}
\begin{equation}
\Psi_\delta=\left(\frac{2^{\delta-1}3^{\delta}\delta^{\delta}(\delta-1)^2(\delta+1)^{\delta
+1}}{\sqrt{1-2\delta-2\delta^2}\sqrt{1-2\delta+4\delta^2}}\frac{1}{(2\delta^2+2\delta
-1)^{\delta}}\right)\,.
\end{equation}
\phantom{line}\\
In order to have the BH horizon and a positive Killing temperature,
the range of the parameter $\delta$ in the latter equations has to be restricted
to the values already discussed in \S \ref{2.3}.
Some additional restrictions are also necessary to get a positive entropy.
In the specific, we must require
$\delta>(\sqrt{3}-1)/2$ or $-1<\delta<0$. 
On the other hand, only the solutions of $-1<\delta<0$ give a real value for the temperature.
By taking into account such restrictions, the energy in (\ref{CBEn}) is well defined and positive.
As expected, in the limit  $\delta\rightarrow 0$ it reduces to the
Misner-Sharp mass of General Relativity.

The mechanism works for the entire class of Liefshitz solutions presented in \S~ \ref{Lif}, where only one integration constant which may be identified with the energy is present (see Ref.~\cite{SSSEnergy} for a specific example).

Let us pause to summarize these noticeable results. 
Making use of definition (\ref{differentialform}), we are able to give a reasonable expression for BH mass in topological $F(R)$-gravity: this expression results to be proportional to the integration constant of SSS solutions and positive defined for all the physical cases.

\subsection{The dS/AdS-Schwarzshild solution in the $R^2$-model\label{R^2}}

\paragraph{} Let us consider the model\\
\phantom{line}
\begin{equation}
I=\frac{1}{32\pi}\int_{\mathcal{M}} d^4x\sqrt{-g}\,R^2\,.
\end{equation}
\phantom{line}\\
It is easy to directly verify that such Lagrangian
admits the topological Schwarzschild dS/AdS solution
\begin{equation}
B(r)=k-\frac{c_1}{r}-\frac{\lambda r^2}{3}\,,\quad R=4\lambda\,.
\end{equation}
The interesting point is that in this case $\lambda$ is not a fixed parameter and the solution depends on two integration constants. Moreover, the entropy reads
\begin{equation}
S_K=\frac{\mathcal A_H}{4}\left(2R_H\right)=2\mathcal A_H\lambda\,,
\end{equation}
and in principle we cannot derive the First Law of thermodynamics from the EOMs, 
since the variation of the entropy with respect to $r_H$ only is not a thermodynamical differential.

However, we observe that the $R^2$-model
 does not contain dimensional parameters in the action.
Thus, one could think that there exists a  trivial entropy and a vanishing energy,
but, as we have shown above, the solution gives rise to the length scale $L$ 
related to the  integration constant $\lambda=1/L^2$ and, if $\lambda\neq 0$, is possible to get a non trivial entropy.
In this sense, we can consider $\lambda>0$ as a fundamental fixed lenght scale of the model, and the real mass-constant of the solution is $c_1$.
In such a case, by fixing $\lambda$, the First Law can be derived from the EOMs and  (\ref{differentialform}) reads
\begin{equation}
T_K=\frac{k-\lambda r_H^2}{4\pi r_H}\,,\quad E_K=\frac{V_K}{\pi}\lambda\left[k r_H-\frac{\lambda r_H^3}{3}\right]=\frac{V_k}{\pi}\lambda c_1\,.
\end{equation} 
Here, for completeness, we have also reported the Killing temperature. Note that, if $\lambda>0$, the BH exists for $k=1$, $c_1>0$ only. In any case, the energy is positive defined.

\section{$\mathcal F (R,G)$-gravity}

\paragraph{}In this and in the following Sections we will try to generalize our results to other classes of modified gravity theories in four dimensional space-time.

Let us analyze the general class of $\mathcal{F}(R,G)$-gravity, whose $F(R)$-gravity is a paricular case. 
Since for the models under consideration $\mathcal L=\mathcal F(R,G)/(2\kappa^2)$, the Wald entropy (\ref{Wald}) becomes\\
\phantom{line}
\begin{eqnarray}
S_W &=& -8\pi \mathcal{A}_H\,
\mathrm{e}^{2\alpha (r_H)}\,\left(\frac{\delta \mathscr L}{\delta R_{0 1 0
1}}\right)\Big\vert_H\nonumber\\
&=&-\frac{8\pi \mathcal{A}_H\mathrm{e}^{2\alpha(r)}}{2\kappa^2}\left(\mathcal F'_R\frac{\delta R}{\delta R_{0 1 0 1}}+\mathcal F'_G\frac{\delta G}{\delta R_{0 1 0 1}}\right)\Big\vert_H\label{waldbis}\,.
\end{eqnarray}
\phantom{line}\\
Here, we have taken into account the form of binormal vector (\ref{binormal}). Thus, since\\
\phantom{line}
\begin{equation}
\frac{\delta R}{\delta R_{\mu \nu \alpha \beta }}=
\frac{1}{2}\left(g^{\alpha \mu}g^{\nu \beta}-g^{\nu \alpha}g^{\mu \beta}  \right)\,,\nonumber
\end{equation}
\begin{equation}
\frac{\delta G}{\delta R_{\mu\nu\xi\sigma}}=
\left[ 2 R^{\mu\nu\xi\sigma} -2 (g^{\mu \xi} R^{\nu \sigma} + g^{\nu
\sigma} R^{\mu \xi} - g^{\mu \sigma} R^{\nu \xi} - g^{\nu \xi} R^{\mu \sigma})+
(g^{\mu \xi} g^{\nu \sigma} - g^{\mu \sigma} g^{\nu \xi}) R \right]\,,
\end{equation}
\phantom{line}\\
by using the horizon condition $B(r_H)=0$, we finally obtain
\phantom{line}\\
\begin{equation}
S_W=\frac{\mathcal A_H}{4 G_N}\left[\mathcal F'_R+\mathcal F'_G\left(\frac{4k}{r^2}\right)\right]\Big\vert_H\,.\label{entropy}
\end{equation}
\phantom{line}\\
As for $F(R)$-gravity, by recasting this expression in the first EOM (\ref{EOM1SphRG}), and by making use of the Killing temperature, we arrive at the following identity valid on the BH horizon,\\
\phantom{line}
\begin{eqnarray}
T_K \frac{\partial S_W}{\partial r_H} &=&
\nonumber\\ \nonumber\\
&&\hspace{-35mm}\mathrm{e}^{\alpha(r_H)}\left(\frac{k\,\mathcal F_R'(R_H, G_H)}{2G_N}-\frac{R_H\mathcal F_R'(R_H, G_H)+G_H\mathcal F'_G (R_H,G_H)-F(R_H, G_H)}{4G_N}r_H^2\right)\frac{V_k}{4\pi}\,.
\end{eqnarray}
\phantom{line}\\
Again, one may try to identify the last term with the thermodynamical variation of the Killing energy, but in Gauss-Bonnet modified gravity, unlike to the case of pure $F(R)$-gravity, the entropy depends on the integration constant of the solution, and in general is not possible to derive the First Law from the EOMs, as it is clear by considering the model (\ref{GBmodel}).
However, we can verify the First Law by hand, namely, we can check if the product between the Killing temperature and the variation of the Wald entropy leads to an exact thermodynamical differential of a quantity proportional to the mass constant of the solution, which can be identified with the energy of the black hole. 

The non trivial model
$\mathcal F(R,G)=R+\sqrt{G}$ of Eq. (\ref{GBmodel}) with the SSS solution $\alpha(r)=0$,  $B(r)=-1+r/c_0$, which corresponds to the topological case $k=1$, satisfies the conditions (\ref{BHconditions}) and may describe a BH.  The event horizon is given by 
$r_H=c_0$ and
the entropy, which explicitly depends on $c_0$ and $r_H$, reads\\
\phantom{line}
\begin{equation}
S_W=\frac{\mathcal A_H}{4G_N}\left[1+\frac{c_0}{r_H}\right]\,.
\end{equation}
\phantom{line}\\
In such a case, 
\begin{equation}
T_K=\frac{1}{4\pi c_0}\,,\quad\Delta S_W=\frac{\pi}{G_N}\left(2 r_H d r_H+c_0 d r_H+r_H d c_0\right)=
\frac{4\pi}{G_N}c_0 dc_0\,,\label{ghgh}
\end{equation}
and the First Law,
\begin{equation}
T_K\Delta S_W=\frac{d c_0}{G_N}=\Delta E_K\,,\quad E_K=\frac{c_0}{G_N}\,,
\end{equation}
permits the identification of the energy with the integration constant of the solution.
Note that in (\ref{ghgh}) we can express the variation of the entropy in terms of $c_0$ due to the fact that in the solution only one parameter appears.  
The solution is strictly related with the existence of the BH, which could be removed only in the limit $c_0\rightarrow 0$. That result is crucial and shows that the solution found is not patological, namely the fact that the mertric changes the signature for $c_0\rightarrow\infty$ is not surpraising, since it means that we are considering a BH with infinite mass.

\section{The Deser-Sarioglu-Tekin topological black hole solutions}

\paragraph*{} In this Section, first we generalize the modified gravity black hole solution 
of Deser et al.~\cite{Deser}, and
then we shall show that for these  metrics
the derivation of the First Law of black hole thermodynamics from the EOMs is valid and the Killing energy is positive and proportional to the constant of integration of the solutions.

For the sake of simplicity we shall restrict ourselves to the four-dimensional case, 
but, since we are interested in black hole with
generalized topological horizon, we have to include a non vanishing 
`cosmological constant' $\Lambda$ (see for example the GR case in Refs.~\cite{Vanzo,Brill,Mann}). 
The $D$-dimensional case as well as the inclusion of electromagnetism presents no difficulties.

To begin with, we write down the action of the model\\
\phantom{line}
\begin{equation}
I = \frac{1}{2\kappa^2} \int_{\mathcal M}\,d^4x\,\sqrt{-g} 
\left(R -2\Lambda + \sqrt{3}\sigma\,\sqrt{C^2}\right) \,, \label{d action}
\end{equation}
\phantom{line}\\
where $\Lambda$ is the cosmological constant, $\sigma$ is a real dimensionless parameter 
and $C^2=C_{\mu\nu\xi\sigma}C^{\mu\nu\xi\sigma}$ is the square of the Weyl tensor, which is an important measure of the curvature of space-time,\\
\phantom{line}
\begin{equation}
C^2=\frac{1}{3}R^{2}-2R_{\mu\nu}R^{\mu\nu}+R_{\mu\nu\xi\sigma}R^{\mu\nu\xi\sigma}\,.\label{Weylsquare} 
\end{equation}
\phantom{line}\\
For $\sigma=0$ the Weyl contribution turns off and GR result is recovered.
This model is a very interesting additive modification 
of GR with cosmological constant. 

A direct computation shows that the noteworthy properties
of the Weyl scalar $F$ discussed in Ref.~\cite{Deser} for $k=1$, 
are still valid for $k=0,-1$. Thus, the unknown functions $\alpha(r)$ and $B(r)$ can be obtained
by imposing the stationary condition $\delta\hat I=0$, where, 
$\hat I$ is the original action evaluated on the metric (\ref{SSS}), 
up to integration by parts and on the `topological' variables \{$\rho,\phi$\}.
It reads\\
\phantom{line}
\begin{equation}
\hat I=\frac{1}{\kappa^2}\int dt\int dr\,\mathrm{e}^{\alpha(r)}\,\left[\left(k-B(r)-r \frac{d B(r)}{d r}\right)
-\sigma\Big\vert k-4B(r)-r \frac{d B(r)}{d r}\Big\vert
-\Lambda r^2\right]\,.\label{hat I}
\end{equation}
\phantom{line}\\
From this equation follow the EOMs\\
\phantom{line}
\begin{equation}
\mathrm{e}^{\alpha(r)}\,\left[(1-\varepsilon\,\sigma)\left(k-B(r)-r \frac{d B(r)}{d r}\right)
+ 3 \varepsilon\sigma B(r)-\Lambda r^2\right]= 0\,, 
\label{EOM1Deser}
\end{equation}
\phantom{line}
\begin{equation}
\mathrm{e}^{\alpha(r)}\left[3\varepsilon\sigma+\frac{d\alpha(r)}{d r}(1-\varepsilon\sigma)r\right]=0\,.
\label{EOM2Deser}
\end{equation}
\phantom{line}\\
The general solution of this system is given by
\begin{equation}
 \alpha(r)= \log\left[\frac{r}{r_0}\right]^{\frac{3\varepsilon\sigma}{\varepsilon\sigma-1}},
\quad
B(r)= k\,\frac{(1-\varepsilon\sigma)}{(1-4\varepsilon\sigma)} - c_0 r^{-\frac{1-4\varepsilon\sigma}{1-\varepsilon\sigma}}
 -\Lambda\,\frac{r^2}{3(1-2\varepsilon\sigma)}\,,\quad\sigma\neq \pm1\,,\pm\frac{1}{4}\,,\label{b}
\end{equation}
where $c_0$  is the integration constant of the solution and $r_0>0$ has to be introduced for dimensional reasons. In the above expressions, the paramter $\varepsilon=\pm 1$ depends on the sign of the argument of the absolute value in the action (\ref{hat I}) and has been added to the original Deser's solution. In principle, this metric is disconnected at some points of the space-time. However, in what follows, we will limit to specify the value of $\varepsilon$ on the event horizon of the BH solution only.

We are assuming  $\sigma\neq\pm1,\pm1/4$ (for $\sigma=\pm1/2$ we can take $\Lambda=0$). For $\sigma=\pm 1$ it is possible to show the existence of the trivially, physically unacceptable solution with $\mathrm{e}^{\alpha(r)}\rightarrow 0$. For $\sigma=\pm 1/4$ and $\Lambda=0$ a simple solution can be found on the sphere $S_2$ ($k=1$). When $\varepsilon=1$, it reads $\alpha(r)=\log [r_0/r]$ and $B(r)=\log [r/c_0]$. 


As usual, the horizon is given by the positive root $r_H$ which satisfies conditions (\ref{BHconditions}).
In particular, the algebraic equation can be easily solved  
and gives\\
\phantom{line}
\begin{equation}
r_H=\left( k\,\frac{1-\varepsilon\sigma}{1-4\varepsilon\sigma} - \Lambda\,\frac{r_H^2}{3(1-2\varepsilon\sigma)}\right)^{\frac{\varepsilon\sigma-1}{1-4\varepsilon\sigma}} 
c_0^{\frac{1-\varepsilon\sigma}{1-4\varepsilon\sigma}}\,.
\label{CDeser}
\end{equation}
\phantom{line}\\
Here, a careful analysis on the metric signarture shows that we must require $c_0>0$ in order to have $r_H>0$ and $d B(r_H)/d r>0$ whatever the choice of the topology $k$.
Then, a direct derivation of the entropy via Wald method has been done in Ref.~\cite{Entropy}. 
Let us compute the Lagrangian
variation, where the constant $\Lambda$ vanishes,\\
\phantom{line}  
\begin{eqnarray}
\delta\mathscr L &=& \frac{1}{2\kappa^2}\left[\delta R + \sqrt{3} \,\sigma \,\delta (\sqrt{C^2})\right]\nonumber\\
&=& \frac{1}{2\kappa^2}\left[\frac{1}{2} (g^{\mu\xi}g^{\nu\sigma}- g^{\mu\sigma}g^{\nu\xi}) \delta R_{\mu\nu\xi\sigma} + \frac{\sqrt{3} \sigma}{2} (C^2)^{-\frac{1}{2}}\right]\, \delta (C^2)\;.
\end{eqnarray}
\phantom{line}\\
Using Eq.~(\ref{Weylsquare}), we get\\
\phantom{line}
\begin{eqnarray}
\frac{\delta \mathscr L}{\delta R_{\mu\nu\xi\sigma}}&=& \frac{1}{2\kappa^2}\left\lbrace \frac{1}{2} (g^{\mu\xi}g^{\nu\sigma}- g^{\mu\sigma}g^{\nu\xi})  + \frac{\sqrt{3} \sigma}{2} (C^2)^{-\frac{1}{2}}\,\times \right.\nonumber \\ \nonumber\\
& &\hspace{-20mm} \left.\left[ 2 R^{\mu\nu\xi\sigma} - (g^{\mu\xi}R^{\nu\sigma}+ g^{\nu
\sigma} R^{\mu\xi}- g^{\mu\sigma}R^{\nu\xi}- g^{\nu\xi}R^{\mu\sigma}) + \frac{1}{3}
(g^{\mu\xi}g^{\nu\sigma} - g^{\mu\sigma}g^{\nu\xi}) R \right]\right\rbrace \;,
\label{variation} 
\end{eqnarray} 
\phantom{line}\\
and in the specific,\\
\phantom{line}
\begin{eqnarray} 
\left(\frac{\delta
\mathscr L}{\delta R_{0 1 0 1}}\right)\Big\vert_H = \frac{1}{4\kappa^2}
\left[ g^{00} g^{11} + \frac{\sqrt{3}
\sigma}{\sqrt{C^2}}\left(2 R^{0 1 0 1} - g^{00} R^{1 1} - g^{1
1} R^{0 0} + \frac{1}{3} g^{0 0} g^{1 1}
R\right)\right]\Big\vert_H \,. \label{wald2} 
\end{eqnarray} 
\phantom{line}\\
For the topological SSS metric we write\footnote{
The trace of the $n$-power of the Weyl tensor with $n>0$ is\\
\phantom{line}
\begin{equation*}
\mathrm{tr} \left(C^2\right)^\frac{n}{2} = \left(-\frac{1}{3}\right)^n [2 +
(-2)^{2-n}] X(r)^n\,, 
\end{equation*}
\phantom{line}\\
where\\
\phantom{line}
\begin{equation*} 
X(r) = \frac{1}{r^2} \left[r^2
\frac{d^2 B(r)}{d r^2} + 2(B(r)-k) -2 r \frac{d B(r)}{d r}\right] + \frac{1}{r a} \left[3 r \frac{d B(r)}{d r} 
\frac{d \alpha(r)}{d r} -2 B(r)
\left(\frac{d \alpha(r)}{d r}
-r\left(\frac{d^2 \alpha(r)}{d r^2}+\left(\frac{d\alpha(r)}{d r}\right)^2\right)\right)
\right]\,. 
\end{equation*}
\phantom{line}\\
In our case, $n=2$.}
\begin{eqnarray}
\sqrt{C^2}\vert_H &=& \frac{1}{\sqrt{3}} \Big \vert \frac{1}{r^2}
\left[r^2 \left(\frac{d^2 B(r)}{d r^2}\right) + 2\left(B(r)-k\right) -2 r \left(\frac{d B(r)}{d r}\right)\right]\nonumber\\ \nonumber\\
&&\hspace{-2cm} + \frac{1}{r} \left[3 r \left(\frac{d B(r)}{d r}\right)\left(\frac{d\alpha(r)}{d r}\right) -2 B(r) \left(\frac{d \alpha(r)}{d r}
-r\left(\frac{d^2 \alpha(r)}{d r^2}+\left(\frac{d\alpha(r)}{d r}\right)^2\right)\right)\right]\Big\vert_{r_H} \,.\label{wald3} 
\end{eqnarray}
Taking together the Wald formula 
~(\ref{Wald}), Eq.~(\ref{wald2}) and Eq.~(\ref{wald3}), for the
solution (\ref{b}), we finally
have that the horizon entropy for the Deser \textit{et al.}
black hole reads\\
\phantom{line} 
\begin{equation} S_W = \frac{\mathcal{A}_H}{4G_N} \left(1 -
\varepsilon \sigma\right) \;,\qquad \mbox{where}\qquad
\varepsilon := \left\lbrace\begin{array}{cc}
-1 ,& \quad \sigma \leq \frac{1}{4} \\
+1,& \quad\sigma >1/4\,,\sigma\neq 1\;
\end{array}\right.\;.\label{entropyresult}
\end{equation}
\phantom{line}\\
Here, we have extended the result to the case $\sigma=1/4$, for which it is easy to see it is still valid and the
entropy function is continuous even if the black hole metric
changes.
\begin{figure}[-!h]
\begin{center}
\includegraphics[angle=0, width=0.55\textwidth]{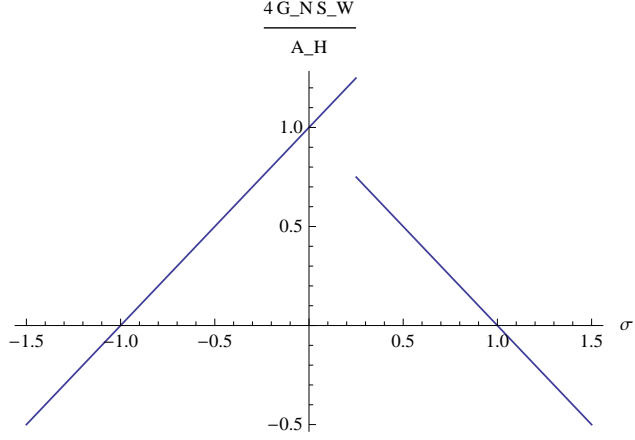}
\end{center}
\caption{Wald's entropy in units of $\mathcal{A}_H/4G_N$ versus $\sigma$ parameter for the Deser \textit{et al.}
BH.\label{Fig2.1}}
\end{figure}

In fact, as shown by Fig.~\ref{Fig2.1}, the entropy of the black hole is positive
only as far as $\sigma \in (-1,1)$. For $\sigma
=-1$, the entropy vanishes suggesting that, for this value of $\sigma$, the
number of microscopic configurations realizing the black hole
is only one. 
For $\sigma \in (-1,0)~\cup~(1/4,1)$, the entropy of Deser's
black hole is always smaller than its value in General
Relativity.
\paragraph{} The first EOM (\ref{EOM1Deser}) can be rewritten on the event horizon ($B(r_H)=0$) as\\
\phantom{line}
\begin{equation}
T_K \frac{\partial S_W}{\partial r_H}=
\frac{V_k}{8\pi G_N}\mathrm{e}^{\alpha(r_H)}\left[(1-\varepsilon)k-\Lambda r_H^2\right]\,,
\end{equation}
\phantom{line}\\
where we have introduced the Killing temperature (\ref{HawkingTemperature}) and the Wald entropy (\ref{entropyresult}). Since the entropy depends on $r_H$ only, we can read this expression as the First Law of thermodynamic for Deser's BH and therefore identify
\begin{equation}
E_K:=\frac{V_k}{8\pi G_N}\int \mathrm{e}^{\alpha(r_H)}\left[(1-\varepsilon)k-\Lambda r_H^2\right]dr_H\,,
\end{equation}
with the Killing energy.
By making use of explicit solution (\ref{b})
and Eq. (\ref{CDeser}),  as a consequence of the First Law, we get 
\begin{equation}
 T_K=\frac{1}{4\pi r_H}\left(k-\Lambda\,\frac{r_H^2}{(1-\varepsilon\sigma)}\right)\left(\frac{r_H}{r_0} \right)^{\frac{3\varepsilon\sigma}{\varepsilon\sigma-1}} \,,\quad E_K=\frac{V_k\left(1 -\epsilon\sigma\right)}{8\pi G_N}\left(\frac{1}{r_0}\right)^{\frac{3\varepsilon\sigma}{\varepsilon\sigma-1}}c_0\,.
\label{t34} 
\end{equation} 
The energy is positive defined for positive values of the Killing temperature and the entropy .
We observe that in this class of modified gravitational models the mass of black hole is particularly simple, 
since the modification is described by 
the dimensionless parameter $\sigma$.

\section{Topological conformal Weyl gravity black hole solutions}

\paragraph*{} In this Section, first we revisit the higher gravity black hole solution 
of Riegert \textit{et al.}~\cite{r,m}, and its topological version~\cite{klem}.
To start, we write down the conformal invariant action of the model, in the form
\begin{equation}
I=3 w \int_{\mathcal M} d^4 x\,\sqrt{-g}\,\,C^2\,,
\label{actionW}
\end{equation}
where $w$  is a dimensionless parameter, usually restricted to the values  $w>0$, and
$C^2$  is the square of the Weyl tensor (\ref{Weylsquare}) again.
The  conformal gravity model has a very interesting future domain and its phenomenology
has been investigated in Ref.~\cite{m21,m22}.
For topological SSS metric (\ref{SSS}), one has
\begin{eqnarray}
C^2=\frac{A(r)^2}{3r^4}\,,
\end{eqnarray}
where
\begin{eqnarray}
A(r)&=&r^2\frac{d^2 B(r)}{d r^2}+\left(3r\frac{d \alpha(r)}{d r}-2\right)r\frac{d B(r)}{d r}\nonumber\\
&&+2\left[r^2\frac{d^2\alpha(r)}{d r^2}+r^2\left(\frac{d \alpha(r)}{d r}\right)^2-r\frac{d \alpha(r)}{d r}+1\right]\,B(r)
-2 k\,.
\label{A(r)}
\end{eqnarray}
The total (effective) action reads now
\begin{equation}
 \hat I=\int dt \int dr\,e^{\alpha(r)}\left[\frac{w\,A(r)^2}{r^2}\right]\,.
\end{equation}
We are dealing  with a higher-order Lagrangian system, because  the Lagrangian depends
on the first and second derivatives of the unknown functions  $\alpha(r)$ and $B(r)$.
The corresponding equations of motion (after simplification) read\\
\phantom{line}
\begin{eqnarray}
w\,\mathrm{e}^{\alpha(r)}\left\{\frac{d^2 A(r)}{d r^2}B(r)+\frac{d A(r)}{d r}\left(\frac{B(r)}{r}+\frac{1}{2}\frac{ d B(r)}{d r}\right)-\frac{1}{4r^2}\,(A(r)^2+4k\,A(r))\right\}=0\,,
\label{bw}\end{eqnarray}
\begin{eqnarray}
w\,\mathrm{e}^{\alpha(r)}\left\{\frac{d^2 A(r)}{d r^2}+\frac{d A}{d r}\left(\frac2r-\frac{d \alpha(r)}{d r}\right)\right\}=0\,.
\label{aw}
\end{eqnarray}
\phantom{line}\\
In order to find explicit solutions, we proceed as in the previous Sections,
first considering the case $\alpha(r)=0$. With this choice, one immediately gets
$A(r)=6c_0+6c_1/r$
and by using Eqs.~(\ref{A(r)}) and (\ref{bw}) we recover the topological Riegert solution
\begin{equation}
B(r)=k+3c_0+\frac{c_1}{r}+b_1\,r+c_2r^2\,,\quad
             c_1\neq0\,,\quad b_1=\frac{c_0(2k+3c_0)}{c_1}\,.
\label{B(r)}
\end{equation}
Here, we have three arbitrary integration constants, namely $c_0\,,c_1\,,c_2$.
In Ref. \cite{Weyl} is possible to find 
the generalization of this metric to the charged case.
By conveniently choosing the integration constants of the Riegert solution, one obtains the Schwarzschild,
Schwarzschild-dS/AdS or the de Sitter metric, correspondingly.

The case $c_1=0$ has to be treated separately, since the limit in (\ref{B(r)}) is singular, and we obtain
\begin{equation}
B(r)=k+b_1\,r+c_2r^2\,,
\end{equation}
$b_1,c_2$ being arbitrary integration constants.

Moreover, if we consider the action of GR plus the Cosmological Constant and we add the contribute of Weyl gravity,
\begin{equation}
I=\int_{\mathcal M} d^4 x\,\sqrt{-g}\,
\left[ \frac{1}{2\kappa^2}\left(R-2\Lambda\right)+3 w C^2\right] \,, \label{EWgravity}
\end{equation}
the topological SSS solution turns out to be the Schwarzshild dS/AdS one, i.e. (\ref{B(r)}) with $c_0=0$ and $c_2=-\Lambda/3$, where $c_2$ (or $\Lambda$) is fixed by the model.\\

In principle, one can try to find other solutions by solving the equations for
 a nonzero (or non constant) $\alpha(r)$, but they are really complicated and so
it is convenient to proceed in a different way, by using the fact that the
Weyl action is conformally invariant.
This implies that, if $ds^2$
is a solution of the field equations, then also
$d\tilde s^2=\Omega^2(x^\mu)ds^2$
is a solution, $\Omega(x^\mu)$ being an arbitrary (smooth) function of the coordinates.
In particular, starting from a topological SSS solution and choosing
$\Omega(x^\mu)=\Omega(r)>0$ to depend on the radial coordinate only, one obtains again
a static and spherically symmetric solution, which can be
set in the form (\ref{SSS}), with the change of radial variable
$\tilde r= r\,\Omega(r)$. In fact,
\begin{equation}
d\tilde s^2=-\Omega^2(r)e^{2\alpha(r)}B(r)\,dt^2+\frac{\Omega^2(r)\,dr^2}{B(r)}+
r^2\Omega^2(r)\,\left(\frac{d\rho^2}{1-k\rho^2}+\rho^2 d\phi^2\right)\,,
\label{ConfTrans}
\end{equation}
and this assumes the form
\begin{equation}
d\tilde s^2=-e^{2\tilde\alpha(r)}\tilde B(r)\,dt^2+\frac{dr^2}{\tilde B(r)}+r^2\,\left(\frac{d\rho^2}{1-k\rho^2}+\rho^2 d\phi^2\right)\,,
\end{equation}
where, for simplicity, we have re-stated the original variable $r$ by
$r\Omega(r)\to r$, $\Omega(r)\to r/\Xi(r)$ and
\begin{equation}
\tilde B(r)=B(\Xi(r))\left[\frac{r}{\Xi(r)}\frac{d\,\Xi(r)}{d r}\right]^{-2}\,,\quad
e^{2\tilde\alpha(r)}\,\tilde B(r)=e^{2\alpha(\Xi(r))}\,B(\Xi(r))\,\left[\frac{r}{\Xi(r)}\right]^2\,.
\end{equation}
In this way we have obtained a class of SSS solutions for the action (\ref{actionW}) specified by the functions $\tilde\alpha(r)\,,\tilde B(r)$.
Any solution in this class is related to the original one
by means of an arbitrary, positive (smooth) function $\Xi(r)$.
As one can immediately see, $\Xi(r)=r$ corresponds to the unitary transformation $\Omega(r)=1$.
On the contrary, starting from Riegert's solution (\ref{B(r)}) and choosing
an arbitrary $\Xi(r)\neq r$ one obtains a class of solutions with
$\tilde \alpha(r)\neq0$ given by means of the equation
\begin{equation}
e^{2\tilde\alpha(r)}=\left[\frac{r}{\Xi(r)}\right]^4\,\left[\frac{d\,\Xi(r)}{d r}\right]^2\,,\quad
\frac{d}{dr}\left[\frac1{\Xi(r)}\right]=-\frac{e^{\tilde\alpha(r)}}{r^2}\,.
\label{alRiegert}
\end{equation}
from which we see that, for fixed $\tilde\alpha(r)$,
one can in principle derive the function $\Xi(r)$ which provides the desired transformation.

\subsection{Riegert-Lifshitz solutions}\label{S:RLS}

\paragraph{} This is a class of solutions similar to the ones obtained for the $F(R)$ case in \S\ref{Lif},
but starting from that of Riegert, performing a conformal transformation as described in the
previous Section, and imposing $\tilde\alpha(r)$ to be of the form \cite{Weyl}
\begin{equation}
\tilde{\alpha}(r)=\log[\gamma\,r^z]\,,\label{B22}
\end{equation}
where $\gamma>0$ is a dimensional constant and $z$ the redshift parameter of Lifshitz solution.

From (\ref{alRiegert}), we immediately get
\begin{eqnarray}
\Xi(r)&=&-\frac{1}{\gamma\log[qr]}\,,\quad\quad\quad\quad\quad z=1\,,
\nonumber\\
\Xi(r)&=&-\frac{1}{q+\gamma r^{z-1}/(z-1)}\,,\quad z\neq1\,,
\end{eqnarray}
$q$ being an integration constant.

In the first case ($z=1$) for $\tilde B(r)$ we easily obtain\\
\phantom{line}
\begin{equation}
\tilde B(r)=\frac{b_1}{\gamma^2}-\frac{b_1}\gamma\,\log[q r]
+(k+3c_0)\,\log^2[q r]-\gamma\,c_1\,\log^3[q r]\,,
\end{equation}
\phantom{line}\\
while, in the second case ($z\neq1$) and for simplicity choosing $q=0$,
for $\tilde B(r)$ we get
\begin{equation}
\tilde B(r)=\tilde c_0+\tilde c_1\,\frac{r^z}{r}
    +\tilde b_1\,\frac{r}{r^z}+\tilde c_2\,\frac{r^2}{r^{2z}}\,,\quad
\left\{\begin{array}{l}
\tilde c_1=\frac{\gamma c_1}{(z-1)^3}\,,\\
\tilde c_0=\frac{k+3c_0}{(z-1)^2}\,,\\
\tilde b_1=-\frac{b_1}{\gamma(z-1)}\,,\\
\tilde c_2=\frac{c_2}{\gamma^2}\,,
\end{array}\right.
\quad z\neq1\,.
\label{tildeCoeff}
\end{equation}

\subsection{Black holes and thermodynamics}

\paragraph{} The Riegert solution (or the solutions derived by it via conformal transformations) describes a BH as soon as conditions (\ref{BHconditions}) are satisfied,\\
\phantom{line}
\begin{equation}
k+3c_0+b_1\,r_H+c_2 r_H^2+\frac{c_1}{r_H}=0\,,\quad
b_1\,+2c_2 r_H-\frac{c_1}{r_H^2}>0\,,\quad r_H>0\,.
\end{equation}
\phantom{line}\\
For example, if $c_1<0$ and $c_2>0$, then
it is easy to show that there always exists a positive root of $B(r_H)=0$,
independently of the values of $c_0$ and $c_2$.

Now we will compute the entropy for these BH solutions.
An important remark is in order.
In general the Wald entropy is not invariant with respect to conformal transformations,
but when the action is conformally invariant, then also  the entropy does not change.
This can be easily seen by recalling that,
under a conformal transformation as in Eq. (\ref{ConfTrans}), one has
\beq
\tilde g_{\mu\nu}=\Om^2(x^{\mu})g_{\mu\nu}\,,\hs\hs
\tilde g^{\mu\nu}=\Om^{-2}(x^{\mu})g^{\mu\nu}\,,\hs
\sqrt{-\tilde g}=\Om^4(x^{\mu})\,\sqrt{-g}\,,\nn
\eeq
\beq
\tilde R_{\mu\nu\xi\sigma}=\Om^2(x^{\mu})\,R_{\mu\nu\xi\sigma}+U(g,\nabla\Om(x^{\mu}))\,,\hs
   \tilde{\mathcal L}=\mathcal L(\tilde R_{\mu\nu\xi\sigma},...)=\Om^{-4}(x^{\mu})\mathcal L\,,
\eeq
where $U(g,\nabla\Om(x^{\mu}))$ is a function which does not depend on the Riemann tensor.
If the action is invariant, both tensors $g_{\mu\nu}$ and
$\tilde g_{\mu\nu}$ related by $\Omega(r)$ are (black hole) solutions and thus, for our usual metric (\ref{SSS}),
we get
\beq
\tilde S_W&=&-2\pi\,\oint_{\begin{array}{c}r=r_H\\t=\mbox{const}\end{array}}\left.
              \left(\sqrt{\tilde g_{00}\tilde g_{11}\tilde g}\,\,
               \frac{\delta\tilde{\mathcal L}}{\delta\tilde R_{0101}}\right)\,
                \right|_{r=\tilde r_H}\,d\rho\,d\phi
\nn\\
  &=&-2\pi\,\oint_{\begin{array}{c}r=r_H\\t=\mbox{const}\end{array}}\left.
           \left(\Om^2(r)\sqrt{g_{00}g_{11}g}\,\,
            \frac{\delta{\mathcal L}}{\delta R_{0101}}\,
             \frac{\delta R_{0101}}{\delta\tilde R_{0101}}\right)\,
              \right|_{r=r_H}\,d\rho\,d\phi
\nn\\
   &=&-2\pi\,\oint_{\begin{array}{c}r=r_H\\t=\mbox{const}\end{array}}\left.
         \left(\sqrt{-g}\,\,\frac{\delta\mathcal L}{\delta R_{0101}}\right)\,
          \right|_{r=r_H}\,d\rho\,d\phi=S_W\,.
\eeq
As it is well known, also the Killing surface gravity and temperature are conformally invariant quantity.
Here, we give a simple proof of the conformal invariance of the Killing/Hawking temperature.
In the metric (\ref{ConfTrans}), since $\tilde{r}=\Omega(r)r$, the event horizon is still located at $\tilde r_H=r_H$. As a consequence, due to the fact that
\begin{equation}
\frac{d\tilde{B}(\tilde r)}{d\tilde r}\Big\vert_{\tilde r_H}=\frac{1}{\Omega^2(r_H)}\frac{d B(r)}{dr}\Big\vert_{r_H}\,,\quad 
e^{\tilde{\alpha}(\tilde r_H)}=\Omega^2(r_H) e^{\alpha(r_H)}\,,
\end{equation}
we easily obtain
\beq
\tilde{T}_K=e^{\alpha(r_H)}\frac{d B(r)}{d r}\Big\vert_{r_H} =T_K\,.
\eeq
This means that all black holes described in \S~\ref{S:RLS}
have the same entropy and Killing temperature.

In Weyl's conformal gravity, since $\mathcal L=3w\,C^2$, making use of  (\ref{Wald}) and (\ref{Weylsquare}), we get\\
\phantom{line}
\beq
S_W&=&-24\pi w\,\mathcal A_He^{2\al(r_H)}\,
   \aq2R^{0101}-(g^{00}R^{11}+g^{11}R^{00})+\frac13\,g^{00}g^{11}R\cq_{r_H}
 \nn\\
    &=&-24\pi w\,\mathcal A_H\,\aq\frac{1}{3}\frac{d^2 B(r)}{d r^2}+\at\frac{d\alpha(r)}{d r}-\frac{2}{3r}\ct\,\frac{d B(r)}{d r}
               -\frac{2k}{3r^2}\cq_{r_H}\,.
\label{Waldgen}
\eeq
\phantom{line}\\
In the particular case of Riegert's solution (\ref{B(r)}), the Killing temperature and the latter equation for the Wald entropy
simplify to
\begin{equation}
T_K=\frac{1}{4\pi}\left(b_1+2c_2 r_H-\frac{c_1}{r_H^2}\right)\,,\quad S_W=-48\pi wV_k\left(\frac{c_1}{r_H}+c_0\right)\,.\label{Entropy}
\end{equation}
In what follows we will consider the thermodynamics of Riegert's solutions only, being temperature, entropy and therefore the Killing energy conformal invariant.\\

We have seen that the Lagrangian of Weyl's
gravity contains only a dimensionless parameter $w$, while the solution $B(r)$
depends on three arbitrary integration constants $c_0\,,c_1\,,c_2$. In this case the  $r_H$ coordinate of the black
hole horizon will depend on several integration constants. 
In the general case, when asymptotically the solution is anti de Sitter,
the energy of the BH may be defined by means of the Euclidean action
and the First Law of black holes thermodynamics holds with additional
thermodynamical potentials~\cite{Pope}. With regard to this issue, let us imagine we are dealing with a BH solution, with $r_H=r_H(c_n)$ being the larger positive solution of $B(r_H)=0$, and $c_n$ the
constants of integration. Then, one has\\
\phantom{line}
\beq
0=\Delta B(r_H)=\sum_n\at\frac{\partial B(r_H)}{\partial c_n}\,dc_n
         +\frac{\partial B(r_H)}{\partial r_H}\,\frac{\partial r_H}{\partial c_n}\,dc_n\ct\,,\quad
\frac{\partial B(r_H)}{\partial r_H}\,\frac{\partial r_H}{\partial c_n}
=-\frac{\partial B}{\partial c_n}\,,
\eeq
\phantom{line}\\
where the $c_n$ are all independent and  $\Delta B(r_H)$ is the (total) variation of $B(r_H)$
on the horizon. Recall that  $T_K$ and $S_W$ are computable, for example, from (\ref{Entropy})
and we get\\
\phantom{line}
\beq
T_K\Delta S_W=12V_k\,(b_1\,dc_0-4c_2dc_1-2c_1dc_2)\,,
        \quad b_1=\frac{c_0(2k+3c_0)}{c_1}\,.
\eeq
\phantom{line}\\
In the particular case when  $c_0$, which is a pure number, and $c_2$, which provides the solution of a lengh scale (see the case of $R^2$-model in \S\ref{R^2}) are held fixed, one recovers the result of Ref. \cite{SSSEnergy}, namely
\beq
E_K=-48V_kc_2  c_1\,.
\eeq
This quantity is positive defined if the temperature and the entropy are also positive. 
In the general case,  all the constants of integration are true thermodynamic variables and, when it is possible, one has to make use  of Euclidean methods  in order to identify the energy. Then the First Law is shown to held true but with additional thermodynamic potentials, as explained in Ref.~\cite{Pope}.
We also observe that for Schwarzshild-dS/AdS solution of Einstein-Weyl gravity  (\ref{EWgravity}) with non vanishing Cosmological Constant $\Lambda$, the Weyl tensor gives a contribute to the energy changing the mass of the black hole as  
\begin{equation*}
E_K=-\frac{c_1 V_k}{8\pi G_N}\rightarrow E_K= -V_k\left(\frac{1}{8\pi G_N}-\frac{48}{3}V_K\Lambda\right)c_1\,,\quad c_1<0\,.
\end{equation*}
In particular, the mass decreases if $\Lambda$ is positive and vice-versa if $\Lambda$ is negative.\\

In conclusion, we have seen that, in all explicit and known examples, 
the  First Law of black hole thermodynamics 
(Clausius relation), that emerges from equations of motion, 
gives a reasonable value for the energy, which results proportional to the integration constant of the SSS solutions.


\chapter{The finite-time future singularities in $\mathcal{F}(R,G)$-modified gravity}

\paragraph*{} Many of $\mathcal{F}(R,G)$-modified gravity models suffer from the fact that they
bring the future universe evolution to finite-time singularities. It means, there is a finite time, for which some physical quantity (like the scale factor, the effective energy density/pressure of the universe or, more simplicity, some derivatives of Hubble parameter and therefore the components of Riemann Tensor) becomes singular rendering unphysical the solution.  
Some of these singularities are softer than other and not all physical quantities  necessarily diverge in rip time. Since singular solutions correspond to accelerated universe, they may appear as the final
attractor in realistic models which mimic the de Sitter universe where we live, leading to various instabilities in
the universe and destroying the feasibility of the models. 
Thus, before analyzing in the following Chapters the viable conditions of modified gravity, it is of some interest to explore in detail the $\mathcal{F}(R,G)$-gravity models realizing future time singularities. A detailed study of singularities in $F(R)$-gravity can be found in Ref.~\cite{OdWork}, here we will follow Refs.~\cite{GBSingularities},~\cite{ProceedingDue}.
In principle, Dark Energy could be described by scalar fields,
quintessence or phantom fluids, and so on. Any of such DE-models (including modified gravity) may be represented as the effective fluid with corresponding
characteristics.
Otherwise, we will see that, unlike to convenient DE-fluids which may be singular or
not, modified gravity suggests an universal scenario to
cure the finite-time future singularities.

\section{Four types of the finite-time future singularities\label{3.1}}

\paragraph*{} In general, in FRW Universe described by the metric (\ref{metric}), singularities appear during cosmological evolution when the Hubble parameter is expressed as
\begin{equation}
H(t)=\frac{h_0}{(t_{0}-t)^{\beta}}+H_{0}\,,
\label{Hsingular}
\end{equation}
where $h_0$, $t_{0}$ and $H_{0}$ are positive constants,
$\beta$ is a generic parameter which describes the type of singularity, and $t<t_{0}$ because it has to be for expanding Universe.
We can see that if
$\beta>0$, $H$ becomes singular in the limit $t\rightarrow t_{0}$.
Hence, $t_{0}$ is the time when a singularity appears.
On the other hand,
if $\beta<0$, even for non-integer values of $\beta$ some derivative of $H$,
and therefore the curvature or some combination of curvature invariants, becomes singular.
We assume $\beta\neq 0$ because $\beta=0$ corresponds to de Sitter space,
which has no singularity. Since $H_0$ is not a dynamical term, in the next Sections we will often put it equal to zero.

The finite-time future singularities can be classified
in the following way~\cite{classificationSingularities}:
\begin{itemize}
\item Type I (Big Rip \cite{Rip1}-\cite{Rip19}): for $t\rightarrow t_{0}$, $a(t)\rightarrow\infty$,
$\rho_\mathrm{{eff}}\rightarrow\infty$ and
$|p_\mathrm{{eff}}|\rightarrow\infty$.
It corresponds to $\beta=1$ and $\beta>1$.
\item Type II (sudden \cite{sudden}):
for $t\rightarrow t_{0}$, $a(t)\rightarrow a_{0}$,
$\rho_\mathrm{{eff}}\rightarrow\rho_{0}$ and $|p_\mathrm{{eff}}|
\rightarrow\infty$.
It corresponds to $-1<\beta<0$.
\item Type III: for $t\rightarrow t_{0}$, $a(t)\rightarrow a_{0}$,
$\rho_\mathrm{{eff}}\rightarrow\infty$ and
$|p_\mathrm{{eff}}|\rightarrow\infty$.
It corresponds to $0<\beta<1$.
\item Type IV: for $t\rightarrow t_{0}$, $a(t)\rightarrow a_{0}$,
$\rho_\mathrm{{eff}}\rightarrow \rho_0$, $|p_\mathrm{{eff}}|
\rightarrow p_0$
and higher derivatives of $H$ diverge.
It corresponds to
$\beta<-1$ but $\beta$ is not any integer number.
\end{itemize}
Here, $a_{0} (\neq 0)$ and $\rho_{0}$, $p_{0}$ are constants.
We remember that $\rho_{\mathrm{eff}}$ and $p_{\mathrm{eff}}$ are referred to the effective energy density and pressure of the universe and for $\mathcal F(R,G)$-gravity are given by Eqs.(\ref{rhoeffRG0})-(\ref{peffRG0}).
Moreover, we call singularities for $\beta=1$ and
those for $\beta>1$ as the `Big Rip' singularities and the `Type I'
singularities, respectively.\\

The Type I, II or III singularity appears when the Ricci scalar diverges and becomes singular. In such a case, 
the cosmological expansion of the model could tend towards this asymptotic solution\footnote{As regard to this point, it is well know that phantom dark energy ($\omega_{\mathrm{eff}}<-1$) reproduces the acceleration of the universe ending in the Big Rip. We will briefly analyze in \S\ref{4.1.1} of the next Chapter a quintessence/phantom (inhomogeneous) fluid with de Sitter solution and final attractor in the Big Rip.}. It is interesting to note that, since singular solution often is an attractor of the system, it can appear and destabilize the model also in the presence of other stable solutions (see also Refs.~\cite{Frolov, Dolgov} where the Starobinsky model~\cite{StaroModel} has been considered). As a qualitative example, we want to consider here the case of a realistic $F(R)$-gravity model, namely the Hu-Sawiki Model~\cite{HuSaw}, able to reproduce the de Sitter solution of dark energy epoch,
\begin{equation}
F(R)=R-\frac{\tilde{m}^{2}c_{1}(R/\tilde{m}^{2})^{n}}{c_{2}(R/\tilde{m}^{2})^{n}+1}=R-\frac{\tilde{m}^{2}c_{1}}{c_{2}}+\frac{\tilde{m}^{2}c_{1}/c_{2}}{c_{2}(R/\tilde{m}^{2})^{n}+1}\,.\label{HuSawModel}
\end{equation}
Here, $\tilde{m}^{2}$ is a mass scale, $c_{1}$ and $c_{2}$ are positive parameters and $n$ is a natural positive number. 
The model is very carefully constructed, such that $c_{1}\tilde{m}^{2}/c_{2}\simeq2\Lambda$, where $\Lambda$ is the Cosmological Constant, and in the high curvature region the physics of $\Lambda$CDM Model can be found. We note that the scalaron $F'(R)$,
\begin{equation}
F'(R)=1-\frac{\tilde{m}^{2}c_{1}/c_{2}}{\left(c_{2}(R/\tilde{m}^{2})^{n}+1\right)^2}(n)\left(\frac{c_{2}}{\tilde{m}^2}\right)\left(\frac{R}{\tilde{m}^2}\right)^{n-1}\,, 
\end{equation}
tends to a constant when $R\rightarrow\pm\infty$. Furthermore, 
one can evaluate the potential $V_{\rm eff}$ of Eq.~(\ref{Veff}) through an integration. By neglecting the contribute of matter, when $R\rightarrow\pm\infty$ one easily gets 
\begin{equation}
V_{\rm eff}(R\rightarrow\pm\infty)\simeq-\frac{\tilde{m}^{2}c_{1}/c_{2}}{3c_2(R/\tilde{m}^2)^n}\left(\frac{c_{2}}{\tilde{m}^2}\right)(n+1)\,.  
\end{equation}
Up to now, we are not able to say if some singular solution appears in this model. In addition, the Hu-Sawiki Model exhibits a stable de Sitter solution, that may be the final attractor of the system. On the other hand, 
if a singular solution 
where $R$ diverges 
exists, 
it is at a finite value of $V_{\rm eff}$ (in particular, it tends to zero) and the scalaron $F'(R)$ can crossover the potential in some point of cosmological evolution and arise the value $F'(R)=0$ for which catastrophic curvature singularity emerges. In the Appendix B the energy conditions related with occurrence of singularities are discussed. In general, it is possible to see that singularities violate the strong energy condition (SEC) describing acceleration. This is the reason for which realistic models for the dark energy could become unstable and fall into a singularity. We will better analyze the singularities in Hu-Sawiki Model in \S~\ref{3.4.1} 
and we will see that for some choices of parameters the model exhibits singularities in expanding universe.\\ 

Finally, the Type IV singularity appears for finite values of $R$. Since in this case only higher derivatives of Hubble parameter
diverge,
then some combination of curvature invariants also diverges and leads to
singularity. As a consequence, the solution becomes unphysical or may cause serious problems at the level of the black holes or stellar astrophysics~\cite{maeda1,maeda2}.\\ 

The study of the singularities is fundamental in order to achieve a correct description of the universe. 
In the next Sections we will reconstruct the specific classes of models which produce finite-time future singularities and the curing terms which protect the theory against singularities. Since near the singularities the Hubble parameter or its derivative diverge, we often analyze the problem in the asymptotic limit, when $t$ is close to $t_0$. We will also reasonably assume that the contribute of matter in expanding universe is avoidable.

\section{Effective parameters and singular solutions\label{3.2}}

\paragraph*{} It could be useful to introduce the (on shell) effective energy density and pressure~(\ref{rhoeffRG})--(\ref{peffRG}) in FRW universe to verify the presence of singularities in $f(R,G)$-modified gravity models, when $\mathcal{F}(R,G)=R+f(R,G)$, as in Eq.~(\ref{actiontwo}). In this case, we can treat modification to gravity like an effective dark energy fluid. 
We get
\begin{equation}
\rho_{\mathrm{eff}} = \frac{1}{2\kappa^{2}}\biggl[(Rf'_{R}
+Gf'_{G}-f)-6H\dot{f}'_{R}
-24H^{3}\dot{f}'_{G}
-6H^{2}f'_R
\biggr]+\rho_{\mathrm{m}}\,,
\label{tic}
\end{equation}
and
\begin{eqnarray}
p_{\mathrm{eff}} &=& \frac{1}{2\kappa^{2}}\biggl[
(f-R f'_{R}-G f'_{G})+4H\dot{f}'_{R}+2\ddot{f}'_{R}\nonumber \\
& &+16H(\dot H+H^{2})\dot{f}'_{G}+8H^{2}\ddot{f}'_{G}
+(4\dot{H}+6H^{2})f'_R                      
\biggr]+p_{\mathrm{m}}\,.
\label{tac}
\end{eqnarray}
Here, $f(R,G)$ has been replaced by $f$ and the subscript `$R$' is the derivative with respect to the Ricci scalar and the subscript `$G$' is the derivative with respect to the Gauss-Bonnet invariant. The point denotes, as usually, the time derivative. 

By combining the EOMs~(\ref{EOM1bis})--(\ref{EOM2bis}),
we obtain
\begin{equation}
\mathcal{G}(H,\dot{H}...)=
-\frac{1}{\kappa^2}\left[ 2\dot{H}+3(1+w)H^2 \right]\,,
\label{prime}
\end{equation}
where
\begin{equation}
\mathcal{G}(H,\dot{H}...) =
p_{\mathrm{eff}}-\omega\rho_{\mathrm{eff}}\label{G(HdotH)}\,.
\end{equation}
In the above expressions, $\omega=p_{\mathrm{m}}/\rho_{\mathrm{m}}$ is the (constant) EoS parameter of matter.
When a cosmology is given by Hubble parameter $H=H(t)$, the right-hand side of
Eq.~(\ref{prime}) is described by a function of $t$.
If the function $\mathcal{G}(H,\dot{H}...)$ in Eq.~(\ref{G(HdotH)}),
which is the combination of $H$, $\dot{H}$, $\ddot{H}$ and
the higher derivatives of $H$,
reproduces the above function of $t$, this cosmology could be realized\footnote{A remark is in order. When matter is taken into account, we need two EOMs, so that the using of one equation only is not enough. However, we will consider singular solutions in vacuum.}.
Hence, the function $\mathcal{G}(H,\dot{H}...)$ can be used to
judge whether the particular cosmology could be realized or
not~\cite{OdWork}.
The form of $\mathcal{G}(H,\dot{H}...)$ is determined by the
gravitational theory which one considers.
In the case of $f(R,G)$-gravity,
by substituting Eqs.~(\ref{tic}) and (\ref{tac}) into
Eq.~(\ref{G(HdotH)}), we find\\
\phantom{line}
\begin{eqnarray}
\hspace{-5mm}
\mathcal{G}(H,\dot{H}...) &=&
\frac{1}{2\kappa^{2}} \biggl\{ (1+\omega)(f
-Rf'_{R}-Gf'_{G})
+f'_R
\left[6H^2(1+\omega)+4\dot{H}\right]
\nonumber \\
\hspace{-5mm}
& &
+ H{\dot{f}}'_{R}(4+6\omega)
+8H{\dot{f}}'_{G}\left[2\dot{H}+ H^{2}(2+3\omega)\right]
+2{\ddot{f}}'_{R}+8 H^{2}{\ddot{f}}'_{G} \biggr\}\,.
\label{Feldwebel}
\end{eqnarray} 
\phantom{line}\\
If any singularity occurs, Eq.~(\ref{prime}) behaves as\\
\phantom{line}
\begin{equation}
\mathcal{G}(H, \dot{H}...)\simeq
\begin{cases}
-\frac{3(1+\omega)h_0^{2}+2\beta h_0}{\kappa^{2}}(t_{0}-t)^{-2} &
\mbox{($\beta=1$) Big Rip}\,;\\
-\frac{3(1+\omega)h_0^{2}}{\kappa^{2}}(t_{0}-t)^{-2\beta} &
\mbox{$\beta>1$ (Type I)}\,;\\
-\frac{2\beta h_0}{\kappa^{2}}(t_{0}-t)^{-\beta-1} &
\mbox{$\beta<1$ (Types II, III, IV )}\,.
\end{cases}
\label{casistiche}
\end{equation}
\phantom{line}\\
Here, we have used the singular form of $H$ in Eq.~(\ref{Hsingular}) with $H_0=0$ and we have considered the limit $t\rightarrow t_0$. 

In order to check the presence of singularities 
in a specific model of $f(R,G)$-gravity, it could be useful to verify the consistence of Eq.~(\ref{Feldwebel}) with Eq.~(\ref{casistiche}).
The behavior
of Eq.~(\ref{Feldwebel}) takes two different asymptotic forms
which depend on the parameter $\beta$ as follows:
\begin{itemize}
\item
Case of $\beta \geq 1$:
In the limit $t\rightarrow t_{0}$, one has\\
\phantom{line}
\begin{eqnarray}
\mathcal{G}(H, \dot{H}...) &\sim& \alpha
\left[ f+\frac{{f}'_{R}}{(t_{0}-t)^{2\beta}}+
\frac{{f}'_{G}}{(t_{0}-t)^{4\beta}} \right]
+\gamma\frac{{\dot{f}}'_{R}}{(t_{0}-t)^{\beta}}\nonumber \\
&&
+\delta\frac{{\dot{f}}'_{G}}{(t_{0}-t)^{3\beta}}
+\epsilon{\ddot{f}}'_{R}
+\zeta\frac{{\ddot{f}}'_{G}}{(t_{0}-t)^{2\beta}}\,,
\label{Achtung}
\end{eqnarray}
\phantom{line}\\
where $\alpha$, $\gamma$, $\delta$, $\epsilon$ and $\zeta$ are constants.
To realize a I Type singularity, we must verify the consistence with the first cases of Eq.~(\ref{casistiche}).
Hence,
if for $G\sim 1/(t_{0}-t)^{4\beta}$ and $R\sim 1/(t_{0}-t)^{2\beta}$
with $\beta \geq 1$, the highest term of Eq.~($\ref{Achtung}$) is
proportional to $1/(t_{0}-t)^{2\beta}$, it is possible to have an (asymptotic) Type I
singularity.
This condition is necessary and not sufficient.
Another very important condition that must be satisfied is the concordance
of the signs in Eq.~(\ref{casistiche}),
which depends on the parameters of the model.
\item
Case of $\beta<1$:
In the limit $t\rightarrow t_{0}$, one has\\
\phantom{line}
\begin{eqnarray}
\mathcal{G}(H, \dot{H}...)&\sim&
\alpha \left[ f
+\frac{{f}'_{R}}{(t_{0}-t)^{\beta+1}}
+\frac{{f}'_{G}}{(t_{0}-t)^{3\beta+1}} \right] 
+\gamma\frac{{\dot{f}}'_{R}}{(t_{0}-t)^{\beta}}\nonumber\\
& & 
+\delta\frac{{\dot{f}}'_{G}}{(t_{0}-t)^{2\beta+1}}
+\epsilon{\ddot{f}}'_{R}
+\zeta\frac{{\ddot{f}}'_{G}}{(t_{0}-t)^{2\beta}}\,.
\label{Hilfe}
\end{eqnarray}
\phantom{line}\\
To realize this kind of singularities, the last case of Eq.(\ref{casistiche}) has to be verified.
Thus,
if for $G\sim 1/(t_{0}-t)^{3\beta+1}$ and $R\sim 1/(t_{0}-t)^{\beta+1}$
with $\beta<1$, the highest term of Eq.~($\ref{Hilfe}$) is proportional to
$1/(t_{0}-t)^{\beta+1}$, it is possible to have a Type II, III or IV
singularity. Also this condition is necessary and not sufficient.
\end{itemize}
In the next Sections, we will reconstruct the typical forms of modify gravity which could lead to singularities.

\section{The reconstruction of singular $f(G)$-gravity\label{3.3}}

\paragraph*{} In this Section, as an explicit example of $\mathcal{F}(R,G)$-gravity, we reconstruct the $f(G)$-gravity models where
finite-time future singularities may occur.
The action is given by 
\begin{equation}
I=\int_\mathcal{M}d^4 x\sqrt{-g}\left[R+f(G)\right]\,.
\end{equation}
It means that the modification to GR is represented by the function $f(G)$ of the Gauss-Bonnet invariant only.
To find the $f(G)$-models which realize the form of $H$ given by Eq.~(\ref{Hsingular}), we adopt
the reconstruction method of modified gravity~\cite{OdWork,Reconstruction1,Reconstruction2,Reconstruction3,Reconstruction4}.
By using proper functions $P(t)$ and $Q(t)$ of a scalar field $t$
which we identify with the cosmic time,
we can write the action (in vacuum) as
\begin{equation}
I=\frac{1}{2\kappa^2}\int_{\mathcal{M}} d^{4}x \sqrt{-g} \left[
R+P(t)G+Q(t)\right]\,.
\label{azionef(G)bis}
\end{equation}
The variation with respect to $t$ yields 
\begin{equation}
\dot{P}(t)G+\dot{Q}(t)=0\,,
\label{PQ}
\end{equation}
from which, in principle, we can get $t$ as a function of $G$, $t=t(G)$.
By substituting $t=t(G)$ into Eq.~(\ref{azionef(G)bis}), we
find the action in terms of $f(G)$,
\begin{equation}
f(G)=P(t(G))G+Q(t(G))\,.
\label{f(G)}
\end{equation}
We describe the scale factor as
\begin{eqnarray}
a(t)= a_0 \exp \left[ g(t) \right]\,,
\label{eq:Scale-Factor}
\end{eqnarray}
where $a_0$ is a constant and $g(t)$ is a proper function of $t$.
By using the explicit form of the EOMs (\ref{EOM1})--(\ref{EOM2}), and by writing $f(G)$ as in Eq.~(\ref{f(G)}) and the scale factor into the Hubble parameter ($H(t)=\dot{g}(t)$), and by using the
matter conservation law (\ref{conservationlaw}) and then neglecting the
contribution from matter,
we get the differential equation
\begin{equation}
2 \frac{d}{d t} \left[ \dot{g}^2(t)  \dot{P}(t) \right] -2
\dot{g}^{3}(t) \dot{P}(t) + \ddot{g}(t)=0\,.
\label{P}
\end{equation}
From the first EOM (\ref{EOM1}),
$Q(t)$ is derived as
\begin{equation}
Q(t)= -24 \dot{g}^{3}(t)\dot{P}(t)-6\dot{g}^2(t)\,.
\label{Q}
\end{equation}

\subsubsection{Big Rip singularity}

\paragraph*{} First, we examine the Big Rip singularity.
If $\beta=1$ in Eq.~(\ref{Hsingular}) with $H_{0}=0$,
$H$ and $G$ are given by
\begin{equation}
\beta=1\,,\quad
\left\{\begin{array}{l}
H(t) = \frac{h_0}{(t_{0}-t)}\,,\\ \\
G(t)= \frac{24h_0^3}{(t_{0}-t)^4}(1+h_0)\,.
\end{array}
\right.\label{HGBigRip}
\end{equation}
The scale factor results
\begin{equation}
a(t)=\frac{a_0}{(t_0-t)^h}\,.\label{aBigRip} 
\end{equation}
From Eqs.~(\ref{P})--($\ref{Q}$), we get for $h_0\neq 1$,\\
\phantom{line}
\begin{equation}
h_0\neq 1\,,\quad
\begin{cases}
P(t)=\frac{1}{4h_0(h_0-1)}(2t_{0}-t)t+c_{1}\frac{(t_{0}-t)^{3-h_0}}{(3-h_0)}+c_{2}\,,
\\ \\
Q(t)=-\frac{6h_0^{2}}{(t_{0}-t)^2}-\dfrac{24 h_0^{3} \left[
(t_{0}-t)/(2h_0(h_0-1))-c_{1}(t_{0}-t)^{2-h_0} \right]}{(t_{0}-t)^3}\,.
\end{cases}
\end{equation}
\phantom{line}\\
Here, $c_{1}$ and $c_{2}$ are generic constants. We can take $c_{1}=0$ in the particular case of $h_0=3$. 

Furthermore, from Eq.~(\ref{PQ}) we correctly obtain
\begin{equation}
t= \left[ \dfrac{24(h_0^{3}+h_0^{4})}{G} \right]^{1/4}+t_{0}\,.
\end{equation}
By solving Eq.~(\ref{f(G)}), we reconstruct
the most general form of $f(G)$ which realizes the Big Rip singularity,
\begin{equation}
f(G)=\frac{\sqrt{6h_0^{3}(1+h_0)}}{h_0(1-h_0)}\sqrt{G}
+c_{1}G^{\frac{h_0+1}{4}}+c_{2}G\,.
\label{Garr}
\end{equation}
This is an exact solution of Eq.~(\ref{prime}) in the case of Eq.~(\ref{HGBigRip}).
In the model
$R+\alpha G^{1/2}$, where
$\alpha (\neq 0)$ is a constant,
the Big Rip singularity for $G\rightarrow+\infty$ could appear realizing any value of $h_0\neq 1$.
In the above expression, $G^{(1+h_0)/4}$ is an invariant with
respect to the Big Rip solution and $G$ is the (trivial) topological invariant and does not contribute to the dynamic.

In the case of $h_0=1$, it is possible to find
another exact solution of $P(t)$ and $Q(t)$, namely
\phantom{line}
\begin{equation}
h_0= 1\,,\quad
\begin{cases}
P(t)=-\frac{1}{4c_2}(t_{0}-t)^{2}\ln \left[ c_1 (t_{0}-t)^{c_2} \right]\,,
\\ \\
Q(t)=-\frac{12}{(t_{0}-t)^{2}}\ln \left[ c_1 (t_{0}-t) \right]\,.
\end{cases}
\end{equation}
\phantom{line}\\
The final form of $f(G)$ is given by
\begin{equation}
f(G)= \frac{\sqrt{3}}{2}\sqrt{G} \ln[c_1 G]\,.
\end{equation}
This is another exact solution of Eq.~(\ref{prime}) for $H(t)=1/(t_{0}-t)$.
In general,
in the model
$R+\alpha\sqrt{G} \ln[\gamma G]$
with $\alpha\,,\gamma$ positive constant parameters,
this Big Rip singularity could appear.

\subsubsection{Other types of singularities}

\paragraph*{} Next, we investigate the other types of singularities. 
We restrict our investigation to the case of $H_0=0$ in Eq. (\ref{Hsingular}). For $\beta>1$ we get in the asymptotic limit
\begin{equation}
\beta> 1\,,\quad
\left\{\begin{array}{l}
H(t) = \frac{h_0}{(t_{0}-t)^{\beta}}\,,\\ \\
G(t) \simeq \frac{24h_0^4}{(t_{0}-t)^{4\beta}}\,.
\end{array}
\right.\label{HGsingular1}
\end{equation}
When $\beta\neq 1$, Eq.~(\ref{Hsingular}) 
implies that the scale factor $a(t)$ behaves as
\begin{equation}
a(t)= a_0\exp \left[ \frac{h_0(t_{0}-t)^{1-\beta}}{\beta-1} \right]\,.\label{ageneric}
\end{equation}
Thus, by evaluating Eq.~(\ref{P}) in the limit $t\rightarrow t_{0}$, we obtain 
\begin{equation}
P(t)\simeq-\frac{1}{4h_02}(t_{0}-t)^{2\beta}\,,\quad f(G)=-12\sqrt{\frac{G}{24}}\,.\label{primo}
\end{equation}
Hence, in the model $R-\alpha\sqrt{G}$
with $\alpha>0$, a Type I singularity for $G\rightarrow+\infty$ could appear.\\

When $\beta<1$, the forms of $H$ and $G$ are given by
\begin{equation}
\beta< 1\,,\quad
\left\{\begin{array}{l}
H(t) =  \frac{h_0}{(t_{0}-t)^{\beta}}\,,\\ \\
G(t) \simeq \frac{24h_0^3\beta}{(t_{0}-t)^{3\beta+1}}\,.
\end{array}
\right.\label{HGsingular2}
\end{equation}
An asymptotic solution of Eq.~(\ref{P}) in the limit $t\rightarrow t_{0}$ is
given by
\begin{equation}
P(t)\simeq \frac{1}{2h_0(1+\beta)}(t_{0}-t)^{1+\beta}\,,\quad
f(G)=\frac{6h_0^{2}}{(\beta+1)}(3\beta+1)\left(\frac{|G|}{24h_0^{3}|\beta|}
\right)^{2\beta/(3\beta+1)}\,.\label{secondo}
\end{equation}
Hence, if $f(G)$ behaves as
\begin{equation}
f(G)\simeq\alpha |G|^{\gamma}\,,
\quad
\gamma = \frac{2\beta}{3\beta+1}\,,
\label{trentatre}
\end{equation}
with $\alpha>0$ and $0<\gamma <1/2$,
we find $0<\beta<1$ and a Type III singularity for $G\rightarrow+\infty$ could emerge.

If $\alpha>0$ and $-\infty<\gamma<0$,
we find $-1/3<\beta<0$ and
a Type II (sudden) singularity for $G\rightarrow-\infty$ could appear.
Moreover,
if $\alpha<0$ and $1<\gamma<\infty$,
we obtain $-1<\beta<-1/3$ and a Type II singularity with $G\rightarrow 0^{-}$ could occur.

Finally, if $\alpha>0$ and $2/3<\gamma<1$,
we obtain $-\infty<\beta<-1$ and a Type IV singularity for $G\rightarrow 0^{-}$ could appear.
We also require that $\gamma\neq2n/(3n-1)$,
where $n$ is a natural number, in order to exclude the non singular solutions with $\beta<-1$ integer numbers.

We can generate all the possible Type II singularities
as shown above except for the case $\beta=-1/3$, namely
\begin{equation}
H(t)=\frac{h_0}{(t_0-t)^{1/3}}\,,\quad G(t)=-8h_0^{3}+24h_0^{4}(t_{0}-t)^{4/3}\,.\label{-1/3}
\end{equation}
To find $t$ in terms of $G$, we must
consider the whole expression of $G$ by taking into account also the low dynamical
term
of $(t_{0}-t)$. We finally obtain
\begin{equation}
f(G)\simeq \frac{1}{4 \sqrt{6}h_0^{3}}G|G+8h_0^{3}|^{1/2}
+\frac{2}{\sqrt{6}}|G+8h_0^{3}|^{1/2}\,.
\end{equation}
As a consequence,
the specific model $R+\gamma_1 G|G+\gamma_{3}|^{1/2}+\gamma_2
|G+\gamma_{3}|^{1/2}$,
with $\gamma_{1,2,3}$
positive constants, can generate the Type II singularity where $G$ tends to the negative constant $-\gamma_3(=-8h_0^3)$ as in~(\ref{-1/3}).\\

All the asymptotic solutions we have found satisfy Eq.~(\ref{casistiche}) in the corresponding cases.

\subsection{Example of realistic $f(G)$-models generating singularities}

\paragraph*{} Here, we study the presence of singularities in the following realistic models of $f(G)$-gravity
which reproduce the dark energy epoch, namely~\cite{Nojiri:2007bt}\\
\phantom{line}
\begin{eqnarray}
f_{1}(G) &=& \frac{a_{1}G^{n}+b_{1}}{a_{2}G^{n}+b_{2}}\,;
\label{uno} \\ \nonumber\\
f_{2}(G) &=& \frac{a_{1}G^{n+N}+b_{1}}{a_{2}G^{n}+b_{2}}\,;
\label{due} \\ \nonumber\\
f_{3}(G) &=& a_{3} G^{n}(1+b_{3} G^{m})\,;
\label{terzo}\\ \nonumber\\
f_{4}(G) &=& \left(G^m\right) \frac{a_{1}G^{n}+b_{1}}{a_{2}G^{n}+b_{2}}\,.
\label{quarto}
\end{eqnarray}
\phantom{line}\\
Here, $a_{1,2,3}$, $b_{1,2,3}$, $n$, $m$ and $N$
are constants.

We start from the model~(\ref{due}), which is a generalization of (\ref{uno}). When $n>0$,
Types I, II and III singularities may be present. In fact, for  $N=1/2$, one
could have Big Rip singularity,
since in this case, in the asymptotic Big Rip limit of large $G$, Eq.~(\ref{due}) gives
$f_2(G)\simeq\alpha G^{1/2}$ with $\alpha\lessgtr 0$. 
Moreover, again with
$N=1/2$, if $a_1/a_2 <0$, Eq.~(\ref{due})
for large value of $G$ leads to $f_2(G)\simeq-\alpha G^{1/2}$ with $ \alpha >0$ and
thus
Type I singularity could appear.
If $n$ and $N$ are integers and $n+N >0$, for large and negative value of
$G$, $f_2(G)\sim a_1/(a_2 G^{-N})$.
As a result, a Type II singularity could
appear, when $-n <N<0$, $N$ even and $a_1/a_2>0$ or $N$ odd and   $a_1/a_2
<0$ (see Eq.~(\ref{trentatre}) and the related discussion). If $0<N< 1/2$
and $a_{1}/a_{2}>0$, we have the Type III singularity for large and positive values of $G$, such that $f_2(G)\sim (a_1/a_2)G^N$. Finally, when $G\rightarrow 0^{-}$, we do not recover
any example of singularity of
the preceding analysis.\\

If there exists any singularity solution, the consistence of Eq.~(\ref{Feldwebel}) and Eq.~(\ref{casistiche}) has to be verified, as we have already discussed in \S\ref{3.2}.

We see that our model in Eq.~(\ref{due}) with
$n>0$ and $N>0$
could also produce Type II singularity for $0<\beta<-1/3$, or Type IV singularity for $\beta<-1$, when $G\rightarrow0^-$. 
We get
\begin{equation*}
f_{2}(G)\sim\frac{b_{1}}{b_{2}}\,,
\quad
\frac{d f_{2}(G)}{d G}\sim-n\frac{b_{1}a_{2}}{b_{2}^{2}}G^{n-1}\,.
\end{equation*}
We are assuming $b_1/b_2$ very small and avoidable in Eq.~(\ref{Feldwebel}), otherwise we have to consider $H_0\neq 0$ in Eq.~(\ref{Hsingular}) and a different analysis of Eq.~(\ref{Feldwebel}) has to be done.
It can be shown that, under the requirement $n>2/3$ (the relation between $n$
and $\beta$ is $n=2\beta/(3\beta+1)$), the asymptotic behavior of
Eq.~(\ref{Hilfe}) when $G\simeq 24 h_0^{3}\beta/(t_{0}-t)^{3\beta+1}$ is
proportional to $1/(t_{0}-t)^{\beta+1}$ and
therefore it is possible to realize the Type II or IV singularity. Here, we include some examples:
\begin{enumerate}
\item For $N=1$ and $n=2$, $\mathcal{G}(H,\dot H...)\simeq
-[(24 h_0^{5})b_{1}a_{2}/(\kappa^2 b_{2}^{2})](t_0-t)^{-1/2}$ when $\beta=-1/2$.
Hence,
if $b_{1}a_{2}<0$, the model can become singular when $G\rightarrow 0^{-}$
(Type II singularity);
\item
For $N=1$ and $n=3$, $\mathcal{G}(H,\dot H...)\simeq
[b_{1}a_{2}/(\kappa^2 b_{2}^{2})](t_0-t)^{-4/7}$ when $\beta=-3/7$.
Thus, if $b_{1}a_{2}>0$, the model can become singular when
$G\rightarrow 0^{-}$ (Type II singularity);
\item For $N=1$ and $n=8/9$, $\mathcal{G}(H,\dot H...)\simeq
[2(8/9)^2 (32^{-1/9})h_0^{5/3}b_{1}a_{2}/(\kappa^2 b_{2}^{2})](t_0-t)^{1/3}$ when $\beta=-4/3$.
Hence,
if $b_{1}a_{2}>0$, the model can become singular when $G\rightarrow 0^{-}$
(Type IV singularity).
\end{enumerate}
 
The model $f_{1}(G)$ in Eq.~(\ref{uno}) is a particular
case of the one in Eq.~(\ref{due}) just analized. For large values of $G$, 
it is easy to see that $\mathcal{G}(H, \dot{H}...)$ in Eq.~(\ref{Achtung}) or in Eq.~(\ref{Hilfe}) tends to a constant,
so that it is impossible to find singularities.
Nevertheless, by performing a similar analysis as above,
Type II or III singularities can occur when $G\rightarrow 0^{-}$
for $n>2/3$.
For example,
if $n=2$, and therefore $\beta=-1/2$, one finds $\mathcal{G}(H,\dot H)\simeq
[(24 h_0^{5}/(\kappa^2 b_{2}^{2}))(a_{1}b_{2}-a_{2}b_{1})](t_0-t)^{-1/2}$. If $(a_{1}b_{2}-a_{2}b_{1})>0$, the model can become singular
when $G\rightarrow 0^{-}$ (Type II singularity).\\

With regard to $f_{3}(G)$ in Eq.~(\ref{terzo}), it is interesting to find
the conditions on $m$, $n$, $a_{3}$ and $b_{3}$ for
which we
do not have any type of singularities. When $G\rightarrow \pm\infty$ or
$G\rightarrow 0^{-}$, the model assumes the form
$f(G)\simeq\alpha G^{\gamma}$, $\alpha$ and $\gamma$ being constants, which we have investigated on in the
first part of this Section.
We do not consider the trivial case $n=m$.
The no-singularity conditions follow directly from the
preceding results as complementary  conditions to the singularity ones:
\begin{enumerate}
\item
Case (1):
$n>0$, $m>0$, $n\neq 1$ and $m\neq 1$.
We avoid any singularity if $0<n+m<1/2$ and $a_{3}b_{3}<0$; $n+m>1/2$, $n>1$
and $a_{3}>0$; $n+m>1/2$, $2/3<n<1$ and $a_{3}<0$; $n+m>1/2$,
$0<n \leq 2/3$; $n=1/2$, $a_{3}>0$.
\item
Case (2):
$n>0$, $m<0$ and $n\neq 1$.
We avoid any singularity if $0<n<1/2$ and
$a_{3}<0$; $n>1/2$, $n+m> 1$ and $a_{3}b_{3}>0$; $n>1/2$, $2/3<n+m < 1$
and $a_{3}b_{3}<0$; $n>1/2$, $n+m \leq 2/3$; $n+m=1/2$,
$a_{3}b_{3}>0$.
\item
Case (3):
$n<0$, $m>0$ and $m\neq 1$.
We avoid any singularity if $m+n>1/2$; $m+n<1/2$ and
$a_{3}b_{3}<0$.
\item
Case (4):
$n<0$ and $m<0$.
We avoid any singularity if $a_{3}<0$.
\end{enumerate}

We end this Subsection considering the last realistic model of Eq.~(\ref{quarto}), again for
$n>0$.
Since for large $G$, one has $f_{4}(G) \simeq (a_1/a_2)G^m $ and for
small $G$, one has
$f_{4}(G) \simeq (b_1/b_2)G^m $, the preceding analysis leads
to the absence of any type of singularities
for
\begin{equation}
\frac{1}{2} < m \leq \frac{2}{3}\,.
\label{sf}
\end{equation}
In fact, for this range of values, the asymptotic behavior of the right-hand
side of Eq.~(\ref{prime}) is different
from the asymptotic behavior of its left-hand side on the singularity
solutions. Thus, Eq.~(\ref{quarto}) provides an example of realistic model
free of all possible singularities when Eq.~(\ref{sf}) is satisfied,
independently on the coefficients. Moreover, this model suggests the
universal scenario to cure finite-time future singularities. In \S~\ref{curing} we will see that 
adding $\alpha G^{m}$, $\alpha$ being a constant and $1/2<m\leq2/3$, to any singular dark energy, results in combined
non-singular model. 

\section{The reconstruction of singular $\mathcal{F}(R,G)$-gravity\label{3.4}}

\paragraph*{} In this Section, we reconstruct the generic $\mathcal{F}(R,G)$-gravity models producing
finite-time future singularities.

We rewrite the action (\ref{azione}) in vacuum by using 
proper functions $Z(t)$, $P(t)$ and $Q(t)$ of a
scalar field which is identified with the time $t$,
\begin{equation}
I=\frac{1}{2\kappa^2}\int_{\mathcal{M}} d^{4}x \sqrt{-g} \left[Z(t)R+P(t)G+Q(t)\right]\,.
\label{azionemodificata}
\end{equation}
By the variation with respect to $t$, we obtain 
\begin{equation}
\dot{Z}(t)R+\dot{P}(t)G+\dot{Q}(t)=0\,,
\label{t}
\end{equation}
from which in principle it is possible to get $t$ as a function of $R$ and $G$, namely $t=t(R,G)$.
By substituting $t=t(R,G)$ into Eq.~(\ref{azionemodificata}),
we derive the action in terms of $\mathcal{F}(R,G)$,
\begin{equation}
\mathcal{F}(R,G)=Z(t(R,G))R+P(t(R,G))G+Q(t(R,G))\label{F(R,G)}\,.
\end{equation}
By using the conservation law and the EOMs, and then neglecting the contribution from matter, we obtain the differential
equation\\
\phantom{line}
\begin{equation}
\ddot{Z}(t)+4 \dot{g}^{2}(t)\ddot{P}(t)-\dot g(t)\dot{Z}(t)+[8 \dot{g}(t) \ddot{g}(t)
-4 \dot{g}^{3}(t)]\dot{P}(t)+2 \ddot{g}(t)Z(t)=0\,,
\label{Pi}
\end{equation}
\phantom{line}\\
where we have putted the scale factor $a(t)=a_0\exp[g(t)]$
as in Eq.~(\ref{eq:Scale-Factor})
and the Hubble parameter $H(t)=\dot{g}(t)$.
From the first EOM again, $Q(t)$ becomes
\begin{equation}
Q(t)=-24\dot{g}^{3}(t)\dot{P}(t)-6\dot{g}^{2}(t)Z(t)-6\dot{g}(t)\dot{Z}(t)\,.
\label{Qu}
\end{equation}

\subsubsection{Big Rip singularity}

\paragraph*{} First, we investigate the Big Rip singularity.
By putting $\beta=1$ in Eq.~(\ref{Hsingular}) with $H_{0}=0$,
we have\\
\phantom{line}
\begin{equation}
\beta=1\,,\quad
\left\{\begin{array}{l}
H(t) = \frac{h_0}{(t_{0}-t)}\,,\\\\
R(t)= \frac{6 h_0}{(t_{0}-t)^{2}}(2h_0+1)\,,\\\\
G(t) = \frac{24 h_0^{3}}{(t_{0}-t)^{4}}(1+h_0)\,.
\end{array}
\right.
\end{equation}
\phantom{line}\\
Let us see for some solutions of Eqs.~(\ref{Pi})--(\ref{Qu}). A simple (trivial) solution is given by\\
\phantom{line}
\begin{equation}
\left\{\begin{array}{l}
Z(t) = c_{1}(t_{0}-t)^{z_{+}}+c_{2}(t_{0}-t)^{z_{-}}\,,\\\\ 
P(t) = c_{0}(t_{0}-t)^{3-h_0}\,,\\\\
Q(t)=\dfrac{24h_0^{3}c_0(3-h_0)}{(t_{0}-t)^{h_0+1}}+\dfrac{6h_0c_{1}(z_{+}-h_0
)}{(t_{0}-t)^{2-z_{+}}}+\dfrac{6h_0c_{2}(z_{-}-h_0)}{(t_{0}-t)^{2-z_{-}}}\,.
\end{array}
\right.
\end{equation}
\phantom{line}\\
Here, $c_{0,1,2}$ are generic constant and $z_{\pm}$ read\\
\phantom{line}
\begin{equation}
z_{\pm}
=\dfrac{1-h_0\pm \sqrt{h_0^{2}-10 h_0+1}}{2}\,.
\end{equation}
Under the condition $0<h_0<5-2\sqrt{6}$ or $h_0>2+\sqrt{6}$, the solution of
$\mathcal{F}(R,G)$ takes the form
\begin{equation}
\mathcal{F}(R,G)=\alpha_{1}R^{1-\frac{z_{+}}{2}}+\alpha_{2}R^{1-\frac{z_{-}}{2}}
+\delta\,G^{\frac{h_0+1}{4}}\,,
\label{trivial}
\end{equation}
where $\alpha_{1,2}$ and $\delta$ are constants.
If $\delta=0$, we find a $F(R)$-model realizing Big Rip according with Ref.~\cite{OdWork}.
$G^{\frac{h_0+1}{4}}$, combined with $R$, is an invariant of the Big Rip solution
in the $f(G)$-models $R+f(G)$ 
and produces the Big Rip in the general class of $\mathcal{F}(R,G)$-gravity.
Note that
$1-(z_{\pm}/2)\neq 1$, according with the fact that Einstein gravity is free of singularities.

\paragraph*{} Another exact solution of Eqs.~(\ref{Pi})-(\ref{Qu}) is given by
\\
\phantom{line}
\begin{equation}
\left\{\begin{array}{l}
Z(t) = \frac{c_1}{(t_{0}-t)^{z+2}}\,,\\\\ 
P(t) = \frac{c_0}{(t_{0}-t)^{z}}\,,\\\\
Q(t)=-\frac{6h_0}{(t_{0}-t)^{z+4}}\left[4h_0^{2}z\delta+c_1(z+2+h_0)\right]\,.
\end{array}
\right.
\end{equation}
\phantom{line}\\
Here, $c_0$ and $z$ are free constants and $c_1$ is given by
\begin{equation}
c_1=\dfrac{4 h_0^{2}z\,c_0(h_0-z-3)}{[z^{2}+(5-h_0)z+6]}\,.
\end{equation}
From Eq.~(\ref{t}) we get\\
\phantom{line}
\begin{eqnarray}
&& \hspace{-15mm}
(t_{0}-t)
\equiv g(R,G)
\nonumber \\ \nonumber\\
&& \hspace{-15mm}
=\left\{\dfrac{-c_1(z+2)R\pm
\sqrt{c_1^{2}(z+2)^{2}R^{2}+24h_0\left[4h_0^{2}z\,c_0+c_1(z+2+h_0)\right]
(z+4)(z\, c_0)G}}{2(z\,c_0)G}\right\}^{1/2}\,,
\label{radice}
\end{eqnarray}
\phantom{line}\\
with $z\neq 0$ and $c_0\neq 0$.

To have real solutions, we must require that the arguments of the
roots in Eq.~(\ref{radice}) are positive.
Since $h_0>0$, the principal cases are as follows:
\begin{enumerate}
\item
Case (1):
$z>0$, $c_0>0$, $1+z \leq h_0 < z+5+\frac{6}{z}$.
We must use the sign $+$ in Eq.~(\ref{radice});
\item
Case (2):
$-\frac{3}{2} \leq z<0$, $c_0<0$,
$h_0 \geq z+1$.
We must use the sign $+$;
\item
Case (3):
$-4<z<-\frac{3}{2}$, $c_0<0$, $h_0>z+5+\frac{6}{z}$.
We must use the sign $+$;
\item
Case (4):
$z>0$, $c_0<0$, $z+5+\frac{6}{z}>h_0 \geq 1+z$.
We must use the sign $-$;
\item
Case (5):
$-\frac{3}{2} \leq z<0$, $c_0>0$, $h_0 \geq z+1$.
We must use the sign $-$;
\item
Case (6):
$-4<z<-\frac{3}{2}$, $c_0>0$, $h_0>z+5+\frac{6}{z}$.
We must use the sign $-$;
\item
Case (7):
$z=-4$, $c_0>0$. We must use the sign $-$;
\item
Case (8):
$z=-4$, $c_0<0$. We must use the sign $+$.
\end{enumerate}
The solution of $\mathcal{F}(R,G)$ reads\\
\phantom{line}
\begin{equation}
\mathcal{F}(R,G)=\frac{c_1}{(g(R,G))^{z+2}}R+\frac{c_0}{(g(R,G))^{z}}G
-\frac{6h_0}{(g(R,G))^{z+4}}\left[4h_0^{2}z\,c_0+c_1(z+2+h_0)\right]\,,
\label{funzione}
\end{equation}
\phantom{line}\\
where
$g(R,G)$ is given by Eq.~(\ref{radice}).
This is an exact solution of the EOMs 
for the Big Rip case.
We show several examples:
\begin{enumerate}
\item In the case $c_1=1$ and $z=-2$, we find\\
\phantom{line}
\begin{equation}
\mathcal{F}(R,G)=R+\dfrac{\sqrt{6}\sqrt{h_0(1+h_0)}}{(1-h_0)}\sqrt{G}\,,
\quad
h_0\neq 1\,,
\end{equation}
\phantom{line}\\
which is in agreement with the result of the previous Section.
\item
If $c_1=0$ and $z=h_0-3$
(this case corresponds to the cases (1)--(6) presented above),
we find\\
\phantom{line}
\begin{equation}
\mathcal{F}(R,G)=\delta\,G^{\frac{h_0+1}{4}}\,,
\quad
\delta \neq 0\,,
\end{equation}
\phantom{line}\\
which is equivalent to
Eq.~(\ref{trivial}) with $\alpha_{1}=\alpha_{2}=0$.
\item
If $z=-4$, the result is given by\\
\phantom{line}
\begin{equation}
\mathcal{F}(R,G)=
\dfrac{16h_0^{4}c_0}{(1+2h_0^{2})^{2}}\left[(9+21h_0+6h_0^{2})-(1+h_0)^{2}
\frac{R^{2}}{G}\right]\,,
\quad
c_0 \neq 0\,.
\label{Dante}
\end{equation}
\phantom{line}\\
Hence,
if $\mathcal{F}(R,G)=\pm\alpha\mp\delta\cdot(R^{2}/G)$ with
$\alpha>0$ and $\delta>0$, the Big Rip singularity could appear for large values of $R$ and $G$.
\item
If $z=h_0-1$, by absorbing some constant into $\tilde c_0$, the solution
becomes\\
\phantom{line} 
\begin{equation}
\mathcal{F}(R,G)=
\tilde c_0 G \left( \dfrac{R}{G} \right)^{\frac{1-h_0}{2}}\,,
\quad
\tilde c_0 \neq 0\,,
\quad
h_0 \neq 1\,.
\label{Sturmtruppen}
\end{equation}
\phantom{line}\\
Thus,
if $\mathcal{F}(R,G)=\delta\,G^{\gamma}/R^{\gamma-1}$
with $\delta \neq 0$ and $1/2<\gamma<1$ or $1<\gamma<+\infty$,
the Big Rip singularity could appear for large values of $R$ and $G$.\\
\end{enumerate}
To conclude this Subsection, we mention the following form of modified gravity:
\phantom{line}\\
\begin{equation}
\mathcal{F}(R,G)=\delta\,\left(\frac{G^{m}}{R^{n}}\right)\,,
\label{Kriegsmarine}
\end{equation}
\phantom{line}\\
with $\delta$ being a generic constant. It is possible to verify that such model
is a solution of the EOMs (\ref{EOM1})--(\ref{EOM2})
in the case of
the Big Rip singularity ($\beta=1$) for some values of $h_0$. In general, we can obtain solutions for $h_0>0$ if $m>0$, $n>0$ and $m>n$. For
example,
the case $n=2$ and $m=3$
realizes the singularity in $h_0=5$;
the case $n=1$ and $m=3$ realizes the singularity in $h_0=4+\sqrt{19}$
and so forth.
This is a generalization of model~(\ref{Sturmtruppen}).
Note that we do not recover a
physical solution for $m=-1$ and $n=-2$ because
in this case $h_0=-3$: for a similar kind of model, where $\mathcal{F}(R,G)$ is a function of $R^2/G$, namely $\mathcal{F}(R,G)=\mathcal{F}(R^2/G)$,
which produces the Big Rip singularity, see Eq.~(\ref{Dante}).
For $m=0$ or $n=0$, we recover Eq.~(\ref{trivial}) again.

\subsubsection{Other types of singularities}

\paragraph*{} Next, we study the other types of singularities.
We consider the case in which $H$ is given by
\begin{equation}
H(t)=\frac{h_0}{(t_{0}-t)^{\beta}}\,,\quad\beta\neq 1\,.
\end{equation}
An exact solution of Eqs.~(\ref{Pi})-(\ref{Qu}) is\\
\phantom{line}
\begin{equation}
\left\{\begin{array}{l}
Z(t) = -c_0(4h_0^{2})(t_{0}-t)\,,\\\\ 
P(t) = c_0(t_{0}-t)^{2\beta+1}\,,\\\\
Q(t)=\dfrac{24h_0^{4}c_0}{(t_{0}-t)^{2\beta-1}}+\dfrac{48h_0^{3}\beta}{(t_{0
}-t)^{\beta}}\,.
\end{array}
\right.
\end{equation}
\phantom{line}\\
As usually, $c_0$ is a generic constant. 
For $\beta=1$, we find a special case of Eq.~(\ref{funzione}). For
$\beta>1$,
we obtain the following asymptotic real solution of Eq.~(\ref{t}),
\begin{equation}
(t_{0}-t)
=g(R,G)=2^{1/2\beta}\left[\dfrac{h_0^{2}R+\sqrt{h_0^{4}R^{2}+6h_0^{4}(4\beta^{2}-1
)G}}{(1+2\beta)G}\right]^{1/2\beta}\,.
\end{equation}
The form of $\mathcal{F}(R,G)$ is expressed as\\
\phantom{line}
\begin{equation}
\mathcal{F}(R,G)=-4h_0^{2}c_0 [g(R,G)]R+c_0
[g(R,G)^{1+2\beta}]G+24h_0^{4}c_0 [g(R,G)^{1-2\beta}]\,,
\quad
\beta>1\,.
\end{equation}
\phantom{line}\\
In the case $\beta \gg 1$, the form of $\mathcal{F}(R,G)$ can be written as\\
\phantom{line}
\begin{equation}
\mathcal{F}(R,G)
\simeq
R-\dfrac{\alpha G}{R+\sqrt{R^{2}+\gamma G}}\,,
\quad
\alpha>0\,,
\quad
\gamma>0\,.
\end{equation}
\phantom{line}\\
This is the behavior of a $\mathcal{F}(R,G)$-model in which
a ``strong'' Type I singularity ($\beta \gg 1$) could appear for $R,G\rightarrow+\infty$ and asymptotically solves Eq.~(\ref{prime}).\\

To find other models, we can consider the results of \S~\ref{3.3}.
The Type I singularities ($\beta>1$) correspond to the asymptotic limits for $R$ and $G$,
\\
\phantom{line}
\begin{equation}
\beta>1\,,\quad
\left\{\begin{array}{l}
R(t)\simeq\frac{12h_0^2}{(t_0-t)^{2\beta}}\,,\\\\
G(t)\simeq\frac{24h_0^4}{(t_0-t)^{4\beta}}\,.
\end{array}
\right.
\end{equation}
\phantom{line}\\
These are two functions of the Hubble parameter only, so that
\begin{equation}
\lim_{t\rightarrow t_{0}} 24\left( \frac{R}{12} \right)^{2}
= \lim_{t\rightarrow t_{0}} G\,.
\label{Limite}
\end{equation}
If we substitute $G$ for $R$ in the model~(\ref{primo}) by taking into account
Eq.~(\ref{Limite}), we obtain a
zero function (this is because model~(\ref{primo}) is zero on the singularity
solution).
Howevere, if we substitute $G$ for $G/R$, we obtain the following model,\\
\phantom{line}
\begin{equation}
\mathcal{F}(R,G)= R-\dfrac{6G}{R}\,.
\label{zap}
\end{equation}
\phantom{line}\\
This is an asymptotic solution of Eq.~(\ref{prime}).
Thus, there appears Type I singularity with $R,G\rightarrow+\infty$ in the class of models
$\mathcal{F}(R,G)= R-\alpha(G/R)$, with $\alpha>0$.\\

In the case of $H(t)=h_0/(t_{0}-t)^{\beta}$ with $\beta<1$, it is not possible to
write $G$ and $R$ like functions of the same variable ($H$ or the same
combination of $H$ and $\dot H$). Nevertheless, if we
examine the asymptotic behavior of $G$ and $R$, we have
\\
\phantom{line}
\begin{equation}
\beta<1\,,\quad
\left\{\begin{array}{l}
R(t)\simeq\frac{6h_0\beta}{(t_{0}-t)^{\beta+1}}\,,\\\\
G(t)\simeq\frac{24 h_0^{3}\beta}{(t_{0}-t)^{3\beta+1}}\,.
\end{array}
\right.
\end{equation}
\phantom{line}\\
As a consequence, one finds
\begin{equation}
\frac{G}{R}\sim G^{\frac{2\beta}{3\beta+1}}\,.\label{Todt}
\end{equation}
If we use $G/R$ for $G$ in the model~(\ref{secondo}) as in Eq.~(\ref{Todt}), we
see that Eq.~(\ref{prime}) is asymptotically verified for $\beta<1$.
Under this consideration, by setting some parameters, it is possible to derive
a general $\mathcal{F}(R,G)$-gravity theory from
Eq.~(\ref{secondo}) as
\phantom{line}\\
\begin{equation}
\mathcal{F}(R,G)=R+\frac{3}{2}\frac{G}{R}\,,\label{zapzap}
\end{equation}
\phantom{line}\\
in which the other types of singularities appear.
Thus, in the model $\mathcal{F}(R,G)= R+\alpha(G/R)$ with $\alpha>0$,
the Type II, III and IV singularities could appear.
Then, by substituting $G$ for $R$ we get\\
\phantom{line}
\begin{equation}
\mathcal{F}(R,G)
\simeq
R-\delta\frac{(1+\beta)}{(\beta-1)}
|R|^{\frac{2\beta}{1+\beta}}\,,
\quad
\delta>0\,.\label{singularf(R)}
\end{equation}
\phantom{line}\\
This result\footnote{Note that in the Big Rip case we
have found exact solutions. This kind of reasoning is therefore
inapplicable.} is according with Ref.~\cite{OdWork}. In the model
$F(R)=R+\alpha R^{\gamma}$,
with $0<\gamma<1$ and $\alpha>0$,
a Type III singularity could appear for $R\rightarrow+\infty$.
In the model $F(R)= R+\alpha |R|^{\gamma}$,
with $-\infty<\gamma<0$ and $\alpha>0$,
a Type II singularity could appear for $R\rightarrow -\infty$.
In the model $F(R)=R+\alpha |R|^{\gamma}$,
with $2<\gamma<+\infty$ ($\gamma\neq 2n/(n-1)$, where $n$ is a natural
number)
and $\alpha<0$, a Type IV singularity could appear for $R\rightarrow 0^{-}$.

In the next Subsection we will analyze an example of realistic $F(R)$-gravity
generating singularity.

\subsection{Example of realistic singular-$F(R)$-model: the Hu-Sawicki Model\label{3.4.1}}

\paragraph*{} Let us return to Hu-Sawicki Model of Eq.~(\ref{HuSawModel}). 
The Hu-Sawiki Model could become singular when $R$ diverges. In particular, it shows a Type II singularity when $H$ behaves as:    
\begin{equation}
H(t)=\frac{h_0}{(t_{0}-t)^{\beta}}+H_{0}\,,\quad-1<\beta<0\,,
\end{equation}
where we have reintroduced the positive constant $H_0$. As usually, the constant $h_0$ has to be positive.
In the asymptotic limit, Eq.~(\ref{prime}) with Eq.~(\ref{Feldwebel}) are verified by putting:
\begin{equation}
\beta=-\frac{n}{n+2}\,,\quad H_{0}=\sqrt{\frac{c_{1}\tilde{m}^{2}}{6c_{2}}}\label{accazero}\,, \quad
h_0= \left[\frac{6n^{2}(n+1)}{(n+2)^{2}}\left (\frac{2+n}{-6n}\right)^{n+2}\left(\frac{c_{1}}{c_{2}^{2}}\phantom{s}\left(\tilde{m}^2\right)^{(n+1)}\right)\right]^{n+2}\,.
\end{equation}
Here, $h_0$ is positive if $n$ is an even number and the model may show the Type II singularity in expanding universe (if $n$ is an odd number, this kind of singularity could appear for contracting universe, as the Big Crunch). Note that $H_{0}$ is the constant Hubble parameter $H_{\mathrm{dS}}$ of the de Sitter universe, $H_0=H_{\mathrm{dS}}$. We have just discussed in \S~\ref{3.1} the problems generated by the possibility to have singular solutions in the cosmological scenario described by Hu-Sawiki Model. Let us have a look for the strategy to use in order to cure singularities occurrence.  

\section{Curing the finite-time future singularities\label{curing}}

\paragraph*{} In this last Section, we discuss a possible way to cure the finite-time future
singularities in modified gravity whose lagrangian is in the form $\mathcal{F}(R,G)=R+f(R,G)$.
We will see some simple curing terms, that is, some power functions of $R$ or $G$, to add into the theory in order to prevent the singularities. 
In the last Subsection, the quantum effects in the range of high curvatures are also discussed.

\subsection{Power terms of $R$ and $G$\label{curingf(R)}}

\paragraph*{} First, we consider $f(G)$-modified gravity.
If any singularity occurs, $\mathcal{G}(H, \dot{H}...)$ evaluated on the singular form of $H$ of Eq.~(\ref{Hsingular}) with $H_0=0$, behaves as in Eq.~(\ref{casistiche}).

The singularities appear in two cases: (a) $G\rightarrow \pm\infty$ (Big Rip, Type I and Type III singularities and Type II singularities with $-1/3<\beta<0$ ); (b) $G\rightarrow 0^{-}$ (Type IV singularities and Type II singularities with $-1<\beta<-1/3$)\footnote{Note that, if $H$ tends to a non avoidable constant $H_0$, the Gauss-Bonnet diverges for any value of $-1<\beta<0$, i.e. for any kind of Type II singularity, as the Ricci scalar $R$.}.

\begin{enumerate}
\item Case of $G\rightarrow\pm\infty$.\\ 
Let us consider the $f^*(G)$ curing term
\begin{equation}
f^*(G)=\gamma G^{m}\,,
\quad
m\neq 1\,,
\label{eins}
\end{equation}
with $\gamma \neq 0$ and $m$ being a constant.
One way to prevent a singularity appearing could be
that the function $\mathcal{G}(H, \dot H...)$ becomes inconsistent with the
behavior of
Eq.~(\ref{casistiche}). In general, $\mathcal{G}(H, \dot H...)$ must tend to
infinity faster than Eq.~(\ref{casistiche}). For
$H=h_0/(t_{0}-t)$, this is the
Big Rip, the (additive) contribute of $f^*(G)$ to $\mathcal{G}(H, \dot H...)$ is given by
\begin{equation}
\mathcal{G}^*(H, \dot H...)\sim\frac{\alpha}{(t_{0}-t)^{4m}}\,.
\end{equation}
Here, $\alpha$ is a generic constant. Hence, if $m > 1/2$, we avoid the singularity. Nevertheless, there is one
specific case in which the Big Rip singularity could still occur. If $m=(1+h_0)/4$,
$\mathcal{G}^*(H, \dot H...)$ is exactly equal to zero, so that (for example)
the following specific model, with $m>1/2$, admits the Big Rip singularity,
\begin{equation}
R+f(G)=R+\frac{\sqrt{24 m (4m-1)^{3}}}{2h_0(1-2m)}G^{1/2}+\gamma G^{m}\,.
\label{esempio}
\end{equation}
This is because the power function $G^{m}$ is an invariant
with respect to the Big Rip singularity generated by $G^{1/2}$-term.
If we have the model $R+\alpha\sqrt{G}$, we can eliminate the Big Rip
singularity with a power function $\gamma G^{m}$ ($m>1/2$) only if
$\alpha>0$ (such that the configuration of Eq.~(\ref{esempio}) cannot be realized).

For $H=h_0/(t_{0}-t)^{\beta}$ with $\beta>1$, this is the Type I singularity, the curing term in Eq.~(\ref{eins}) leads to
\begin{equation}
\mathcal{G}^*(H, \dot H...)\sim\dfrac{\alpha}{(t_{0}-t)^{4\beta m}}\,.
\end{equation}
Also in this case, if $m>1/2$, we avoid the singularity.\\
For example, the model
$R+\alpha\sqrt{G}+\gamma G^{2}$ with $\alpha>0$ is free of
Type I singularities,
while if $\alpha<0$
the Big Rip singularity could still appear.

For $H=h_0/(t_{0}-t)^{\beta}$ with $0<\beta<1$, this is the Type III singularity,
the curing term in Eq.~(\ref{eins}) leads to
\begin{equation}
\mathcal{G}^*(H, \dot H...)\sim
\frac{\alpha}{(t_{0}-t)^{m(3\beta+1)+(1-\beta)}}\,.\label{again}
\end{equation}
If $m> 2\beta/(3\beta+1)$, i.e. $m> 1/2$, we avoid the singularity.

Also for $H=h_0/(t_{0}-t)^{\beta}$ with $-1/3<\beta<0$, this is the case of
Type II singularity when $G\rightarrow-\infty$, we have to require the same condition.
For example, the model
$R+\alpha |G|^{m}+\gamma G^{2}$ with $m<1/2$, is free of
Type I, II (with $-1/3<\beta<0$) and III singularities.

\item Case of $G\rightarrow0^-$.\\
For $H=h_0/(t_{0}-t)^{\beta}$ with $\beta<-1/3$
(Type II and IV singularities), the curing term in Eq.~(\ref{eins}) leads to Eq.~(\ref{again}) again, which
diverges and hence becomes inconsistent with Eq.~(\ref{casistiche})
if $m\leq 2/3$.
For example, the model
$R+\alpha |G|^{\zeta}+\gamma G^{-1}$ with $\zeta>2/3$ is free of Types IV
singularities.
\end{enumerate}

As a result,
the term $\gamma G^{m}$ with $m> 1/2$ and $m\neq 1$ cures the singularities
occurring when $G\rightarrow \pm\infty$.
Moreover, the term $\gamma G^{m}$ with $m \leq 2/3$ cures the singularities occurring when $G\rightarrow 0^{-}$.

In $f(R)$-gravity ($F(R)=R+f(R)$), 
by using the term $\gamma R^{m}$, the same consequences are found. 
The term $\gamma R^{m}$ with $m>1$ cures the Type I, II and III 
singularities occurring when $R\rightarrow\pm\infty$. 
On the other hand, the term $\gamma R^{m}$ with $m\leq2$ cures the Type IV
singularity occurring when $R\rightarrow0^-$.

Note that $\gamma G^{m}$ or $\gamma R^{m}$ are invariants
with respect to the Big Rip solution (see Eq.~(\ref{trivial})), so that in this cases it is
necessary to pay attention to the whole form of the
theory.

A general important result is the following: the terms like $\gamma R^n$ or $\gamma G^m$ with $1<n\leq 2$ or $1/2<m\leq2/3$ respectively, avoid any types of singularities in $f(R,G)$-gravity or in the presence of dark energy fluid producing singularities (with regard to $R^2$ curing term see Refs.~\cite{OdWork,Abdalla:2004sw,Nojiri:2008fk}).

\subsection{Combinations of $R$ and $G$}

\paragraph*{} Within the framework of $f(R,G)$-gravity, that is when the Ricci scalar and the Gauss-Bonnet invariant are combined in the modification to gravity, we can use terms like
\begin{equation}
f^*(R,G)=\gamma\frac{G^{m}}{R^{n}}\,,
\label{R+F}
\end{equation}
with $\gamma \neq 0$ and $m$, $n$ constants, to cure the singularities.

The singularities appear in the following three cases:
(a)
$R\rightarrow \pm\infty$, $G\rightarrow \pm\infty$ (Type I and Type III singularities and Type II singularities for $-1/3<\beta<0$);
(b)
$R\rightarrow -\infty$,
$G\rightarrow 0^{-}$
(Type II singularities for $-1<\beta<-1/3$),
and
(c)
$R\rightarrow 0^{-}$, $G\rightarrow 0^{-}$ (Type IV singularities).

We investigate general possibilities.\\

In the case of the Big Rip singularity,
the contribute of $f^*(R,G)$ to $\mathcal{G}(H,\dot{H}...)$ in Eq.~(\ref{Feldwebel})
diverges as
\begin{equation}
\mathcal{G}^*(H,\dot{H}...)\sim
\frac{1}{(t_{0}-t)^{4m-2n}}\,.
\end{equation}
Thus, if $m > (n+1)/2$, we avoid the singularity. Nevertheless,
there is the possibility that
$\mathcal{G}^*(H, \dot{H}...)$ is exactly equal to zero and the
curing term does not protect the theory against the Big Rip (see Eq.~(\ref{Sturmtruppen}) and Eq.~(\ref{Kriegsmarine})
in the case of $m=n+1$, where such combination of $R$ and $G$ alone produces the Big Rip, and therefore is trivial in $R+f(R,G)$ models).
Hence, the whole form of $\mathcal{F}(R,G)$
as well as its form in the asymptotic limit must be examined.\\

In the case of
Type I singularities, $\mathcal{G}^*(H,\dot{H}...)$ diverges as
\begin{equation}
\mathcal{G}^*(H,\dot{H}...)\sim
\frac{1}{(t_{0}-t)^{4\beta m-2\beta n}}\,.
\end{equation}
Also in this case, if $m>(n+1)/2$, we avoid the singularity.\\

When $\beta<1$, $\mathcal{G}^*(H, \dot{H}...)$ behaves as\\
\phantom{line}
\begin{equation}
\mathcal{G}^*(H,\dot{H}...)\sim\frac{\alpha}{(t_{0}-t)^{(3\beta+1)m-(\beta+1)n+(1-\beta)}}\,.
\end{equation}
\phantom{line}\\
As a result, $\mathcal{G}^*(H, \dot{H}...)$ diverges faster than $(t_{0}-t)^{-\beta-1}$ and therefore
the Type III singularity ($0<\beta<1$) is avoided if $m\,,n>0$ such that $m>(1+n)/2$ (for example, one can choose $n=1$ and $m=2$).
The Type II singularity for $-1/3<\beta<0$ is avoided if $m>0$ and $n<0$. 
The Type II singularity for $-1<\beta<-1/3$ is avoided if $m<0$ and $n<0$. 
Finally, the Type IV singularity ($\beta<-1$) is avoided if $n>0$ and $m<0$.

\subsection{Quantum effects} 

\paragraph*{} In the high curvature limit, quantum effects could become relevant and they have to be taken into account.
Consider next the quantum contribution to the conformal anomaly. The complete energy density $\rho_{\mathrm{tot}}$ and pressure $p_{\mathrm{tot}}$ of matter are:
\begin{equation}
\rho_{\mathrm{tot}}=\rho_{\mathrm{m}}+\rho_{\mathrm{A}}\,,
\quad
p_{\mathrm{tot}}=p_{\mathrm{m}}+p_{\mathrm{A}}\,.
\end{equation}
Here, $\rho_{\mathrm{m}}$ and $p_{\mathrm{m}}$ are, as usually, the standard contributes of matter and $\rho_{\mathrm{A}}$ and $p_{\mathrm{A}}$ are given by quantum effects. Taking the trace $\mathrm{T}_A$ of the conformal anomaly energy-momentum tensor,
\begin{equation}
\mathrm{T}_{A}=-\rho_{\mathrm{A}}+3p_{\mathrm{A}}\,,
\end{equation}
plus observing the energy conservation law,
\begin{equation}
\dot{\rho_{\mathrm{A}}}+3H(\rho_{\mathrm{A}}+p_{\mathrm{A}})=0\,,
\end{equation}
we find that
\begin{equation}
p_{\mathrm{A}}=-\rho_{\mathrm{A}}-\frac{\dot{\rho_{\mathrm{A}}}}{3H}\,.
\end{equation}
Thus we obtain for the conformal anomaly energy density~\cite{conformalanomaly1,conformalanomaly2}:\\
\phantom{line}
\begin{eqnarray}
\rho_{\mathrm{A}}&=&-\frac{1}{a(t)^{4}}\int a(t)^{4}H \left(\mathrm{T}_{A}\right)dt\nonumber\\ \nonumber \\
&=& -\frac{1}{a(t)^{4}}\int a(t)^{4}H\Bigl\{-12b\dot{H}^{2}+24b_1(-\dot{H}^{2}+H^{2}\dot{H}+H^{4})-\nonumber\\ \nonumber \\ 
& &\hspace{10mm}(4b_0+6b_2)(\dddot{H}+7H\ddot{H}+4\dot{H}^{2}+12H^{2}\dot{H})\Bigr\}d t\,.\label{equazioneanomaly}
\end{eqnarray}
\phantom{line}\\
Here, $b_{0,1,2}$ are constants, occuring in the expression for the conformal trace anomaly
\begin{equation}
\mathrm{T}_{\mathrm{A}}=b_0(C^2+\frac{2}{3}\Box R)+b_1G+b_2R\,,
\end{equation}
where $C^2$ is the square of the Weyl Tensor and $G$ the Gauss-Bonnet invariant. Explicitly, if there are $N$ scalars, $N_{1/2}$ spinors, $N_{2}$ gravitons and $N_{HD}$ higher derivative conformal scalars, one has for $b_0$ and $b_1$ the following expressions:\\
\phantom{line}
\begin{equation}
b_0=\frac{N+6N_{1/2}+12N_{1}+611N_{2}-8N_{HD}}{120(4\pi)^{2}}\,,\nonumber
\end{equation}
\phantom{line}
\begin{equation}
b_1=\frac{N+11N_{1/2}+62N_{1}+1411N_{2}-28N_{HD}}{360(4\pi)^{2}}\,,
\end{equation}
\phantom{line}\\
whereas $b_2$ is an arbitrary constant whose value depends on the regularization.

The quantum corrected EOM (\ref{EOM1bis}) results to be
\begin{equation}
\rho_{\mathrm{eff}}+\rho_{\mathrm{A}}=\frac{3}{\kappa^2}H^{2}\label{EOM1corrected}\,.
\end{equation}
Quantum effects become relevant for large values of curvature $R$ and when the effective energy density of the universe is not too much large. In particular, this is the case of Type II singularities, when $H=h_0/(t_0-t)^{\beta}$ with $-1<\beta<0$. 
Eq.~(\ref{equazioneanomaly}) gives
\begin{equation}
\rho_{\mathrm{A}}\simeq \frac{\alpha}{(t_{0}-t)^{\beta+2}}\,.
\end{equation}
Here, $\alpha$ is a number. In some scenario, quantum effects have to be taken into account. In this case, $\rho_{\mathrm{A}}$ diverges in Eq.(\ref{EOM1corrected}) faster than $H^2$, so that the Type II singularity is not realized.


\chapter{Viscous fluids and singularities}

\paragraph*{} Here, we study some features of inhomogeneous viscous fluids, especially related with future-time singularities. Fluids in general have been considered as candidate to dark energy into the context of GR, since the evolution of cosmological parameters is not defined with precise accuracy, except for the current values with 3-5\% error at least, and the observations do not exclude the possibility to have dark energy with a dynamical EoS parameter. 
Furthermore, modified gravity has an equivalent description as effective (viscous) fluid.
In this Chapter, as a prosecution of the previous one, we analyze the behaviour of dark energy fluids in singular theories of modified gravity, investigating
how the singularities may change or disappear, due to the contribution of these fluids.
After that, a Section is devoted to the study of (viscous) dark energy (DE) fluids coupled with dark matter (DM). 

\section{Viscous fluids and modified gravity}

\paragraph*{} The most general form of inhomogeneous viscous fluid in FRW background is given by the Equation of State~\cite{fluidsOd1,fluidsOd2}
\begin{equation}
p_{\mathrm{F}}=\omega(\rho_{\mathrm{F}})\rho_{\mathrm{F}}+B(\rho_{\mathrm{F}},a(t),H, \dot{H}...)\,,\label{start}
\end{equation}
where $p_{\mathrm{F}}$ and $\rho_{\mathrm{F}}$ are the pressure and energy density of fluid, respectively, and the thermodynamical variable $\omega(\rho_{\mathrm{F}})$ is an arbitrary function of the density $\rho_{\mathrm{F}}$. The bulk viscosity $B(\rho_{\mathrm{F}},a(t),H, \dot{H}...)$ is a function of the density $\rho_{\mathrm{F}}$, the scale factor $a(t)$, and the Hubble parameter $H$ and its derivatives. 
The motivation in considering this general form of time-dependent bulk viscosity comes from the modification of gravity, which can be treated as a fluid in this form. For example, in the framework of $\mathcal{F}(R,G)$-gravity, if we define the effective energy density and pressure 
as in Eqs.~(\ref{rhoeffRG0})--(\ref{peffRG0}), we may take $B(\rho_{\mathrm{F}},a(t),H, \dot{H}...)=0$ and $\omega(\rho_{\mathrm{F}})=\omega_{\mathrm{eff}}$ given by Eq.~(\ref{omegaeffdef}). In this way, the EoS parameter depends on $H$ and its derivatives. Otherwise, in \S~\ref{3.2}, we have used an other fluid representation for modified gravity with constant $\omega$.
In this case, we obtain Eq.~(\ref{start}) by identifying $B(\rho_{\mathrm{F}},a(t),H, \dot{H}...)$ with $\mathcal{G}(H,\dot{H}...)$ of Eq.~(\ref{Feldwebel}).

Generally speaking, we refer to the quintessence when $-1<\omega(\rho_{\mathrm{F}})<-1/3$ and  phantom fluid when $\omega(\rho_{\mathrm{F}})<-1$.

\subsection{Example of realistic fluid model generating the Big Rip\label{4.1.1}}

\paragraph*{} In principle DE-fluids -as modified gravity- may bring the future universe evolution to become singular. Let us analyze in some detail an interesting inhomogeneous non viscous fluid introduced in Ref.~\cite{BigRipfluid}, whose EoS is
\begin{equation}
p_{\mathrm{F}}=-\rho_{\mathrm{F}}+f(\rho_{\mathrm{F}})\,, 
\end{equation}
where
\begin{eqnarray}
\left\{\begin{array}{lll}
f(\rho_{\mathrm{F}})&=&+\frac{2\rho_{\mathrm{F}}}{3n}\left(1-\frac{4n}{\delta}\left(\frac{3\tilde{m}^2}{\kappa^2\rho_{\mathrm{F}}}\right)^{\frac{1}{2}}\right)^{\frac{1}{2}}\,,\quad t \leq t_0\,, \\
f(\rho_{\mathrm{F}})&=&- \frac{2\rho_{\mathrm{F}}}{3n}\left(1-\frac{4n}{\delta}\left(\frac{3\tilde{m}^2}{\kappa^2\rho_{\mathrm{F}}}\right)^{\frac{1}{2}}\right)^{\frac{1}{2}}\,,\quad t >t_0\,.
\end{array}\right.
\end{eqnarray} 
Here, $n\geq 1$ and $\delta$ are constant positive parameters, $\tilde{m}^2$ is a mass scale and $t_0$ is the fixed time for which $f(\rho_{\mathrm{F}})$ assumes the smallest value and it is equal to zero.  
The EoS parameter $\omega(\rho_{\mathrm{F}})=p_{\mathrm{F}}/\rho_{\mathrm{F}}$ reads
\begin{equation}
\omega(\rho_{\mathrm{F}})=-1+\sigma(t)\frac{2}{3n}\left(1-\frac{4n}{\delta}\left(\frac{3\tilde{m}^2}{\kappa^2\rho_{\mathrm{F}}}\right)^{\frac{1}{2}}\right)^{\frac{1}{2}}\,, 
\end{equation}
where $\sigma(t)=1$ when $t\leq t_0$ and $\sigma(t)=-1$ when $t>t_0$. We note that $t=t_0$, such that $f(\rho_{\mathrm{F}})=0$, corresponds to the transition point between quintessence ($-1<\omega(\rho_{\mathrm{F}})<-1/3$) and phantom ($\omega_{\mathrm{F}}<-1$) region, so that $\omega(\rho_{\mathrm{F}})=-1$.  
In particular, when $t<t_0$, $-1<\omega(\rho_{\mathrm{F}})<-1+2/(3n)\leq-1/3$, and when  $t>t_0$, $-5/3\leq-1-2/(3n)<\omega(\rho_{\mathrm{F}})<-1$. 

This model may be used to correctly reproduce the matter era and the present accelerated epoch at the time $t=t_0$. 
The fluid energy conservation law reads
\begin{equation}
\dot{\rho}_{\mathrm{F}}+3Hf(\rho_{\mathrm{F}})=0\,, 
\end{equation}
which leads to\\
\phantom{line}
\begin{equation}
\rho_{\mathrm{F}}=\frac{3\tilde{m}^2\left(\frac{a(t)}{n}\right)^{\frac{2}{n}}\left(4n+c_0^{-\left(\frac{1}{2}\right)}\left(\frac{a(t)}{n}\right)^{-\frac{1}{n}}\right)^4c_0}{16\delta^2\kappa^2}\,.\label{generalFBigRip}
\end{equation}
\phantom{line}\\
Here, $a(t)$ is the scale factor of the universe and $c_0>0$ is an integration constant. Furthermore, we put $a(t_0)=1$ and indicate the fluid energy density at the present time $t_0$ with $\rho_{\mathrm{F}(0)}$. 

If the mass scale $\tilde{m}^2$ corresponds to the energy density of matter at the present time $\rho_{\mathrm{m}(0)}$, i.e. $\rho_{\mathrm{m}(0)}=3\tilde{m}^2/\kappa^2$, by imposing $\rho_{F(0)}/\rho_{\mathrm{m}(0)}=\Lambda/(3\tilde{m}^2)$, such that $\Lambda/\kappa^2$ is the observed dark energy density in our universe, $\Lambda$ being the Cosmological Constant, and $\dot{\rho}_{\mathrm{F}(0)}=0$, which is the condition to have $\omega(\rho_{\mathrm{F}(0)})=-1$, one derives:\\
\phantom{line}    
\begin{equation}
\left\{\begin{array}{l}
c_0=\frac{1}{16}\left(n^{1-\frac{1}{n}}\right)^{-2}\,, \\ \\ 
\frac{16n^2}{\delta^2}=\frac{\Lambda}{3\tilde{m}^2}\,. \label{Ccondition}
\end{array}\right.
\end{equation}
\phantom{line}\\
It is easy to see that, for $t\ll t_0$, since matter evolves as $\rho_m\sim a(t)^{-3}$, its energy density grows up in the past faster than the one of the fluid and we have the matter era, but since for $t=t_0$, $\rho_{\mathrm{F(0)}}>\rho_{\mathrm{m(0)}}$, there is a point in the past when the energy density of fluid overtakes the energy density of matter and an accelerated epoch driven in a first step by quintessence fluid (for $t<t_0$) and therefore by phantom fluid (for $t>t_0$) takes place. The solution of equation of motion $\rho_{\mathrm{F}}=3H^2/\kappa^2$ reads
\begin{equation}
H(t)=\frac{n\left(\frac{\delta}{\sqrt{\tilde{m}^2}}\right)}{(t_s-t)\left(t-t_s+\frac{\delta}{\sqrt{\tilde{m}^2}}\right)}\,,\quad t<t_s\,,\label{HBigRipfluid}
\end{equation}
where $t_s>0$ is a fixed time parameter. Here, we have used the first condition in (\ref{Ccondition}). We observe that the Hubble parameter diverges at finite-future time when $t\rightarrow t_s$, and the Big Rip singularity appears. Therefore, $t_s$ corresponds to the life time of the universe. Thus, in order to get an expanding universe ($H(t)>0$), $\delta/\sqrt{\tilde{m}^2}$ has to be larger than $t_s$. The de Sitter solution at the present corresponds to $t_0=t_s-(\delta/2\sqrt{\tilde{m}^2})$.

In conclusion, we have seen that fluid exits from de Sitter-phase involving in a phantom region. The de Sitter solution is not a final attractor of the system, which becomes singular. If we want to remove such singularity, we can use some power functions of $R$ or some power functions of $G$ in the framework of $f(R,G)$-modified gravity via scenario suggested in Chapter 3. Now, we will see how (vice-versa) inhomogeneous fluids can cure singularities in $f(R,G)$-gravity.

\section{Viscous fluids in singular universe\label{4.2}}

\paragraph*{} In this Section, we analyze the results obtained in Ref.~\cite{Viscousfluidsingularities} and we take a simple theory of modified gravity where $\mathcal{F}(R,G)=R+f(R,G)$ as in Eq.~(\ref{actiontwo}). Moreover, we consider the presence of a viscous fluid, whose Equation of State is a simple formulation of Eq.~(\ref{start}) and it is given by 
\begin{equation}
p_{\mathrm{F}}=\omega(\rho_{\mathrm{F}})\rho_{\mathrm{F}}-3 H\zeta(H)\,,\label{eq.state}
\end{equation}
where $\zeta(H)$ is the bulk viscosity and it depends on the Hubble parameter $H$ only. On thermodynamical grounds, in order to have the positive sign of the entropy change in an irreversible process, $\zeta(H)$ has to be a positive quantity, so we assume $\zeta(H)>0$~\cite{Alessia, Alessia(2)}. For the stress-energy tensor of fluid $T_{\mu\nu}^{\mathrm{(fluid)}}$, one has :
\begin{equation}
T_{\mu\nu}^{\mathrm{(fluid)}}=\rho_{\mathrm{F}} u_{\mu}u_{\nu}+\left(\omega(\rho_{\mathrm{F}})\rho-3H\zeta(H)\right)(g_{\mu\nu}+u_{\mu}u_{\nu})\,, 
\end{equation}
where $u_{\mu}=(1,0,0,0)$ is the four velocity vector. 
Now in the effective parameters of Eqs.~(\ref{rhoeffRG})--(\ref{peffRG}), by avoiding the ordinary matter and radiation, we want to take into account the presence of viscous fluid, such that
\begin{equation}
\rho_{\mathrm{eff}}=\rho_{\mathrm{MG}}+\rho_{\mathrm{F}}\,,
\quad
p_{\mathrm{eff}}=p_{\mathrm{MG}}+p_{\mathrm{F}}\,,
\end{equation} 
where the suffix 
`MG' indicates the `modified gravity' contributes given by
\begin{eqnarray}
\rho_{\mathrm{MG}} &=& \frac{1}{2\kappa^{2}}\biggl[(Rf'_{R}
+Gf'_{G}-f)-6H\dot{f}'_{R}
-24H^{3}\dot{f}'_{G}
-6H^{2}f'_R
\biggr]\,,
\nonumber
\\ \nonumber\\
p_{\mathrm{MG}} &=& \frac{1}{2\kappa^{2}}\biggl[
(f-R f'_{R}-G f'_{G})+4H\dot{f}'_{R}+2\ddot{f}'_{R}\nonumber \\
& &+16H(\dot H+H^{2})\dot{f}'_{G}+8H^{2}\ddot{f}'_{G}
+(4\dot{H}+6H^{2})f'_R                      
\biggr]\,.
\label{tactac}
\end{eqnarray}
The equations of motion finally read 
\begin{equation}
\rho_{\mathrm{MG}}+\rho_{\mathrm{F}}=\frac{3}{\kappa^2}H^{2}\,,\quad
p_{\mathrm{MG}}+p_{\mathrm{F}}=-\frac{1}{\kappa^2}\left(2\dot{H}+3H^2\right)\label{EOM1fluid}\,.
\end{equation}
As a consequence, we obtain the fluid energy conservation law, 
\begin{equation}
\dot{\rho_{\mathrm{F}}}+3H\rho_{\mathrm{F}}(1+\omega(\rho_{\mathrm{F}}))=9H^{2}\zeta(H) \label{conservationlawfluid}\,.
\end{equation}
In what follows, we will concentrate again on the singular form of Hubble parameter as in Eq.~(\ref{Hsingular}), namely $H(t)=h_0/(t_0-t)^{\beta}+H_0$, such that the scale factor behaves as\\
\phantom{line}
\begin{eqnarray}
\left\{\begin{array}{lll}
a(t)&=& \frac{a_0}{(t_0-t)^h}\,,\quad\quad\quad\quad\quad\quad\quad\quad\quad\beta=1\quad\text{(Big Rip)}\,;\\ \\
a(t)&=& a_0\exp\left[\frac{h_0(t_0-t)^{1-\beta}}{\beta-1}\right]\,,\quad\beta(\neq 1)>0\quad\text{(Type I, III singularities)}\,;\\ \\
a(t)&=& a_0\exp\left[\frac{h_0(t_0-t)^{1-\beta}}{\beta-1}+H_0\right]\,,\quad\beta<0\quad\text{(Type II, IV singularities)}\label{asingularforms}\,.
\end{array}\right.
\end{eqnarray} 
As usually, $a_0$ and $h_0$ are positive constants and $t_0$ is the finite time for which singularity appears. Here, the positive constant $H_0\neq 0$ is considered in the significant cases of Type II and IV singularities only.

By using Eq.~(\ref{conservationlawfluid}), we will check the solution of the fluid energy density when $H$ is singular. We will see how changes the total effective energy density (and, as a consequence, the total effective pressure) of the universe due to the fluid contribute in the case of singular theories of $f(R,G)$-modified gravity, and if the singularities are still realized. In particular, we are interested in the quintessence ($-1<\omega(\rho_{\mathrm{F}})<-1/3$) and phantom ($\omega(\rho_{\mathrm{F}})<-1$) regions.

We investigate the cases of $\omega(\rho_{\mathrm{F}})$ constant and of $\omega(\rho_{\mathrm{F}})$ dependent on energy density.

\subsection{$\omega(\rho_{\mathrm{F}})$ constant}

\paragraph*{}Let us start considering the simple case when $\omega(\rho_{\mathrm{F}})$ is a constant, such that $\omega(\rho_{\mathrm{F}})=\omega_{\mathrm{F}}$, where $\omega_{\mathrm{F}}$ is the constant EoS parameter of fluid. We take different choices of bulk viscosity $\zeta(H)$.

\subsubsection{Non-viscous case}

\paragraph*{} In the non-viscous case $\zeta(H)=0$ (perfect fluid), the solution of Eq.~(\ref{conservationlawfluid}) assumes the classical form
\begin{equation}
\rho_{\mathrm{F}}=\rho_{0}a(t)^{-3(1+\omega_{\mathrm{F}})}\,,
\end{equation}
where $\rho_{0}$ is a positive constant and $a(t)$ the scale factor. As a consequence, when singularities occur,
from~(\ref{asingularforms}) we see that $\rho_{\mathrm{F}}$ behaves as\\
\phantom{line} 
\begin{eqnarray}
1)\,\,\rho_{\mathrm{F}}&=& \rho_{0}(t_{0}-t)^{3h_0(1+\omega_{\mathrm{F}})}\,,\quad\quad\quad\quad\quad\quad\beta=1\,;\label{tric}\\ \nonumber\\
2)\,\,\rho_{\mathrm{F}}&=&\rho_{0}\mathrm{e}^{\frac{3h_0(1+\omega_{\mathrm{F}})(t_{0}-t)^{1-\beta}}{1-\beta}}\,,\,\,\quad\quad\quad\quad\quad\beta(\neq 1)>0\,;\label{trac}\\ \nonumber\\
3)\,\,\rho_{\mathrm{F}}&=&\rho_{0}\mathrm{e}^{3(1+\omega_{\mathrm{F}})(t_{0}-t)\left(H_{0}-\frac{h_0(t_{0}-t)^{-\beta}}{\beta-1}\right)}\,,\,\,\,\beta<0\,.\label{truc}
\end{eqnarray}
\phantom{line}\\
For $\beta=1$ (Big Rip) and $\beta>1$ (Type I singularity), $\rho_{\mathrm{F}}$ grows up and becomes relevant when $t$ is close to $t_{0}$ only if $\omega_{\mathrm{F}}<-1$. It means that phantom fluids increases the effective density and pressure of the universe in the case of Big Rip and Type I singularities, whereas quintessence fluid becomes negligible and do not influence the asymptotic behaviour of $f(R,G)$-models that realize this kind of singularities. For this reason, for modified gravity which produces Type I singularities, we will examine the case of phantom fluid only.

In Einstein's gravity ($f(R,G)=0$), Eq.~(\ref{EOM1fluid}) and Eq.~(\ref{tric}) admit the solution
\begin{equation}
H(t)=-\frac{2}{3(1+\omega_{\mathrm{F}})}\frac{1}{(t_{0}-t)}\,,\label{oraetlabora}
\end{equation}
and we can see that the phantom fluid produces the Big Rip for $H(t)=h_0/(t_{0}-t)$, where $h_0=-2/3(1+\omega_{\mathrm{F}})$.  

In general, by considering $f(R,G)$-modified gravity in the presence of phantom fluid, the asymptotically Big Rip singularity could appear if $\rho_{\mathrm{MG}}$  diverges less than $H^{2}$ ($\sim (t_{0}-t)^{-2}$) on the singular solution of Eq.~(\ref{oraetlabora}), namely the modified gravity becomes negligible with respect to the fluid contribute in the first EOM of~(\ref{EOM1fluid}). On the other hand, if a $f(R,G)$ model realizes the Big Rip for a certain value of $h_0$, the fluid energy density $\rho_{\mathrm{F}}$ of Eq.~(\ref{tric}) becomes negligible on this singular solution if $\omega_{\mathrm{F}}>-(1+2/(3h_0))$, since in this case it diverges less than $H^{2}$.

When $\beta>1$, the energy density $\rho_{\mathrm{F}}$ of phantom fluid exponentially diverges in Eq.~(\ref{trac}), so that the EOMs (\ref{EOM1fluid}) become inconsistent and the Type I singularity is never realized.
 
When $0<\beta<1$, $\rho_{\mathrm{F}}$ tends to $\rho_{0}$ with time in Eq.~(\ref{trac}), and it is asymptotically negligible with respect to $H^{2}$ ($\sim (t_{0}-t)^{-2\beta}$). In this case, the behaviour of $f(R,G)$-model realizing Type III singularity is not influenced by perfect fluids.

For Type II and IV singular models ($\beta<0$), the presence of quintessence or phantom fluids can make the singularities more difficult to realize. Note that $H^{2}$ of Type II and IV singularities tends to the constant $3H_{0}^{2}/\kappa^2$ like $\sim(t_{0}-t)^{-\beta}$, while $\rho_{\mathrm{F}}$ in Eq.~(\ref{truc}), after the expansion in power series, tends to $\rho_{0}$ like $\sim(1+\omega_{\mathrm{F}})(t_{0}-t)$. 

In the case of $-1<\beta<0$, the presence of fluid may change the numerical value of $H_{0}$ for which the singularity appears in $f(R,G)$-gravity, but does not necessarily avoid  the singularity. 

In the case of $\beta<-1$, since $\rho_{\mathrm{F}}$ behaves as $(t_{0}-t)$ and asymptotically dominates the dynamic of $H^2$, the EOMs~(\ref{EOM1fluid}) could become inconsistent. In particular, the softest Type IV singularities with $|\beta|\gg1$ are very difficult to realize in the presence of phantom or quintessence perfect fluids.\\
\\
Examples:
\begin{itemize}
\item In the model $R-\alpha\sqrt{G}$, where $\alpha$ is a positive constant, the Type I singularity or the Big Rip for some values of $h_0>1$ could occur (see Eq.~(\ref{Garr}) and Eq.~(\ref{primo}) together). If we add a phantom fluid ($\omega_{\mathrm{F}}<-1$), the Type I singularity is avoided, while the Big Rip could still appear.\\ 
If $\omega_{\mathrm{F}}<-5/3$ (namely, $\omega_{\mathrm{F}}<-[1+2/(3h_0)]$ for any value of $h_0>1$), the fluid energy density of Eq.~(\ref{tric}) grows up faster than $H^{2}$ in the case of the Big Rip, and the Big Rip with $h_0>1$ is not realized. On the other hand, the phantom fluid could produce the Big Rip for some value of $0<h_0<1$, when $h_0=-2/3(1+\omega_{\mathrm{F}})$ like in Eq.~(\ref{oraetlabora}). However, it is possible to verify, by using (\ref{tactac}), that $\rho_{\mathrm{MG}}$ of this model, when $0<h_0<1$, diverges still like $H^2$, but is negative. If the effective energy density of the universe becomes negative, the Big Rip is not a physical solution.  

\item The model $R+\alpha R^{\gamma}$, where $\alpha$ is a constant, could realize the Type II singularity when $\gamma<0$ or the Type IV singularity when $2<\gamma$ (see Eq.~(\ref{singularf(R)})). In both cases we assume $H_{0}$ negligible in Eq.~(\ref{Hsingular}).\\ 
The presence of quintessence or phantom fluids does not avoid the Type II singularity, because the numerical value of $H_{0}$ changes on the singular solution ($H_{0}=\sqrt{\kappa^2\rho_{0}/3}$), but the dynamical behaviour of the modified function $f(R)$ keeps the same, due to the fact that $R$ tends to infinity and the constant $H_{0}$ is avoidable. Moreover, if we use a phantom fluid, there is the possibility that the Type II singularity is changed into the Big Rip in the form of Eq.~(\ref{oraetlabora}), since in this case it is easy to verify that $\rho_{\mathrm{MG}}$ of the model tends to zero, so that the fluid is dominant and makes the future singularity stronger.\\
The Type IV singularity could be avoided by phantom or quintessence fluids, especially if $\gamma$ parameter is very close to two (it means, $|\beta|\gg 1$). As a consequence, other future scenarios for the universe are possible. For example, if $\gamma=3$, the model admits an unstable de Sitter solution with $R_{dS}=\sqrt{1/\alpha}$ (see Eq.~(\ref{RdeSitter})), or the phantom fluid may produce an accelerating phase. 

\item The model $R-\alpha G^{\gamma}$, where $\alpha>0$ and $\gamma>1$, shows the Type II singularity with $H_{0}=0$ and $-1<\beta<-1/3$ (see Eq.~(\ref{trentatre}) and the following discussion). Now, the presence of phantom or quintessence fluids with suitable boundary conditions on $\rho_{0}$, avoids the Type II singularity. Unlike the preceding example, the value of $H_{0}$ and the dynamical behaviour of $f(G)$ change together, because in the case of $H_{0}=0$, when $-1<\beta<-1/3$, $G$ tends to zero, but if $H_{0}\neq 0$, $G$ diverges to infinitive and the EOMs~(\ref{EOM1fluid}) for this kind of model become inconsistent on the Type II singularity.
\end{itemize}

\subsubsection{Constant viscosity}

\paragraph*{}Now, we introduce bulk viscosity in cosmic fluid. Note that viscous fluids belong to more general inhomogeneous EoS fluids introduced in Refs.~\cite{BigRipfluid, Capozielloetal}.

Suppose to have the bulk viscosity equal to a constant $\zeta_0$, i.e. $\zeta(H)=\zeta_{0}$. Eq.~(\ref{conservationlawfluid}) yields:
\begin{equation}
\rho_{\mathrm{F}}=\rho_{0}a^{-3(1+\omega_{\mathrm{F}})}+9\zeta_{0} a^{-3(1+\omega_{\mathrm{F}})}\int^{t} a(t')^{1+3\omega_{\mathrm{F}}}\dot{a}(t')^{2}dt'\,.
\end{equation}
For the Big Rip ($\beta=1$), $\rho_{\mathrm{F}}$ behaves as  
\begin{equation}
\rho_{\mathrm{F}}=\rho_{0}(t_{0}-t)^{3h_0(1+\omega_{\mathrm{F}})}+\frac{9h_0^{2}\zeta_{0}}{(t_{0}-t)(1+3h_0+3h_0\omega_{\mathrm{F}})}\,.\label{zig}
\end{equation}
In this case, in Einstein's framework ($f(R,G)=0$), the solution of Eq.~(\ref{EOM1fluid}) becomes~\cite{Alessia}:
\begin{equation}
H(t)=\frac{\sqrt{3\kappa^2\rho_{0}} \mathrm{e}^{(3\kappa^2\zeta_{0}/2)t}}{3+\left[\frac{3}{\zeta_0}(1+\omega_{\mathrm{F}})\sqrt{\frac{\rho_0}{3\kappa^2}}(\mathrm{e}^{(3\kappa^2\zeta_{0}/2)t}-1)\right]}\,.\label{exp}
\end{equation}
$H$ shows a finite-time future singularity when $t$ tends to $t_{0}$, where 
\begin{equation}
t_0=\frac{2}{3\kappa^2\zeta_0}\ln\left[1-\sqrt{\frac{3\kappa^2}{\rho_0}}\frac{\zeta_0}{(1+\omega_{\mathrm{F}})}\right]\,. 
\end{equation}
If we expand the exponential functions around $t_{0}$, we obtain:
\begin{equation}
H(t)\simeq-\frac{2}{3(1+\omega_{\mathrm{F}})}\frac{1}{(t_{0}-t)}+\frac{\kappa^2}{1+\omega_{\mathrm{F}}}\zeta_0+\mathcal{O}(t_{0}-t)\,,\label{oraetlaborabis}
\end{equation}
that corresponds to Eq.~(\ref{Hsingular}) with $\beta=1$ (Big Rip), $h_0=-2/(3+3\omega_{\mathrm{F}})$, where $\omega_{\mathrm{F}}<-1$, and $H_{0}=\kappa^2\zeta_{0}/(1+\omega_{\mathrm{F}})$. The viscosity $\zeta_{0}$ is not relevant in the asymptotic singular limit of $H$ (here, $H_{0}$ is negative, but the first positive term of $H$ is much larger), and we recover Eq.~(\ref{oraetlabora}), that is valid for phantom perfect fluids, and the related discussion already done is still valid.\\

In order to study the effects of the viscosity on Type I, II, III and IV singular models, it is worth considering the asymptotic behaviour of the conservation law in Eq.~(\ref{conservationlawfluid}). We require that the left part diverges like the right part on the singular solutions,
\begin{equation}
\dot{\rho_{\mathrm{F}}}+3\rho_{\mathrm{F}}(1+\omega_{\mathrm{F}})\left(\frac{h_0}{(t_{0}-t)^{\beta}}+H_{0}\right)\simeq \frac{9 h_0^{2}\zeta_{0}}{(t_{0}-t)^{2\beta}}+\frac{18 h_0 H_{0}\zeta_{0}}{(t_{0}-t)^{\beta}}+9H_{0}^{2}\zeta_{0}\,,\label{claw}
\end{equation}
where we take $H_{0}=0$ if $\beta>0$. In what follows, we neglect the homogeneous solutions, already discussed above. 

The following asymptotic solutions of Eq.~(\ref{claw}) are found:
\begin{eqnarray}
1)\,\rho_{\mathrm{F}}&\simeq& \frac{3 h_0 \zeta_{0}}{(1+\omega_{\mathrm{F}})(t_{0}-t)^{\beta}}\,,\quad\quad\quad\quad\quad\,\,\,\beta>1\,;\label{zagzag}
\\ \nonumber\\
2)\,\rho_{\mathrm{F}}&\simeq& \frac{9\zeta_{0}h_0^{2}}{(2\beta-1)(t_{0}-t)^{2\beta-1}}\,,\quad\quad\quad\quad\,\, 1>\beta>0\,;\label{otto}
\\ \nonumber\\
3)\,\rho_{\mathrm{F}}&\simeq&\frac{9h_0 H_{0}\zeta_{0}}{(\beta-1)(t_{0}-t)^{\beta-1}}+\frac{3H_{0}\zeta_{0}}{1+\omega_{\mathrm{F}}}\,,\quad\beta<0\,,H_{0}\neq 0\,.\label{zag}
\end{eqnarray}

In the first case ($\beta>1$), it is possible to see that fluid energy density diverges more slowly than $H^{2}$ in~(\ref{EOM1fluid}), so that viscous fluid does not influence the (asymptotically) behaviour of Type~I singularity in $f(R,G)$ models, due to the constant viscosity.

Also in the second case ($0<\beta<1$), viscous fluid is asymptotically avoidable in the case of Type III singularity in $f(R,G)$ models, since fluid energy density diverges less than $H^{2}$ again.
  
In the end, we consider fluid which tends to a constant when $\beta<0$. Large bulk viscosity $\zeta_{0}$ becomes relevant in the EOM and, if $\omega_{\mathrm{F}}<-1$, the effective energy density, due to the fluid contribute, could become negative avoiding the Type II and IV singularities.\\
\\
Example:
\begin{itemize}
\item We have seen in Chapter 3 that the Hu-Sawicki Model in some cases produces the Type II singularity for a certain positive value of $H_0=H_{\mathrm{dS}}$ as in Eq.~(\ref{accazero}).\\
A fluid with $\omega_{\mathrm{F}}>-1$ and constant viscosity $\zeta_{0}$ large with respect to $H_{dS}$, may change the value of $H_{0}$ for which singularity appears, but does not avoid it.\\
On the other hand, a fluid with $\omega_{\mathrm{F}}<-1$ and $\zeta_{0}\gg H_{dS}$, makes the singularity unphysical, since the solution of the EOMs~(\ref{EOM1fluid}) with Eq.~(\ref{zag}) leads to $H_{0}$ imaginary and the singularity does not appear. 
\end{itemize}

\subsubsection*{Viscosity proportional to H}

\paragraph*{} This is the case $\zeta(H)=3H\tau$. As $\zeta$ is assumed to be positive, the constant $\tau$ has to be positive. Eq.~(\ref{conservationlawfluid}) yields:
\begin{equation}
\rho_{\mathrm{F}}=\rho_{0}a^{-3(1+\omega_{\mathrm{F}})}+27\tau a^{-3(1+\omega_{\mathrm{F}})}\int^{t} dt'a(t')^{3\omega_{\mathrm{F}}}\dot{a}(t')^{3}\,.
\end{equation}
For the Big Rip ($\beta=0$), $\rho_{\mathrm{F}}$ behaves as:  
\begin{equation}
\rho_{\mathrm{F}}=\frac{27 h_0^{3}\tau}{(t_{0}-t)^{2}(2+3h_0+3h_0\omega_{\mathrm{F}})}\,,\label{zurp}
\end{equation}
In Einstein's gravity, from~(\ref{EOM1fluid}) with~(\ref{zurp}) we get
\begin{equation}
H(t)=\frac{2}{[9\kappa^2\tau-3(1+\omega_{\mathrm{F}})]}\frac{1}{(t_{0}-t)}\,,  \label{zip}
\end{equation}
and realize the Big Rip for $H=h_0/(t_{0}-t)$, where $h_0=2/[9\kappa^2\tau-3(1+\omega_{\mathrm{F}})]$. We have that $h_0$ is positive if~\cite{Alessia}:
\begin{equation}
(1+\omega_{\mathrm{F}})-3\kappa^2\tau<0\,.\label{zizizip}
\end{equation}

It means that phantom fluid or fluid in the quintessence region with sufficiently large bulk viscosity could produce the Big Rip. On the other hand, if $(1+\omega_{\mathrm{F}})-3\kappa^2\tau>0$, the fluid does not realize the Big Rip for expanding universe.   

The other asymptotic solutions of Eq.~(\ref{conservationlawfluid}) are:
\begin{eqnarray}
1)\,\rho_{\mathrm{F}}&\simeq& \frac{9 h_0^{2} \tau}{(1+\omega_{\mathrm{F}})(t_{0}-t)^{2\beta}}\,,\phantom{spa}\quad\quad\quad\quad\,\,\,\,\beta>1\,;\label{uno1}
\\ \nonumber\\
2)\,\rho_{\mathrm{F}}&\simeq& \frac{27\tau h_0^{3}}{(3\beta-1)(t_{0}-t)^{3\beta-1}}\,,\quad\quad\quad\quad\quad\,\,0<\beta<1\,;\label{due2}
\\ \nonumber\\
3)\,\rho_{\mathrm{F}}&\simeq&\frac{27h H_{0}^{2}\tau}{(\beta-1)(t_{0}-t)^{\beta-1}}+\frac{9H_{0}^{2}\tau}{1+\omega_{\mathrm{F}}}\,,\phantom{sp}\quad\beta<0\,,H_{0}\neq 0\,.\label{tre3}
\end{eqnarray}

For $\beta>1$, $\rho_{\mathrm{F}}$ diverges like $H^2$ if $\omega_{\mathrm{F}}>-1$. Thus, the fluid could asymptotically produce the Type I singularity and generally does not influence the $f(R,G)$-gravity producing such kind of singularity. On the other hand, if $\omega_{\mathrm{F}}<-1$, for large values of viscosity $\tau$, the theory is protected against Type I singularity, since the effective energy density of the universe may become negative.
 
When $0<\beta<1$, since $\rho_{\mathrm{F}}$ diverges less than $H^{2}$, the fluid does not influence the $f(R,G)$-singular models on Type III singularity and can be neglected on singular solutions.
  
When $\beta<0$, the fluid can influence the $f(R,G)$-models producing Type II and IV singularity with $H_{0}\neq 0$, if the viscosity $\tau$ is large. In particular, if $\omega_{\mathrm{F}}<-1$, the fluid energy density becomes negative and may avoid Type II and IV singularities.\\ 
\\
Examples:
\begin{itemize}
\item The model $R-\alpha(G/R)$, where $\alpha$ is a positive constant, shows the Type I singularity (see Eq.~(\ref{zap})).\\
A fluid with $\omega_{\mathrm{F}}>-1$ and energy density in the form of Eq.~(\ref{uno1}), may influence some feature of singularity, but the Type I singularity is still realized. In addition, if $\tau$ is sufficiently large, Eq.~(\ref{zizizip}) is satisfied and an other possible
scenario is the Big Rip solution.\\
If $\omega_{\mathrm{F}}<-1$, large values of $\tau$ make negative the effective energy density of the universe on the Type I singularity, which could be
changed into the Big Rip.

\item In the model $R+\alpha(G/R)$, where $\alpha$ is a positive constant, the Type II, III and IV singularities could appear (see Eq.~(\ref{zapzap})). The presence of fluids with $\omega_{\mathrm{F}}<-1$ and energy density in the form of Eq.~(\ref{uno1}) and Eq.~(\ref{due2}), does not influence the Type III singularity, but could change the Types II and IV with the Big Rip, like in the previous example.    
\end{itemize}

\subsection{$\omega(\rho_{\mathrm{F}})$ not a constant}

\paragraph*{}In this general case, the fluid EoS parameter $\omega(\rho_{\mathrm{F}})$ explicitly depends on the fluid energy density $\rho_{\mathrm{F}}$. We are interested in some simple case. We consider viscous fluid, whose thermodynamical parameter $\omega(\rho_{\mathrm{F}})$ is given by:
\begin{equation}
\omega(\rho_{\mathrm{F}})=A_{0}\rho_{\mathrm{F}}^{\alpha-1}-1\label{omega}\,,
\end{equation}
where $A_{0}$($\neq 0$) and $\alpha$ are constants. When $\alpha=1$, we find the case when $\omega(\rho_{\mathrm{F}})$ is a constant. Let us suppose the following form of bulk viscosity $\zeta(H)$:
\begin{equation}
\zeta(H)=(3 H)^{n}\tau\,.\label{viscosityinhomogeneus}
\end{equation}
Here, $\tau>0$ and $n$ are constants.

The energy conservation law (\ref{conservationlawfluid}) leads:
\begin{equation}
\dot{\rho_{\mathrm{F}}}+3 H A_{0}\rho_{\mathrm{F}}^{\alpha}=9H^{2}(3H)^{n}\tau\label{super}\,,
\end{equation}
from which we may get the (asymptotic) solutions of the fluid energy density when $H$ is singular. 

In what follows, we consider several examples for $\tau\neq 0$ (viscous case) and $\tau=0$ (non viscous case).

\subsubsection{Viscous case}

\paragraph*{} Let us take $\tau$ positive constant different to zero. For the Big Rip singularity ($\beta=1$), some simple (asymptotic) solutions of Eq.~(\ref{super}) are given by: 
\begin{eqnarray}
\left\{\begin{array}{lll}
\rho_{\mathrm{F}}&=&\frac{3^{n+2}h_0^{n+2}\tau}{(n+1+3 h_0 A_{0})(t_{0}-t)^{n+1}}\,,\quad\alpha=1\,;
\\ \\
\rho_{\mathrm{F}}&\simeq& \left(\frac{3^{n+1} h_0^{n+1}\tau}{A_{0}(t_{0}-t)^{n+1}}\right)^{\frac{1}{\alpha}}\,,\quad\quad\quad\alpha>1\label{rhoin2}\,. 
\end{array}\right.
\end{eqnarray} 
The first expression corresponds to the cases when $\omega(\rho_{\mathrm{F}})$ is a constant. For $n=0,1$, we find Eq.~(\ref{zig}) and Eq.~(\ref{zurp}). When $\alpha=1$ and $n>1$, the fluid energy density diverges faster than $H^{2}$ ($\sim (t_{0}-t)^{-2}$) and the EOMs become inconsistent on the Big Rip. We can say that fluids with $\omega(\rho_{\mathrm{F}})$ constant and bulk viscosity proportional to $H^{n}$, where $n>1$, avoid the Big Rip. The same happens in the presence of this kind of viscous fluids with $n+1>2\alpha$, where $\alpha>1$, as in the second case of (\ref{rhoin2}).

For Type I singularities ($\beta>1$), an asymptotic, simple solution of Eq.~(\ref{super}) is given by
\begin{equation}
\rho_{\mathrm{F}}\simeq \left(\frac{3^{n+1} h_0^{n+1}\tau}{A_{0}(t_{0}-t)^{(n+1)\beta}}\right)^{\frac{1}{\alpha}}\,,\quad\alpha\geqslant 1\label{rhoin3}\,.
\end{equation}
The cases $\alpha=1$ and $n=0,1$ correspond to Eq.~(\ref{zagzag}) and Eq.~(\ref{uno1}). The fluid avoids the Type I singularities if $2\alpha<n+1$ when $\alpha\geqslant 1$, so that its energy density diverges faster than $H^{2}$ in the EOMs. It means that, if the viscosity behaves as a power function of $H$ larger than one, the fluid with $\omega(\rho_{\mathrm{F}})$ constant is able to protect the theory against the Big Rip and Type I singularities together. 

Note that the viscosity is introduced in the EOMs by the fluid pressure of Eq.~(\ref{eq.state}). On the Big Rip and Type I singularities, the curvature $R$ behaves as $H^{2}$. Motivated by fact that the correction term $\gamma R^{m}$, with $\gamma$ constant and $m>1$, cures Big Rip and Type I singularities in $f(R,G)$ gravity (see Chapter 3), we may directly conclude that the term $-3H\zeta(H)$ proportional to $H^{1+n}$, with $n>1$, shows the same effect, like we have just verified.
    
For Type III singularities ($0<\beta<1$), an asymptotic solution of Eq.~(\ref{super}) is 
\begin{equation}
\rho_{\mathrm{F}}\simeq \frac{3^{n+2} h_0^{n+2}\tau}{(2\beta+n\beta-1)(t_{0}-t)^{2\beta+n\beta-1}}\,,\quad1/2< \alpha\leqslant 1\label{rhoin4}\,.
\end{equation}

The cases $\alpha=1$ and $n=0,1$ correspond to Eq.~(\ref{otto}) and Eq.~(\ref{due2}). The fluid energy density diverges faster than $H^{2}$ when $n>1/\beta$. In principle, if a $f(R,G)$-theory shows the Type III singularity for a certain value of $\beta$, the presence of a fluid with viscosity proportional to $H^{n}$, where $n>1/\beta$ (and, as a consequence, always $n>1$), can make inconsistent the EOMs and avoid this kind of singularity. Otherwise, it could appear a new Type III singularity realized by fluid for $H(t)=h_0/(t_{0}-t)^{1/n}$, so that $\rho_{\mathrm{F}}\sim H^{2}$, solving in some cases the EOMs~(\ref{EOM1fluid}).
We will see a nice example in the end of the Section.

For Type II and IV singularities ($\beta<0$), if $H_{0}\neq 0$, an asymptotic solution of Eq.~(\ref{super}) reads
\begin{equation}
\rho_{\mathrm{F}}\simeq \frac{3^{n+2}H_{0}^{n+1}h_0\tau}{(\beta-1)(t_{0}-t)^{\beta-1}}+\left(\frac{3^{n+1}H_{0}^{n+1}\tau}{A_{0}}\right)^{\alpha}\,,\quad\alpha\geqslant 1\label{rhoin5}\,.
\end{equation}

The cases $\alpha=1$ and $n=0,1$ correspond to Eq. (\ref{zag}) and Eq.~(\ref{tre3}). In general, this kind of fluid influences the feature of Type II and IV singularities in $f(R,G)$-gravity, but not necessarily avoids they.

\subsubsection{Non viscous case}

\paragraph*{} If the viscosity is equal to zero, i.e. $\tau=0$, Eq.~(\ref{super}) yields\\
\phantom{line}
\begin{equation}
\rho_{\mathrm{F}}=\left[(\alpha-1)\left(3A_{0}\ln \frac{a(t)}{a_{0}}\right)\right]^{\frac{1}{1-\alpha}}\,,
\end{equation}
\phantom{line}\\
where $a(t)$ is, as usual, the scale factor, $a_{0}$ is a positive parameter and $\alpha\neq 1$ (non perfect fluid). We may take $A_0(\alpha-1)$ positive, so that, in general, $\rho_{\mathrm{F}}$ is also positive. 
In addition, we set\\
\phantom{line}
\begin{equation}
\rho_{\mathrm{F}}=\frac{H_{0}^{2}}{\kappa^2}\left[\ln \frac{a(t)}{a_{0}}\right]^{\frac{1}{1-\alpha}}\,,
\quad
H_0^2=\kappa^2[3A_{0}(\alpha-1)]^{\frac{1}{1-\alpha}}\,.
\label{Lorenzo}
\end{equation}
\phantom{line}\\
In Einstein's gravity ($\rho_{\mathrm{MG}}=0$), the first EOM of (\ref{EOM1fluid}) reads\\
\phantom{line}
\begin{equation}
a(t)=\left(a_{0}\right)\text{Exp}\left\{ 6^{\frac{2-2\alpha}{2\alpha-1}}\left[\pm \dfrac{(2\alpha-1)(\sqrt{3}H_{0} t)}{\alpha-1}\right]^{2(\alpha-1)/(2\alpha-1)}\right\}\label{zirp}\,.
\end{equation}
\phantom{line}\\
Note that for large values of $\alpha$, the fluid energy density tends to $H_{0}^{2}/\kappa^2$, and Eq.~(\ref{omega}) leads to $\omega(\rho_{\mathrm{F}})\simeq -1$ and $a(t)\simeq a_{0}\mathrm{e}^{H_{0}t/3}$ (de Sitter universe).

Moreover, one can see that Eq.~(\ref{zirp}) produces the following form of Hubble parameter,\\
\phantom{line}
\begin{equation}
H(t)=\frac{h_0}{(t_0-t)^{\frac{1}{(2\alpha-1)}}}\,,\quad h_0>0\,.
\end{equation}
\phantom{line}\\
In principle, this inhomogeneous non viscous fluid can generate any type of singularity with $\beta=1/(2\alpha-1)\,,\alpha\neq 1$ (except the Big Rip, for which one has to consider the perfect fluid case).\\

We conclude with a special non-viscous fluid with $\omega(\rho_{\mathrm{F}})$ non-constant, namely the Chaplygin gas~\cite{Ciappinski}, which also has been studied as a candidate for dark energy in many works and whose Equation of State reads\\
\phantom{line}
\begin{equation}
p_{\mathrm{F}}=-\frac{A_{0}}{\rho_{\mathrm{F}}}\,,
\end{equation}
\phantom{line}\\
where $A_{0}$ is a positive constant. Eq.~(\ref{conservationlawfluid}) leads to\\
\phantom{line}
\begin{equation}
\rho_{\mathrm{F}}=\sqrt{A_{0}+\frac{1}{a(t)^{6}}}\label{rhoChap}\,.
\end{equation}
\phantom{line}\\
Since $a(t)$ diverges or tends to a constant for Big Rip, Type I and Type III singularities, it is easy to see that the Chaplygin gas does not influence in the EOMs (\ref{EOM1fluid}) the asymptotic behaviour of $f(R,G)$-models where such kind of singularities appear. It only may influence $f(R,G)$-models where Type II and IV singularities are realized, but not necessarily prevents the singularities.\\
\\
Example:
\begin{itemize}

\item In the model $R+\alpha R^{1/2}$, with $\alpha$ positive constant, the Type III singularity for $\beta=1/3$ can appear (see Eq.~(\ref{singularf(R)})). A fluid with $\omega(\rho_{\mathrm{F}})$ constant and energy density in ther form of Eq.~(\ref{rhoin4}), where $n>3$, avoids this kind of singularity. It is also interesting to see that in such a case, since if $\beta=1/n$, $\rho_{\mathrm{MG}}$ of~(\ref{tactac}) diverges faster than $H^{2}$ in the first EOM of~(\ref{EOM1fluid}), the fluid does not produce a new Type III singularity, due to the contribute of modified gravity. Moreover, the model is free of any type of singularity, being the theory protected against singularities of $f(R)$-gravity by fluid and against fluid singularities by modified gravity itself.  
\end{itemize}

\section{Viscous fluids coupled with Dark Matter}

\paragraph*{} In this Section, we consider viscous fluid coupled with dark matter~\cite{Viscousfluids} .
Their energy conservation laws are given by:
\begin{equation}
\dot \rho_{\mathrm{F}} +3H(\rho_{\mathrm{F}}+p_{\mathrm{F}})=-Q_0\rho_{\mathrm{F}}\,,\label{ECL1}
\end{equation}
\begin{equation}
\dot \rho_{\mathrm{DM}} +3H\rho_{\mathrm{DM}}=Q_0\rho_{\mathrm{F}}\,.\label{ECL2}
\end{equation}
Here, $Q_0$ is the coupling constant, $\rho_{\mathrm{DM}}$ is the energy density of dark matter (the corresponding pressure is equal to zero), whereas $\rho_{\mathrm{F}}$ and $p_{\mathrm{F}}$ are, as usually, the energy density and pressure of viscous fluid. The fluid pressure $p_{\mathrm{F}}$ is written as in Eq.~(\ref{eq.state}).

The equations of motion simply read
\begin{equation}
\rho_{\mathrm{F}}+\rho_{\mathrm{DM}}=\frac{3}{\kappa^2}H^{2}\,,
\quad
p_{\mathrm{F}}=-\frac{1}{\kappa^2}\left(2\dot{H}+3H^2\right)\label{EOM1fluidDM}\,.
\end{equation}
We will motivate this study by showing how this coupling may solve the coincidence problem and remove singular solutions of DE-fluids. 

\subsection{$\omega(\rho_{\mathrm{F}})$ constant}

\paragraph*{} Suppose to have $\omega(\rho_{\mathrm{F}})=\omega_{\mathrm{F}}$ constant for the fluid and bulk viscosity in the form of Eq.~(\ref{viscosityinhomogeneus}), namely $\zeta(H)=\tau (3H)^n$, $\tau>0$ and $n$ being constants.
In this case, the general solution of Eq.~(\ref{ECL1}) is\\
\phantom{space}
\begin{equation}
\rho_{\mathrm{F}}=\rho_{\mathrm{F(0)}}\frac{\mathrm{e}^{-Q_0 t-3\omega_{\mathrm{F}} \log a(t)}}{a(t)^3}+\frac{\tau 3^{2+n} \mathrm{e}^{-Q_0 t-3\omega_{\mathrm{F}} \log a(t)}}{a(t)^3}\int^{t} \mathrm{e}^{Q_0 t'+3\omega_{\mathrm{F}} \log a(t')}a(t')\dot{a}(t')^2\left(\frac{\dot{a}(t')}{a(t')}\right)^n dt'\,,\label{unobis}
\end{equation}
\phantom{space}\\
where $\rho_{\mathrm{F(0)}}$ is a positive constant of integration.

One possible solution is the de Sitter space, where $H=H_{\mathrm{dS}}$ is a constant. One may identify the Hubble parameter $H_{\mathrm{dS}}$ with the present value of accelerated universe. In this case, Eq.~(\ref{unobis}) can be solved as
\begin{equation}
\rho_{\mathrm{F}}=\rho_{\mathrm{F(0)}}\mathrm{e}^{-t(Q_0+3H_{\mathrm{dS}}(1+\omega_{\mathrm{F}}))}+\frac{(3H_{\mathrm{dS}})^{n+2} \tau}{(Q_0+3H_{\mathrm{dS}}(1+\omega_{\mathrm{F}}))}\,.\label{duebis}
\end{equation}
It follows the solution of Eq.~(\ref{ECL2}) for dark matter
\begin{equation}
\rho_{\mathrm{DM}}=\rho_{\mathrm{DM(0)}}\mathrm{e}^{-3H_{\mathrm{dS}}t}-\rho_{F(0)}\frac{Q_0}{Q_0+3H_{\mathrm{dS}}\omega_{\mathrm{F}}}\mathrm{e}^{-t(Q_0+3H_{\mathrm{dS}}(1+\omega_{\mathrm{F}}))}+\frac{(3H_{\mathrm{dS}})^{n+1}Q_0\tau}{(Q_0+3H_{\mathrm{dS}}(1+\omega_{\mathrm{F}}))}\,,\label{trebis}
\end{equation}
where $\rho_{\mathrm{DM(0)}}$ is a positive constant. It is easy to see that, if $\tau\neq 0$, the EOMs (\ref{EOM1fluidDM}) are satisfied only if $\rho_{\mathrm{F(0)}}=\rho_{\mathrm{DM(0)}}=0$. Therefore, we note that, if the de Sitter solution is an attractor and it is able to describe our universe today, we can require
\begin{equation}
\frac{\rho_{\mathrm{DM}}}{\rho_{\mathrm{F}}}=\frac{Q_0}{3H_{\mathrm{dS}}}=\frac{1}{3}\,,
\end{equation}
and the coincidence problem is solved by setting
\begin{equation}
Q_0=H_{\mathrm{dS}}\,.\label{Qcoupling} 
\end{equation}
The ratio of DM and fluid is approximately $1/3$, almost independent from initial conditions. By evaluating 	the second EOM of~(\ref{EOM1fluidDM}) on the de Sitter solution, one has the relation between $\omega_{\mathrm{F}}$ and $\tau$, namely 
\begin{equation}
 \omega_{\mathrm{F}}=-\frac{4}{3}+4\kappa^2 (3H_{dS})^{n-1}\tau\,.\label{omegafluidDM}
\end{equation}
Here, Eq.~(\ref{Qcoupling}) has been used. Note that $\rho_{\mathrm{F}}$ of Eq.~(\ref{duebis}) results to be positive. For example, a DE-fluid with $\omega_{\mathrm{F}}=-1$ admits the de Sitter solution for $H=H_{dS}$ if its bulk viscosity is given by
\begin{equation*}
\zeta(H)=\frac{(3H)^{n}}{12\kappa^2 (3H_{\mathrm{dS}})^{n-1}}\,, 
\end{equation*}
and the coupling constant with DM is $Q_0=H_{\mathrm{dS}}$. We stress that, if from one side with
the coupling between dark matter and viscous fluid we can solve the
coincidence problem, on the other side the energy density of dark matter
in accelearted universe does not follow the usual law $\rho_{\mathrm{DM}}\sim a(t)^{-3}$.

This is a generalization of the result achieved in Ref.~\cite{Odintsovcouplingfluids} for coupled non viscous DE-fluid with DM. If $\tau=0$, it is easy to see that Eqs.~(\ref{duebis})--(\ref{trebis}) are de Sitter solutions of the EOMs if $Q_0=-3(1+\omega_{\mathrm{F}})H_{\mathrm{dS}}$ and $\rho_{DM(0)}=0$, so that the coincidence problem is solved by putting
\begin{equation}
\frac{\rho_{\mathrm{DM}}}{\rho_{\mathrm{F}}}=-(1+\omega_{\mathrm{F}})\sim \frac{1}{3}\,, 
\end{equation}
which leads to the condition of phantom fluid
\begin{equation}
\omega_{\mathrm{F}}=-\frac{4}{3}\,. 
\end{equation}
Let us return to the case of $\tau\neq 0$. In order to investigate if the de Sitter solution is an attractor or not, we consider the perturbation as 
\begin{equation}
H(t)=H_{\mathrm{dS}}+\Delta(t)\,.\label{perturbazione}
\end{equation}
Here, $\Delta(t)$ is a function of the cosmic time $t$ and it is assumed to be small. The second EOM of (\ref{EOM1fluidDM}) gives\\
\phantom{line}  
\begin{equation}
2\dot{\Delta}(t)+6H_{\mathrm{dS}}\Delta(t)\simeq 3H_{\mathrm{dS}}(n+1)\Delta(t)\,,
\end{equation}
\phantom{line}\\
where we have used Eq.~(\ref{duebis}) and Eq.~(\ref{omegafluidDM}). By assuming $\Delta(t)=\mathrm{e}^{\lambda t}$, we find
\begin{equation}
 \lambda+3H_{\mathrm{dS}}-\frac{3}{2}H_{\mathrm{dS}}(n+1)\simeq 0\,,
\end{equation}
that is
\begin{equation}
 \lambda\simeq \frac{3}{2}H_{\mathrm{dS}}(n-1)\,.
\end{equation}

Then, if $n<1$, the de Sitter solution is stable and the coupling of viscous fluid and dark matter at last generates a stable accelerated universe with a constant rate of DM and DE-fluid. If $n>1$, the de Sitter solution is not stable and other future scenarios are possible. 

We have seen in \S\ref{4.2} that phantom (viscous) fluid ($\omega_{\mathrm{F}}<-1$) can generate the Big Rip singularity. On the other hand, the coupling with DM seems to avoid such scenario, being constant the value of fluid energy density in stable de Sitter universe.

\subsection{$\omega(\rho_{\mathrm{F}})$ not a constant}

\paragraph*{} To complete this Section, let us consider a more general case, when the thermodynamical parameter $\omega(\rho_{\mathrm{F}})$ of viscous fluid is not a constant. A simple example is given by Eq.~(\ref{omega}), namely $\omega(\rho_{\mathrm{F}})=\left[A_0\rho_{\mathrm{F}}^{\alpha-1}-1\right]$,
$A_{0}$ and $\alpha$ being constant parameters. The energy conservation law (\ref{ECL1}) of viscous fluid becomes\\
\phantom{line}
\begin{equation}
\dot\rho_{\mathrm{F}}+3H A_{0}\rho_{\mathrm{F}}^{\alpha}+Q_0\rho_{\mathrm{F}}=9 H^2(3H)^n\tau\,. 
\end{equation}
\phantom{line}\\
Here, we suppose the bulk viscosity proportional to $H^{n}$, namely $\zeta(H)=\tau (3H)^{n}$ as in Eq.~(\ref{viscosityinhomogeneus}), $\tau>0$ and $n$ being constants. If we assume $\alpha\gg1$, on the de Sitter solution $H=H_{\mathrm{dS}}$, we obtain 
\begin{equation}
\rho_{\mathrm{F}}\simeq \left(\frac{\tau(3H_{\mathrm{dS}})^{n+1}}{A_{0}}\right)^\frac{1}{\alpha}\,.\label{duebisbis}
\end{equation}
By using Eq.~(\ref{ECL2}), the energy density of dark matter reads
\begin{equation}
 \rho_{\mathrm{DM}}\simeq\frac{Q_0}{3H_{\mathrm{dS}}}\rho_{\mathrm{F}}\,,
\end{equation}
and in order to solve the coincidence problem we have to require $Q_0=H_{\mathrm{dS}}$. 

From the EOMs (\ref{EOM1fluidDM}), by assuming that the fluid drives the accelerated expansion of the universe, it follows 
\begin{equation}
A_{0}\simeq \tau(3H_{\mathrm{dS}})^{n+1}\left(\frac{\kappa^2}{3H_{\mathrm{dS}}^{2}}\right)^\alpha\,,\label{omegabis}
\end{equation}
and for $\omega(\rho_{\mathrm{F}})$ in the de Sitter space one has
\begin{equation}
\omega(\rho_{\mathrm{F}})\simeq -1+3(3H_{\mathrm{dS}})^{n-1}\kappa^2\tau\,,\label{sturmtruppen}
\end{equation}
being $\rho_{\mathrm{F}}$ constant. 

In order to investigate if the de Sitter solution is an attractor or not, we consider the perturbation as in Eq.~(\ref{perturbazione}). Thus, the second EOM  gives  
\begin{equation}
2\dot{\Delta}(t)+6H_{\mathrm{dS}}\Delta(t)\simeq H_{\mathrm{dS}}\left(\frac{n+1}{\alpha}\right)\Delta(t)\,,
\end{equation}
where we have used Eq.~(\ref{duebisbis}) and Eq.~(\ref{omegabis}). By assuming $\Delta(t)=\mathrm{e}^{\lambda t}$, we find
\begin{equation}
 \lambda+3H_{\mathrm{dS}}-\frac{1}{2}H_{\mathrm{dS}}\left(\frac{n+1}{\alpha}\right)\simeq0\,,
\end{equation}
that is
\begin{equation}
 \lambda\simeq H_{\mathrm{dS}}\left(\frac{1}{2}\left(\frac{n+1}{\alpha}\right)-3\right)\,.
\end{equation}
Then, if $(n+1)/\alpha<6$, the de Sitter solution is stable.\\


\chapter{Exponential $F(R)$-gravity models: unified theories for inflation and current acceleration}

\paragraph*{} Here, we review viable conditions of realistic $F(R)$-gravity able to reproduce the universe where we live.
A suitable modification to Einstein's gravity is given by 
the class of models $F(R)=R+f(R)$, where the function $f(R)$ plays the role of an effective dark energy fluid. The final goal of modified gravity is to unify the early time with the late time cosmic acceleration. In this sense, exponential gravity models represent an interesting proposal.

\section{Viability conditions in $F(R)$-gravity \label{5.1}}

\paragraph*{} In this and in the next Chapter we will concentrate on $F(R)$-modified gravity, whose action is given by Eq.~(\ref{actionF(R)}). 
The viability conditions~\cite{Viableconditions} follow from the fact that 
the theory has to be consistent with the results of General Relativity and with the important goals arisen with $\Lambda$CDM Model, whose Lagrangian is given by $R-2\Lambda$, $\Lambda$ being the Cosmological Constant, in the description of the universe and the Solar System. 

If $R=0$ it is reasonable to recover the Special Relativity  and the condition $F(0)=0$ permits to obtain the Minkowski's solution of flat space. 
Furthermore,
recall that, in order to avoid anti-gravity effects, it is required that
$F'(R)>0$, namely  the positivity of the effective gravitational coupling $G_{\mathrm{eff}}$, where $G_{\mathrm{eff}}=G_N/F'(R)$, at least when $R$ assumes
the curvature values of present and past universe (in general, when $R\geq4\Lambda$).

\subsection{Existence of a matter era and stability of cosmological
perturbations \label{mattersection}}

\paragraph*{} On the critical points of the theory, one has $\dot{F'}(R)=0$ (see Eq.~(\ref{criticalpoint})). In particular, during matter era, modified gravity has to vanish, so that $\rho_{\mathrm{eff}}=\rho_{\mathrm{m}}$ and $p_{\mathrm{eff}}=p_{\mathrm{m}}= 0$ in Eqs.~(\ref{EOM1bis})--(\ref{EOM2bis}), and the Ricci scalar~(\ref{R}) turns out to be $R=3H^2$.
As a consequence, from Eqs.~(\ref{rhoeffRG})--(\ref{peffRG}) we obtain the conditions on critical point of matter era~\cite{Amendola}:
\begin{equation}
\frac{RF'(R)}{F(R)}=1\,,\quad
F'(R)=1\,.\label{matterpoint2} 
\end{equation}
In order to reproduce the results of the Standard Model, where
$R=\kappa^2\rho_{\mathrm{m}}$ when matter drives the cosmological expansion, a
$F(R)$-theory is acceptable if the modified gravity contribution
vanishes
during this era and $F'(R)\simeq 1$.
However, another condition is required on the second derivative of
$F(R)$:
it has to be positive~\cite{Faraoni1,Faraoni2}. This last
condition
arises from the stability of the cosmological perturbations. If we
consider a
small region of space-time in the weak-field regime, so that the
curvature
is approximated by $R= R^{(0)}+\delta R$, were $R^{(0)}=-\kappa^2 T^{\mathrm{(matter)}}$ is the matter solution, we get Eq.~(\ref{completeperturbationEq}).   
By using Eq.~(\ref{criticalpointbis}), it is easy to see that, since $F'(R)>0$, the solution is stable when
\begin{equation}
F''(R)>0\,,\label{F''(R)>0}
\end{equation}
during matter era. However, a more detailed investigation on dark energy perturbations during this epoch will be carried out in Chapter {\bf 7}.

\subsection{Existence and stability of a late-time de Sitter point}

\paragraph*{} A reasonable theory of modified gravity which reproduces the current
acceleration of the universe needs to show an accelerating solution
for $R_{\mathrm{dS}}=
4\Lambda$, $\Lambda$ being the Cosmological Constant and typically $\Lambda\simeq 10^{-66}\text{eV}^2$. In principle, it is sufficient
to
require that the EoS parameter $\omega_{\mathrm{eff}}$ in
Eq.~(\ref{omegaeff}) is smaller than $-1/3$, but all
the
cosmological data confirm that its value is actually very close to $-1$. The
possibility of the effective quintessence/phantom dark energy and different future scenarios
of the universe evolution, such as the so-called `Little Rip cosmology'~\cite{lRip1,lRip2, lRip3} (for observational bounds and viable models of Little Rip cosmology see Ref.~\cite{viableRip1,viableRip2,viableRip3}), are not exluded, but 
we have seen in Chapter {\bf 4} that the presence of singularities in the framework of modify gravity 
can destroy the feasibility of the models. As a consequence,
the most realistic solution for our current universe in modified gravity is an
(asymptotically) stable de Sitter solution given by Eq.~(\ref{dScondition}) under condition (\ref{dSstability}).

\subsection{Newton Law and the stability on a planet's surface}

\paragraph*{} The results of GR were first confirmed by local tests
at the level of the Solar System. A theory of modified gravity has to
admit a
static spherically symmetric solution of the type of the Schwarzshild solution
(\ref{S2}) or, more in general, the Schwarzshild-de Sitter solution (\ref{S2bis}) with $\Lambda$ very small. The typical value of the curvature in
the Solar
System far from sources is $R=R^{*}$, where $R^*\simeq 10^{-61}
\text{eV}^2$
(it corresponds to one hydrogen atom per cubic centimeter). If a Scwarzshild-de Sitter solution exists,
it will be stable provided by Eq.~(\ref{dSstability}) evaluated on $R_{\mathrm{dS}}=R^*$.
The stability of the solution is necessary in order to find the
post-Newtonian parameters in GR~\cite{SolarSystemconstraints}.
As regard this point, we recall that some (realistic) models of $F(R)$-gravity may lead to 
significant Newton law corrections at large cosmological scales~\cite{matterinstabilityplanet}. 
From the trace of the field equations (\ref{scalaroneq}), by
performing a 
variation with respect to $R=R_0+\delta R$, where 
$R_0$
is the constant background such that $2F(R_0)-R_0 F'(R_0)= 0$, 
and supposing the presence of a matter point source (like a planet), that is, 
$T^{\mathrm{(matter)}}=T_0\,\delta(x)$, where $\delta(x)$ is the Dirac's 
distribution, 
we find, to first order in $\delta R$, 
\begin{equation}
\left(\Box-m^2\right)\delta R=\frac{\kappa^2}{3F''(R_0)}T_0\delta(x)\,,
\quad
m^2=\frac{1}{3}\left(\frac{F'(R_0)}{F''(R_0)}-R_0\right)\,.
\end{equation}
The solution is given by 
\begin{equation}
\delta R=\frac{\kappa^2}{3F''(R_0)}T_0\,G(m^2,|x|)\,,\quad \left(\Box-m^2\right)G(m^2,|x|)=\delta(x)\,.
\end{equation}
Here, $G(m^2,|x|)$ is the correlation function which satisfies the last equation on the right.
Hence, if $m^2 < 0$, there appears a tachyon and thus there could be some 
instability. Even if $m^2 > 0$, when $m^2$ is small compared with $R_0$, 
$\delta R\neq 0$ at long ranges, which generates the large correction to 
the Newton Law.
For this reason, we must require
the stability condition (\ref{dSstability})
for gravitational systems like Solar System.\\ 

Concerning the matter instability \cite{
matterinstabilityplanet}, this might
also occur when the curvature is rather large, as on a planet
($R\simeq
10^{-38} \text{eV}^2$), as compared with the average curvature of the
universe today ($ R\simeq 10^{-66} \text{eV}^2$). In order to arrive to a stability condition,
we
can start from Eq.~(\ref{completeperturbationEq}), where $R^{(0)}=R_{b}$ is
the curvature of the planet surface and $\delta R$ is a perturbation due to
the
difference of the curvature between the internal and the external solution.
We assume that the curvature $R_b=-\kappa^2 T^{\mathrm{(matter)}}$ depends on the radial coordinate $r$ of the planet. Since we are interested in the change of the perturbation $\delta R$ in the time, we get
\begin{eqnarray}
-\partial_{t}^2(\delta R)\sim U(R_{b})\delta R\,,
\end{eqnarray}
where
\begin{eqnarray}
U(R_{b})&=&\left[\left(\frac{F'''(R_{b})}{F''(R_{b})}\right)^2
-\frac{F'''(R_{b})}{F''(R_{b})}\right]g^{rr}\nabla_{r}R_{b}\nabla_{r}R_{b}
-\frac{R_{b}}{3}+\frac{F'(R_{b})}{3F''(R_{b})}\nonumber\\ \nonumber \\
&&\frac{F'''(R_{b})}{3(F''(R_{b}))^2}(2F(R_{b})-R_{b}F'(R_{b})-R_{b})\,.\label{U}
\end{eqnarray}
Here, $g_{\mu\nu}$ is the diagonal metric which describes the planet.
If $U(R_{b})$ is negative, then the perturbation $\delta R$ grows up and the system becomes unstable. Thus, the
planet
stability condition reads
\begin{equation}
U(R_{b})>0\,.\label{Ustab}
\end{equation}
This expression has to be evaluated for typical values of $R_{b}\simeq
10^{-38}\text{eV}^2$. 

\subsection{Existence of an early-time acceleration and future
singularities}

\paragraph*{} In order to reproduce the early-time acceleration of our universe,
namely
the inflation epoch, the modified gravity models have to admit a
solution
for $\omega_{\mathrm{eff}}$
in~(\ref{omegaeff}) smaller than $-1/3$. An important point is
that this
solution should be unstable.
For example, if the model admits the de Sitter solution when
$R_{\mathrm{dS}}\simeq
10^{20-38} \text{GeV}^2$ (this is the typical curvature value at
inflation),
we have to require the violation of condition~(\ref{dSstability}). 
In principle, other scenarios for the very early universe are possible instead the standard cosmic inflation, such as the ekpyrotic one~\cite{ekpyrotic1,ekpyrotic2}, which also accommodates the physics of the Big Bang.\\

We have seen in Chapter {\bf 4} and in Chapter {\bf 5} that many
DE-models, including modified gravity, bring the future
universe evolution to a finite-time singularity.
The presence of a finite-time future singularity may cause serious
problems to the cosmological evolution or to the corresponding black hole and stellar
astrophysics. Thus, it is always necessary to avoid such scenario in realistic models of modified gravity. It is remarkable that modified gravity actually provides a very natural way to cure such singularities by adding, for instance, higher-power term of $R$ (see \S~\ref{curingf(R)}) in the framework of modified gravity. Simultaneously with the removal of any possible
future singularity, the addition of this term supports the early-time inflation
caused by modified gravity (it may be the case of $R^2$-term, which protects the theory against singularities and could produce inflation~\cite{Starobinsky:1980te}). Remarkably, even in the case inflation were not an element
of the alternative gravity dark energy model considered, it eventually occurs after
adding such higher-power term. Hence, the removal of future singularities is a natural prescription for the unified description of inflation and current acceleration.

\section{Unified description of early- and late-time acceleration\label{viablemodels}}

\paragraph*{} In Refs.~\cite{StaroModel,HuSaw,Battye} several versions of viable
modified
$F(R)$-gravity have been proposed, namely so-called `one-step' models, which
reproduce the current acceleration of the universe.
In these models, 
a correction term 
to the Hilbert-Einstein action is added 
as $F(R)=R+f(R)$, 
so that 
the dark energy epoch can be reproduced in a simple and intuitive way. 
Namely, 
a vanishing (or fast decreasing) cosmological constant in the flat limit 
of $R\rightarrow 0$ is incorporated, and a suitable, constant asymptotic 
behavior for large values of $R$ is exhibited.
These models can be collected in the following class of $F(R)$-gravity toy models~\cite{Twostepmodels0}:\\
\phantom{line}
\begin{equation*}
F(R)=R+f(R)\,, 
\end{equation*}
\phantom{line}
\begin{equation}
f(R)=-2\Lambda\, \theta(R-R_0)\,.\label{toymodel}
\end{equation}
\phantom{line}\\
Here, $\theta(R-R_0)$ is Heaviside's step distribution and $\Lambda$ is the Cosmological Constant.
Models in this class are characterized by the existence of one
transition scalar curvature $R_0$. For $R=0$, $f(0)=0$ and we recover the limit of Special Relativity. When $R\gg R_0$, $f(R)\simeq -2\Lambda$ and we mimic the $\Lambda$CDM Model. 

These models
contain a sort of `switching on' of the Cosmological Constant as a function
of the scalar curvature $R$. The simplest version of this kind reads 
\begin{equation}
f(R)=-2\Lambda(1-\mathrm{e}^{-\frac{R}{R_0}})\,,\label{OneStepModel}
\end{equation}
where the transition appears around $R_0$.
\begin{figure}
\begin{center}
\unitlength=1mm
\begin{picture}(100,60)
\put(10,10){\vector(1,0){80}}
\put(10,10){\vector(0,1){40}}
\put(95,10){\makebox(0,0){$R$}}
\put(10,55){\makebox(0,0){$-f(R)$}}
\put(35,5){\makebox(0,0){$R_0$}}
\put(5,41){\makebox(0,0){$2\Lambda$}}
\thicklines
\qbezier(10,10)(34,11)(35,25)
\qbezier(35,25)(36,39)(50,40)
\qbezier(50,40)(64,41)(90,41)
\thinlines
\put(35,25){\line(0,-1){15}}
\put(10,41){\line(1,0){80}}
\end{picture}
\end{center}
\caption{\label{Fig1} Typical behavior of $f(R)$-gravity for dark energy epoch.}
\end{figure}
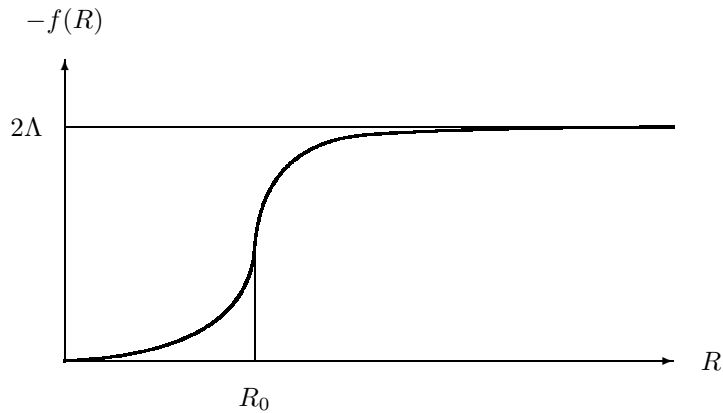

\begin{figure}
\begin{center}
\unitlength=1mm
\begin{picture}(130,80)
\put(10,10){\vector(1,0){110}}
\put(10,10){\vector(0,1){65}}
\put(125,10){\makebox(0,0){$R$}}
\put(10,80){\makebox(0,0){$-f(R)$}}
\put(25,5){\makebox(0,0){$R_0$}}
\put(5,30){\makebox(0,0){$2\Lambda$}}
\put(70,5){\makebox(0,0){$R_i$}}
\put(5,70){\makebox(0,0){$2\Lambda_{\rm eff}$}}
\thicklines
\qbezier(10,10)(24.5,10.5)(25,20)
\qbezier(25,20)(25.5,29.5)(40,30)
\qbezier(40,30)(69,31)(70,45)
\qbezier(70,45)(71,69)(120,70)
\thinlines
\put(25,20){\line(0,-1){10}}
\put(40,30){\line(-1,0){30}}
\put(70,45){\line(0,-1){35}}
\put(120,70){\line(-1,0){110}}
\end{picture}
\end{center}
\caption{\label{Fig2} Typical behavior of $f(R)$-gravity for inflation and dark energy epoch.}
\end{figure}
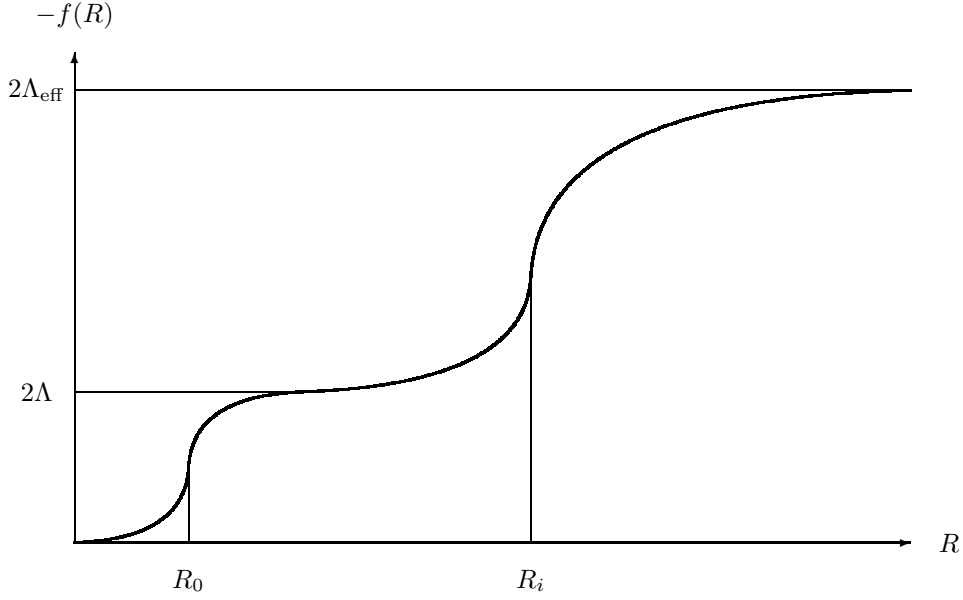

Moreover, models in the form of Eq.~(\ref{toymodel}) may be combined in a natural
way, if one is also interested in the phenomenological description of the
inflationary epoch. For example, a `two-steps' model may be the smooth version of \\
\phantom{line}
\begin{equation*}
F(R)=R+f(R)\,, 
\end{equation*}
\phantom{line}
\begin{equation}
f(R)=-2\Lambda\, \theta(R-R_0)\,-\Lambda_{\rm i} \, \theta(R-R_i)\,.\label{toymodel2}
\end{equation}
\phantom{line}\\
Here, $R_i$ is the transition scalar curvature at inflationary scale, and $\Lambda_{\rm i}$ is a suitable Cosmological Constant producing the acceleration of inflation, when $R\gg R_i$. Thus, the effective Cosmological Constant $\Lambda_{\rm eff}$ at inflation results to be 
\begin{equation}
\Lambda_{\rm eff}=\Lambda+\Lambda_{\rm i}/2\,.\label{Lambdaeff}
\end{equation}
The typical behavior of $f(R)$ associated with (\ref{toymodel}) and (\ref{toymodel2}) models is
given in Fig.~\ref{Fig1} and Fig.~\ref{Fig2}, respectively.
The main problem which may affect these sharp models is the appearance of
possible antigravity regime around the transition between the inflation and the $\Lambda$CDM region and
antigravity in a past epoch, what is not phenomenologically acceptable. Furthermore, adding some terms would be necessary in order for 
inflation to end. 
The exponential gravity seems to give a viable possibility to unify early- and late-time acceleration as in Eq.~(\ref{toymodel2}) and in \S~\ref{inflation} we will study two applications of it to achieve such unified description.

\paragraph*{} We conclude this Section by showing other two (more complicate) exponential models which can be used to reproduce the Cosmological Constant at high curvatures, namely\\
\phantom{line}  
\begin{equation}
f_2(R)=2\Lambda\left(\frac{1+\mathrm{e}^{-\beta R_0}}{1+\mathrm{e}^{-\beta(R_0-R)}}-1\right)\,,
\end{equation}
\phantom{line}
\begin{eqnarray}
 f_3(R)&=&-2\Lambda\left[\tanh\left(\frac{\beta}{2}(R-R_0)\right)+\tanh\left(\frac{\beta}{2} R_0\right)\right]\nonumber\\ \nonumber\\
&=&- 2\Lambda\left(\frac{\mathrm{e}^{\beta\left(R-R_0\right)} - 1}{\mathrm{e}^{\beta\left(R-R_0\right)} + 1}
+ \frac{\mathrm{e}^{\beta R_0} - 1}{\mathrm{e}^{\beta R_0} + 1}\right)\,.
\end{eqnarray}
\phantom{line}\\
Here, $\beta$ is a positive parameter which regulates the amplitude of the transition between the region $R<R_0$ and the region $R_0<R$. 
The advantage of this models is given by the possibility to pass in an analytical way to the scalar tensor framework~{\cite{Twostepmodels0}}.

\section{Realistic exponential gravity\label{exponential}}

\paragraph*{} Let us start by analyzing the exponential model of Eq.~(\ref{OneStepModel}) for dark energy epoch\\
\phantom{line} 
\begin{equation}
F(R)=R-2\Lambda \left(1-\mathrm{e}^{-R/R_0}\right)\,,\label{model}
\end{equation}
\phantom{line}\\
where the curvature parameter $R_0$ is on the same order of the Cosmological Constant,
$R_{0}\sim \Lambda(\simeq 10^{-66}\text{eV}^2)$. In flat space $F(0)=0$
and one recovers the Minkowski's solution. For $R \gg R_{0}$, $F(R)\simeq
R-2\Lambda$, and the theory mimics the $\Lambda$CDM Model. 
Then, we have:
\begin{equation}
F'(R)=1-2\frac{\Lambda}{R_{0}}\mathrm{e}^{-R/R_0}\,,
\quad
F''(R)=2\frac{\Lambda}{R_{0}^2}\mathrm{e}^{-R/R_0}\,.
\end{equation}
It is remarkable that exponential model in Eq.~(\ref{model}) corresponds
to a
polynomial modification of gravity without a true cosmological
constant. One
can write
\begin{equation}
F(R)=R+2\Lambda\sum_{k=1}^{+\infty}\frac{(-1)^{k}}{k!}\left(\frac{R}{R_{0}}\right)^k\,,
\end{equation} 
and modified gravity can be viewed as a correction to the Einstein's gravity given by a sum of power terms which become relevant at different scales of energy.

Since $F'(R \gg R_{0})>0$, the model is protected against
anti-gravity during the
cosmological evolution until the de Sitter solution ($R_{\mathrm{dS}}= 4\Lambda$)
of
today's universe is reached.
For large values of the curvature, $F(R \gg R_{0})\simeq R$ and we
can
reconstruct the matter-dominated era as in GR. In particular, $F''(R)>0$, and we do not have any
instability problems related to the matter epoch, obtaining matter stability
on a planet's surface of Eq.~(\ref{Ustab}), where $U(R_b)\sim1/[3F''(R_b)]$, and at the Solar System scale.

\begin{figure}[-!h]
\begin{center}
\includegraphics[angle=0, width=0.5\textwidth]{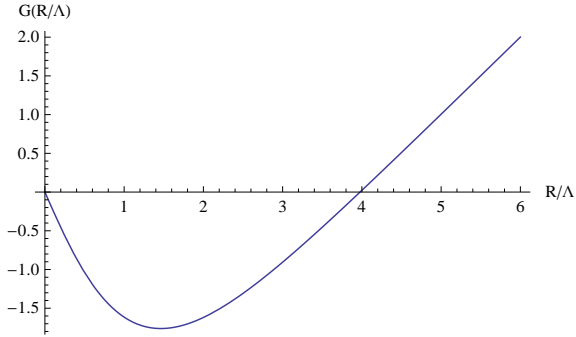}
\caption{Plot of $G(R/\Lambda)$ of exponential model for $R_0=0.6\Lambda$.
The ``zeros'' of these graphics indicate the de Sitter solutions of the model.\label{Fig3}}
\end{center}
\end{figure}

In order to study the de Sitter era, it is convenient to introduce the following function, 
\begin{equation}
G(R)=2F(R)-RF'(R)\,,\label{G(R)}
\end{equation}
whose zeros correspond to the de Sitter solutions for accelerated expansion
(see Eq.~(\ref{dScondition})). 
In our case,
since $G(0)=0$, we immediatly get the trivial de Sitter/Minkowski solution at $R=0$.
Consider now
\begin{equation}
G'(R)=F'(R)-RF''(R)\,.
\end{equation}
If $G'(0)>0$, the function $G(R)$ becomes positive and it is quite simple to see that any non-trivial zero (i.e. de Sitter solution) exists. In order to obtain the de Sitter solution describing universe today, we must require
\begin{equation}
R_{0}<2\Lambda\,,\quad G'(0)<0\,,\label{R0condition}
\end{equation}
so that $G(R)$ becomes negative and starts to
increase
after $R= R_{0}$.
For $R=4\Lambda$, $F(R)\simeq R-2\Lambda$, $F'(R)\simeq 1$ and
$F''(R)\simeq
0^+$.
It means that $G(4\Lambda)\simeq 0$ and we get the de Sitter
solution of dark energy epoch.
Then, $G(R>4\Lambda)\simeq R$ is positive and we do
not find
other de Sitter solutions. Note that dark energy epoch 
is stable, since stability condition (\ref{dSstability}) leads to
\begin{equation}
\left(\frac{R_0}{\Lambda}\right)^2\frac{e^{\left(\frac{4\Lambda}{R_0}\right)}}{2}-\left(\frac{R_0}{\Lambda}\right)>4\,. 
\end{equation}
This condition always is satisfied and $G'(4\Lambda)>0$.
On the other hand, as a consequence of (\ref{R0condition}) which leads to $G'(0)<0$, the Minkowski space is unstable, but it is not a problem in respect of the viable representation of the universe. In Fig.~\ref{Fig3} the graphic of $G(R/\Lambda)$ for the case $R_0=0.6\Lambda$ is shown. Summing up,
we have two FRW-vacuum solutions, which correspond to the Minkowski's space at $R=0$ and to the stable de Sitter point at $R=4\Lambda$.

Finally, we have to consider the existence of spherically symmetric
solutions. At $R=0$ we get the Schwarzschild solution, which is an
unstable solution. On
the other hand, the physical Schwarzschild-de Sitter solution, which also describes our Solar System, is
obtained
at $R \gg R_{0}$. For example, at $R^*\simeq
10^{-61}
\text{eV}^2$, 
$F(R^*)\simeq R^*-2\Lambda$, being $R^*\gg R_0(\sim\Lambda)$.  Thus, stability condition
(\ref{dSstability}) is verified for $R=R^*$, and the
solution results
stable without leading to any significant modification to the Newton Law.

The description of the cosmological evolution in exponential gravity
has
been carefully studied in Refs.~\cite{Linder,Bamba,Twostepmodels, altri}, where it has been
explicitly demonstrated that the late-time cosmic acceleration
following the
matter-dominated stage, as final attractor of the universe, can
indeed be
realized. By carefully fitting the value of $R_{0}$, the correct
rate between matter and dark energy of the current universe
follows. We will analyze the dynamic of exponential models in 
Chapter {\bf 7}.
Here, we mention that in Ref.~\cite{Yang:2011cp}, 
the gravitational waves in viable $F(R)$ models have been studied, 
and that the observational constraints on exponential gravity 
have also been examined in Ref.~\cite{Yang:2010xq}.
As our next step, we want to generalize the model in order to
describe inflation.

\section{Exponential gravity to describe the inflation\label{5.3.1}\label{inflation}}

\paragraph*{} A simple modification of the `one-step' model which incorporates the
inflationary era is given by a combination of the function discussed
in the previous Section
with another `one-step' function reproducing the cosmological constant
during
inflation, as in Eq. (\ref{toymodel2}). 

Following the first proposal of Ref.~\cite{Twostepmodels}, 
we start with the following model,
\begin{equation}
F(R)=R-2\Lambda\left(1-\mathrm{e}^{-\frac{R}{R_0}}\right)
-\Lambda_\mathrm{i}\left[1-\mathrm{e}^{-\left(\frac{R}{R_\mathrm{i}}\right)^n}\right]
+\bar{\gamma} \left(\frac{1}{\tilde R_\mathrm{i}^{\alpha-1}}\right)R^\alpha\,.
\label{total}
\end{equation}
Here, $R_{i}$ and $\Lambda_{\rm i}$ assume the typical values of the
curvature
and expected cosmological constant during inflation, namely $R_{i}$,
$\Lambda_{\rm i}$($\simeq 10^{100-120}\Lambda$) $\sim 10^{20-38} \text{eV}^2$ (note that, since $\Lambda\ll\Lambda_{\mathrm{i}}$, Eq.~(\ref{Lambdaeff}) leads to $\Lambda_{\mathrm{eff}}\simeq\Lambda_{\mathrm{i}}/2$), while $n$ is a natural
number
larger than one. 
In this equation, 
the last term  
$\bar{\gamma}(1/\tilde R_\mathrm{i}^{\alpha-1}) R^\alpha$, 
where $\bar{\gamma}$ 
is a positive dimensional constant and $\alpha$ is a real number, 
works at the inflation scale $\tilde R_\mathrm{i}$, and 
is actually necessary in order to realize an exit from inflation. 

We also introduce another nice inflation model based on the good behavior of exponential function described as \cite{malloppone}
\begin{equation}
F(R)=R-2\Lambda\left(1-\mathrm{e}^{-\frac{R}{R_0}}\right)-\Lambda_\mathrm{i}\frac{\sin\left(\pi\,\mathrm{e}^{-\left(\frac{R}{R_\mathrm{i}}\right)^n}\right)}{\pi\,\mathrm{e}^{-\left(\frac{R}{R_\mathrm{i}}\right)^n}}+\bar{\gamma} \left(\frac{1}{\tilde R_\mathrm{i}^{\alpha-1}}\right)R^\alpha\,.
\label{sin}
\end{equation}
Here, the parameters have the same roles of the corresponding ones 
in the model in Eq.~(\ref{total}). 
We note that the second term of the model vanishes when $R\ll R_i$ and 
tends to $\Lambda_\mathrm{i}$ when $R_\mathrm{i}\ll R$. 
We analyze these models, i.e., Model I in Eq.~(\ref{total}) and Model II 
in Eq.~(\ref{sin}), and explore the possibilities to reproduce 
the phenomenologically acceptable inflation. 

\subsection{Inflation in Model I\label{inflation1}}

\paragraph{} First, we investigate the model in Eq.~(\ref{total}). 
For simplicity, we describe a part of it as 
\begin{equation}
f_\mathrm{i}(R) \equiv 
-\Lambda_\mathrm{i}\left(1-\mathrm{e}^{-\left(\frac{R}{R_\mathrm{i}}\right)^n}\right)+\bar{\gamma}\left(\frac{1}{\tilde R_\mathrm{i}^{\alpha-1}}\right)R^\alpha\,,\quad
f_{i}(0)=0\,.
\end{equation}
We note that, 
if $n>1$ and $\alpha>1$, we
avoid the effects
of inflation during the matter era,
when $R\ll R_{\mathrm i}\,,\tilde R_\mathrm{i}$, due to the fact that
\begin{equation}
R \gg \left|f_\mathrm{i}(R)\right| \simeq \left|-\frac{R^{n}}{R_\mathrm{i}^{n-1}}+\bar{\gamma}\frac{R^\alpha}{\tilde R_\mathrm{i}^{\alpha-1}}\right|\,,
\quad R\ll R_{\mathrm i}\,,\tilde R_\mathrm{i}\,.
\end{equation}
We also find 
\begin{eqnarray}
f_\mathrm{i}'(R) &=& -\frac{\Lambda_\mathrm{i} n
R^{n-1}}{R_\mathrm{i}^n}\mathrm{e}^{-\left(\frac{R}{R_\mathrm{i}}\right)^n}+\bar{\gamma}\alpha\,\left(\frac{R}{\tilde R_\mathrm{i}}\right)^{\alpha-1}\,, \nonumber\\
f_\mathrm{i}''(R) &=& \left[-\frac{\Lambda_\mathrm{i} 
n(n-1)R^{n-2}}{R_\mathrm{i}^n}
+\Lambda_i\left(\frac{n R^{n-1}}{R_\mathrm{i}^n}\right)^2\right]
\mathrm{e}^{-\left(\frac{R}{R_\mathrm{i}}\right)^n}+\bar{\gamma}\alpha(\alpha-1)\frac{R^{\alpha-2}}{\tilde R_\mathrm{i}^{\alpha-1}}\,.
\label{secondderivative}
\end{eqnarray}
Since when $R=R_\mathrm{i}\left[(n-1)/n\right]^{1/n}$ 
the negative term of $f_\mathrm{i}'(R)$ has a minimum, 
in order to avoid the anti-gravity effects, it means, in order to have $|f'_\mathrm{i}(R)|<1$,
it is sufficient to require 
\begin{equation}
R_\mathrm{i}>\Lambda_\mathrm{i} n
\left(\frac{n-1}{n}\right)^{\frac{n-1}{n}}\mathrm{e}^{-\frac{n-1}{n}}\,.
\label{unol}
\end{equation}
It is necessary for the modification of gravity describing inflation 
not to have any influence on the stability of the matter dominated era 
in the small curvature limit. 
When $R \ll R_\mathrm{i}\,,\tilde R_i$, the second derivative of $f_\mathrm{i}''(R)$, 
given by 
\begin{equation}
f''_\mathrm{i}(R)\simeq
\frac{1}{R}\left[-n(n-1)\left(\frac{R}{R_\mathrm{i}}\right)^{n-1}
+\bar{\gamma} \alpha(\alpha-1)\left(\frac{R}{\tilde R_\mathrm{i}}\right)^{\alpha-1}\right]\,,
\quad R\ll R_i\,,\tilde R_i\,,
\label{zwei}
\end{equation}
must be positive, that is, 
\begin{equation}
n>\alpha\,.
\end{equation}
Now we require the existence of the de Sitter critical point
$R_{\mathrm{dS}}$
which describes early time acceleration at the scale of inflation, when\\
\phantom{line}
\begin{eqnarray}
&&f_\mathrm{i}(R_{\mathrm{dS}})\simeq -\Lambda_\mathrm{i}+\bar{\gamma} \left(\frac{1}{\tilde R_\mathrm{i}^{\alpha-1}}\right)R^\alpha\,,\quad
f'_\mathrm{i}(R_{\mathrm{dS}})\simeq
\bar{\gamma}\,\alpha\left(\frac{R}{\tilde R_{\mathrm{i}}}\right)^{\alpha-1}\,,\nonumber\\
&&f''_\mathrm{i}(R_{\mathrm{dS}})\simeq
\bar{\gamma}\,\alpha(\alpha-1)\left(\frac{1}{\tilde R_{\mathrm{i}}^{\alpha-1}}\right)R^{\alpha-2}\,,\quad\quad\quad\,\,\,
\left(\frac{R_{\mathrm{dS}}}{R_\mathrm{i}}\right)^n\gg 1\,.
\end{eqnarray}
\phantom{line}\\
In this region, the
role of
the first term in Eq.~(\ref{total}) is negligible. For simplicity, we shall
assume that 
\begin{equation}
\tilde R_\mathrm{i}=R_{\mathrm{dS}}\,.
\end{equation}
The function $G(R)$ of Eq.~(\ref{G(R)}) now reads
\begin{equation}
G(R)\simeq R-2\Lambda_{\mathrm{i}}+
\bar{\gamma}\frac{(2-\alpha)}{R_{\mathrm{dS}}^{\alpha-1}}R^\alpha
+\Lambda_\mathrm{i} n
\left(\frac{R}{R_{\mathrm{i}}}\right)^n\mathrm{e}^{-\left(\frac{R}{R_\mathrm{i}}\right)^n}\,,\label{ughino}
\end{equation}
and has to vanish on the de Sitter solution of the inflation, when the exponential term can be dropped out. As a consequence, we get
\begin{equation}
R_{\mathrm{dS}}=\frac{2\Lambda_\mathrm{i}}{\bar{\gamma}(2-\alpha)+1}\,,
\quad
\left(\frac{R_{\mathrm{dS}}}{R_\mathrm{i}}\right)^n \gg 1\,.
\label{duel}
\end{equation}
The last two conditions have to be satisfied simultaneously. 
By using Eq.~(\ref{unol}), we also acquire 
\begin{equation}
\frac{2}{\bar{\gamma}(2-\alpha)+1}>n
\left(\frac{n-1}{n}\right)^{\frac{n-1}{n}}\mathrm{e}^{-\frac{n-1}{n}}\,.
\label{unouno}
\end{equation}
In order to check the instability of the inflation, 
we must violate condition (\ref{dSstability}) 
on
the de
Sitter solution,
which leads to
\begin{equation}
\alpha\,\bar{\gamma}(\alpha-2)>1\,, 
\label{duedue}
\end{equation}
for our model. From Eqs.~(\ref{unouno})(\ref{duedue}), we observe 
\begin{equation}
2+1/\bar{\gamma}>\alpha>2\,.
\end{equation} 
Let us have a look on the evolution of the function $G(R)$
in Eq.~(\ref{ughino}) as we did in the previous Section for dark energy sector.
When $R=0$, we find the Minkowski solution and $G(0)=0$.
For the first derivative of $G(R)$, one has\\
\phantom{line}
\begin{equation}
G'(R)\simeq 1
+\bar\gamma\alpha(2-\alpha)\frac{R^{\alpha-1}}{R_{\mathrm{dS}}^{\alpha-1}}
+\Lambda_{\mathrm i}
\left[\frac{
n(n-1)R^{n-1}}{R_\mathrm{i}^n}
-\left(\frac{n R^{2n-1}}{R_\mathrm{i}^{2n}}\right)
-\frac{ n
R^{n-1}}{R_\mathrm{i}^n}
\right]
\mathrm{e}^{-\left(\frac{R}{R_\mathrm{i}}\right)^n}\,.
\label{ugo}
\end{equation}
\phantom{line}\\
Since $G'(0)>0$, $G(R)$ increases. Then, due to the fact that the term in the square brackets starts being
negative for $R>R_\mathrm{i}[(n-1)n]^{1/n}$, and
$0>(2-\alpha)$,
it is easy to see that $G(R)$ begins to decrease at around $R=R_i$
and 
vanishes at $R=R_{\mathrm{dS}}$.
After this point, $G'(R>R_{\mathrm{dS}})<0$ and we do not recover other
de
Sitter solutions.
On the other hand, it is possible to have a fluctuation of $G(R)$
along the
$R$-axis just before the de Sitter point describing inflation takes
over. In
order to avoid other de Sitter solutions (i.e., possible final
attractors
for the system), we need to verify the fulfillment of the following
condition:
\begin{equation}
G(R)>0\,,\quad0<R<R_{\mathrm{dS}}\,.
\end{equation}
Precise analysis of this condition leads to a transcendental
equation.
Below, we will limit ourselves to a graphical
evaluation.
In general, it will be sufficient to choose $n$ sufficiently large in
order
to avoid such effects.\\

\begin{figure}[!h]
\subfigure[]{\includegraphics[width=0.3\textwidth]{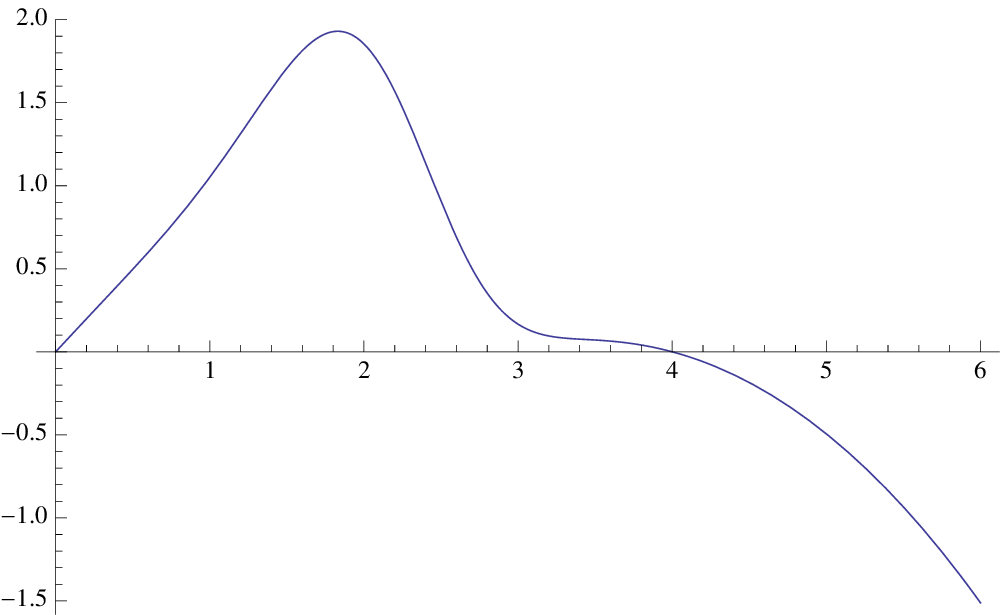}}
\centering
\subfigure[]{\includegraphics[width=0.3\textwidth]{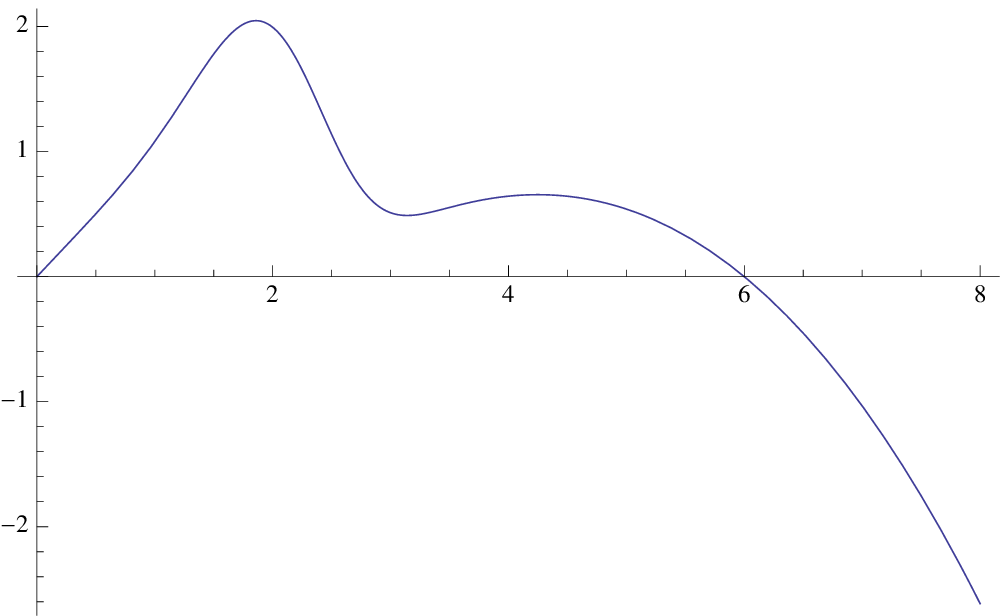}}
\centering
\subfigure[]{\includegraphics[width=0.3\textwidth]{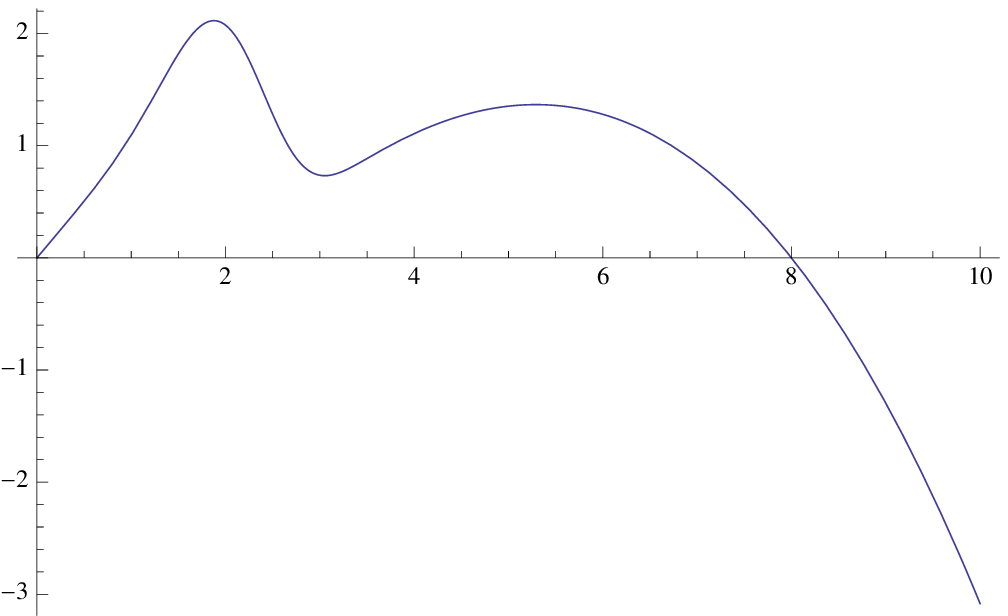}}
\caption{
Plot of $G(R/\Lambda_{\mathrm i})$
for Model I with $\alpha=5/2$ (a), $\alpha=8/3$ (b) and $\alpha=11/4$ (c). The ``zeros'' of these graphics indicate the de Sitter solutions of the model. 
\label{00}}
\end{figure}

After the analysis carried out in the first part of this Subsection, we finally propose some viable choices of the parameters for Model I (\ref{total}).  
Here, we summarize the conditions for inflation already stated:
\begin{eqnarray}
&&
R_\mathrm{i}>\Lambda_\mathrm{i} n
\left(\frac{n-1}{n}\right)^{\frac{n-1}{n}}\mathrm{e}^{-\frac{n-1}{n}}\,,\quad\quad\quad\quad\quad\quad\quad\,\,\,\,\text{(no antigravity effects);}\nonumber\\ 
&&
\tilde R_\mathrm{i}=R_{\mathrm{dS}}\,,\quad\alpha\bar{\gamma}(\alpha-2)>1\,,\quad\left(\frac{R_{\mathrm{dS}}}{R_\mathrm{i}}\right)^n\gg 1\,,\quad\text{(existence of unstable dS solution).} \nonumber
\end{eqnarray}
In this expressions, $n>1$, $2+1/\bar{\gamma}>\alpha>2$ and $R_{\mathrm{dS}}=2\Lambda_\mathrm{i}/\left[\bar{\gamma}(2-\alpha)+1\right]$. Since $\bar{\gamma}$ and $\alpha$ are combined in $\bar\gamma(\alpha-2)$, we can fix $\bar{\gamma}=1$, so that $R_{\mathrm{dS}}=2\Lambda_\mathrm{i}/(3-\alpha)$ and $3>\alpha>2$. 
Hence, 
in order to reproduce and study the phenomenology of (different) realistic inflationary scenarios, 
one may examine the effects of the variation of $\alpha$ parameter (and, as a consequence, that of $\tilde R_\mathrm{i}$). 
Thus, a reasonable choice is to take $n=4$ and $R_\mathrm{i}=2\Lambda_\mathrm{i}$, which satisfy the condition for no antigravity well. 
In Chapter {\bf 7} we will analyze three different cases of $\alpha = 5/2$, $8/3$, and $11/4$. 
In these cases, we have 
$R_{\mathrm{dS}}=4\Lambda_\mathrm{i}$, $6\Lambda_\mathrm{i}$, and $8\Lambda_\mathrm{i}$, respectively. 
For the sake of completeness,
in Fig.~\ref{00} we plot the cosmological evolution of 
$G(R/\Lambda_{\mathrm{i}})$
in the three cases and with $R_0=0.6\Lambda$ ($\Lambda$ is taken as $\Lambda=10^{-100}\Lambda_\mathrm{i}$). The ``zeros'' of $G(R/\Lambda_\mathrm{i})$ correspond to the de Sitter points of the model. We can recognize the unstable de Sitter solutions of inflation and the attractor in zero (the de Sitter point of current acceleration is obviously out of scale).
The system gives rise to the de Sitter solution where the universe
expands in an accelerating way but, suddenly, it exits from inflation and tends towards
the minimal attractor at $R=0$, unless the theory develops a singularity solution for $R\rightarrow+\infty$. In such a case, the model could exit from inflation and move in the
wrong direction, where the curvature would grow up and diverge, and a singularity would appear. In \S~\ref{sing} singularities
will be considered in the context of exponential gravity. We will see that the theory is free of such future-time singularities.

\subsection{Inflation in Model II\label{inflation2}}

\paragraph{} Next, we study the inflation model in Eq.~(\ref{sin}). 
By performing a similar analysis to that in the previous Subsection, 
we find that also in this case, if $\alpha>1$ and $n>1$, we avoid the effects 
of inflation at small curvatures and it does not influence the stability of 
the matter dominated era. 
The de Sitter point exists if $\tilde R_\mathrm{i}=R_{\mathrm{dS}}$ and 
it reads as in Eq.~(\ref{duel}) under the condition $(R/R_\mathrm{i})^n\gg 1$. 
Thus, the inflation is unstable if the condition in Eq.~(\ref{duedue}) is 
satisfied. 
The bigger difference between the two models exists in those behaviors 
in the transition phase between the small curvature region 
(where the physics of the $\Lambda$CDM Model emerges) 
and the high curvature region. 
This means that the no antigravity condition is different in the two models 
and such a condition becomes more critical in the transition region. 
Therefore, we are able to 
make different choices of parameters in the two models. 
In the specific, in order to satisfy the condition for no antigravity 
we may choose $n=3$ and $R_\mathrm{i}=2\Lambda_\mathrm{i}$. 
Then, we can take $\bar{\gamma}=1$ as in the Model I and analyze the cases
$\alpha=5/2\,, 13/5\,, 21/8$, for which we will
execute the numerical evaluation in Chapter {\bf 7}.
The corresponding de Sitter curvatures of inflation are 
$R_{\mathrm{dS}}=4\Lambda_\mathrm{i}\,, 5\Lambda_\mathrm{i}\,, 16\Lambda_\mathrm{i}/3$.

We note that 
since dark energy sector of the above models only originates from the first part of (\ref{total})--(\ref{sin}), 
all qualitative results in terms of the behavior of the 
dark energy component in exponential gravity 
found in \S~{\ref{exponential}} remain to be valid.

\section{Singularities in exponential gravity\label{sing}}

\paragraph*{} In \S~\ref{curingf(R)} we have seen that the terms of the type $\gamma R^\alpha$ with $\alpha>1$ protect the theory against singularities occurring when $R\rightarrow\pm\infty$, it means Type I, II, III singularities, and the Big Rip. As a consequence, our models in Eqs.~(\ref{total})--(\ref{sin}) are free of catastrophic divergences in the curvature, due to the contribution of the last power term necessary to exit from inflation. A remark is in order. Since in `one-step' models for large values of curvature the high derivatives of $F(R)$ tend to zero, it is easy to see that effective energy density never diverges, and the Big Rip and the Type I and III singularities are not realized. Nevertheless, the Type II singularity may affect this kind of models. For example, in \S~\ref{3.4.1}, we explicit found the possibility to realize the Type II singularity in the Hu-Sawiki Model. This fact suggests the presence of high power terms of the curvature, which prevents such singularity and induces also inflationary effects.
 
In the next Section we will see that, in the very asymptotic limit $R\gg R_i$, our models may  exhibit a (disconnected) singularity. Since the de Sitter solution of inflation is unstable, we can ask if the models could move toward such extreme limit. We will provide an argument to exclude this possibility.   

Thus, we have found that our theories are free from singularities where curvature diverges. In
particular, when we are in the region of inflation, catastrophic
Types I, II or III singularities do not occur. When inflation ends, it is reasonable to suppose that
the models move to the attractors at $R\rightarrow0$. In this
way, the small curvature regime arises, the first term of
Eqs.~(\ref{total})--(\ref{sin})
becomes dominant and the physics of the $\Lambda$CDM model is
reproduced.
 
However, 
it is important to remark that 
in these models of $F(R)$-gravity, when singularity appears 
at high curvature,
the solution for the problem is adding the higher 
powers of $R$ so that the behavior at large curvatures can be soften, 
but this argument is applicable to 
the Type I, II and III finite-time 
future singularities only. The Type IV singularities, where only some derivatives of Hubble parameter diverge,
occur 
at small curvatures, 
and the higher power of $R$ are avoidable
in this range of detection: for this reason, it may be necessary
to introduce in the theory lower power term of $R$.
This problem
will be analyzed in Chapter {\bf 7} in the context of DE-oscillations in matter era.

\section{Asymptotic behavior}

\paragraph*{} As the last issue, we will analyze the solutions of our models when $R$ is
very large in comparison with the curvature $R_i$ of inflation.
This means that in Eqs.~(\ref{total})--(\ref{sin}) the last term is dominant due to the viability conditions before analyzed and we get
\begin{equation}
F(R)\simeq \bar{\gamma} \left(\frac{1}{\tilde R_\mathrm{i}^{\alpha-1}}\right)R^\alpha\,,\quad
R\gg R_\mathrm i, \tilde R_{\mathrm i}\,.
\label{approx}
\end{equation}
By using Eq.~(\ref{EOM1}) and Eq.~(\ref{R}) for $F(R)$-gravity in vacuum, one can deal with a first order differential autonomous system in $R$ and $H$ as\\
\phantom{line}
\begin{equation}
\left\{\begin{array}{l}
\dot{R}=-\frac{1}{F''(R)}\left(H F'(R)+\frac{F(R)-RF'(R)}{6H}\right)\,,\\ \\
\dot{H}=\frac{R}{6}-2H^2\,.
\end{array}
\right.
\end{equation}
\phantom{line}\\
For the model of
Eq.~(\ref{approx}), at the limit $t\rightarrow 0^+$, we can find the following solutions:
\begin{equation}
H(t)\simeq\frac{H_{0}}{(t_0-t)^{\beta}}\,,
\quad
R(t)\simeq 12\frac{H_0^2}{(t_0-t)^{2\beta}}\,.
\end{equation}
Here, $t<t_0$ and $H_{0}$ is a large positive constant. Moreover, $\beta$ is a positive
parameter
so that $\beta=1$ or $\beta>1$, and the solution corresponds to a Type I (or Big Rip) singularity.

This result shows that in the limit $R\rightarrow+\infty$
the model exhibits a past singularity, which could be identified with
the Big Bang one if we consider $t_0\rightarrow-\infty$ or we replace $(t_0-t)\rightarrow t$. However, we may also have a future-time singularity solution.  
It is important to stress that this kind of solution is disconnected
from the de Sitter phase of inflation, where the term $R$ is of the
same order of $\bar\gamma R^\alpha/\tilde R_{\mathrm i}^{\alpha-1}$ and is therefore not negligible as in Eq.~(\ref{approx}).
Furthermore, in the very asymptotic limit, the scalaron $F'(R)$ results to be
\begin{equation}
F'(R)\simeq \bar{\gamma}\,\alpha \left(\frac{R}{\tilde R_\mathrm{i}}\right)^{\alpha-1}\,,\quad
R\gg R_\mathrm i, \tilde R_{\mathrm i}\,.
\end{equation}
We can also evaluate the potential $V_{\rm eff}$ of Eq.~(\ref{Veff}), through integration of $\partial V_{\rm eff}(R)/\partial R\equiv F''(R)[\partial V_{\rm eff}(R)/\partial F'(R)]$. By neglecting the contribute of matter, one easily finds 
\begin{equation}
V_{\rm eff}(R)\simeq\frac{\gamma^2\alpha(\alpha-1)(2-\alpha)}{3(2\alpha-1)}\left(\frac{1}{\tilde R_i^{\alpha-1}}\right)^2 R^{(2\alpha-1)}\,,
\quad
R\gg R_\mathrm i, \tilde R_{\mathrm i}\,.
\end{equation}
We observe that, in order to reach the singularity, the scalaron has to crossover an infinite potential barrier ($V_{\rm eff}(R\rightarrow\infty)\rightarrow\infty$) and go to infinity ($F'(R\rightarrow\infty)\rightarrow\infty$), but clearly this
dynamical behavior is forbidden. 

We may safely assume that, just after the Big Bang, a Planck epoch takes over where
physics is not described by GR and where quantum gravity effects are
dominant. When the universe exits from the Planck epoch, its
curvature is bound to be the characteristic curvature of inflation
and the unstable de Sitter solution takes over.


\chapter{Cosmic history of viable $F(R)$-gravity: dark energy
oscillations, future evolution and growth index}

\paragraph*{} 
In this Chapter, we study the feature of dark energy in the fluid representation of viable $F(R)$-modified gravity.
Many viable models lead to an oscillating behaviour of the dark energy during de Sitter and matter epoch. 
In particular, in matter dominated era, the
large frequency oscillations may influence 
the behavior of higher derivatives of the Hubble parameter 
with the risk to produce some singular unphysical solutions at high redshift. 
This behavior is typical of realistic $F(R)$-gravity, as we will explicit show
by considering exponentail gravity and 
Hu-Sawiki model (as an example of power  
form model). 
To stabilize such oscillations, we may introduce an additional modification of 
the models via a correction term which does not destroy 
the viability properties. 
In the second part of the Chapter, a detailed analysis on the future evolution of the universe and 
the evolution history of the growth index of the matter density perturbations 
are performed via numerical evaluation.

\section{Dark energy in the de Sitter universe}

\paragraph*{} Here, we are interested in the analysis of cosmological behavior of realistic $F(R)$-models of modified gravity describing dark energy epoch. The tag `realistic' has been defined in Chapter~{\bf 6} and has to do with the feasibility of the models in view the all the most recent and accurate observational data. 
In particular, we will still consider modified gravity in the form $F(R)=R+f(R)$, by explicitly separating the contribution of GR from its modification. Up to now, with this class of models, it is possible to turn out the most realistic reproduction of our universe, as in the case of exponential gravity previously analized.   

Let us consider the effects of modified gravity and matter together as the ones of an effective fluid with energy density and pressure given by Eqs.~(\ref{rhoeffRG})--(\ref{peffRG}). Now, for $F(R)$-gravity, we define the dark energy density $\rho_{\mathrm{DE}}$ and the dark energy pressure $p_{\mathrm{DE}}$ as 
\begin{equation}
\rho_{\mathrm{DE}}=\rho_{\mathrm{eff}}-\rho_{\mathrm{m}}\,,\quad
p_{\mathrm{DE}}=p_{\mathrm{eff}}-p_{\mathrm{m}}\,,\label{rhoDE}
\end{equation}
by explicitly separate the contribute of matter. In this way, we obtain a fluid representation of $F(R)$-gravity. Then, we introduce the variable~\cite{HuSaw, Bamba}
\begin{equation}
y_H (z)\equiv\frac{\rho_{\mathrm{DE}}}{\rho_{\mathrm{m}(0)}}=\frac{H^2}{\tilde{m}^2}-(z+1)^3-\chi
(z+1)^{4}\,,\label{y}
\end{equation}
as a function of the redshift parameter 
$z=1/a(t)-1$, such that at the present time the scale factor $a(t)=1$ and $z=0$.
Furthermore, $\rho_{\mathrm{m}(0)}$ is the energy density of matter at the present, while
$\tilde{m}^2$ and $\chi$ are a suitable mass scale and the today's radiation abundance\footnote{Here, we have used the data of Ref.~\cite{WMAP}.}, respectively, and they are defined as
\begin{equation}
\tilde{m}^2\equiv\frac{\kappa^2\rho_{\mathrm{m}(0)}}{3}\simeq 1.5 \times
10^{-67}\text{eV}^2\,,
\quad
\chi\equiv\frac{\rho_{\mathrm{r}(0)}}{\rho_{\mathrm{m}(0)}}\simeq 3.1 \times
10^{-4}\,.\label{scale}
\end{equation}
In the last expression, $\rho_{\mathrm{r}(0)}$ is the energy density of radiation at the present.
The EoS parameter for the dark energy reads
\begin{equation}
\omega_{\mathrm{DE}}\equiv\frac{p_{\mathrm{DE}}}{\rho_{\mathrm{DE}}}=-1+\frac{1}{3}(z+1)\frac{1}{y_H(z)}\frac{d y_H(z)}{d (z)}\,.\label{oo}
\end{equation}
By combining Eq.~(\ref{R}) with Eq.~(\ref{EOM1bis}), and then using
Eq.~(\ref{y}), one gets
\begin{equation}
\frac{d^2 y_H(z)}{d z^2}+J_1\frac{d y_H(z)}{d z}+J_2
\left[y_H(z)\right]+J_3=0\,,\label{superEq}
\end{equation}
where\\
\phantom{line}
\begin{equation}
\left\{\begin{array}{l}
J_1=\frac{1}{(z+1)}\left(-3-\frac{1}{y_H+(z+1)^{3}+\chi (z+1)^{4}}\frac{1-F'(R)}{6\tilde{m}^2
F''(R)}\right)\,,\\ \\
J_2=\frac{1}{(z+1)^2}\left(\frac{1}{y_H+(z+1)^{3}+\chi (z+1)^{4}}\frac{2-F'(R)}{3\tilde{m}^2
F''(R)}\right)\,, \\ \\
J_3=-3 (z+1)-\frac{(1-F'(R))((z+1)^{3}+2\chi (z+1)^{4})
+(R-F(R))/(3\tilde{m}^2)}{(z+1)^2(y_H+(z+1)^{3}+\chi
(z+1)^{4})}\frac{1}{6\tilde{m}^2
F''(R)}\,.
\end{array}
\right.
\end{equation}
\phantom{line}\\
Thus, the Ricci scalar reads\\
\phantom{line}
\begin{equation}
R=3\tilde{m}^2 \left[4y_H(z)-(z+1)\frac{d y_H(z)}{d z}+(z+1)^{3}\right]\,.\label{Ricciscalar}
\end{equation}
\phantom{line}\\
We remember that $d/dt\equiv-(z+1)[H(z)] d/d z=[H(t)] d/d[\ln a(t)]$, where $H$ could be an explicit function of the red shift, $H(z)$, or an explicit function of the time, $H(t)$.
In general, Eq.~(\ref{superEq}) can be solved in a numerical way, once we write the explicit form of the $F(R)$-model. 

Let us study the perturbations around the de Sitter solution $R_{\mathrm{dS}}$ given by Eq.~(\ref{dScondition}), to see that we are able to recover the stability condition~(\ref{dSstability}). We restrict our analysis to homogeneous perturbations. The behavior of general, linear, inhomogeneous perturbations has been discussed in Ref.~\cite{fara}, where the equivalence between the two approaches has been
explicitly shown: as regards this point, see also the independent proof contained in Ref.~\cite{cogno}.

The starting point will be\\
\phantom{line}
\begin{equation}
y_{H}(z)\simeq y_{0}+y_1(z)\,,\label{yexpansion}
\end{equation}
\phantom{line}\\
where $y_0=R_{\mathrm{dS}}/12\tilde{m}^2$ is the constant dark energy of the dS-universe and $|y_1(z)/y_0|\ll 1$. Eq.~(\ref{Ricciscalar}) leads to\\
\phantom{line}
\begin{equation}
R=3\tilde{m}^2 \left[4y_0+4y_1(z)-(z+1)\frac{d y_1(z)}{d z}+(z+1)^3\right]\,.\label{RdS}
\end{equation}
\phantom{line}\\
In this case, by neglecting the contribution of radiation and assuming the matter one to be much smaller than $y_0$, Eq.~(\ref{superEq}) becomes, at first order in $y_1(z)$,\\
\phantom{line}
\begin{equation}
\frac{d^2 y_1(z)}{d z^2}+\frac{\alpha}{(z+1)}\frac{d y_1(z)}{d z}+\frac{\beta}{(z+1)^2}
y_1(z)=4\zeta(z+1)\,,\label{superEqbis}
\end{equation}
\phantom{line}\\
where\\
\phantom{line}
\begin{equation}
\alpha=-2\,,\quad
 \beta= -4+\frac{4F'(R_{\mathrm{dS}})}{RF''(R_{\mathrm{dS}})}\,,\quad
 \zeta=1+\frac{1-F'(R_{\mathrm{dS}})}{R_{\mathrm{dS}}F''(R_{\mathrm{dS}})}\,.\label{zeta}
\end{equation}
\phantom{line}\\
In this derivation, we have used the de Sitter condition (\ref{dScondition}). The solution of Eq.~(\ref{superEqbis}) reads\\
\phantom{line}
\begin{equation}
y_1(z)=C_0(z+1)^{\frac{1}{2}\left(1-\alpha\pm\sqrt{(1-\alpha)^2-4\beta}\right)}+
\frac{4\zeta}{\beta}(z+1)^3\,,\label{result}
\end{equation}
\phantom{line}\\
where $C_0$ is a generic constant. It is easy to see that $y_1(z)\rightarrow 0$ when $z\rightarrow -1^{+}$, and, therefore, the de Sitter solution is stable, provided by Eq.~(\ref{dSstability}), i.e. $F'(R_{\mathrm{dS}})/R_{\mathrm{dS}}F''(R_{\mathrm{dS}})>1$.
Furthermore, we have two possible behaviors of dark energy density  for a  stable de Sitter universe~\cite{starodS}. If\\
\phantom{line}
\begin{equation}
\frac{25}{16}>\frac{F'(R_{\mathrm{dS}})}{R_{\mathrm{dS}}F''(R_{\mathrm{dS}})}>1\,,\label{gamma>0}
\end{equation}
\phantom{line}\\
the solution approaches the de Sitter point as a power function of $(z+1)$, that is $y_1(z)\sim (z+1)^{\gamma}$, $\gamma>0$. Otherwise, if\\
\phantom{line}
\begin{equation}
\frac{F'(R_{\mathrm{dS}})}{R_{\mathrm{dS}}F''(R_{\mathrm{dS}})}>\frac{25}{16}\,,\label{discriminant}
\end{equation}
\phantom{line}\\
the discriminant in the square root of Eq.~(\ref{result}) is negative and the
dark energy density possesses an oscillatory behavior whose amplitude decreases as $(z+1)^{3/2}$ when $z\rightarrow -1^{+}$.
As a consequence, we can write $y_H(z)$ as\\
\phantom{line}
\begin{eqnarray}
\label{oscillatorysolution}&& y_H(z)= \frac{R_{\mathrm{dS}}}{12\tilde{m}^2}+\left(\frac{1}{F'(R_{\mathrm{dS}})-R_{\mathrm{dS}}F''(R_{\mathrm{dS}})}-1\right)(z+1)^3+
(z+1)^{\frac{3}{2}}\times\\ \nonumber\\ \nonumber
&&\hspace{-10mm}\left[A_0\cos\left(\sqrt{\left(\frac{4F'(R_{\mathrm{dS}})}{R_{\mathrm{dS}}F''(R_{\mathrm{dS}})}-
\frac{25}{4}\right)}\log(z+1)\right)+B_0\sin\left(\sqrt{\left(\frac{4F'(R_{\mathrm{dS}})}{R_{\mathrm{dS}}F''(R_{\mathrm{dS}})}-
\frac{25}{4}\right)}\log(z+1)\right)\right]\,,
\end{eqnarray}
\phantom{line}\\
$A_0$ and $B_0$ being constants which depend on the boundary conditions.
Using Eq.~(\ref{oo}), we can also evaluate the EoS DE-parameter
\begin{equation}
 \omega_{\mathrm{DE}}=-1+\frac{4\zeta}{\beta}\frac{(z+1)^3}{y_0}+\frac{1}{3}\gamma\frac{(z+1)^\gamma}{y_0}\,,
\end{equation}
where
\begin{equation}
 \gamma=\frac{1}{2}\left(1-\alpha\pm\sqrt{(1-\alpha)^2-4\beta}\right)\,.\label{gamma}
\end{equation}
In the case of oscillating models which satisfy Eq.~(\ref{discriminant}), one has\\
\phantom{line}
\begin{eqnarray}
\label{omegaoscillating}&&\omega_{\mathrm{DE}}=-1+\frac{12\tilde{m}^2}{R_{\mathrm{dS}}}\left(\frac{1}{F'(R_{\mathrm{dS}})-
R_{\mathrm{dS}}F''(R_{\mathrm{dS}})}-1\right)(z+1)^3+4\tilde{m}^2\frac{(z+1)^{\frac{3}{2}}}{R_{\mathrm{dS}}}\times\\\nonumber\\\nonumber
&&\hspace{-10mm}\left[\tilde A_0\cos\left(\sqrt{\left(\frac{4F'(R_{\mathrm{dS}})}{R_{\mathrm{dS}}F''(R_{\mathrm{dS}})}-\frac{25}{4}\right)}
\log(z+1)\right)+\tilde B_0\sin\left(\sqrt{\left(\frac{4F'(R_{\mathrm{dS}})}{R_{\mathrm{dS}}F''(R_{\mathrm{dS}})}-
\frac{25}{4}\right)}\log(z+1)\right)\right]\,,
\end{eqnarray}
\phantom{line}\\
with
\begin{equation}
\tilde A_0=\frac{3}{2}A_0+\sqrt{\left(\frac{4F'(R_{\mathrm{dS}})}{R_{\mathrm{dS}}F''(R_{\mathrm{dS}})}-\frac{25}{4}\right)}B_0\,,\quad
\tilde B_0=\frac{3}{2}B_0-\sqrt{\left(\frac{4F'(R_{\mathrm{dS}})}{R_{\mathrm{dS}}F''(R_{\mathrm{dS}})}-\frac{25}{4}\right)}A_0\,. \label{u}
\end{equation}
We observe that $\omega_{\mathrm{DE}}$ exhibits the same oscillation period of $y_H(z)$ and that its amplitude is amplified by its frequency, encoded in the coefficients $\tilde A_0$ and $\tilde B_0$.

\subsection{Time evolution}

\paragraph*{} Let us consider the stable de Sitter solution of Eq.~(\ref{yexpansion}) and Eq.~(\ref{result}) when the condition (\ref{gamma>0}) is satisfied,
\begin{equation}
 y_H(z)=y_0+ C_0(z+1)^\gamma+\frac{4\zeta}{\beta}(z+1)^3\,,\quad\gamma>0\,.
\end{equation}
Here, $\beta$, $\zeta$ and $\gamma$ are given by (\ref{zeta}) and~(\ref{gamma}),
$y_0=H_{\mathrm{dS}}^2/\tilde{m}^2=R_{\mathrm{dS}}/(12\tilde{m}^2)$, and $C_0$ is a constant. We assume $C_0>0$.
The first EOM (\ref{EOM1bis}) leads to\\
\phantom{line}
\begin{equation}
 \frac{H^2}{\tilde{m}^2}=y_H(z)+(z+1)^3=y_0+ C_0(z+1)^\gamma + \left(\frac{1}{F'(R_{\mathrm{dS}})-R_{\mathrm{dS}}F''(R_{\mathrm{dS}})}\right)(z+1)^3\,.
\end{equation}
\phantom{line}\\
We will explicitly solve $H$ as a function of the cosmic time $t$. By writing $z+1$ as $1/a(t)$, and
by omitting the matter contribution,
one gets
\begin{equation}
\left(\frac{\dot{a}(t)}{a(t)}\right)^2=H_{\mathrm{dS}}^2+(C_0\tilde{m}^2)\left(\frac{1}{a(t)}\right)^\gamma\,.
\end{equation}
By taking $t>0$, the general solution for the expanding universe is
\begin{equation}
 a(t)=\left(\frac{C_0\tilde{m}^2}{H^2_{dS}}\right)^\frac{1}{\gamma}\left[\sinh\left(\frac{H_{\mathrm{dS}}}{2}\gamma t+\phi\right)\right]^{\frac{2}{\gamma}}\,,
\end{equation}
being $\phi$ a positive constant. It is then easy to obtain
\begin{equation}
 a(t)=a_0e^{H_{\mathrm{dS}}t}\left[1-e^{-\left(\frac{H_{\mathrm{dS}}}{2}\gamma t+\phi\right)}\right]^\frac{2}{\gamma}\,,
\end{equation}
where $a_{0}$ is a constant which depends on $\phi$. As $\gamma>0$, we get $a(t)\simeq a_0e^{H_{\mathrm{dS}} t}$. For the Hubble parameter, one has
\begin{equation}
H(t)=H_{\mathrm{dS}}\coth\left(\frac{1}{2}H_{\mathrm{dS}}\gamma t+\phi\right)\,,
\end{equation}
and, in general, this expression leads to $H\simeq H_{\mathrm{dS}}$.

Consider now the case of an oscillatory behavior followed by the condition (\ref{discriminant}). Thus,\\
\phantom{line}
\begin{equation}
H(t)=\left(\frac{\dot{a}(t)}{a(t)}\right)=\sqrt{H_{\mathrm{dS}}^2+
\tilde{m}^2a(t)^{-\frac{3}{2}}\left[A\cos\left[\mathcal F_{\mathrm {dS}}\log(a(t)^{-1})\right]+
B\sin\left[\mathcal F_{\mathrm {dS}}\log(a(t)^{-1})\right]\right]}\,,\label{bip}
\end{equation}
\phantom{line}\\
where we have used Eq.~(\ref{oscillatorysolution}), omitting matter contribution, and the frequency $\mathcal F_{\mathrm dS}$ is given by
\begin{equation}
\mathcal F_{\mathrm dS}=\sqrt{\frac{4F'(R_{\mathrm{dS}})}{R_{\mathrm{dS}}F''(R_{\mathrm{dS}})}-\frac{25}{4}}\,.
\end{equation}
If we assume $a(t)\simeq \exp(H_0 t)$, Eq.~(\ref{bip}) yields\\
\phantom{line}
\begin{equation}
H(t)\simeq\sqrt{H_{\mathrm{dS}}^2+\tilde{m}^2 \left(\frac{1}{e^{H_0 t}}\right)^{\frac{3}{2}}\left[-A\cos\left(\mathcal F_{\mathrm {dS}} H_{0}t\right)+B\sin\left(\mathcal F_{\mathrm {dS}} H_0 t\right)\right]}\,.
\end{equation}
\phantom{line}\\
Also in this case, $H\simeq H_{\mathrm{dS}}$.

\section{Dark energy in the matter era}

\paragraph*{} The critical points associated with matter dominated era for $F(R)$-gravity have been briefly discussed in \S~\ref{mattersection}.
An important viable condition is the positivity of the second derivative of $F(R)$, namely $F''(R)>0$, during this phase. When $F''(R)<0$ perturbations grow up and, as a consequence, the theory becomes strongly unstable. Now, a detailed analysis of such perturbations will be carried out following Ref. \cite{omegaDE}. Since in the late-time matter era is not stable (raugly speaking, $\dot F (R)\neq 0$), it will be necessary to introduce some physical assumptions in order to solve the corresponding equations. 
 
At first, we assume that $y_H(z)\ll(1+z)^3$ and neglect the contribute of radiation. Eq.~(\ref{superEq}) becomes, to first order in $y_H(z)/(z+1)^3$,\\
\phantom{line}
\begin{eqnarray}
\label{Trento}y''_H(z)-\frac{y'_H(z)}{(z+1)}\left(3\right)+
\frac{y_H(z)}{(z+1)^2}\left(\frac{4F'(R)-3}{R F''(R)}\right)&=&\\\nonumber
&&\hspace{-30mm}(z+1)\left[3+\frac{1}{2F''(R)R}\left((1-F'(R))+\frac{(R-F(R))}{R}\right)\right]\,,
\end{eqnarray}
\phantom{line}\\
where $R$ is written in full form, as in Eq.~(\ref{Ricciscalar}), and we have used the conditions (\ref{matterpoint2}), which usually read
\begin{equation}
R-F(R)\simeq 6\tilde{m}^2y_H(z)\,,
\quad
F'(R)\simeq 1\,.\label{matterpoint4} 
\end{equation}
In the first expression, we have considered the case of realistic models mimicing an effective cosmological constant: in general, we may say that $R-F(R)\sim\tilde{m}^2y_H(z)$.
Now, Eq.~(\ref{Trento}) can be expanded at first order in $y_H(z)$ as\\
\phantom{line}
\begin{eqnarray}
\label{strong}y_H''(z)+y'_H(z)\frac{1}{(z+1)}\left(-\frac{7}{2}-\frac{(1-F'(R))F'''(R)}{2F''(R)^2}\right)&+&
\\\nonumber\\\nonumber y_H(z)\frac{1}{(z+1)^2}\left(2+\frac{1}{R F''(R)}+\frac{2(1-F'(R))F'''(R)}{F''(R)^2}\right)&=&\left(3+\frac{2-F'(R)-F(R)/R}{2RF''(R)}\right)(z+1)\,.
\end{eqnarray}
\phantom{line}\\
We are completely neglecting the effects of the dark energy in the Ricci scalar which simply reads
\begin{equation}
 R=3\tilde{m}^2(z+1)^3\,.\label{111}
\end{equation}
In order to solve Eq.~(\ref{strong}) we can set $z=z_0+(z-z_0)$, where $|(z-z_0)/z_0|\ll 1$, and perform a variation with respect to $z$. To first order in $(z-z_0)$, we find
\begin{eqnarray}
&&\nonumber\\
&&y_H''(z)+y'_H(z)\frac{1}{(z_0+1)}\left(-\frac{7}{2}-\frac{(1-F'(R_0))F'''(R_0)}{2F''(R_0)^2}\right)+
\\\nonumber\\\nonumber&&\hspace{30mm}y_H(z)\frac{1}{(z_0+1)^2}\left(2+\frac{1}{R_0 F''(R_0)}+\frac{2(1-F'(R_0))F'''(R_0)}{F''(R_0)^2}\right)=\nonumber\\\nonumber\\&&
\left(3+\frac{2-F'(R_0)-F(R_0)/R_0}{2R_0F''(R_0)}\right)(z_0+1)+\nonumber\\\nonumber
&&\hspace{10mm} 3\left(\frac{1}{2}+\frac{5F(R_0)/R_0-F'(R_0)-4}{6R_0F''(R_0)}-
\frac{(2-F'(R_0)-F(R_0)/R_0)F'''(R_0)}{2F''(R_0)^2}\right)(z-z_0)\,,
\end{eqnarray}
\phantom{line}\\
where 
\begin{equation}
R_0=3\tilde{m}^2(z_0+1)^3\,.
\end{equation}
The solution of this equation is
\begin{equation}
y_H(z)= a+b(z-z_0)+ C_0\cdot e^{\frac{1}{2(z_0+1)}\left(\alpha\pm\sqrt{\alpha^2-4\beta}\right)(z-z_0)}\,, \label{resultmatter}
\end{equation}
where $C_0$ is a constant, $a$ and $b$ are given by\\
\phantom{line}
\begin{equation}
\left\{\begin{array}{l}
a=\left(\frac{1}{6\tilde{m}^2}\right)\frac{6R_0^2F''(R_0)+(2-F'(R_0))R_0-F(R_0)}{1+2R_0F''(R_0)
+2(2-F'(R_0)-F(R_0)/R_0)R_0 F'''(R_0)/F''(R_0)}+\\ \\
\hspace{0.6cm}\left(\frac{R_0^2}{4\tilde{m}^2}\right)\frac{7F''(R_0)^2+(2-F'(R_0)-F(R_0)/R_0)
F'''(R_0)}{[2R_0F''(R_0)^2+F''(R_0)+2R_0(2-F'(R_0)-F(R_0)/R_0)F'''(R_0)]^2}\times\\ \\
\left[R_0 F''(R_0)^2+(5F(R_0)/R_0-F'(R_0)-4)F''(R_0)/3- R_0(2-F'(R_0)-F(R_0)/R_0)F'''(R_0)\right]
\,,\\ \\
b=\frac{R_0}{2\tilde{m}^2(z_0+1)}
\frac{R_0F''(R_0)^2+(5F(R_0)/R_0-F'(R_0)-4)F''(R_0)/3
-(2-F'(R_0)-F(R_0))R_0 F'''(R_0))}{2R_0F''(R_0)^2+F''(R_0)+2(2-F'(R_0)-F(R_0)/R_0)R_0 F'''(R_0)}\,, \label{bb}
\end{array}
\right.
\end{equation}
\phantom{line}\\
and finally\\
\phantom{line}
\begin{equation}
\left\{\begin{array}{l}
\alpha=\frac{7}{2}+\frac{(1-F'(R_0))F'''(R_0)}{2F''(R_0)^2}\,,\\ \\
\beta=2+\frac{1}{R_0 F''(R_0)}+\frac{2(1-F'(R_0))F'''(R_0)}{F''(R_0)^2}\,.\label{2a}
\end{array}
\right.
\end{equation}
\phantom{line}\\
Let us now analyze this result. Since in the expanding universe $(z-z_0)<0$, it turns out that the matter solution is stable around $R_0$ if $\alpha>0$ and $\beta>0$. This means that in viable matter era we must require\\
\phantom{line}
\begin{equation}
\frac{(1-F'(R))F'''(R)}{2F''(R)^2}>-\frac{7}{2}\,,
\quad
\frac{1}{R F''(R)}>12\,.\label{qq}
\end{equation}
\phantom{line}\\
This conditions are in perfect agreement with~(\ref{F''(R)>0}).
We can thus have an oscillatory behavior of the dark energy if the discriminant of the square root in Eq.~(\ref{resultmatter}) is negative.
We will analyze oscillations in matter era and their consequences in \S~\ref{matteroscillations}.

\subsection{Late-time matter era}

\paragraph*{} In realistic models of modified gravity, the de Sitter universe follows the matter era.
The effects of dark energy could be relevant at a late-time matter era, near the transition between the matter and de Sitter epochs (thypically, around $z=1$). In this case, we cannot do an expansion of the $F(R)$-functions in terms of $y_H(z)$, as we did before. On the other hand, in realistic models of modified gravity, $y_H(z)$ tends to a constant value, as in Eq.~(\ref{yexpansion}), namely $y_H(z)=y_0+y_1(z)$, where $y_0\simeq R_{dS}/12\tilde{m}^2$ is related to the de Sitter solution and $|y_1(z)/y_0|\ll 1$ (in this way, we reproduce the correct dynamical evolution of the universe, as in the $\Lambda$CDM Model). As a consequence, we can actually perform the variation of Eq.~(\ref{Trento}) with respect to $y_1(z)$, to obtain\\
\phantom{line}
\begin{eqnarray}
&&\hspace{-10mm}y_1''(z)+y_1'(z)\frac{1}{(z+1)}\left[-\frac{7}{2}-\frac{(1-F'(R))F'''(R)}{2F''(R)^2}\right]+
\frac{y_0+y_1(z)}{(z+1)^2}\left(\frac{4F'(R)-3}{RF''(R)}\right)=\\\nonumber\\\nonumber
&&\hspace{40mm}(z+1)\left[3+\frac{1}{2F''(R)R}\left((1-F'(R))+\frac{(R-F(R))}{R}\right)\right]\,,
\end{eqnarray}
\phantom{line}\\
where
\begin{equation}
R=3\tilde{m}^2\left[(z+1)^3+4y_0\right]\,.\label{222}
\end{equation}
Also in this case, we can take $z=z_0+(z-z_0)$, with $|(z-z_0)/z_0|\ll 1$, and doing the variation with respect to $z$, we find, up to first order in $(z-z_0)$,\\
\phantom{line}
\begin{eqnarray}
&&y_1''(z)+y_1'(z)\frac{1}{(z_0+1)}\left[-\frac{7}{2}-\frac{(1-F'(R_0))F'''(R_0)}{2F''(R_0)^2}\right]+
\frac{y_0+y_1(z)}{(z_0+1)^2}\left(\frac{1}{R_0 F''(R_0)}\right)=\nonumber\\\nonumber\\
&&\hspace{-10mm}(z_0+1)\left[3+\frac{1}{2F''(R_0)R_0}\left(1-F'(R_0)+\frac{R_0-F(R_0)}{R_0}\right)\right]+\nonumber\\ \nonumber\\
&&\hspace{30mm}3\left[\frac{1}{2}-\frac{1-F'(R_0)}{2F''(R_0)^2}F'''(R_0)+
\frac{1-F'(R_0)}{6F''(R_0)R_0}\right](z-z_0)\,,
\end{eqnarray}
\phantom{line}\\
where\\
\phantom{line} 
\begin{equation}
R_0=3\tilde{m}^2[(z_0+1)^3+4y_0]\,.
\end{equation}
\phantom{line}\\ 
In the above expression we have used the conditions (\ref{matterpoint4}). Owing to the fact that $y_0\tilde{m}^2\ll R_0$, we have considered terms at least of first order in $(y_0\tilde{m}^2)/R_0$. The solution of the last equation is
\begin{eqnarray}
&&y_0=a\,,\nonumber\\
&&y_1(z)= b\,(z-z_0) + C_0\cdot e^{\frac{1}{2(z_0+1)}\left(\alpha\pm\sqrt{\alpha^2-4\beta}\right)(z-z_0)}\,, \label{resultmatter2}
\end{eqnarray}
where $C_0$ is a constant, $a$ and $b$ read\\
\phantom{line}
\begin{equation}
\left\{\begin{array}{l}
a\simeq\frac{R_0}{6\tilde{m}^2(4F'(R_0)-3)}(6F''(R_0)R_0+2-F'(R_0)-F(R_0)/R_0)+\\ \\
\hspace{0.6cm}\frac{R_0^2}{4\tilde{m}^2}(7F''(R_0)^2+2-F'(R_0)-F(R_0)/R_0)F'''(R_0))\times\\ \\
\hspace{0.6cm}(R_0F''(R_0)^2-(1-F'(R_0))R_0F'''(R_0)+(1-F'(R_0))F''(R_0)/3)\,,\\ \\
b=\frac{3(z_0+1)^2(R_0F''(R_0)^2-(1-F'(R_0))R_0F'''(R_0)+(1-F'(R_0))F''(R_0)/3)}{2\, F''(R_0)}\,,
\end{array}
\right.
\end{equation}
\phantom{line}\\
and finally
\phantom{line}\\
\begin{equation}
\left\{\begin{array}{l}
\alpha=\frac{7}{2}+\frac{(1-F'(R_0))F'''(R_0)}{2F''(R_0)^2}\,,\\ \\
\label{Nvier}\beta=\frac{1}{R_0 F''(R_0)}\,.
\end{array}\right.
\end{equation}
\phantom{line}\\
The solution is stable around $R_0$ if $\alpha>0$ and $\beta>0$. This means that in late-time matter era one has to verify
\begin{equation}
\frac{(1-F'(R))F'''(R)}{2F''(R)^2}>-\frac{7}{2}\,,\quad
 \frac{1}{RF''(R)}>0\,,\label{xx}
\end{equation}
otherwise the model is stongly unstable.
The oscillatory behavior of the dark energy occurs when the discriminant of the square root of Eq.~(\ref{resultmatter2}) is negative.
We observe that the expression (\ref{222}) is more accurate than (\ref{111}). In general, if the conditions (\ref{xx}) are satisfied for $R_0=3\tilde{m}^2(z_0+1)^3+12\tilde{m}^2y_0$, the conditions (\ref{qq}) will be also satisfied, provided it is possible to use the approximation $R_0=3\tilde{m}^2(z_0+1)^3$.

\section{DE-oscillations in viable matter era\label{matteroscillations}}

\paragraph{} In this Section, 
we consider viable $F(R)$-gravity representing 
a realistic scenario for the dark energy epoch, in particular, the class of models presented in \S~\ref{viablemodels}. As representative exemples, we take exponential gravity and a power form model. 
As the result of the analysis carried out in the previous Section, we show that for these models, large frequency oscillation of dark energy 
in the matter dominated era appears. 
This behaviour may influence the
higher derivatives of the Hubble parameter with the risk to produce some divergence.
Therefore, we suggest a way to stabilize such frequency oscillation.

Let us start with the exponential model of \S~\ref{exponential}. By replacing $R_0\rightarrow b\,\Lambda$, we get
\begin{equation}
F(R)=R-2\Lambda\left[1-\mathrm{e}^{-R/\left(b\,\Lambda\right)}\right]\,,
\label{modell}
\end{equation}
where $2>b>0$ is a free parameter which satisfies condition (\ref{R0condition}) for stable de Sitter universe. 
Then, we examine the Hu-Sawicki model in Eq. (\ref{HuSawModel}). 
For our treatment, we reparameterize this model by 
putting $c_{1}\tilde{m}^{2}/c_{2}=2\Lambda$ and 
$(c_2)^{1/n}\,\tilde{m}^2=b\,\Lambda$, with $2>b>0$, and we obtain 
\begin{equation}
F(R)=R-2\Lambda\left\{1-\frac{1}{\left[R/\left(b\,\Lambda\right)\right]^{n}+1}\right\}\,, 
\quad n=4\,.
\label{model2l}
\end{equation}
We observe that, 
in both of these models, the term $b\,\Lambda$ corresponds to 
the curvature for which the Cosmological 
Constant is ``switched on''. 
In the model of Eq.~(\ref{model2l}), since 
$n$ has to be sufficiently large in order to mimic the $\Lambda$CDM model, 
we have assumed $n=4$ and we keep free the $b$ parameter only.
At high curvatures, during matter era, this models accurately reproduce the Cosmological Constant. It is easy to verify that conditions (\ref{qq}), or, for late-time matter era, (\ref{xx}), are well satisfied and, 
owing to the fact that $F''(R)$ is very close to $0^+$ for both of the models, the discriminant in the square root of Eq.~(\ref{resultmatter}) is negative and the dark energy oscillates 
as\\
\phantom{line}
\begin{equation}
y_H(z)\simeq\frac{\Lambda}{3\tilde{m}^2}+\mathrm{e}^{\frac{\alpha_{1,2}(z-z_0)}{2(z_0+1)}}\left[A_0\sin\left(\frac{\sqrt{\beta_{1,2}}}{(z_0+1)}(z-z_0)\right)+
B_0\cos\left(\frac{\sqrt{\beta_{1,2}}}{(z_0+1)}(z-z_0)\right)\right]\,.
\label{matterDE}
\end{equation}
\phantom{line}\\
Here, $A_0$ and $B_0$ are constants and $\alpha_{1,2}$ and $\beta_{1,2}$ are given by (\ref{2a}). In the specific, $\alpha_1=3$ for the model in Eq. (\ref{modell}) and $\alpha_2\simeq 29/10$ for the model in Eq. (\ref{model2l}), while $\beta_{1,2}\simeq 1/[R_0F''(R_0)]$ in the both cases, namely 
%
\begin{equation}
\beta_1\simeq\left(\frac{b^2\Lambda\mathrm{e}^{\frac{R_0}{\tilde{R}}}}{2 R_0}\right)\,,  
\end{equation}
for exponential model (\ref{modell}) and 
\begin{equation}
\beta_2\simeq\frac{R_0\left[1+\left(\frac{R_0}{b\Lambda}\right)^n\right]^3\left(\frac{b\Lambda}{R_0}\right)^n}{2\Lambda\,n\left\{1+n\left[\left(\frac{R_0}{b\Lambda}\right)^n-1\right]+\left(\frac{R_0}{b\Lambda}\right)^n\right\}}\simeq\frac{R_0}{2\Lambda\,n(n+1)}\left(\frac{R_0}{b\Lambda}\right)^n\,, 
\end{equation}
for the power-law model~(\ref{model2l}). 
We see that
the frequency of dark energy oscillations increases with the curvature (and redshift). Moreover, the effects of such oscillations are amplified in 
the derivatives of the dark energy density, namely, 
\begin{equation}
\left| \frac{d^n}{d t^n}y_H(t_0)\right|\propto 
\left[\mathcal{F}(z_0)\right]^n\,,\quad\mathcal F (z)\simeq \frac{1}{(z+1)\sqrt{R F''(R)}}\,.
\label{frequency}
\end{equation}
Here, $\mathcal{F}(z)$ is the oscillation frequency of dark energy and 
$t_0$ is the cosmic time corresponding to the redshift $z_0$. 
This is for example the case of the dark energy EoS parameter 
(\ref{oo}):
for large values of the redshift, the dark energy density oscillates 
with a high frequency and also its derivative becomes large, showing a different feature of the dark energy EoS parameter in the models (\ref{modell})--(\ref{model2l}) compared with the case of $\Lambda$CDM Model. 
Then, during matter era, the Hubble parameter behaves as 
\begin{equation}
H(z)\simeq \sqrt{\tilde{m}^2}\left[(z+1)^{3/2}+\frac{y_H(z)}{2(z+1)^{3/2}}\right]\,,\quad \Big\vert\frac{y_H(z)}{(z+1)^3}\Big\vert\ll 1\,,
\label{eq:FR5-15-3.14}
\end{equation}
and it is simply to understand that, if the frequency $\mathcal{F}(z)$ of Eq.~(\ref{frequency}) is extremely large, the derivatives of dark energy density could become dominant in some higher derivatives of the Hubble parameter 
which may approach an effective (Type IV) singularity and therefore make 
the solution unphysical. 
We see it for specific cases.
Exponential gravity as well as Hu-Sawicki model can correctly reproduce 
the late-time cosmic acceleration 
following the matter dominated epoch, in agreement with astrophysical 
data~\cite{Linder,Bamba,Twostepmodels, altri}.
A reasonable choice is to take $b=1$ for both these models and
put $\Lambda = 7.93 \tilde{m}^2$, where $\tilde m^2$ is given by (\ref{scale}), according with data of Ref.~\cite{WMAP}. 
We can solve Eq.~(\ref{superEq}) numerically\footnote{We have used 
Mathematica 7 \textcopyright.} by taking the following boundary conditions at 
$z=z_i$, where $z_i\gg 0$, 
\begin{eqnarray}
\left\{\begin{array}{l}
\frac{d y_H(z)}{d (z)}\Big\vert_{z_i} = 0\,, \\ 
y_H(z)\Big\vert_{z_i} = \frac{\Lambda}{3\tilde{m}^2}\,. 
\end{array}\right.
\end{eqnarray}
Here, we have taken into account the fact that at a high redshift the universe should be 
very close to the $\Lambda$CDM model. 
We have set $z_i=2.80$ for the model in Eq.~(\ref{modell}) and $z_i=4.5$ for the model in Eq.~(\ref{model2l}), such that $R\,F''(R)\sim 10^{-8}$ at $R=3\tilde m^2(z_i+1)^3$. 
We observe that it is hard to extrapolate 
the numerical results to the higher redshifts because of the large frequency of dark energy oscillations. 

\begin{figure}[!h]
\subfigure[]{\includegraphics[width=0.3\textwidth]{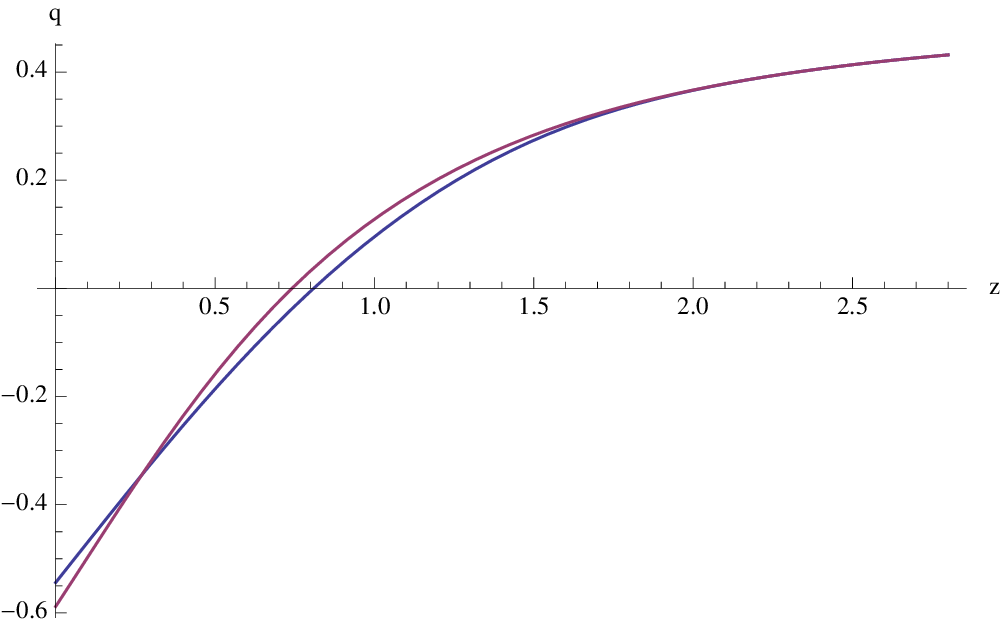}}
\qquad
\subfigure[]{\includegraphics[width=0.3\textwidth]{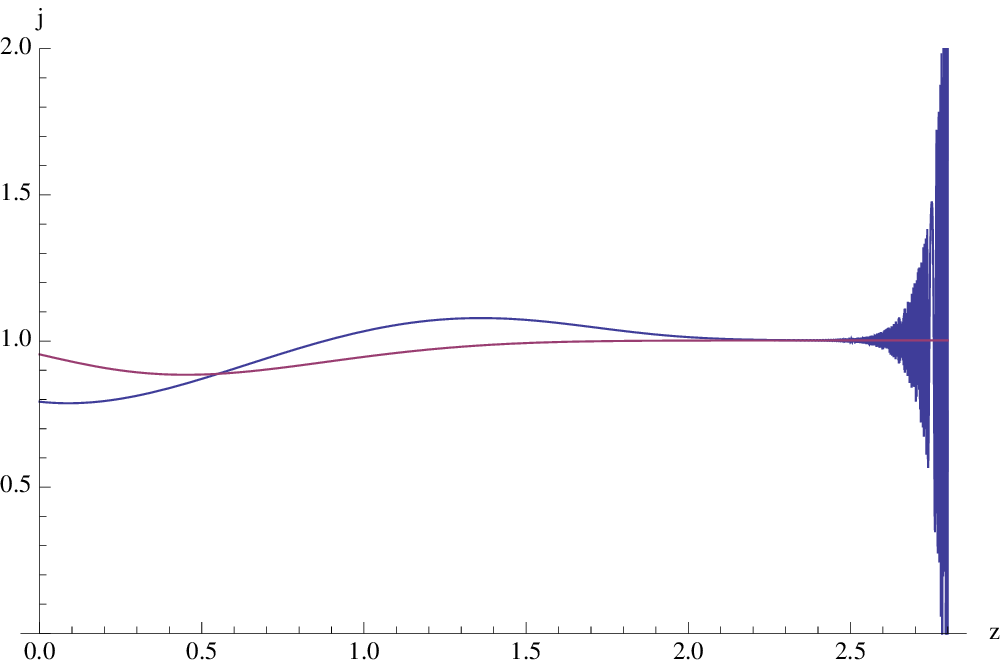}}
\qquad
\subfigure[]{\includegraphics[width=0.3\textwidth]{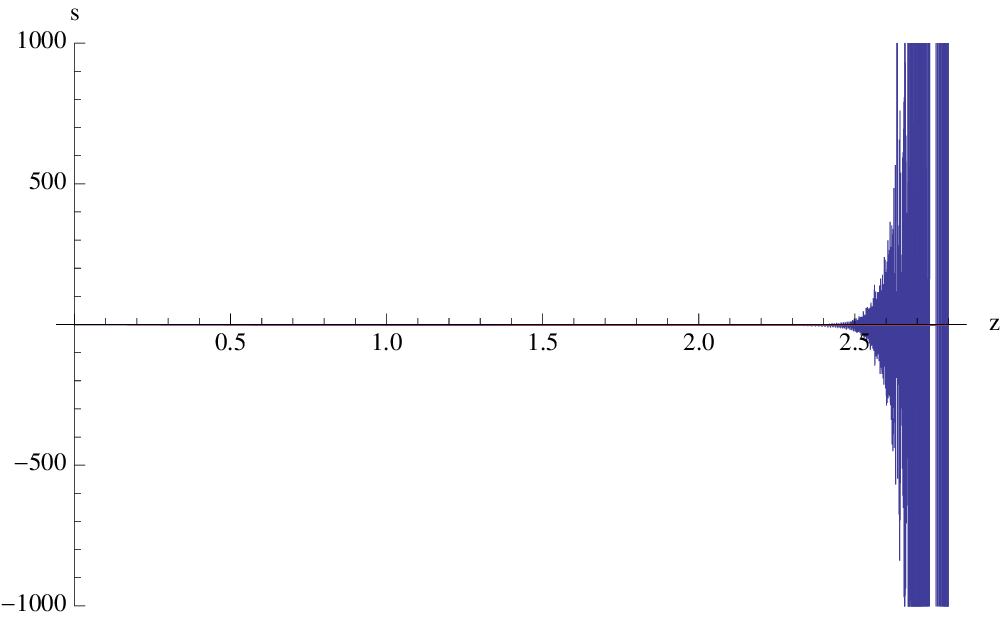}}
\qquad
\subfigure[]{\includegraphics[width=0.3\textwidth]{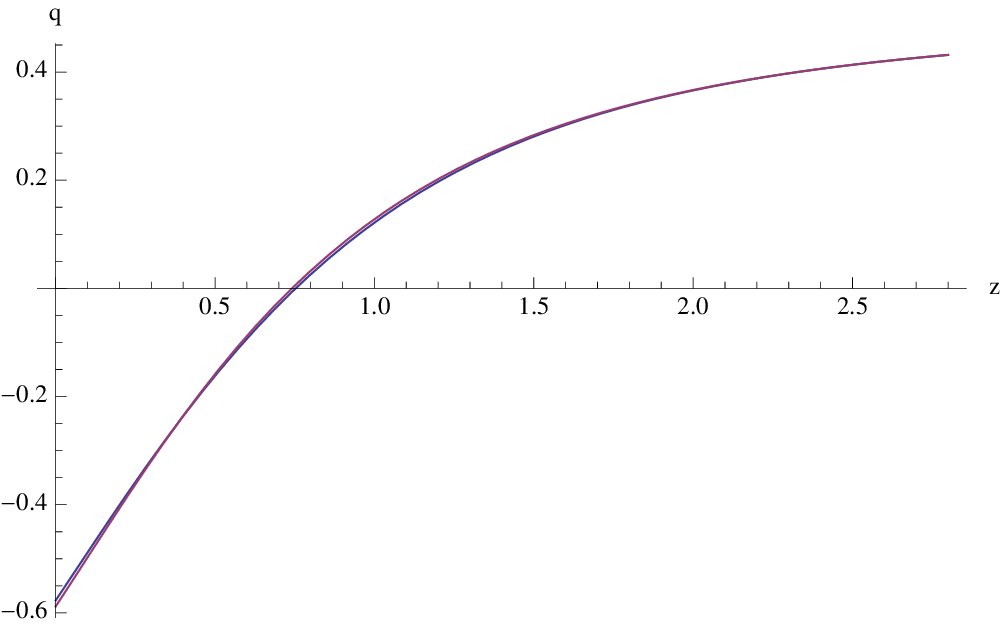}}
\qquad
\subfigure[]{\includegraphics[width=0.3\textwidth]{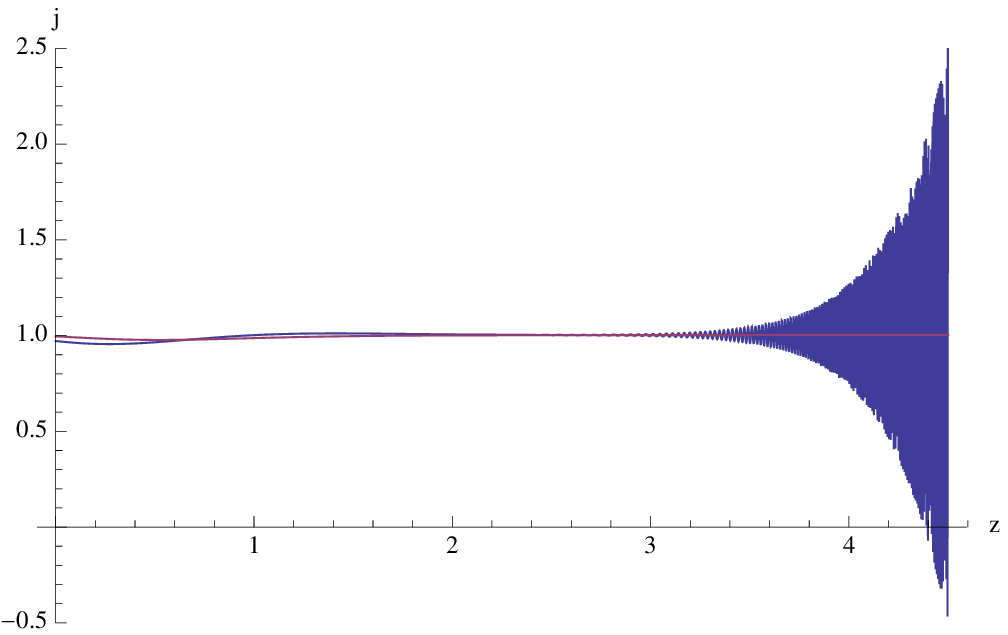}}
\qquad
\subfigure[]{\includegraphics[width=0.3\textwidth]{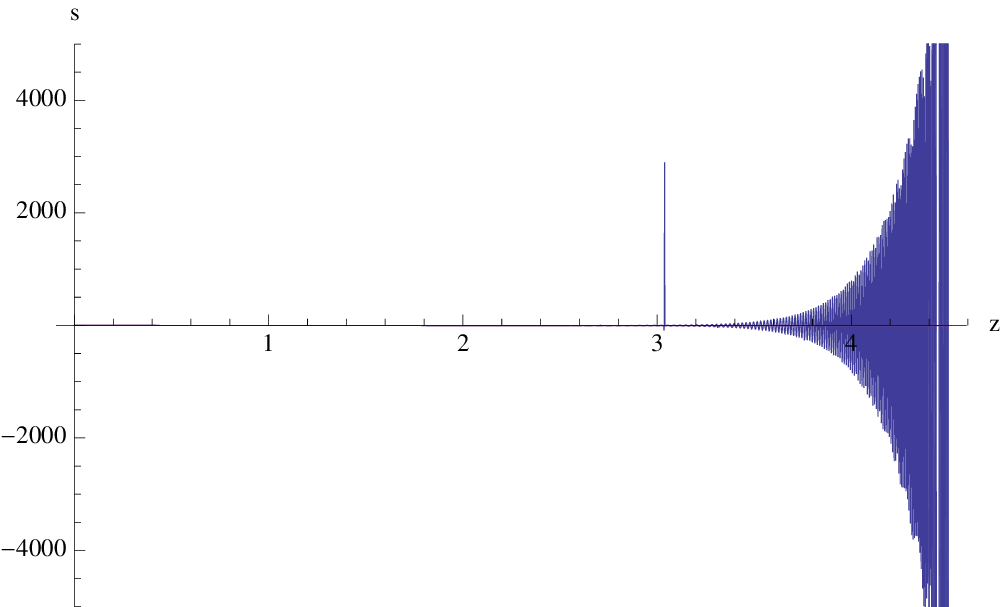}}
\caption{
Cosmological evolutions of $q(z)$ [(a) and (d)], $j(z)$ [(b) and (e)] and $s(z)$ [(c) and (f)] parameters as functions of the redshift $z$ for Model I 
[(a)--(c)] and Model II [(d)--(f)] in the region of $z>0$. 
The overlapped regions with $\Lambda$CDM Model are shown.
\label{parameters}}
\end{figure}

Using Eq.~(\ref{oo}), we also can derive $\omega_\mathrm{DE}$ and
from Eq.~(\ref{Ricciscalar}) we obtain $R$ as a function of the redshift. 
Furthermore, we extrapolate 
the behavior of $\Omega_{\mathrm{DE}}$, which is given by 
\begin{equation}
\Omega_\mathrm{DE}(z)\equiv\frac{\rho_\mathrm{DE}}{\rho_\mathrm{eff}}
=\frac{y_H}{y_H+\left(z+1\right)^3+\chi\left(z+1\right)^4}\,.
\end{equation} 
For the present universe, we get the 
following results: 
for the model in Eq.~(\ref{modell}), 
$y_H(0)=2.736$, $\omega_{\mathrm{DE}}(0)=-0.950$, $\Omega_{\mathrm{DE}}(0)=0.732$ and $R(z=0)=4.365$, 
and for the model~(\ref{model2l}), 
$y_H(0)=2.652$, $\omega_{\mathrm{DE}}(0)=-0.989$, $\Omega_{\mathrm{DE}}(0)=0.726$ and $R(z=0)=4.358$. 
These resultant data are in accordance with the last and very accurate 
observations of current universe~\cite{WMAP}, namely
\begin{eqnarray}
\omega_\mathrm{DE} &=& -0.972^{+0.061}_{-0.060}\,, \nonumber \\ 
\Omega_\mathrm{DE} &=& 0.721\pm 0.015\,.
\label{data}
\end{eqnarray}
Next, 
we introduce the deceleration $q$, jerk $j$ and snap $s$ 
parameters~\cite{Chiba:1998tc, Sahni:2002fz},\\
\phantom{line}
\begin{eqnarray}
q(t) &\equiv& -\frac{1}{a(t)}\frac{d^2 a(t)}{d t^2}\frac{1}{H(t)^2}=-\frac{\dot H}{H^2}-H^2\,;\nonumber\\
j(t) &\equiv& \frac{1}{a(t)}\frac{d^3 a(t)}{d t^3}\frac{1}{H(t)^3}=\frac{\ddot H}{H^3}-3q-2\,;\nonumber\\
s(t) &\equiv& \frac{1}{a(t)}\frac{d^4 a(t)}{d t^4}\frac{1}{H(t)^4}=\frac{\dddot H}{H^4}+4j+3q(q+4)+6\,.
\label{cosmpar}
\end{eqnarray}
\phantom{line}\\
The following values of these cosmological parameters at the present time 
($z=0$) in the two models, 
which we called Model I in Eq.~(\ref{modell}) and Model II 
in Eq.~(\ref{model2l}), and in the $\Lambda$CDM Model are obtained from numerical extrapolation:\\
\phantom{line}
\begin{eqnarray}
q(z=0) &=& -0.650\, (\Lambda \mathrm{CDM})\,, 
\,-0.544\, (\mathrm{Model\, I})\,,
\,-0.577\, (\mathrm{Model\, II})\,;\nonumber\\
j(z=0) &=& 1.000\, (\Lambda \mathrm{CDM})\,, 
\,0.792 (\mathrm{Model\, I})\,, 
\,0.972\, (\mathrm{Model\,II})\,;\nonumber\\
s(z=0) &=& -0.050\, (\Lambda \mathrm{CDM})\,, 
\,-0.171 (\mathrm{Model\, I})\,, 
-0.152\,\, (\mathrm{Model\,II})\,.
\end{eqnarray}
\phantom{line}\\
The deviations of these parameters in Models I and II from those in the 
$\Lambda$CDM Model are very small at the present: 
however, since the parameters depend on the time derivatives of 
the Hubble parameter, it is interesting to analyze their behaviors at high 
redshift (and curvature). 
For this reason, 
in Fig.~\ref{parameters}, we plot the cosmological evolutions of decelaration, jerk and snap as functions of the redshift $z$. 
In this graphics, also the overlapped regions with 
$\Lambda$CDM Model are shown. 

The deceleration parameter in Models I and II is very close to the value in the $\Lambda$CDM Model,
being still negligible the contribute of dark energy to the first derivative of the Hubble parameter,
and the correct cosmological evolution of these models is guaranteed. 
However, it is clear that in the jerk and snap parameters 
the derivatives of the dark energy density become relevant: these parameters 
grow up with an oscillatory behavior and, due to the fact that
the frequency of the oscillations strongly increases with the redshift, 
it is reasonable to expect that some divergence occurs in the past. 
We observe that, 
if from one side at high redshifts 
the exponential Model I is more close to the $\Lambda$CDM Model 
because of the faster decreasing of exponential function in comparison with 
the power function of Model II, from the other side it involves stronger 
oscillations in the matter era. 

It could be stated that, the more similar the model is to 
the $\Lambda$CDM Model, namely as much $F''(R)$ is close to zero, 
the bigger the oscillation frequency of dark energy becomes in the past and, despite the fact that the dynamics of the universe 
depends on the matter and the dark energy density remains avoidable, 
some divergences in the derivatives of the Hubble parameter may occur. 
We can conclude that
the models under consideration,
although the approaching manners to a model with the Cosmological Constant 
are different from each other, 
show a generic feature of realistic $F(R)$-gravity models, in which 
the cosmological evolutions are similar to those of General Relativity with Cosmological Constant.
The corrections to the Einstein's equations in the small curvature regime 
lead to undesired effects in the high curvature regime. 
Thus, we need to investigate additional modifications. 

In Ref.~\cite{omegaDE} the accurate analysis of DE-oscillations, where is possible to appreciate the concordance between predicted oscillation frequency and numerical one, is carried out. 

\subsection{Proposal of a correction term}

\paragraph{} In order to remove the divergences in the derivatives of the Hubble parameter, 
we introduce a function $g(R)$ for which the oscillation frequency of 
the dark energy density~(\ref{matterDE}) acquires a constant value 
$1/\sqrt{\delta}$, where $\delta>0$, for a generic curvature $R\gg b\,\Lambda$, and we try to stabilize the oscillations of dark energy during the matter dominated 
era with the use of a correction term \cite{malloppone, bollettino}. 
Since in the matter era
$(z+1)=\left[R/(3\tilde{m}^2)\right]^{1/3}$, 
we must require
\begin{eqnarray}
\frac{(3\tilde{m}^2)^{2/3}}{R^{5/3}\,g''(R)} = 
\frac{1}{\delta}\,,\quad\quad\quad
g(R) = -\tilde{\gamma}\,\Lambda\left(\frac{R}{3\tilde m^2}\right)^{1/3}\,, 
\quad 
\tilde{\gamma}>0\,, 
\label{eq:FR5-15-3.18}
\end{eqnarray}
where $\tilde{\gamma}\equiv (9/2) \delta (3\tilde m^2/\Lambda)=1.702\,\delta$. 
We explore 
the models in Eqs.~(\ref{modell})--(\ref{model2l}) 
with adding these correction as\\ 
\phantom{line}
\begin{eqnarray}
F_{1}(R) &=& 
R-2\Lambda\left(1-\mathrm{e}^{-\frac{R}{b\,\Lambda}}\right)-\tilde{\gamma}\,\Lambda\left(\frac{R}{3\tilde{m}^2}\right)^{1/3}\,, 
\label{F3exp} \\ 
F_{2}(R) &=& R-2\Lambda\left[1-\frac{1}{(R/b\, \Lambda)^4+1}\right]-\tilde{\gamma}\,\Lambda\left(\frac{R}{3\tilde{m}^2}\right)^{1/3}\,. 
\label{F3HS}
\end{eqnarray}
\phantom{line}\\
We note that in both cases $F_{1,2}(0)=0$, and we still have the
Minkowski's solution of the flat space-time. 
Then, under the requirement $\tilde{\gamma}\ll (\tilde{m}^2/\Lambda)^{1/3}$,
the effects of our last modification vanish in the de Sitter epoch, when 
$R=4\Lambda$, and we recover a model with an effective 
cosmological constant. 
We may also evaluate the dark energy density at high redshifts 
by directly putting $R=3\tilde{m}^2(z+1)^3$ in the definition of 
$\rho_{\mathrm{DE}}=\rho_{\mathrm{eff}}-\rho_\mathrm{m}$ 
given by Eq.~(\ref{rhoeffRG}). We get 
\begin{equation}
y_H(z)\simeq\frac{\Lambda}{3\tilde{m}^2}\left[1+\tilde{\gamma}(1+z)\right]\,. 
\label{densmatter}
\end{equation}
%
We see that
with the reasonable choice $\tilde{\gamma}\sim1/1000$, 
the effects of modification of gravity on the dark energy density 
begin to appear at a very high redshift (for example, 
$y_H(10)= 1.01\times y_H(0)$), and we are very close to the 
$\Lambda$CDM Model. 
However, while the pure models in Eqs.~(\ref{modell}) and (\ref{model2l}) mimic 
an effective cosmological constant, the models in Eqs.~(\ref{F3exp})--(\ref{F3HS}) mimic a quintessence fluid. In fact,
Eq. (\ref{oo}) leads to 
\begin{equation}
\omega_{\mathrm{DE}}(z)\simeq -1+\frac{(1+z)\tilde{\gamma}}{3(1+(1+z)\tilde{\gamma})}\,, 
\end{equation}
so that when $z\rightarrow+\infty$, $\omega_{\mathrm{DE}}(z)\rightarrow -2/3^-$. 

Thus, it is simple to verify that all the cosmological 
constraints are still satisfied. 
Since $|F_{1,2}'(R \gg b\Lambda)-1| \ll 1$, 
the effective gravitational coupling $G_{\mathrm{eff}}=G_N/F_{1,2}'(R)$ is 
positive, and hence the models are protected against the 
anti-gravity during the cosmological evolution. 
Then, due to the fact that $F_{1,2}''(R\gg b\,\Lambda)>0$, 
we do not have any problem in terms of the existence of a stable matter era. 
It should be stressed that the energy density preserves 
its oscillation behavior in the matter dominated era, 
but that, owing to the correction term reconstructed before, 
such oscillations keep a constant frequency 
$\mathcal{F}=\sqrt{1.702/\tilde{\gamma}}$ and do not diverge. 
Despite the small value of $\tilde{\gamma}$, in this way 
the high redshift divergences and possible effective singularities 
can be removed. 


\subsection{Analysis of exponential and power-form models with correction terms 
in the matter dominated era 
\label{matterstudy}}

\begin{figure}[!h]
\subfigure[]{\includegraphics[width=0.3\textwidth]{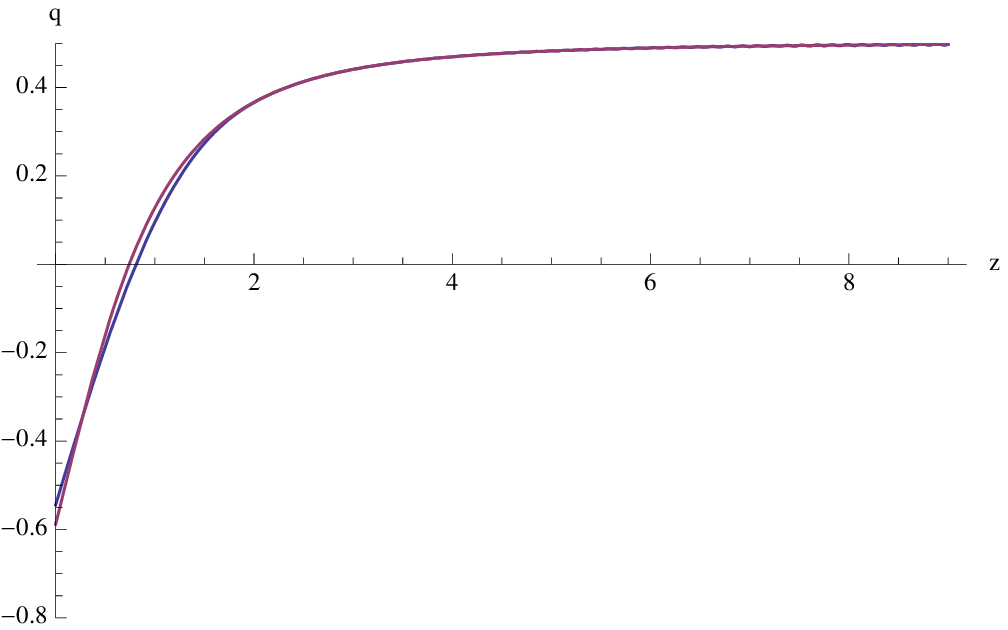}}
\qquad
\subfigure[]{\includegraphics[width=0.3\textwidth]{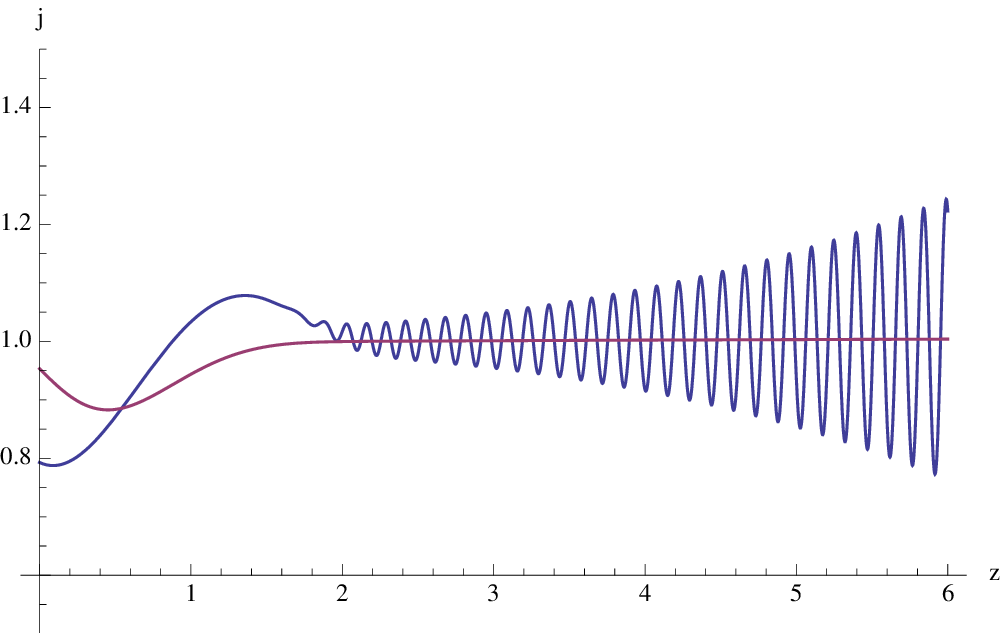}}
\qquad
\subfigure[]{\includegraphics[width=0.3\textwidth]{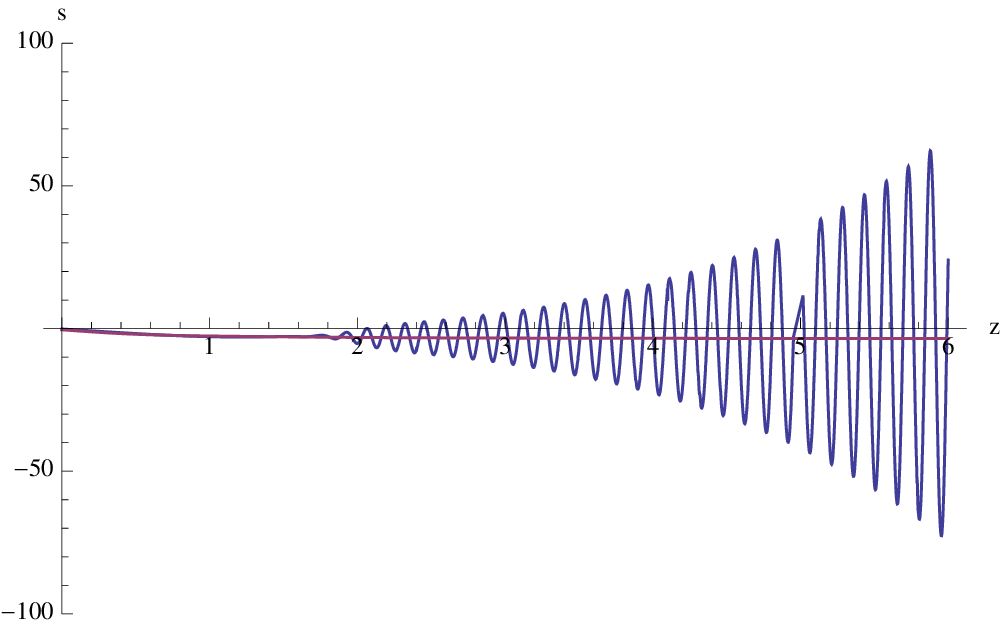}}
\qquad
\subfigure[]{\includegraphics[width=0.3\textwidth]{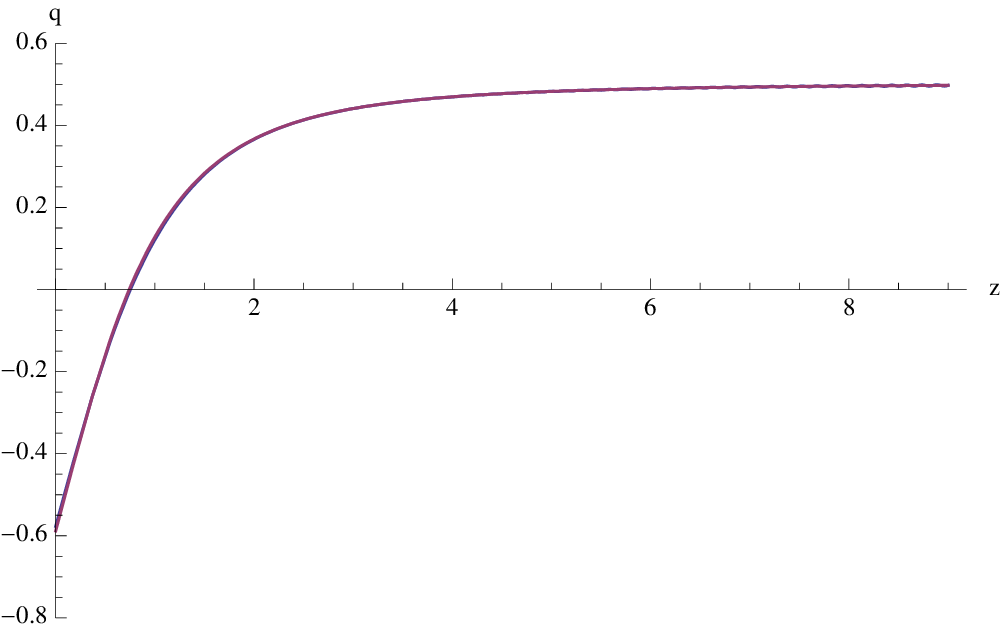}}
\qquad
\subfigure[]{\includegraphics[width=0.3\textwidth]{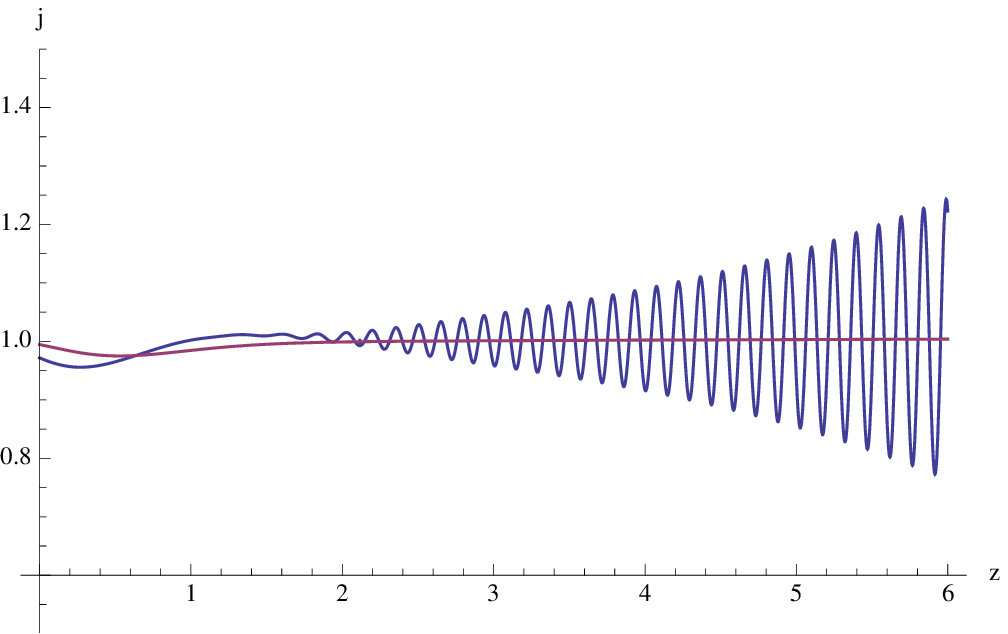}}
\qquad
\subfigure[]{\includegraphics[width=0.3\textwidth]{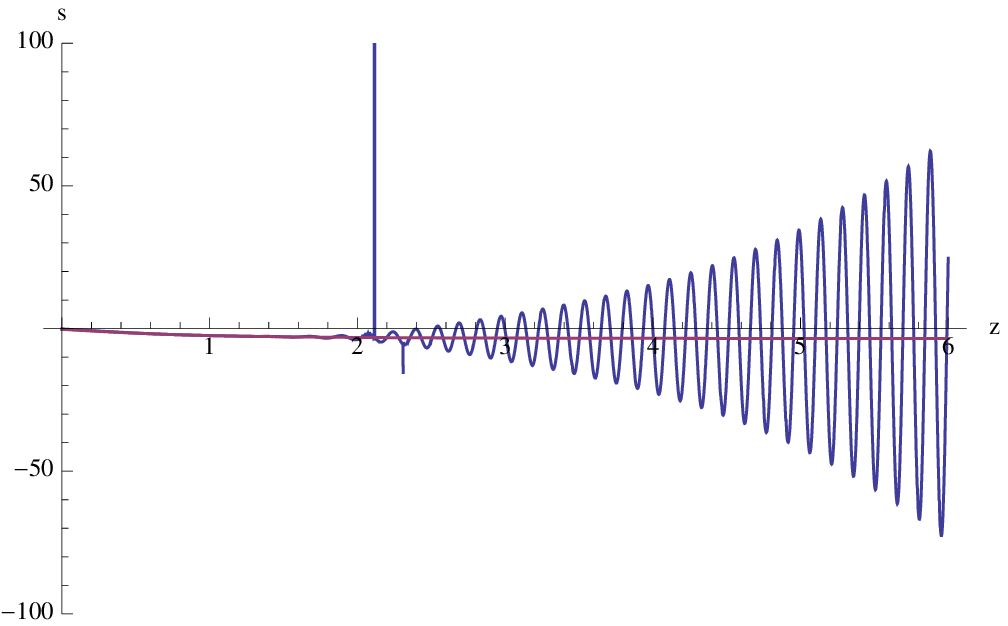}}
\caption{
Cosmological evolutions of $q(z)$ [(a) and (d)], $j(z)$ [(b) and (e)] and $s(z)$ [(c) and (f)] parameters as functions of the redshift $z$ for 
the model $F_1(R)$ [(a)--(c)] and the model $F_2(R)$ [(d)--(f)] 
in the region of $z>0$. 
The overlapped regions with $\Lambda$CDM Model are shown.
\label{3}}
\end{figure}

\paragraph{} In this Subsection, we carry out the numerical analysis of the models 
in Eqs.~(\ref{F3exp}) and (\ref{F3HS}). 
In both cases, 
we assume $b=1$ and $\tilde{\gamma}=1/1000$ and solve 
Eq.~(\ref{superEq}) in a numerical way, 
by taking accurate initial conditions at $z=z_i$,
where $z_i\gg0$. 
By using Eq.~(\ref{densmatter}), we acquire 
\begin{eqnarray}
\left\{\begin{array}{l}
\frac{d y_H(z)}{d (z)}\Big\vert_{z_i} = 
\frac{\Lambda}{3\tilde{m}^2}\tilde\gamma\,,\\ 
y_H(z)\Big\vert_{z_i} = \frac{\Lambda}{3\tilde{m}^2}\left[1+\tilde\gamma\,(z_i+1)\right]\,.
\end{array}\right.
\end{eqnarray}
We have set $z_i=9$. 
The feature of the models~(\ref{F3exp})--(\ref{F3HS}) at the present time is very similar to those of the models in Eqs.~(\ref{modell})--(\ref{model2l}). 
The numerical extrapolation to the current universe 
for the model~(\ref{F3exp}) leads to 
$y_H(0)=2.739$, $\omega_{\mathrm{DE}}(0)=-0.950$, $\Omega_{\mathrm{DE}}(0)=0.732$ and $R(z=0)=4.369$, 
while for the model~(\ref{F3HS}), we obtain 
$y_H(0)=2.654$, $\omega_{\mathrm{DE}}(0)=-0.989$, $\Omega_{\mathrm{DE}}(0)=0.726$ and $R(z=0)=4.361$. 
Let us have a look to those behaviors in the matter dominated era. 
From the initial conditions we get
$y_H(9)=2.670$ and $\omega_{\mathrm{DE}}(9)=-0.997$, and
the universe remains extremely close to the $\Lambda$CDM Model. 
The dynamical correction of 
the Einstein's equation, namely, roughly speaking, 
the fact of having ``a dynamical cosmological constant'', 
introduces the oscillatory behavior of dark energy density, but now,
thanks to the contribution of the correction term, 
we have a constant frequency of oscillation frequency without changing 
the cosmological evolution described by the theory. 
In Fig.~\ref{3}, we show the behaviour of 
the deceleration, jerk and snap parameters as functions of the redshift $z$ 
in these models. 
The overlapped regions in this models with those in the $\Lambda$CDM 
Model are shown. 
We may compare the graphics in Fig.~\ref{3} 
with the corresponding ones in Fig.~\ref{parameters} 
of the models (\ref{modell})--(\ref{model2l}) without the correction 
term.
At high redshifts, 
the deceleration parameter is not influenced by dark energy and
its behaviour
in both of the models~(\ref{F3exp})--(\ref{F3HS}) is close to the one 
of the $\Lambda$CDM Model. 
On the contrary,  
the jerk and snap parameters oscillate, being the derivatives of the dark energy density relevant. 
However, here such oscillations have the constant frequency of the dark energy and do not diverge. 
The predicted value of the oscillation frequency is 
$\mathcal{F}\equiv\sqrt{1.702/\tilde{\gamma}}=41.255$. 
The oscillation period is $T=2\pi/\mathcal{F}\simeq 0.152$. 
Thus, the numerical data are in good accordance with the predicted ones 
(we can appreciate the result in the graphics by taking into account the fact that 
the number of crests per units of the redshift has to be $1/T\simeq7$).\\ 

In conclusion, 
we have shown in both analytical and numerical ways that 
the effects
of dark energy oscillations are evident at high red shift in the higher 
derivative of the Hubble parameter, which may approach an effective singularity. 
It is not a case if all the numerical simulations presented in the 
literature start from small redshifts, since for large values of the redshift this  
problem appears.  
We stress that the average value of the dark energy density remains negligible, 
but the oscillations around this value become huge.  
The analytical results can match with the numerical simulations, and therefore 
all the analyses here presented are consistent. 
This behavior of realistic $F(R)$-gravity models has recently been studied 
also in Ref.~\cite{Lee:2012dk}. Some correction to stabilize such oscillations is required and we have seen
that one possibility is given by the adding of a low-power term of the Ricci scalar.

\section{DS-solution, future universe evolution and growth index}

\paragraph{} We have just seen that the models (\ref{F3exp})--(\ref{F3HS}) with the choice of $b=1$ can be consistent 
with the observational data of the universe. 
Here, we examine all the range of $b$ in which the models are compatible with 
the observations 
and we concentrate on their future evolution. We will show 
that the effective crossing of the phantom divide which characterizes the de Sitter epoch takes place in the very far future. 

Furthermore,
in the way of trying to explain the several aspects of our 
universe, there exists the problem of distinguishing different theories. 
Different theories can achieve the same expansion history, but theories with the same expansion history can have a different cosmic growth history. This fact makes the growth of the large scale structure in the universe, namely the characterization of growth of the matter density perturbations, an important tool in order to discriminate among different models.  In order to execute it, the so-called growth index $\gamma$~\cite{Linder:2005in} is useful. Therefore, in the second part of this Section, we study the evolution of the matter density perturbations in our models.

These results have been presented in Ref. \cite{malloppone}.

\subsection{Cosmological constraints}

\begin{figure}[!h]
\subfigure[]{\includegraphics[width=0.4\textwidth]{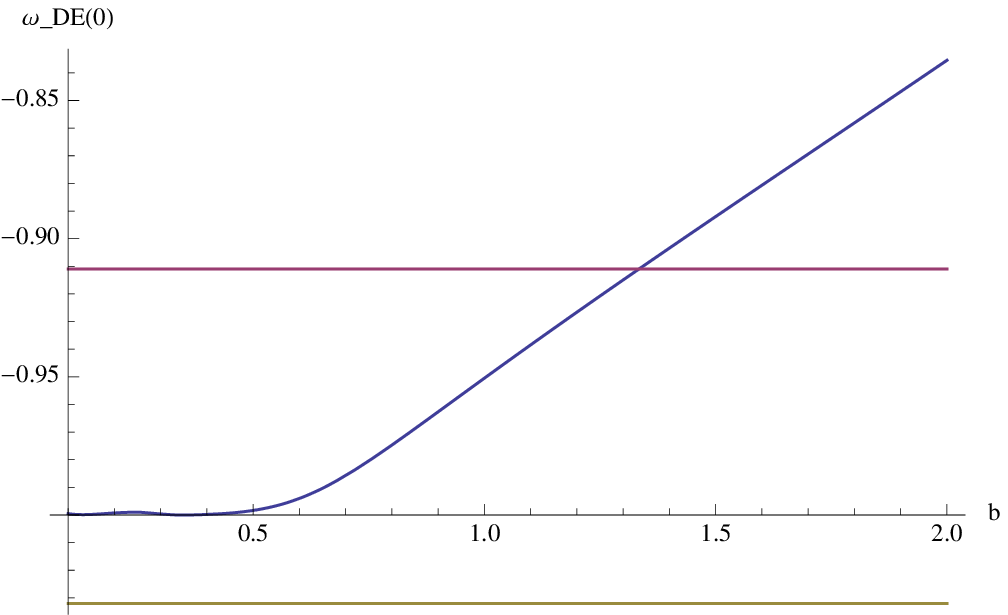}}
\qquad
\subfigure[]{\includegraphics[width=0.4\textwidth]{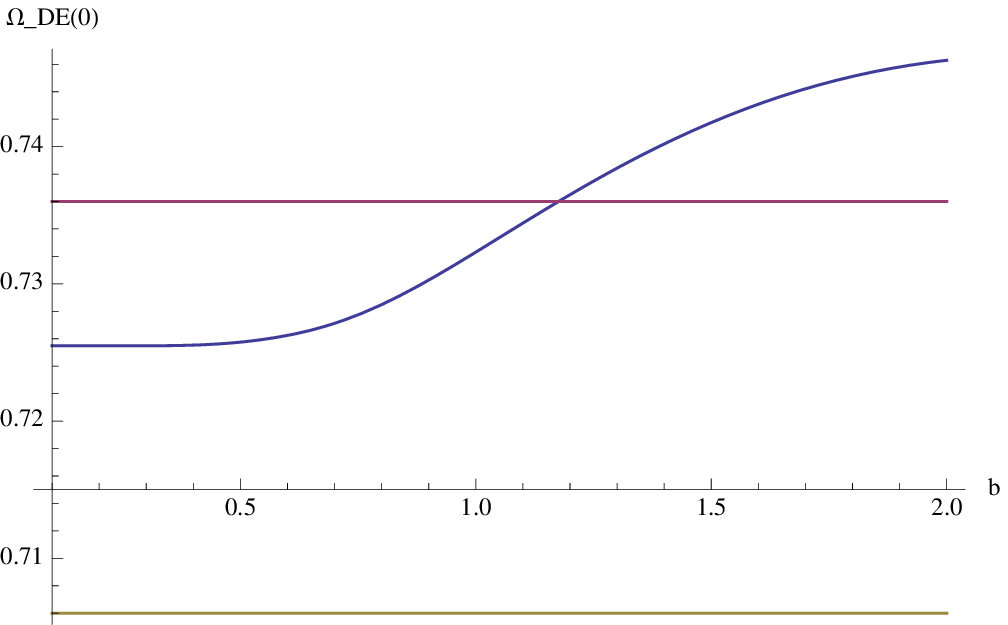}}
\caption{Behaviors 
of $\omega_\mathrm{DE}(z=0,b)$ and of $\Omega_\mathrm{DE}(z=0,b)$ 
as functions of $b$ for exponential model. 
The observational data bounds (horizontal lines) are also shown.
\label{b-1}}
\end{figure}

\begin{figure}[!h]
\subfigure[]{\includegraphics[width=0.4\textwidth]{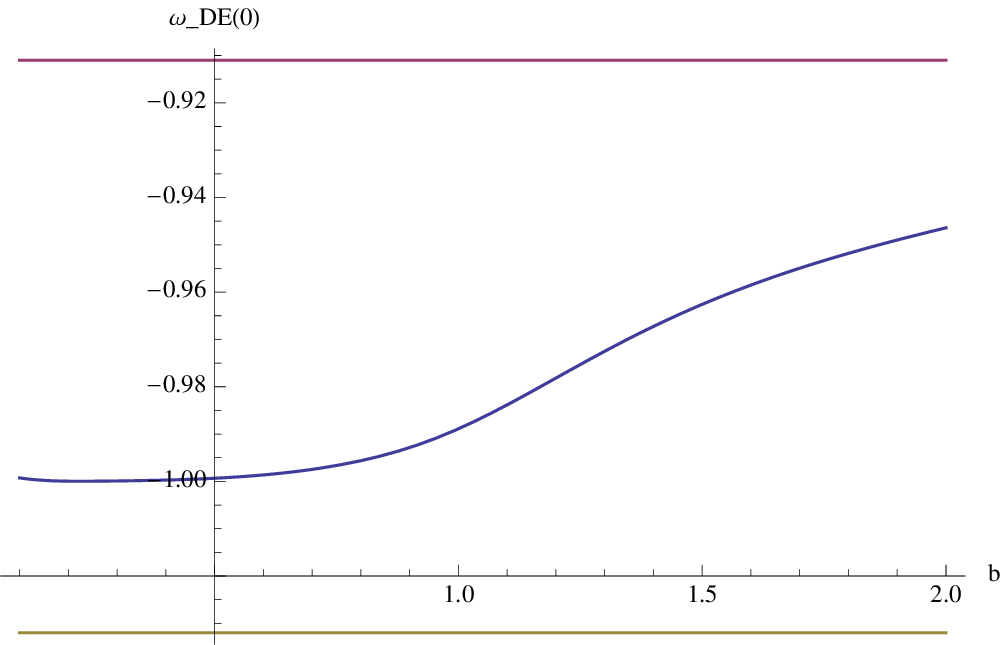}}
\qquad
\subfigure[]{\includegraphics[width=0.4\textwidth]{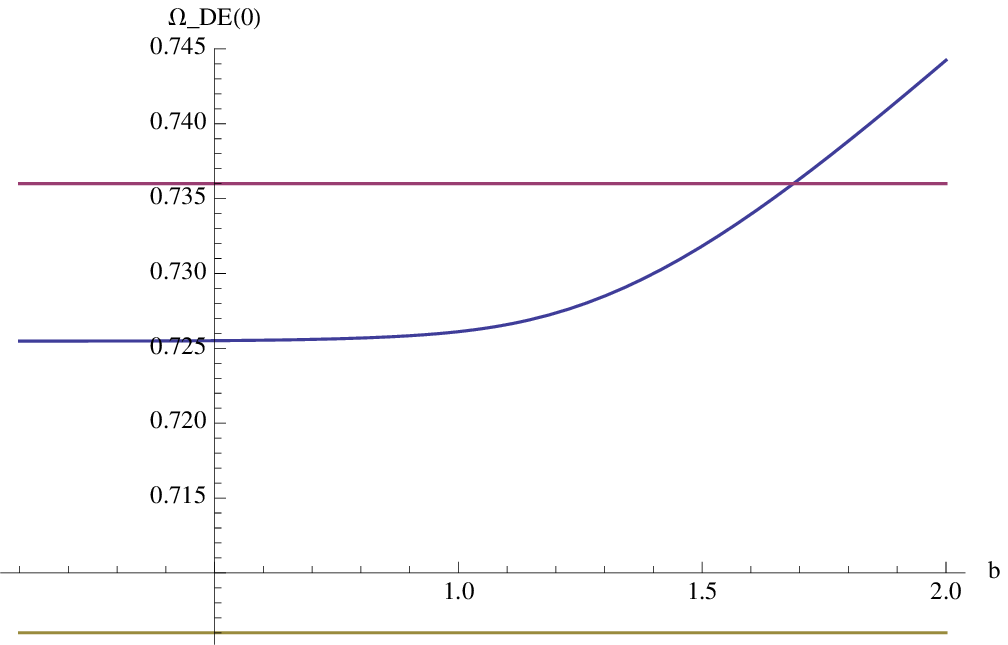}}
\caption{Behaviors 
of $\omega_\mathrm{DE}(z=0,b)$ and of $\Omega_\mathrm{DE}(z=0,b)$ 
as functions of $b$ for power-law model. 
Legend is the same as Fig.~\ref{b-1}. 
\label{b-2}}
\end{figure}

\paragraph{}We take $\tilde{\gamma}=1/1000$ in the models in Eqs.~(\ref{F3exp}) and (\ref{F3HS}), keeping the parameter $b$ free. Now, the dark energy density is a function of $z$ and $b$, namely, $ y_H(z,b)$. 
We can solve Eq.~(\ref{superEq}) numerically, taking the initial conditions at $z_i=9$ as 

\begin{eqnarray}
\left\{\begin{array}{l}
\frac{d y_H(z,b)}{d (z)}\Big\vert_{z_i} = 
\frac{\Lambda}{3\tilde{m}^2}\tilde{\gamma}\,,\\ 
y_H(z,b)\Big\vert_{z_i} = \frac{\Lambda}{3\tilde{m}^2}\left[1+\tilde{\gamma}\,(z_i+1)\right]\,,
\end{array}\right.
\end{eqnarray}
as we did in the previous Section. 
We consider the range $0.1<b<2$. In Figs.~\ref{b-1} and \ref{b-2}, 
we display the resultant values of dark energy EoS parameter $\omega_\mathrm{DE}(z=0,b)$ and 
$\Omega_\mathrm{DE}(z=0,b)$ at the present time as functions of $b$ for the two models. 
We also show the bounds of cosmological data in~(\ref{data}), namely, 
the lines in rose denote the upper bounds, and the lines in yellow denote 
the lower ones. 
From the graphics of every model, 
we find that in order to correctly reproduce the universe where we live with exponential gravity in Eq.~(\ref{F3exp}), $0.1<b<1.174$, with power-law model in Eq.~(\ref{F3HS}), $0.1<b<1.699$. The choices 
in \S~\ref{matterstudy} are consistent with these results.


\subsection{Future universe evolution\label{dS}}

\begin{figure}[!h]
\subfigure[]{\includegraphics[width=0.3\textwidth]{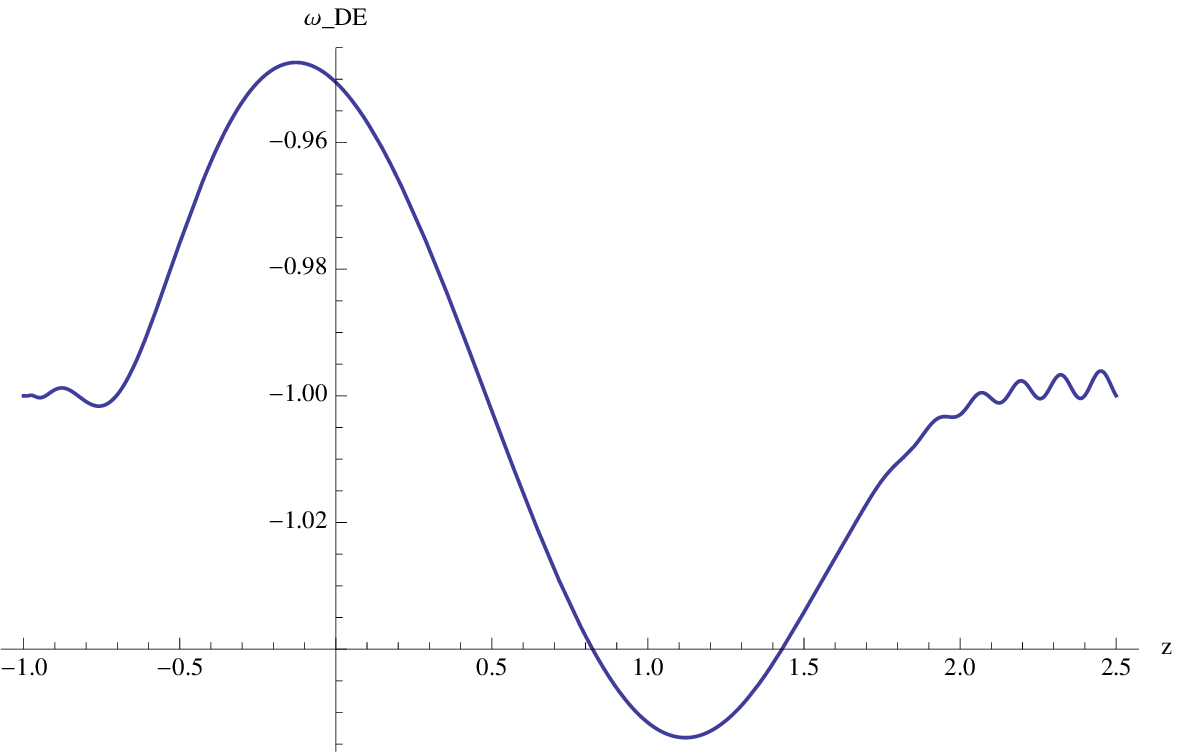}}
\centering
\subfigure[]{\includegraphics[width=0.3\textwidth]{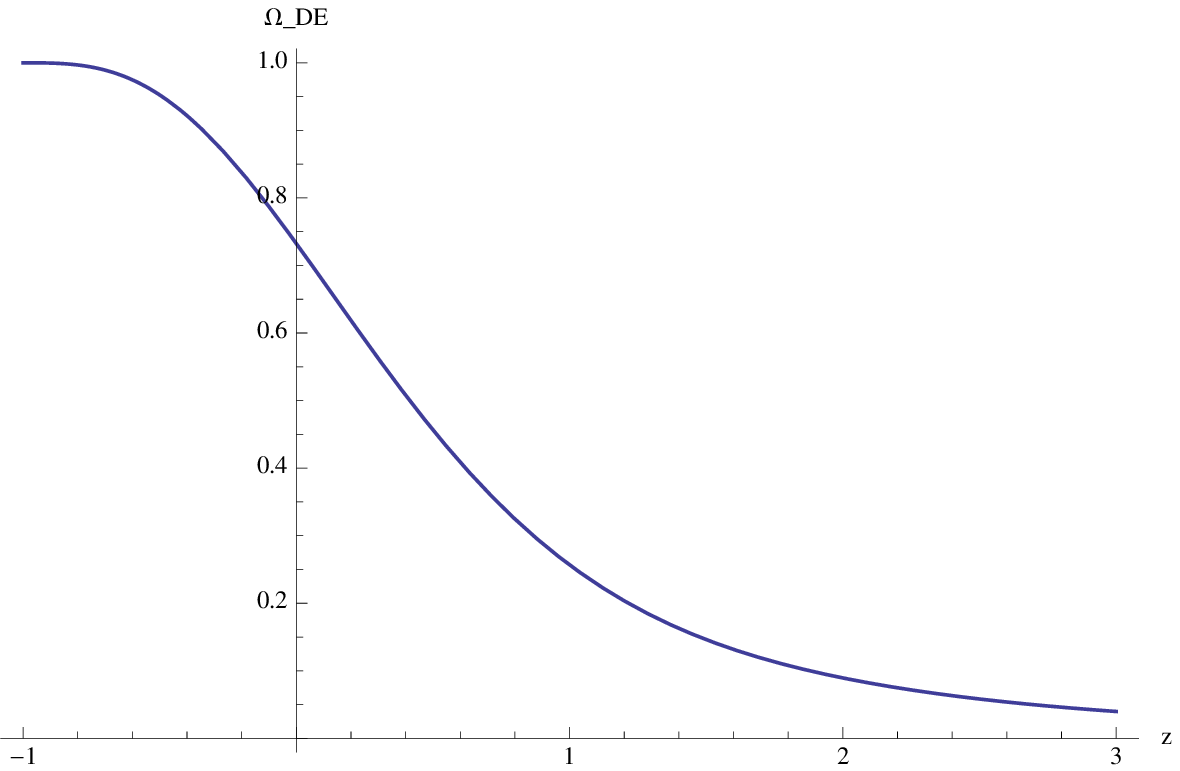}}
\centering
\subfigure[]{\includegraphics[width=0.3\textwidth]{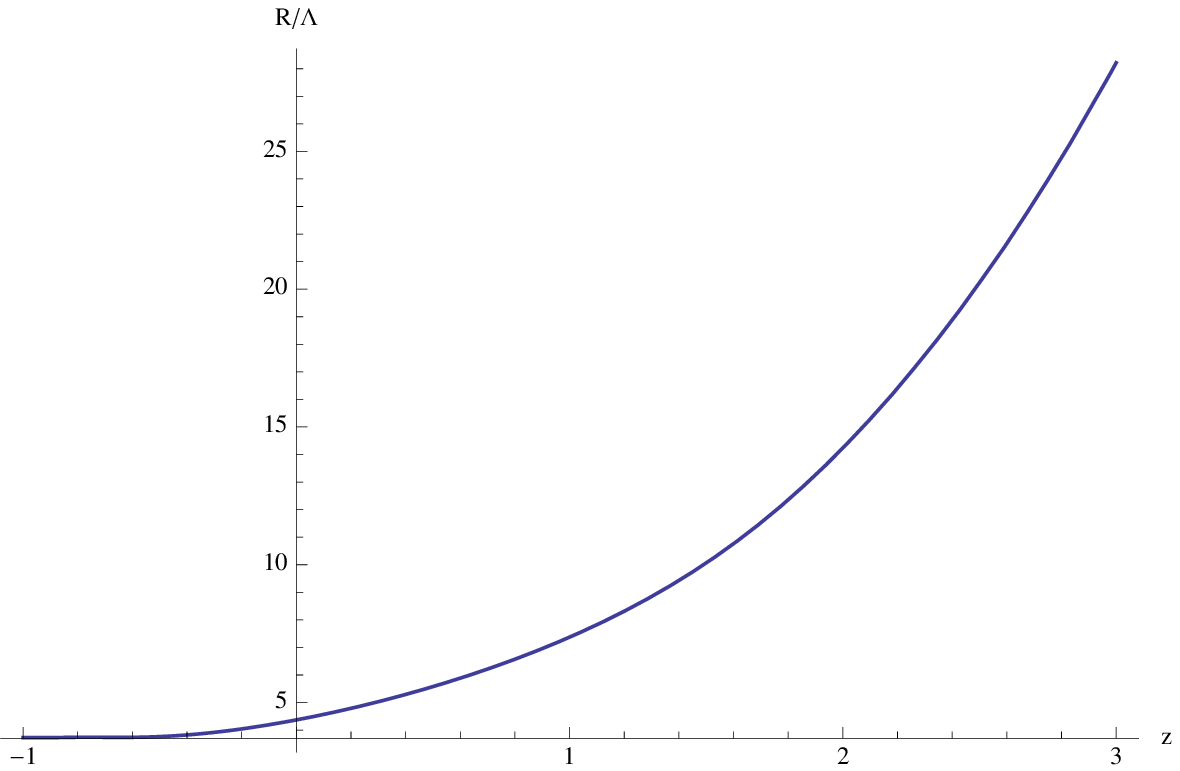}}
\caption{
Plot of $\omega_{DE}$ (a), $\Omega_{DE}$ (b) and $R/\Lambda$ for exponential model (\ref{F3exp}) with $b=1$ and $\tilde\gamma=1/1000$.
\label{BimBumBam}}
\end{figure}

\paragraph{}In the de Sitter universe, we have $R=R_{\mathrm{dS}}$, where $R_{\mathrm{dS}}$ follows from
the constant curvature given by the constant dark energy density $y_H=y_{0}$, 
such that $R_{\mathrm{dS}}=12\tilde{m}^2y_0$. 
Dark energy reads as in Eq. (\ref{yexpansion}) and behaves as in Eq. (\ref{result}).
For models (\ref{F3exp})--(\ref{F3HS}), the de Sitter solution $R_{\mathrm{dS}}=4\Lambda$ is stable and condition (\ref{discriminant}) is well satisfied, such that dark energy has an oscillatory behaviour (\ref{oscillatorysolution}) and, in particular,
oscillates infinitely often around the line of the phantom divide 
$\omega_\mathrm{DE}=-1$, how is clear from Eq. (\ref{omegaoscillating}). 
According to observational data, 
the crossing of the phantom divide may be occurred 
in the near past~\cite{Observational-status1}-\cite{Observational-status4}.
These models possess one crossing in the recent past~\cite{Bamba}, after the end of the matter dominated era, and infinite crossings in the future, but the amplitude of such crossings decreases as $(z+1)^{3/2}$ and this fact does not cause any serious problem to the accuracy of the cosmological evolution during the de Sitter epoch which is in general the final 
attractor of the system~\cite{Bamba, Twostepmodels}. As an example, we can consider exponential model (\ref{F3exp}) with $b=1$ and $\tilde\gamma=1/1000$ (see \S~\ref{matterstudy} for numerical extrapolation).
In Figs.~\ref{BimBumBam}, we plot
$\omega_\mathrm{DE}$, $\Omega_{DE}$ and $R/\Lambda$ as functions of the redshift in the late time matter era/de Sitter region ($-1<z<3$). We note the oscillatory behaviour of $\omega_{\mathrm{DE}}$ around the line of phantom divide. As we previously observed, the values of $\omega_{\mathrm{DE}}$ and $\Omega_{\mathrm{DE}}$ are very close to the cosmological data at the present time. Moreover, in the asymptotic limit $z\rightarrow -1^+$, $R$ tends to $3.727\Lambda$ (the effective cosmological constant of the model is $0.932\Lambda$). As a consequence, the de Sitter solution is a final attractor of the system and describes an eternal accelerating expansion.

\begin{figure}[!h]
\subfigure[]{\includegraphics[width=0.4\textwidth]{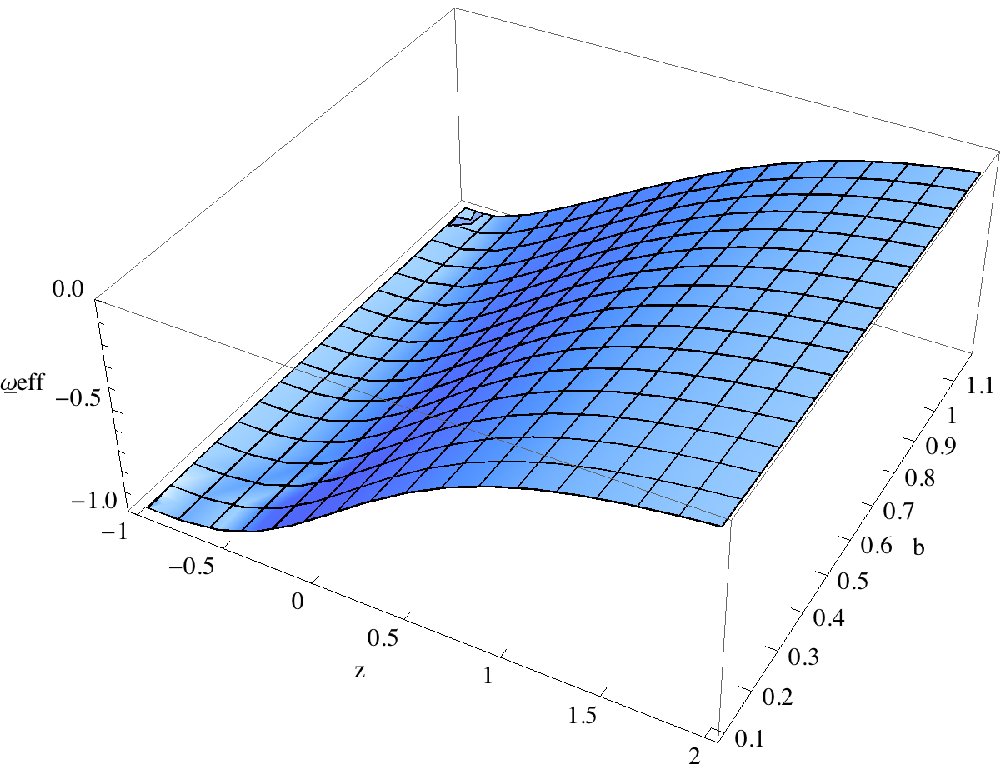}}
\qquad
\subfigure[]{\includegraphics[width=0.4\textwidth]{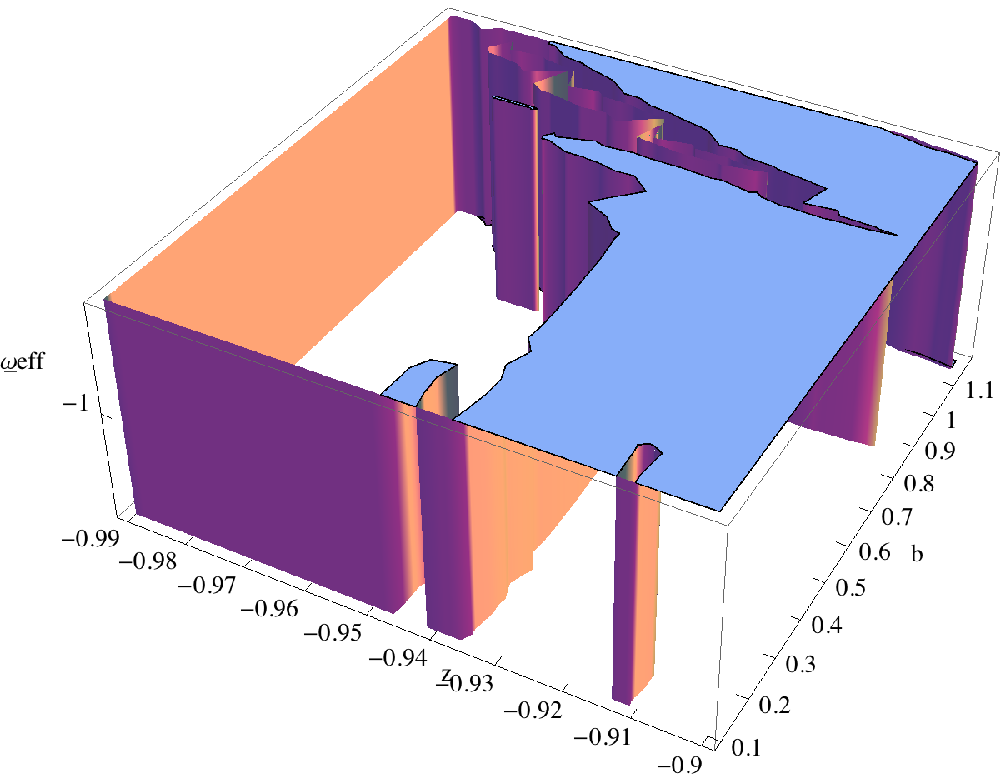}}
\caption{Cosmological evolution of $\omega_{\mathrm{eff}}$ as a function of the red shift $z$ and the $b$ parameter for the model in Eq.~(\ref{F3exp}). 
The left panel plots it for $-1<z<2$ and the right one displays 
around $z=-1$.
\label{33}}
\end{figure}

\begin{figure}[!h]
\qquad
\subfigure[]{\includegraphics[width=0.4\textwidth]{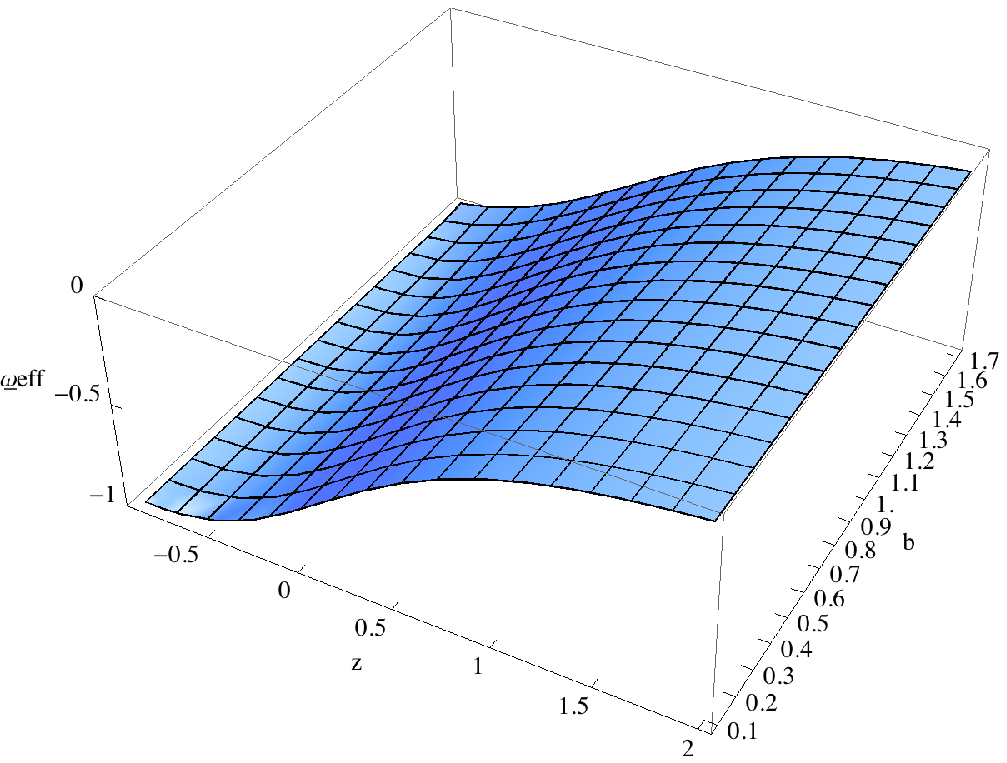}}
\qquad
\subfigure[]{\includegraphics[width=0.4\textwidth]{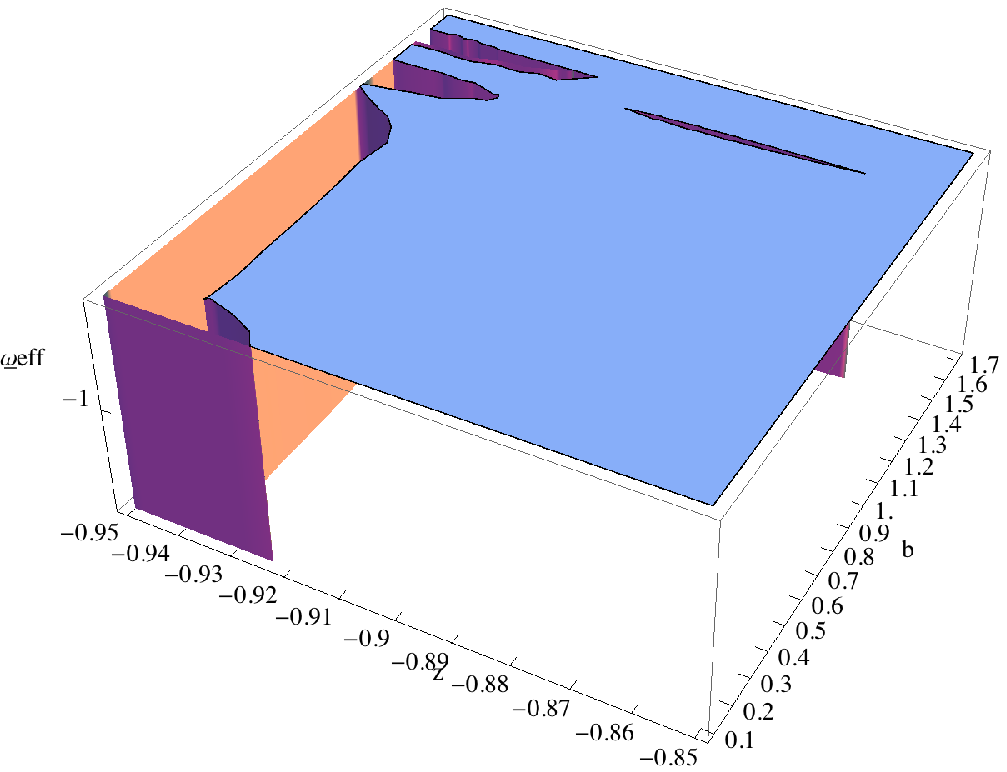}}
\caption{Cosmological evolution of $\omega_{\mathrm{eff}}$ as a function of the red shift $z$ and the $b$ parameter for the model in Eq.~(\ref{F3HS}). 
Legend is the same as Fig.~\ref{33}.
\label{33bis}}
\end{figure}

However, the existence of a phantom phase can give some undesirable effects, like 
the possibility to have the Big Rip~\cite{Caldwell} as an alternative scenario of the universe or 
the disintegration of bound structures which does not necessarily require the final (Big Rip) singularity~\cite{lRip1,lRip2,Rip2}. 
In this Subsection, we show that in the models in Eqs.~(\ref{F3exp})--(\ref{F3HS}) the effective EoS parameter (\ref{omegaeff}) of the universe,
\begin{equation}
\omega_{\mathrm{eff}}\equiv\frac{\rho_{\mathrm{eff}}}{p_{\mathrm{eff}}}=-1+\frac{2(z+1)}{3H(z)}\frac{dH(z)}{dz}\,, 
\end{equation}
never crosses the phantom divide line in the past, and that when $z$ is very close to $-1$ 
it coincides with $\omega_{\mathrm{DE}}$ and the crossings occur. 
We remark that $\rho_{\mathrm{eff}}$ and $p_{\mathrm{eff}}$ correspond 
to the total energy density and pressure of the universe, 
and only when dark energy strongly dominates over ordinary matter, 
we can consider $\omega_{\mathrm{eff}} \approx \omega_{\mathrm{DE}}$. 
In both of the models under investigation, 
we take again $\tilde{\gamma}=1/1000$ and keep the parameter $b$ free, 
such that $0.1<b<1.174$ (model in Eq.~(\ref{F3exp})) and $0.1<b<1.699$ 
(model in Eq.~(\ref{F3HS})), according to the realistic representation 
of current universe. 
From the numerical evaluation of Eq.~(\ref{superEq}), we can also get  
$H(z)$, given by 
\begin{equation}
H(z)=\sqrt{\tilde m^2\left[y_H(z)+(z+1)^3+\chi(z+1)^4\right]}\,,
\end{equation}
and therefore $\omega_{\mathrm{eff}}(z)$. 
We depict the cosmological evolution 
of $\omega_{\mathrm{eff}}$ as a function of the red shift $z$ and 
the $b$ parameter in Fig.~\ref{33} for the model in Eq.~(\ref{F3exp}) 
and in Fig.~\ref{33bis} for the model in Eq.~(\ref{F3HS}). 
On the left panels, we plot the effective EoS parameter for $-1<z<2$. 
In both of the models, independently on the choice of $b$, 
$\omega_{\mathrm{DE}}$ starts from zero in the matter  
era and asymptotically approaches -1 without any appreciable deviation. 
Only when $z$ is very close to $-1$ and the matter contribution to 
$\omega_{\mathrm{eff}}$ is effectively zero, 
we observe the crossing of the phantom divide due to the oscillations of
the 
dark energy. 
On the right panels, we display the behavior of the effective EoS parameter 
around $z=-1$ and 
we focus on the phantom divide line, by excluding the graphic area out of the range $-1.0001<\omega_{\mathrm{eff}}<-0.9999$. 
The blue region indicates that $\omega_{\mathrm{eff}}$ is still in the quintessence phase. We note that especially in the model in Eq.~(\ref{F3HS}), the first crossing of phantom divide is very far in the future. 
For example, with the scale factor $a(t)=\exp \left[H_{\mathrm{dS}}\,t\right]$, 
where $H_\mathrm{dS}\simeq 6.3\times 10^{-34}\text{eV}^{-1}$ is the Hubble parameter of 
the de Sitter universe, $z=-0.90$ 
corresponds to $10^{26}$ years!


\subsection{Growth of the matter density perturbations: growth index}

\begin{figure}[!h]
\subfigure[]{\includegraphics[width=0.45\textwidth]{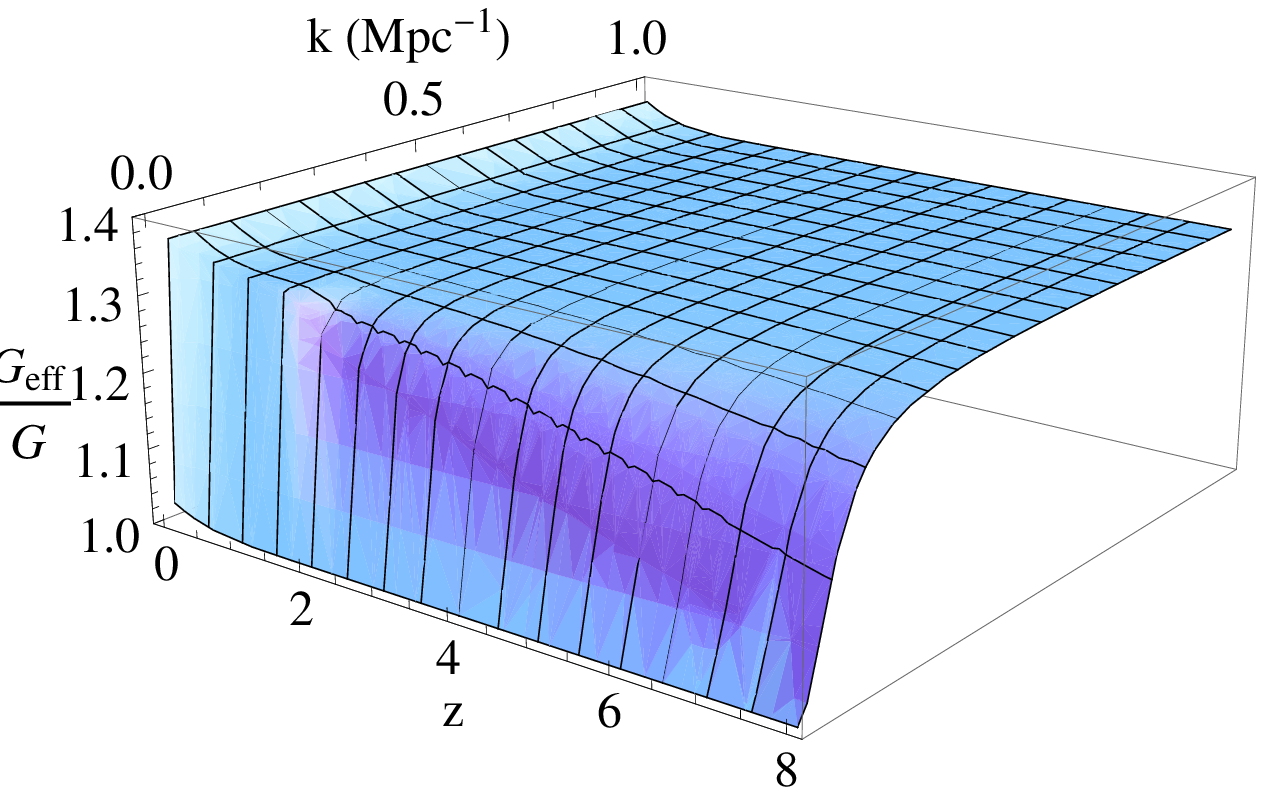}}
\qquad
\subfigure[]{\includegraphics[width=0.45\textwidth]{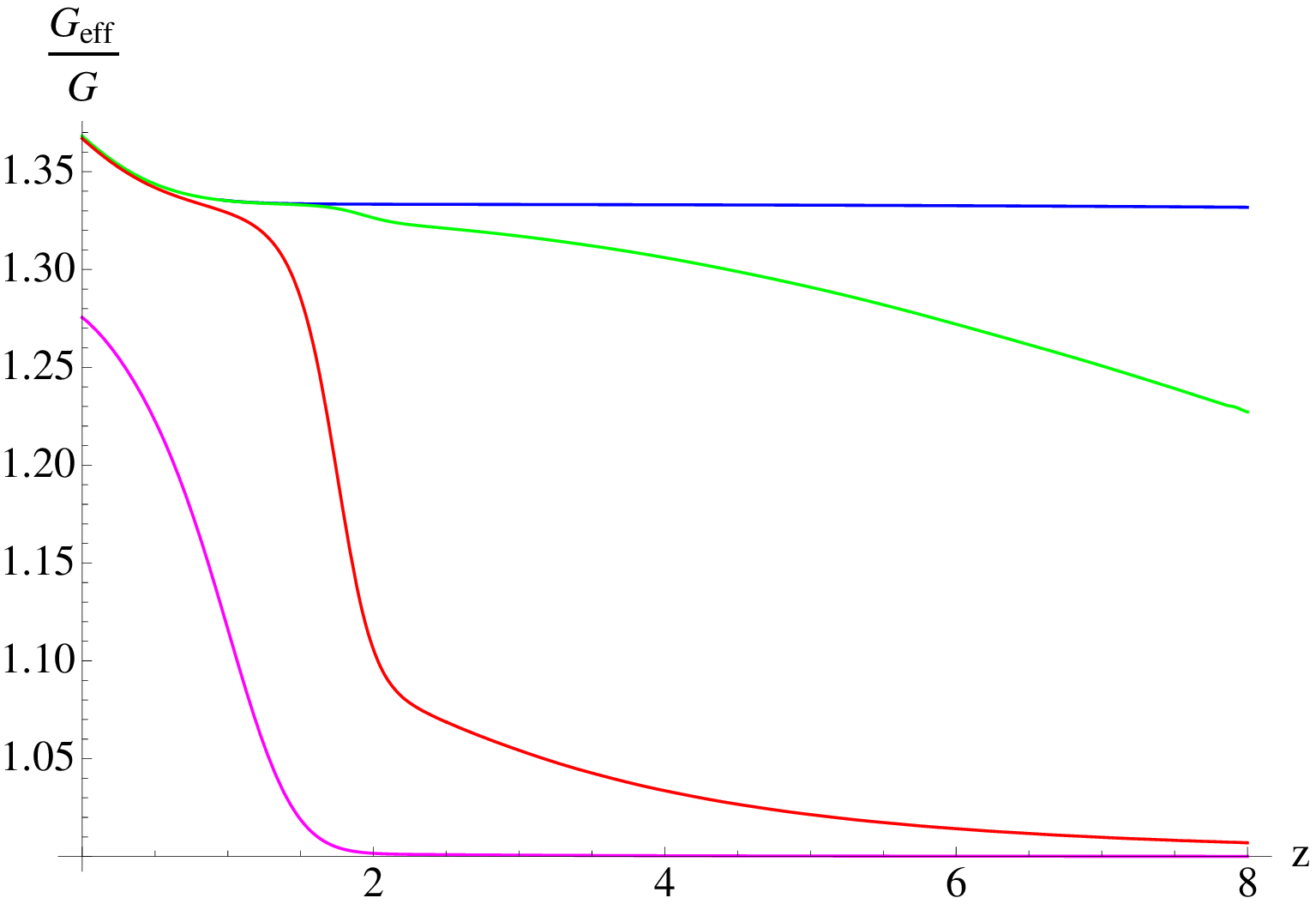}}
\caption{(a) Cosmological evolution as a function of $z$ and the scale dependence on $k$ of the effective gravitational constant $G_\mathrm{eff}/G_N$ 
for the model $F_1(R)$ with $b=1$ and $\tilde{\gamma}=1/1000$. (b) Cosmological evolution of $G_\mathrm{eff}/G_N$ as a function of $z$ in the model $F_1(R)$ with $b=1$ and $\tilde{\gamma}=1/1000$ 
for $k = 1 \mathrm{Mpc}^{-1}$ (blue), $k = 0.1 \mathrm{Mpc}^{-1}$ (green), $k = 0.01 \mathrm{Mpc}^{-1}$ (red) and $k = 0.001 \mathrm{Mpc}^{-1}$ (fuchsia).}
\label{EM_eff_grav_const}
\end{figure}

\begin{figure}[!h]
\subfigure[]{\includegraphics[width=0.45\textwidth]{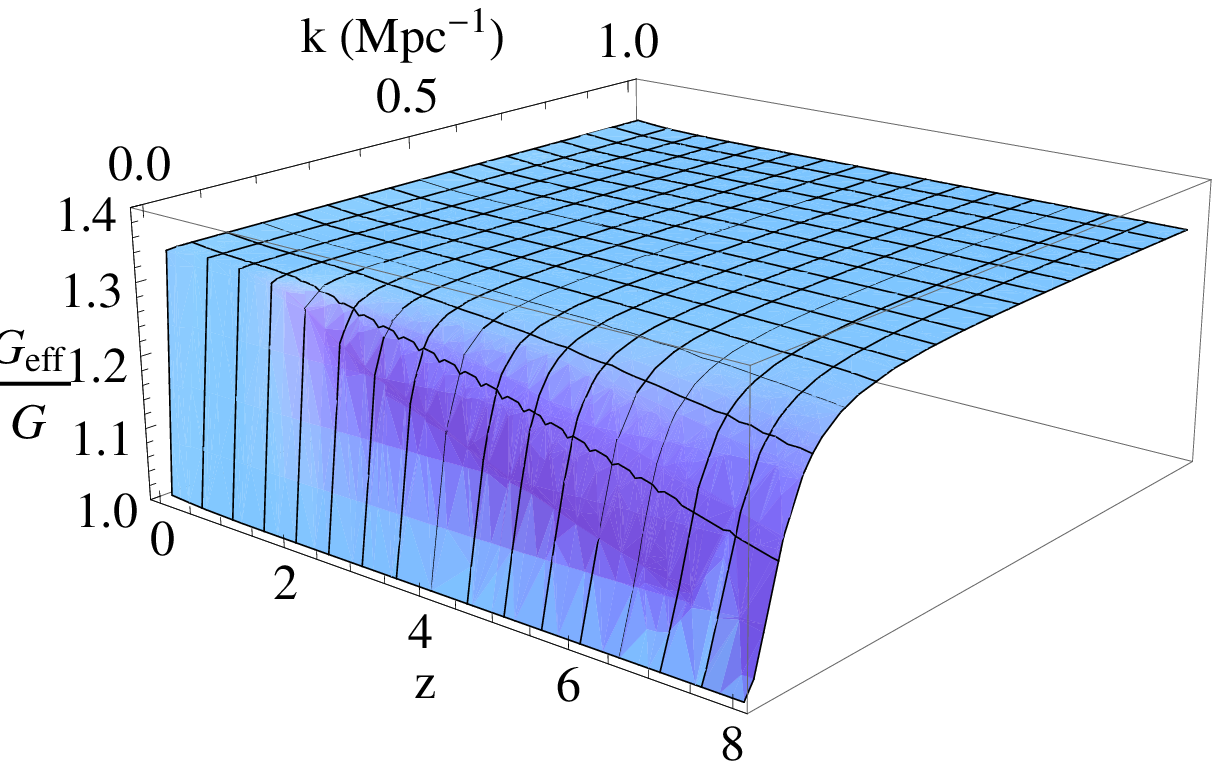}}
\qquad
\subfigure[]{\includegraphics[width=0.45\textwidth]{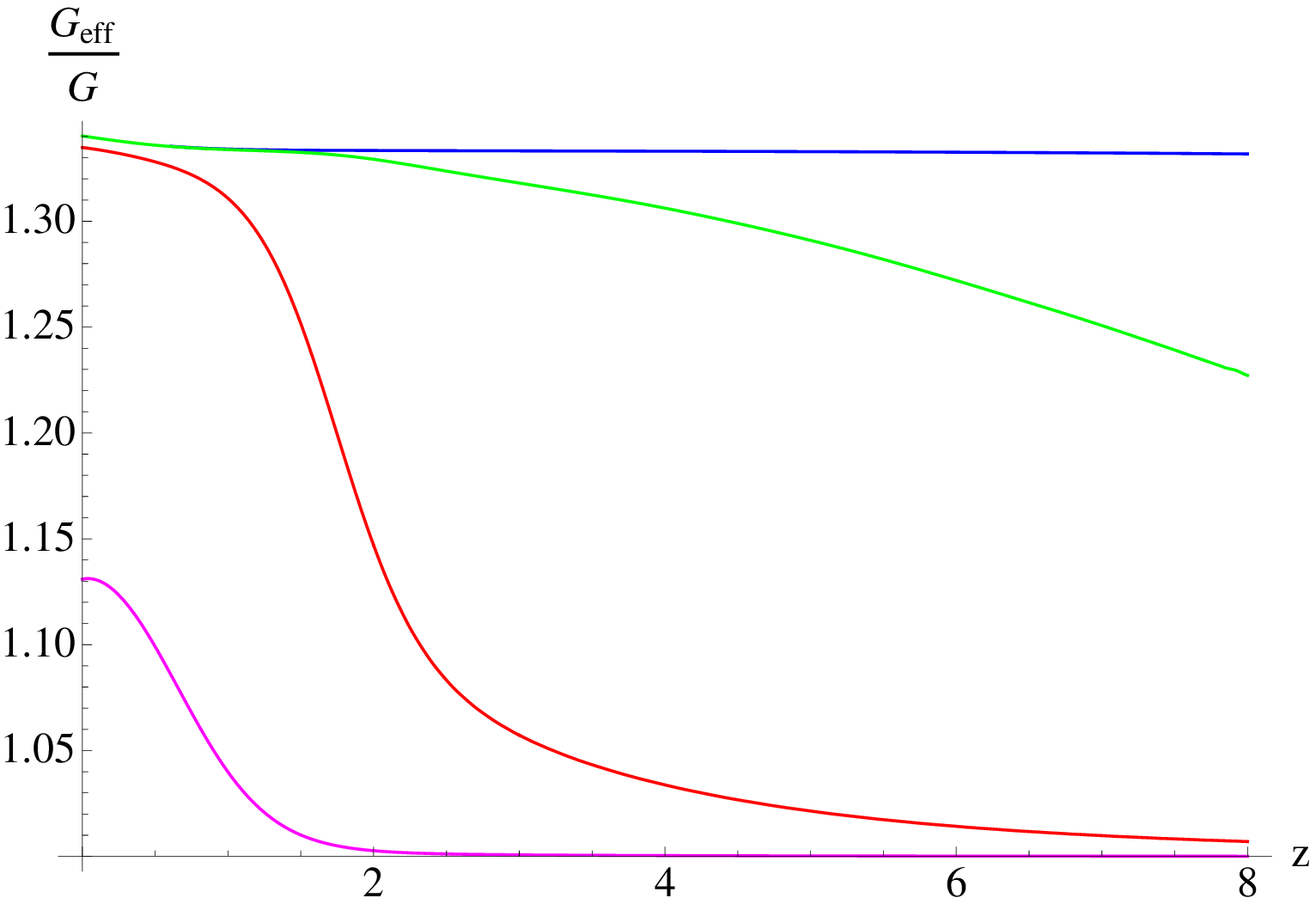}}
\caption{(a) Cosmological evolution as a function of $z$ and the scale dependence on $k$ of the effective gravitational constant $G_\mathrm{eff}/G_N$ 
for the model $F_2(R)$ with $b=1$ and $\tilde{\gamma}=1/1000$. (b) Cosmological evolution of $G_\mathrm{eff}/G_N$ as a function of $z$ for the model $F_2(R)$ with $b=1$ and $\tilde{\gamma}=1/1000$. 
Legend is the same as Fig.~\ref{EM_eff_grav_const}.}
\label{HS_eff_grav_const}
\end{figure}

\paragraph{} In this Subsection, we study the matter density perturbations. 
The equation that governs the evolution of the matter density perturbations in $F(R)$-gravity has been derived in Ref.~\cite{Tsujikawa:2007gd1,Tsujikawa:2007gd2,Tsujikawa:2007gd3}. 
Under the subhorizon approximation, 
the matter density perturbation 
$\delta = \delta \rho_\mathrm{m}/\rho_\mathrm{m}$ 
satisfies the equation
\begin{equation}\label{gmp1}
\ddot \delta \, + \, 2 H \dot \delta \, - \, 4 \pi G_\mathrm{eff}(a,k) \rho_\mathrm{m} \delta = 0\,,
\end{equation}
with $k$ the comoving wavenumber and $G_\mathrm{eff}(a,k)$ 
the effective gravitational ``constant'' given by 
\begin{equation}\label{gmp2}
G_\mathrm{eff}(a,k) = \frac{G_N}{F'(R)} \left[ 1 + \frac{\left( k^2/a^2 \right) \left( F''(R)/F'(R) \right)}{1 + 3 \left( k^2/a^2 \right) \left( F''(R)/F'(R) \right)} \right]\,.
\end{equation}
It is worth noting that the appearance of the comoving wavenumber $k$ in the effective gravitational constant makes the evolution of the matter density perturbations dependent on the comoving wavenumber $k$. It can be easily verified that, by taking $F(R) = R$ in Eq.~(\ref{gmp2}), the evolution of the matter density perturbation does not have this kind of dependence in the case of GR. 
In Fig.~\ref{EM_eff_grav_const}, we depict the cosmological evolution 
as a function of the redshift $z$ and the scale dependence on the comoving wavenumber $k$ of the effective gravitational constant above for the model $F_1(R)$ in Eq.~(\ref{F3exp}), and 
in Fig.~\ref{HS_eff_grav_const} we plot those for the model $F_2(R)$ 
in Eq.~(\ref{F3HS}). 
In both these cases, we have fixed $b=1$ and used $\gamma=1/1000$. 


Another important remark is that in deriving Eq.~(\ref{gmp1}), we have assumed the subhorizon approximation~\cite{EspositoFarese:2000ij}, namely, 
comoving wavelengths $\lambda = a(t)/k$ are considered to be much shorter than the Hubble radius $H^{-1}$ as 
\begin{equation}\label{gmp3}
\frac{k^2}{a^2} \gg H^2\,.
\end{equation}
It means that we must examine the scales of $\log [k] \geq -3$. 
On the other hand, for large $k$ we have to take into account deviations from the linear regime~\cite{Cardone:2012xv}. Thus, we do not consider the scales of $\log [k] > -1$ 
and take $\log [k]$ close to $-1$. 

From Figs.~\ref{EM_eff_grav_const}--\ref{HS_eff_grav_const}, we observe that $G_\mathrm{eff}$ measured today 
can significantly be different from the Newton's constant in the past. 
The Newton's constant should be normalized to the current one 
as $G_\mathrm{eff}/G_N$. 
This implies that the Newton's constant at the decoupling epoch 
can be much lower than what is (implicitly) assumed in CMB codes like CAMB~\cite{Lewis:1999bs, CAMB-code}. 
It may significantly change the CMB power spectrum 
since it changes, for example, 
the relation between the gravitational interaction and the Thomson scattering 
rate. 
Due to the fact that we use the CMB data when we examine whether the theoretical results 
are consistent with the observations in the framework of General Relativity, 
it should be important to take into account this point. 
Therefore, if we compare our results with 
the observations, we have to use the CMB data with the present value of $G_\mathrm{eff}$ 
given by our $F(R)$-modified gravity model, instead of the Newton's constant $G_N$ of GR.

\begin{figure}[!h]
\subfigure[]{\includegraphics[width=0.45\textwidth]{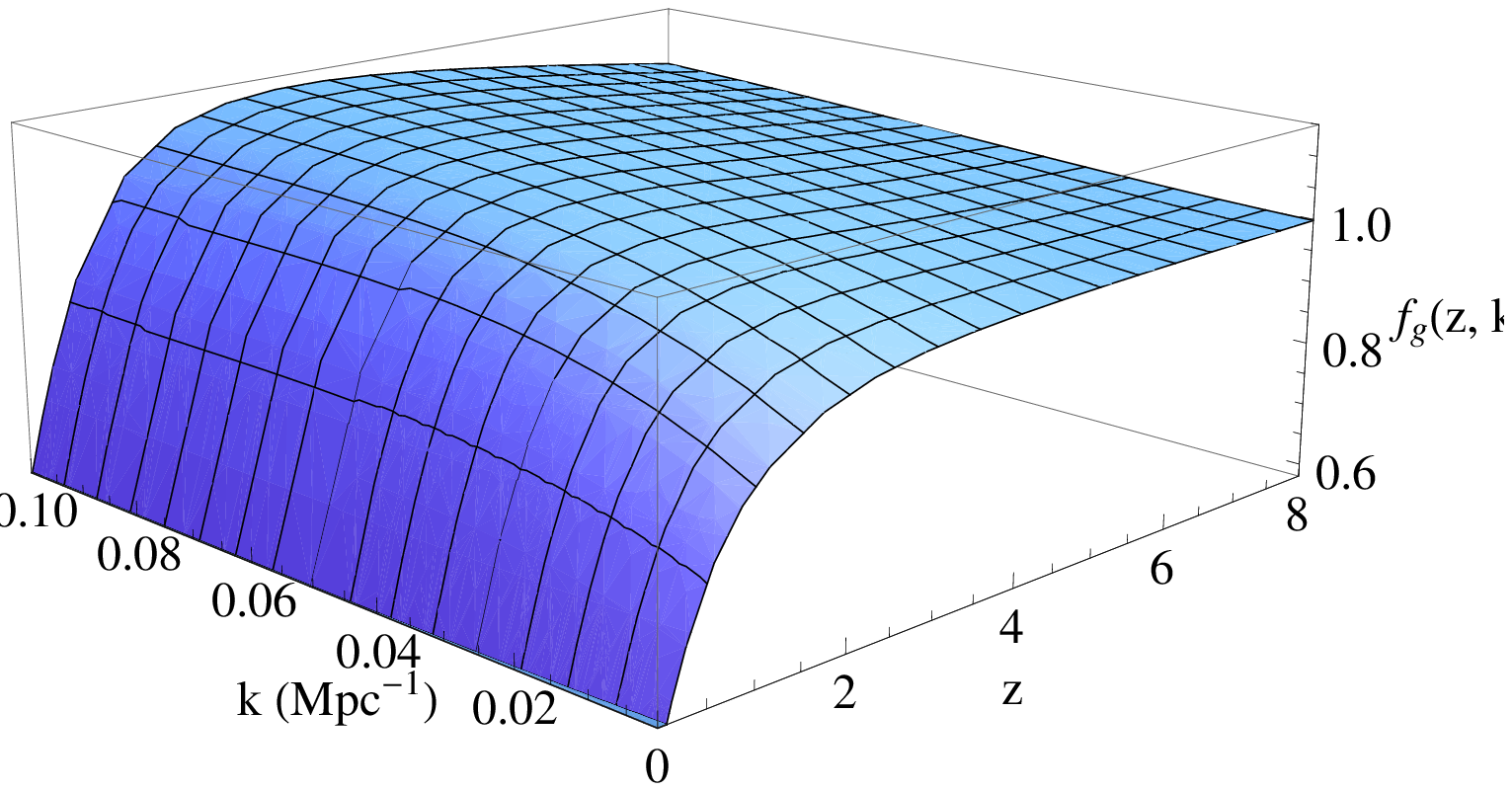}}
\qquad
\subfigure[]{\includegraphics[width=0.45\textwidth]{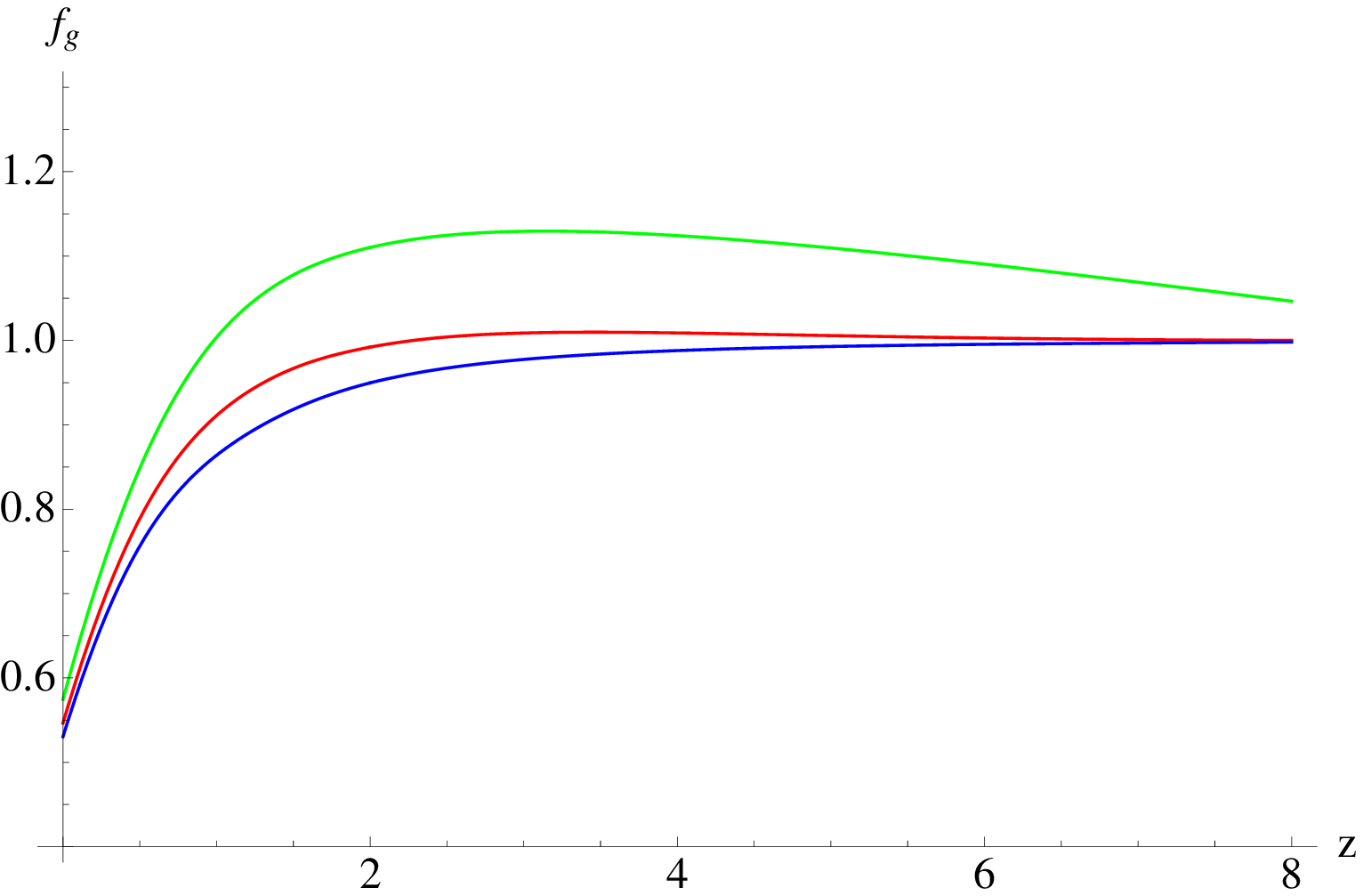}}
\caption{(a) Cosmological evolution as a function of the redshift $z$ and the scale dependence on the comoving wavenumber $k$ of the growth rate 
$f_\mathrm{g}$ for the model $F_1(R)$. (b) Cosmological evolution of the 
growth rate $f_\mathrm{g}$ as a function of $z$ in the model $F_1(R)$ for $k = 0.1 \mathrm{Mpc}^{-1}$ (green), $k = 0.01 \mathrm{Mpc}^{-1}$ (red) and $k = 0.001 \mathrm{Mpc}^{-1}$ (blue).}
\label{EM_growth_rate}
\end{figure}

\begin{figure}[!h]
\subfigure[]{\includegraphics[width=0.45\textwidth]{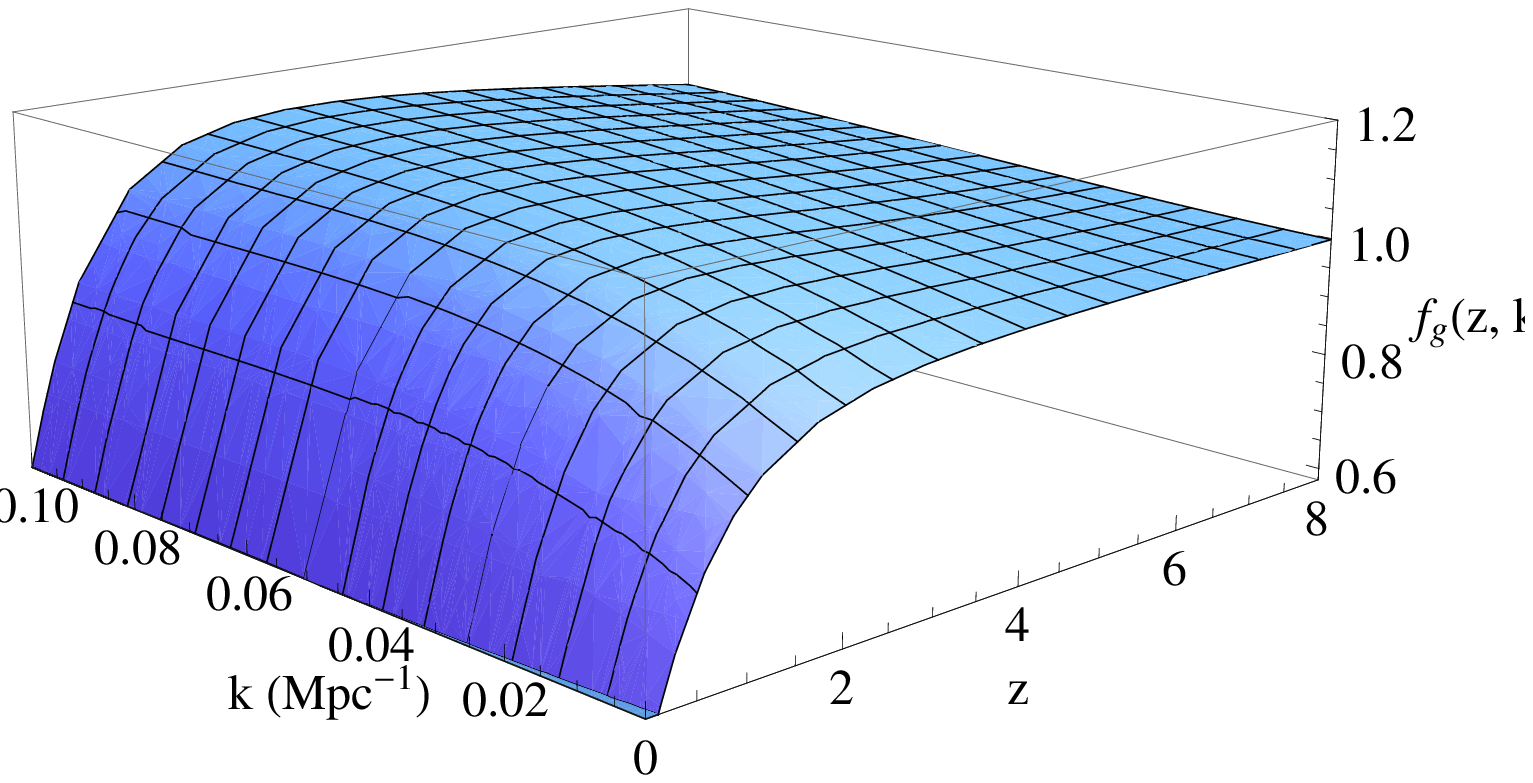}}
\qquad
\subfigure[]{\includegraphics[width=0.45\textwidth]{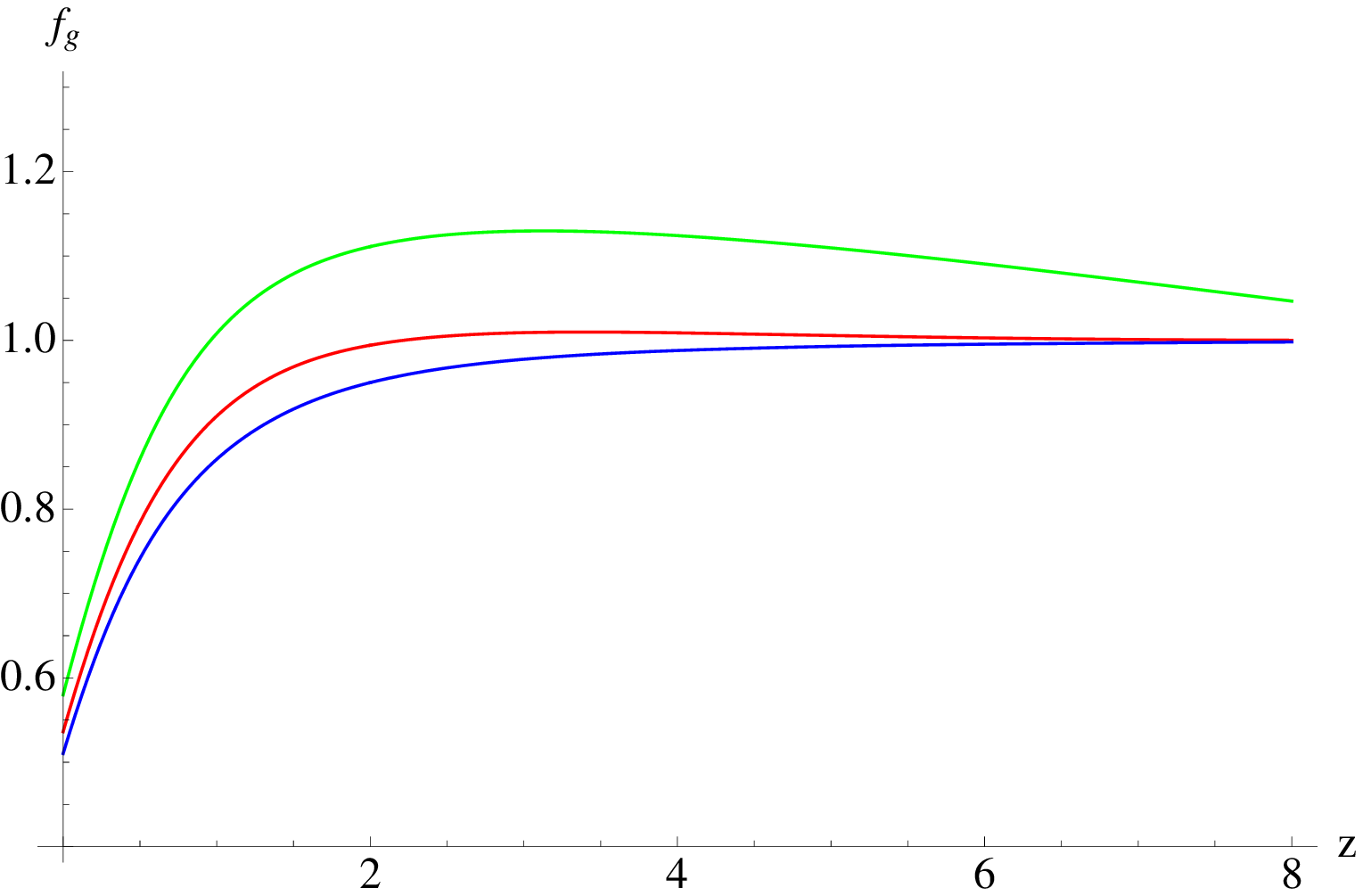}}
\caption{(a) Cosmological evolution as a function of the redshift $z$ and the scale dependence on the comoving wavenumber $k$ of the growth rate 
$f_\mathrm{g}$ for the model $F_2(R)$. (b) Cosmological evolution of 
the growth rate $f_\mathrm{g}$ as a function of $z$ for the model $F_2(R)$. 
Legend is the same as Fig.~\ref{EM_growth_rate}.}
\label{HS_growth_rate}
\end{figure}

Now we introduce the growth rate $f_\mathrm{g} \equiv d \ln{\delta}/d \ln{a}$ and solve the equivalent equation to Eq.~(\ref{gmp1}) for 
the growth rate in terms of the redshift $z$, 
\begin{equation}\label{gmp4}
\frac{df_\mathrm{g}(z)}{dz} \, + \, \left( \frac{1 + z}{H(z)} \frac{dH(z)}{dz} - 2 - f_\mathrm{g}(z) \right) \frac{f_\mathrm{g}(z)}{1 + z} + \frac{3}{2} \frac{\tilde{m}^2 (1 + z)^2}{H^2(z)} \frac{G_\mathrm{eff}(a(z),k)}{G} = 0\,. 
\end{equation}
Unfortunately, this equation cannot be solved analytically for the models $F_1(R)$ and $F_2(R)$, but it may be solved numerically by putting the initial conditions. 
We impose
that at a very high redshift the growth rate becomes that in the $\Lambda$CDM Model. 
In Fig.~\ref{EM_growth_rate}, we illustrate 
the cosmological evolution 
as a function of the redshift $z$ and the scale dependence on the comoving wavenumber $k$ of the growth rate for the model $F_1(R)$, while we plot those of the growth rate for the model $F_2(R)$ in Fig.~\ref{HS_growth_rate}. 


One way of characterizing the growth of the matter density perturbations could be to use the so-called growth index $\gamma$, which is defined as the quantity which satisfy the following equation: 
\begin{equation}\label{gi}
f_\mathrm{g}(z) = \Omega_\mathrm{m}(z)^{\gamma(z)}\,,
\end{equation}
with $\Omega_\mathrm{m}(z) = (8 \pi G_N) \rho_m/[3 H^2]$ being 
the matter density parameter. 

It is known that the growth index $\gamma$ in Eq.~(\ref{gi}) cannot be 
observed directly, but it can be determined from the observational data of the growth factor $f_\mathrm{g}(z)$ and the matter density parameter $\Omega_\mathrm{m}(z)$ at the same redshift. 
Despite the fact that the growth index is not directly observable quantity, it could have a 
fundamental importance in discriminating among the different cosmological models since, for example, the growth factor $f_\mathrm{g}(z)$, which may be estimated from redshift space distortions in the galaxy power spectra at different $z$~\cite{Kaiser:1987qv, Hamilton:1997zq}, cannot be expressed in terms of elementary functions and this fact makes the comparison among the different models difficult. 

Various parameterizations for the growth index $\gamma$ have been proposed in the literature. 
At the beginning, $\gamma$ was taken constant~\cite{Peebles:19841,Peebles:19842}. In the case of dark fluids with the constant EoS parameter
$\omega=\omega_0$ in GR, it is $\gamma = 3 \left(\omega_0 - 1\right)/\left(6 \omega_0 - 5\right)$ (for the $\Lambda$CDM model, the growth index is $\gamma \simeq 0.545$). Although taking $\gamma$ constant is very appropriated for a wide class of dark energy models in the framework of GR (for which $\left| d\gamma(0)/dz \right| < 0.02$), for modified gravity theories $\gamma$ is not a constant~\cite{Gannouji:2008wt, Cardone:2012xv} and the measurement of $\left| d\gamma(0)/dz \right|$ becomes important in distinguishing between different theories. 
For this reason, other parameterizations have been proposed. The case of a linear dependence $\gamma(z) = \gamma_0 + \gamma_1 z$, with $\gamma_{0,1}$ constants, was treated in Ref.~\cite{Polarski:2007rr}. Recently, an Ansatz of the type $\gamma(z) = \gamma_0 + \gamma_1 z/(1 + z)$ with $\gamma_0$ and $\gamma_1$ being constants was explored in Ref.~\cite{Belloso:2011ms} and a generalization given by $\gamma(z) = \gamma_0 + \gamma_1 z/(1 + z)^\alpha$ with $\alpha$ being a constant in Ref.~\cite{Cardone:2012xv}.
In the following, we study some of these parameterizations of the growth index for our models $F_1(R)$ and $F_2(R)$.

\subsubsection{$\gamma = \gamma_0$}

\paragraph{}We start by considering the following Ansatz for the growth index,
\begin{equation} 
\gamma = \gamma_0\,, 
\end{equation}
where $\gamma_0$ is a constant.

\begin{figure}
\subfigure[]{\includegraphics[width=0.45 \textwidth]{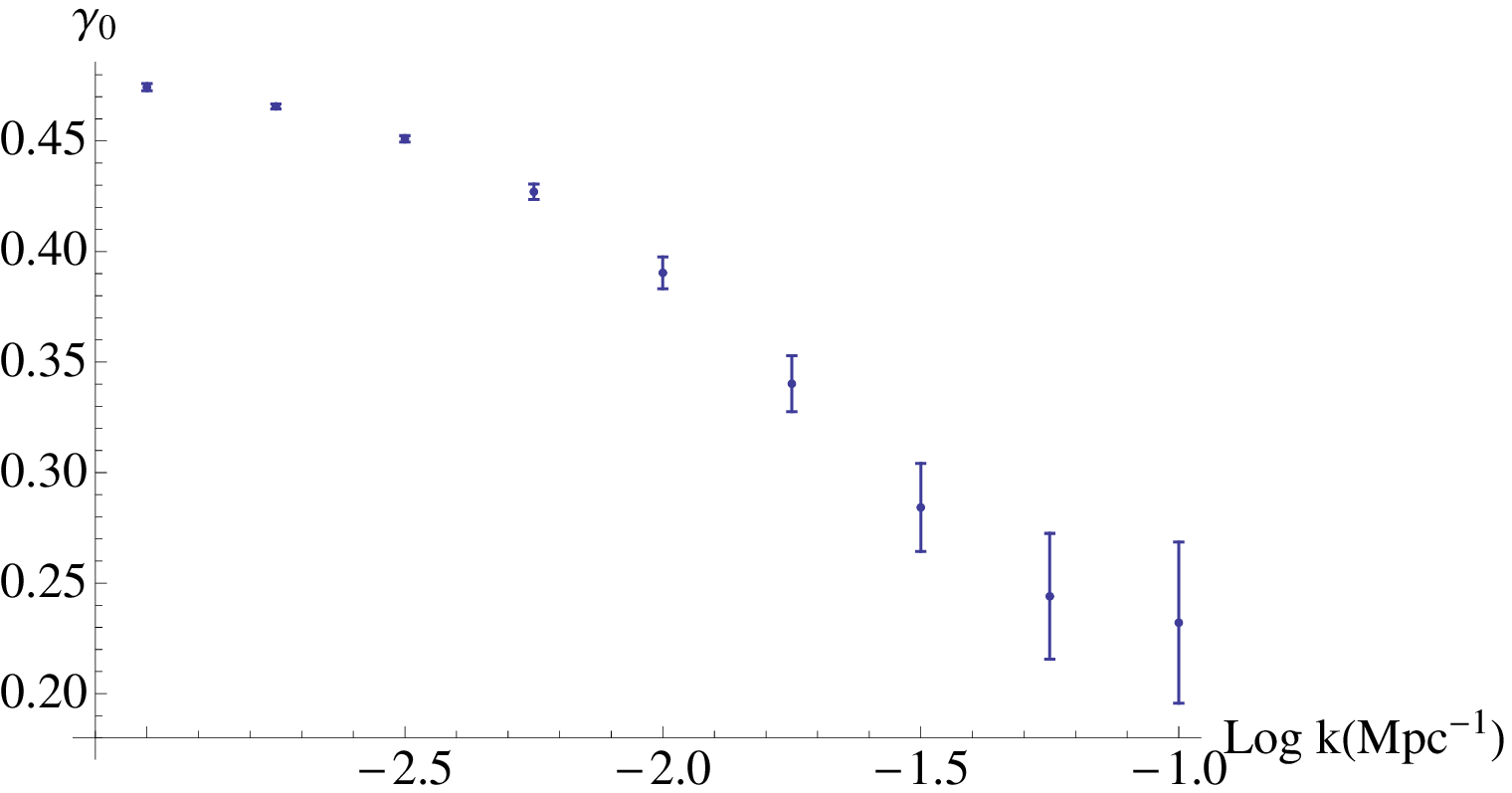}}
\quad
\subfigure[]{\includegraphics[width=0.45 \textwidth]{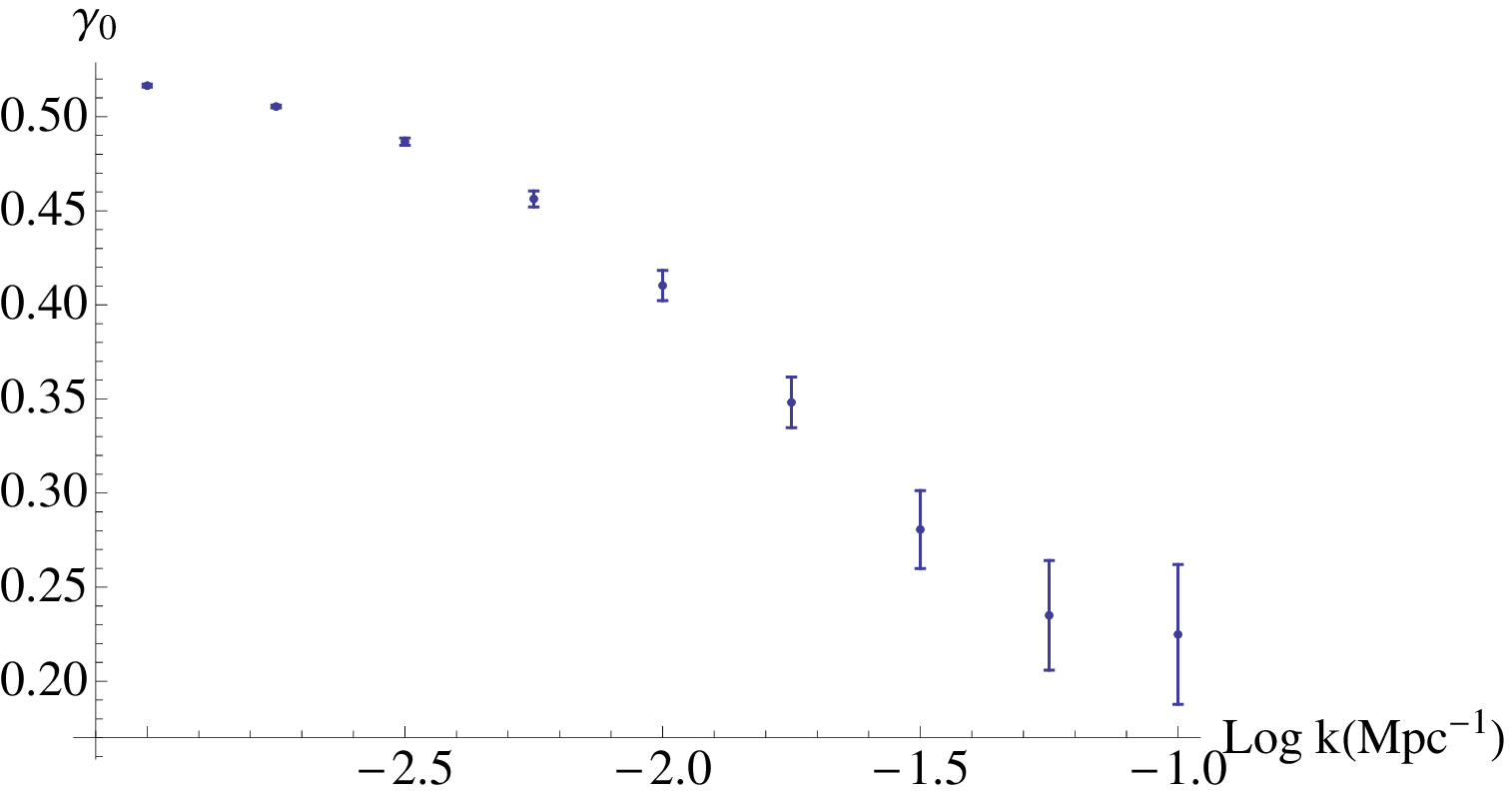}}
\caption{Constant growth index as a function of $\log k$ for the model $F_1(R)$ (a) and for the model $F_2(R)$ (b). The bars express the $68\%$ CL.}
\label{constant_growth_index_vs_logk}
\end{figure}

In Fig.~\ref{constant_growth_index_vs_logk}, we display the results obtained by fitting Eq.~(\ref{gi}) to the solution of Eq.~(\ref{gmp4}) for different values of the comoving wavenumber $k$ for the models $F_1(R)$ and $F_2(R)$. We stress that in these and following plots, the bars express the $68\%$ confidence 
level (CL) and the point denotes the median value. At first, we observe that the value of the growth index has a strong dependence with $\log [k]$ and this scale dependence seems to be quite similar in both of the models. 

\begin{figure}[!h]
\subfigure[]{\includegraphics[width=0.3\textwidth]{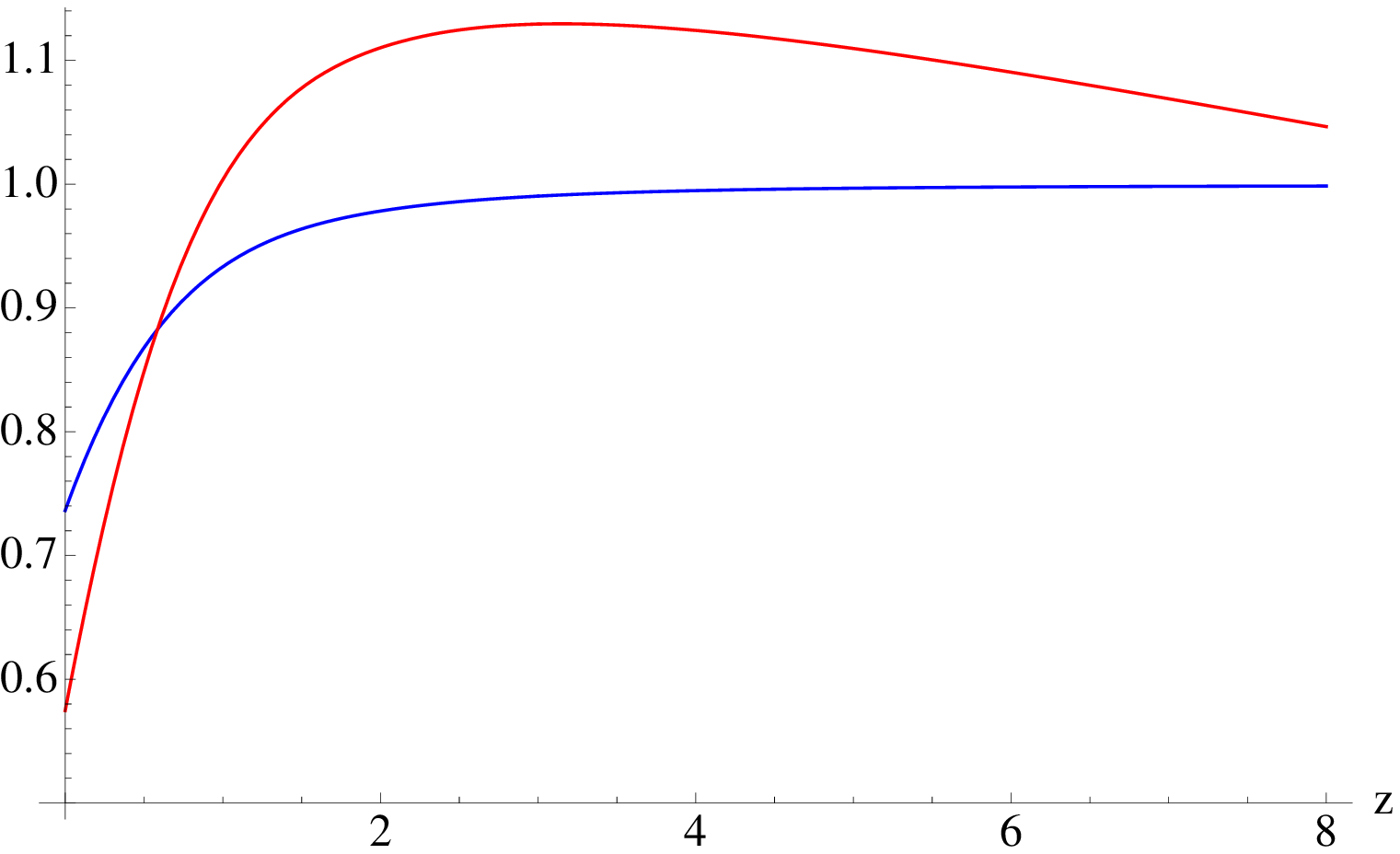}}
\quad
\subfigure[]{\includegraphics[width=0.3\textwidth]{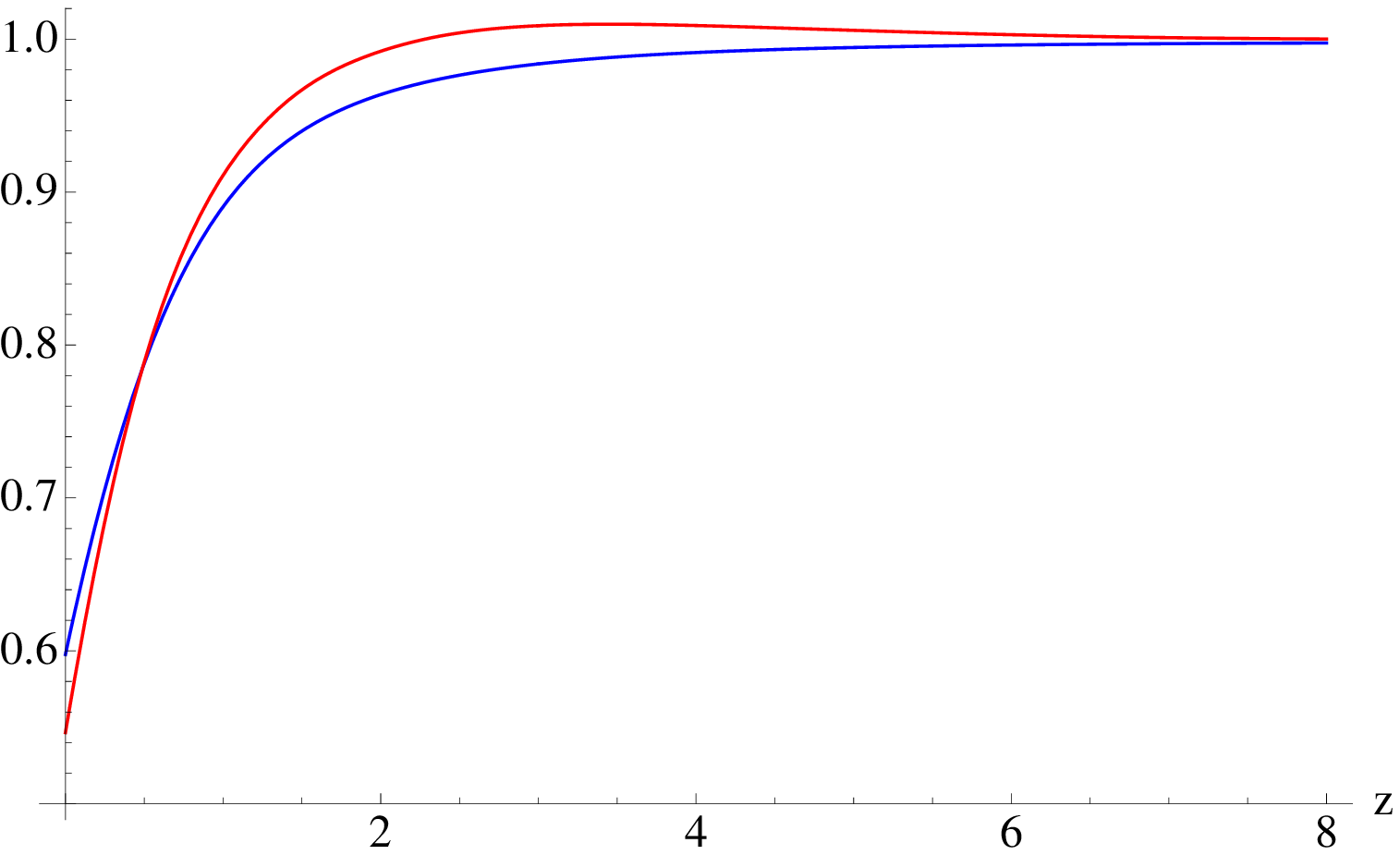}}
\quad
\subfigure[]{\includegraphics[width=0.3\textwidth]{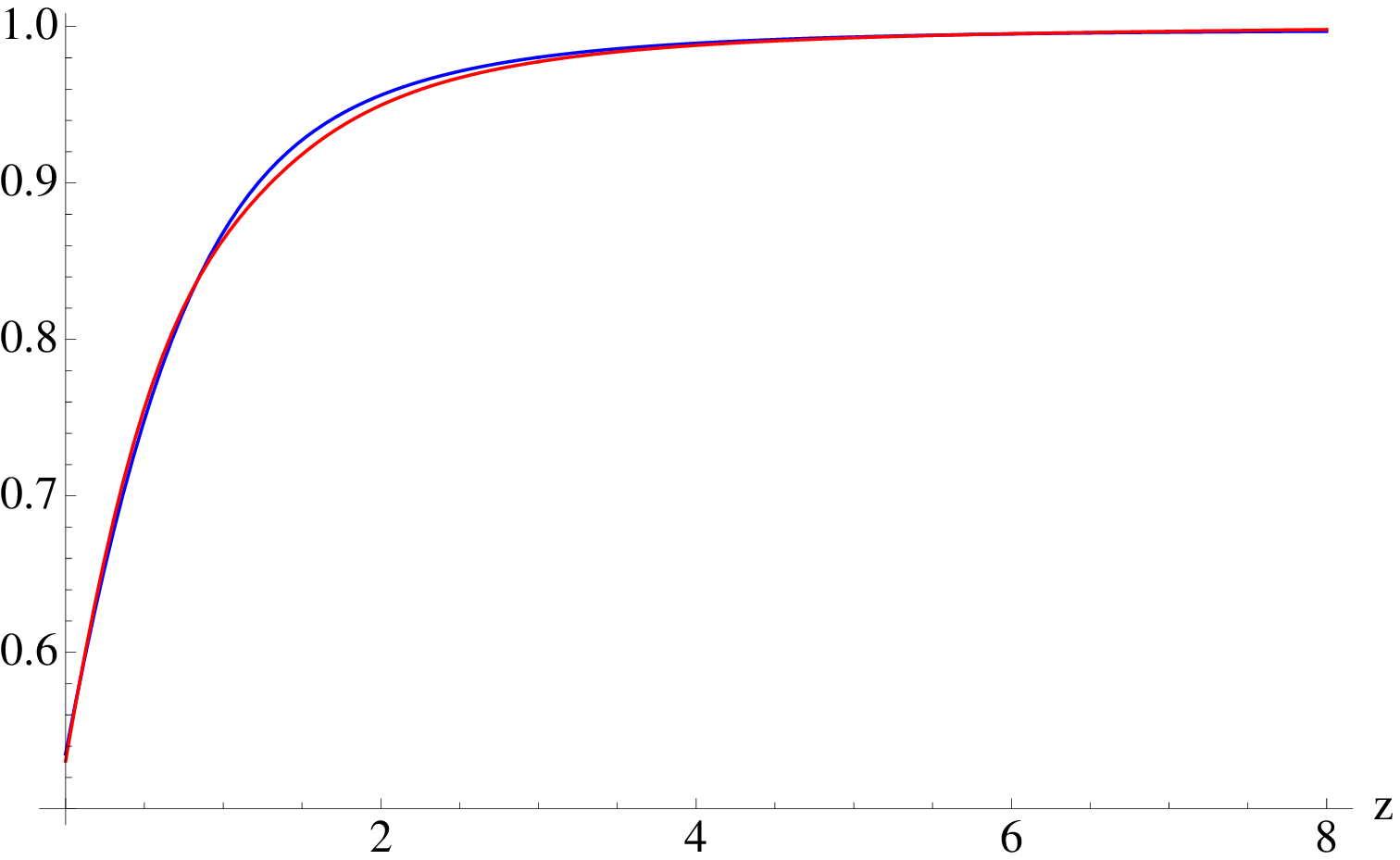}}
\quad
\subfigure[]{\includegraphics[width=0.3\textwidth]{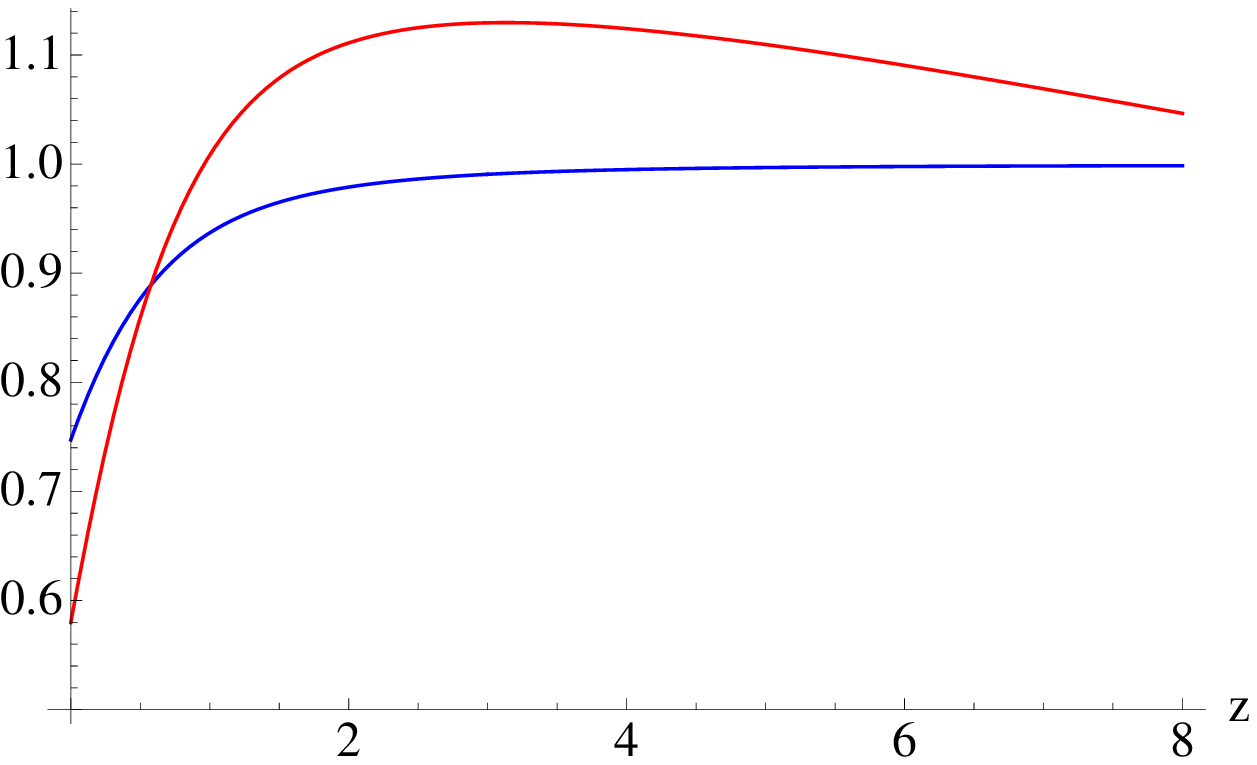}}
\quad
\subfigure[]{\includegraphics[width=0.3\textwidth]{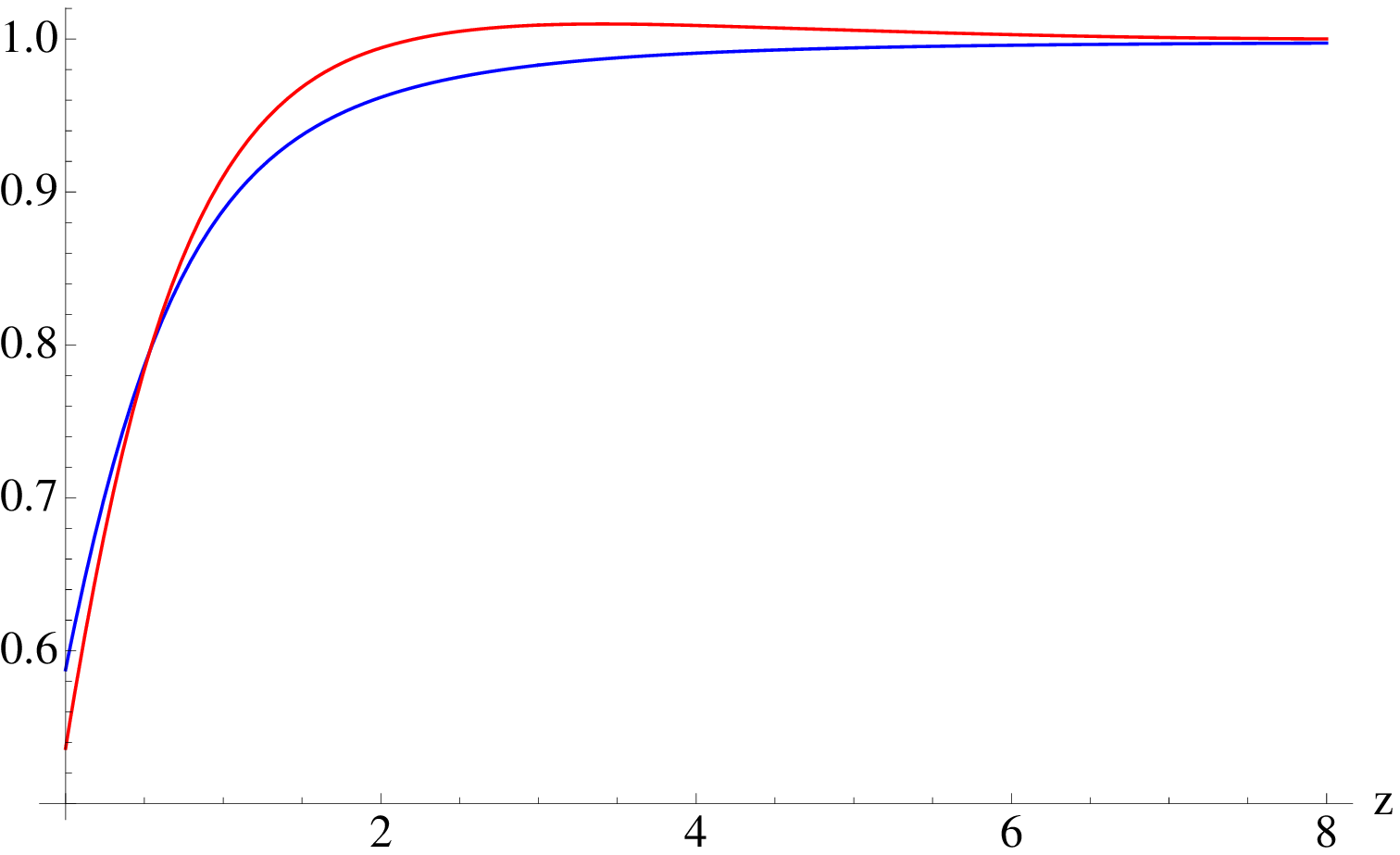}}
\quad
\subfigure[]{\includegraphics[width=0.3\textwidth]{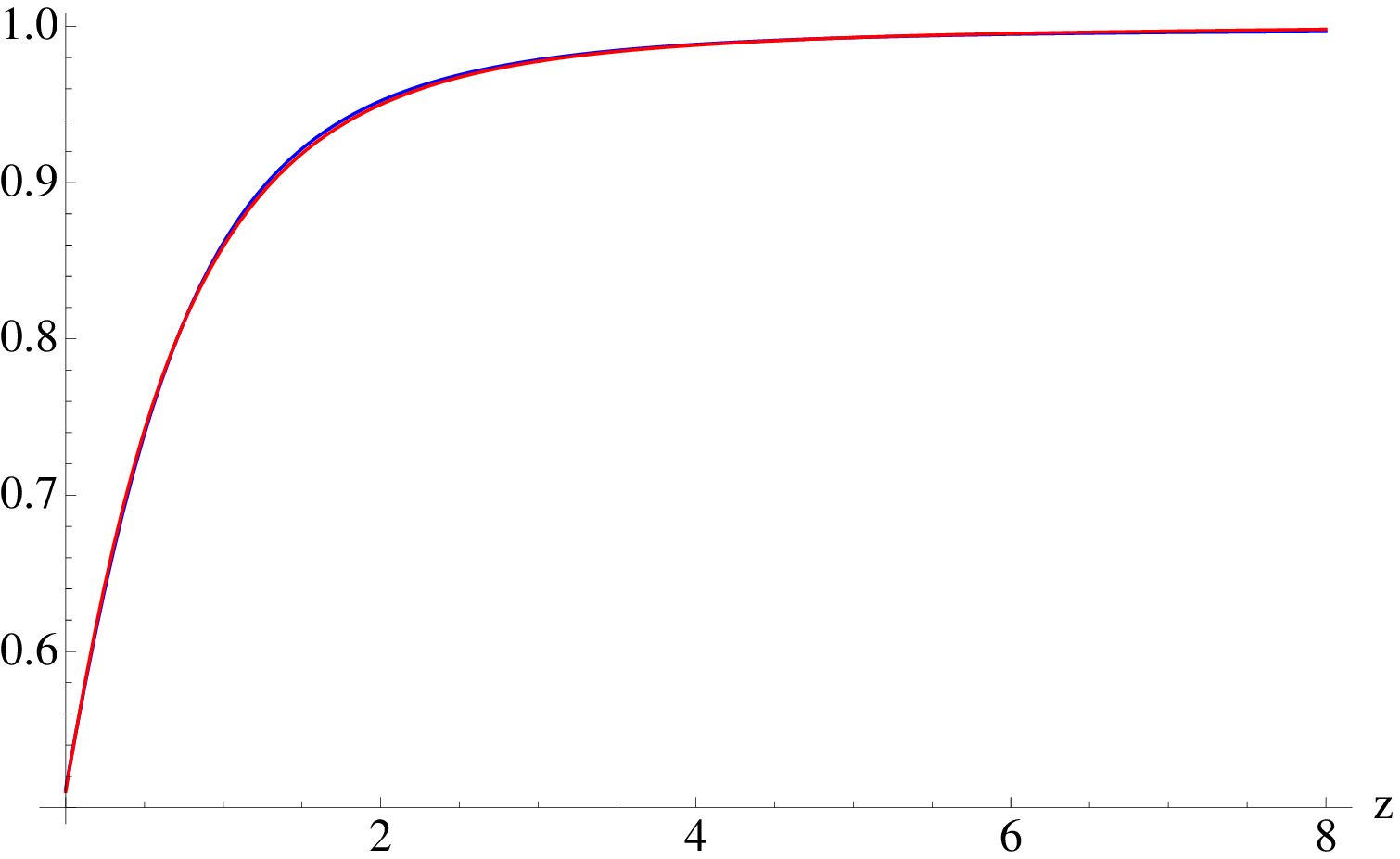}}
\caption{Cosmological evolutions of the 
growth rate $f_\mathrm{g}$ (red) and $\Omega_\mathrm{m}^\gamma$ (blue) with $\gamma = \gamma_0$ as functions of the redshift $z$ in the model $F_1(R)$ for $k = 0.1 \mathrm{Mpc}^{-1}$ (a), $k = 0.01 \mathrm{Mpc}^{-1}$ (b) and $k = 0.001 \mathrm{Mpc}^{-1}$ (c), and those in the model $F_2(R)$ for $k = 0.1 \mathrm{Mpc}^{-1}$ (d), $k = 0.01 \mathrm{Mpc}^{-1}$ (e) and $k = 0.001 \mathrm{Mpc}^{-1}$ (f).}
\label{HS_figure_constant_growth_index}
\end{figure}

\begin{figure}[!h]
\subfigure[]{\includegraphics[width=0.45\textwidth]{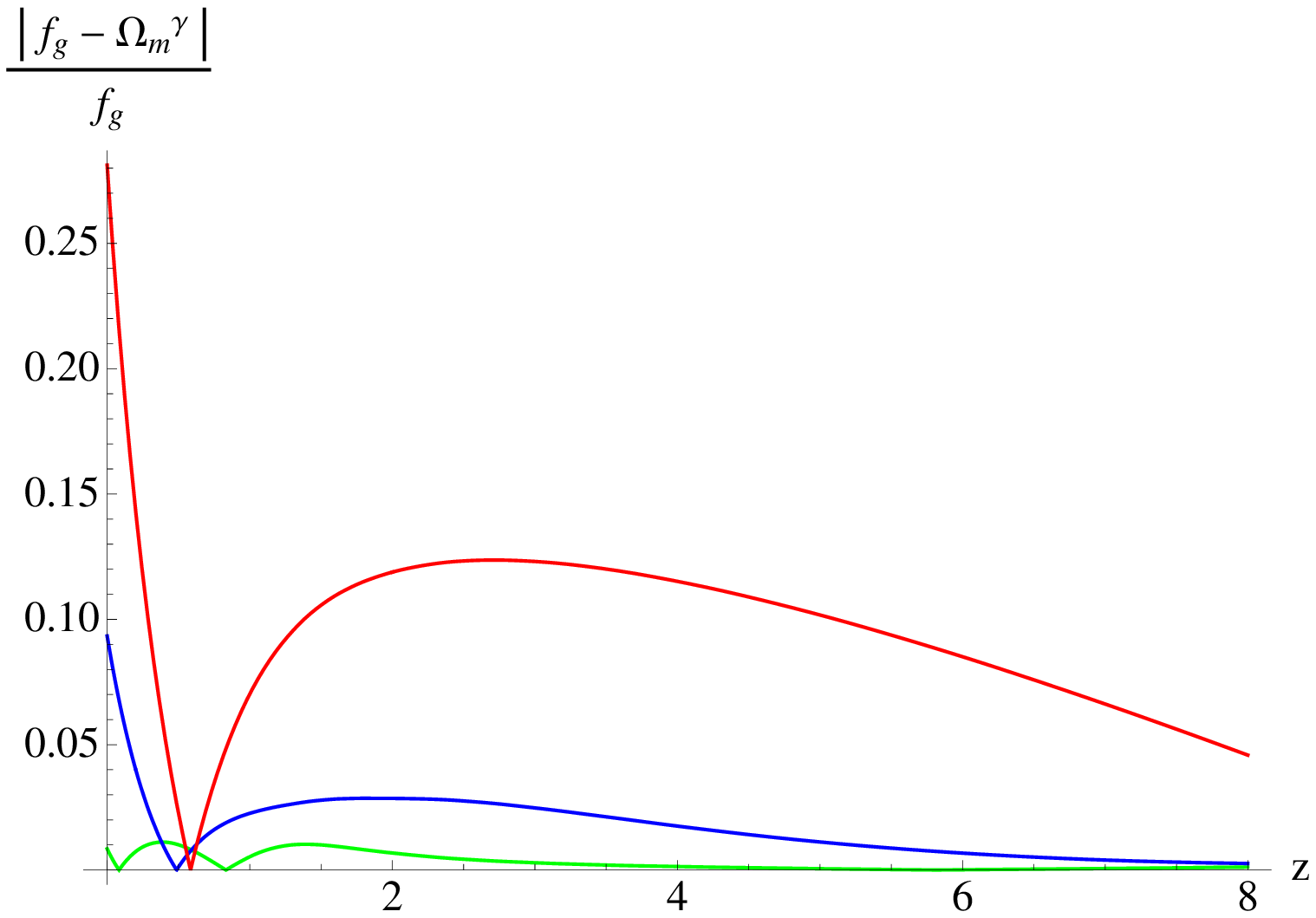}}
\quad
\subfigure[]{\includegraphics[width=0.45\textwidth]{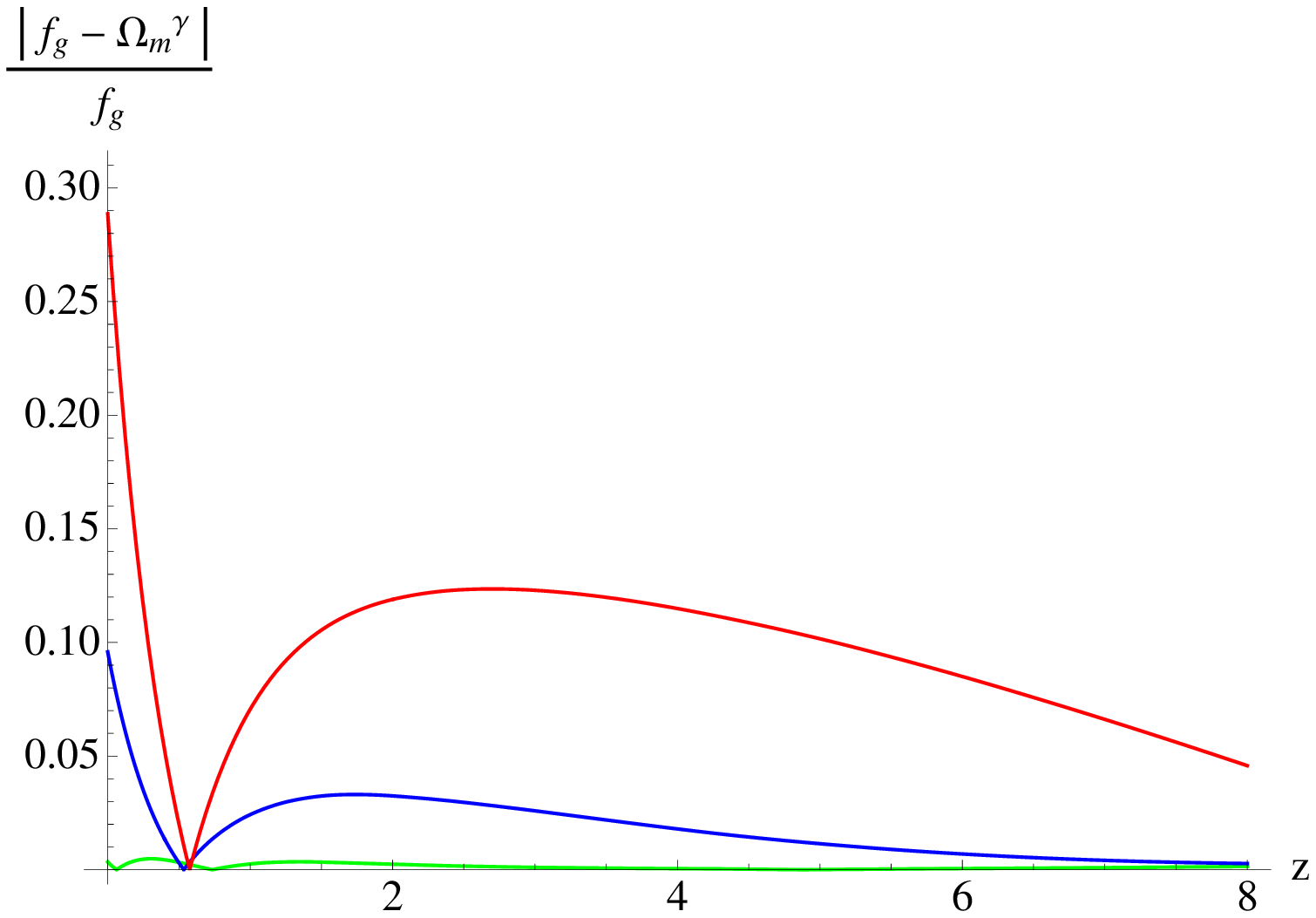}}
\caption{Cosmological evolution of the 
relative difference $\frac{\left| f_\mathrm{g} - \Omega_\mathrm{m}^\gamma \right|}{f_\mathrm{g}}$ with $\gamma = \gamma_0$ for $k = 0.1 \mathrm{Mpc}^{-1}$ (red), $k = 0.01 \mathrm{Mpc}^{-1}$ (blue) and $k = 0.001 \mathrm{Mpc}^{-1}$ (green) in the model $F_1(R)$ (a) and the model $F_2(R)$ (b).}
\label{const_rel_dif}
\end{figure}

In order to check the goodness of our fits, in Fig.~\ref{HS_figure_constant_growth_index} we show cosmological evolutions of 
the growth rate $f_\mathrm{g}(z)$ and $\Omega_\mathrm{m}(z)^{\gamma_0}$ as functions of the redshift $z$ together for several values of the comoving wavenumber $k$ for the models $F_1(R)$ and $F_2(R)$. 
To clarify these results, in Fig.~\ref{const_rel_dif} we also depict the 
cosmological evolution of the relative difference between $f_\mathrm{g}(z)$ and $\Omega_\mathrm{m}(z)^{\gamma_0}$ as a function of $z$ 
for the same values of $k$ in these models. For both models the function $\Omega_\mathrm{m}(z)^{\gamma_0}$ fits the growth rate for large scales (namely, lower $k$) very well, but this is not anymore the case for larger values of $k$. In fact, if we do not consider lower values for $z$ (namely, $z < 0.2$), for $\log [k] = -2$ the relative difference is smaller than $3\%$ for the both models, while for $\log [k] = -1$ can arrive up to almost $13\%$. For $\log [k] = -3$, we see that the relative difference is always smaller than $1.5\%$ for the model $F_1(R)$, and smaller than $1\%$ for the model $F_2(R)$.


\subsubsection{$\gamma = \gamma_0 + \gamma_1 z$}

\paragraph{}With the same procedure used in the previous case, we explore now a linear dependence for the growth index as the following,
\begin{equation}
\gamma = \gamma_0 + \gamma_1 z\,, 
\end{equation}
where $\gamma_{0,1}$ are constants. 

\begin{figure}[!h]
\subfigure[]{\includegraphics[width=0.45\textwidth]{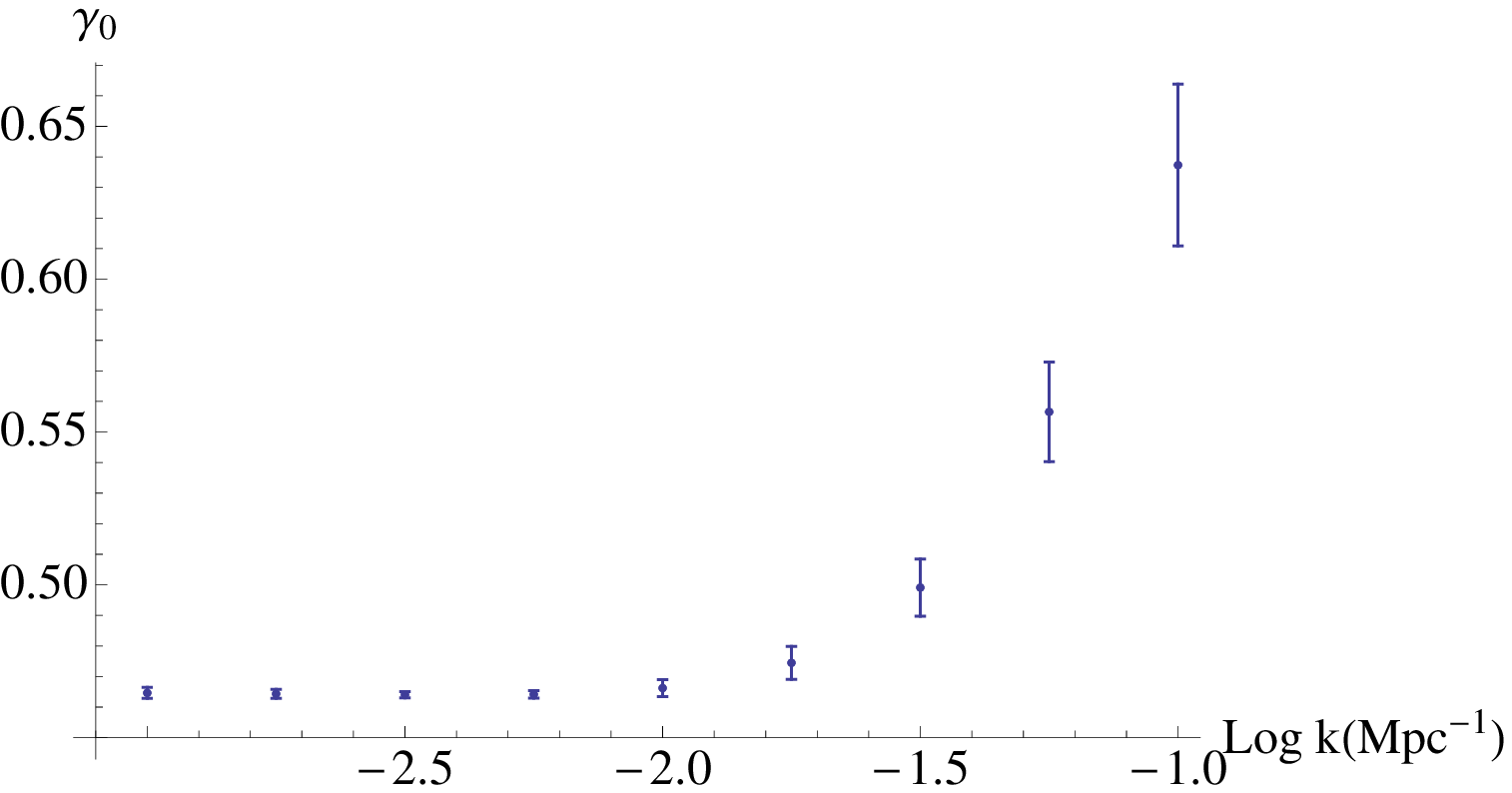}}
\quad
\subfigure[]{\includegraphics[width=0.45\textwidth]{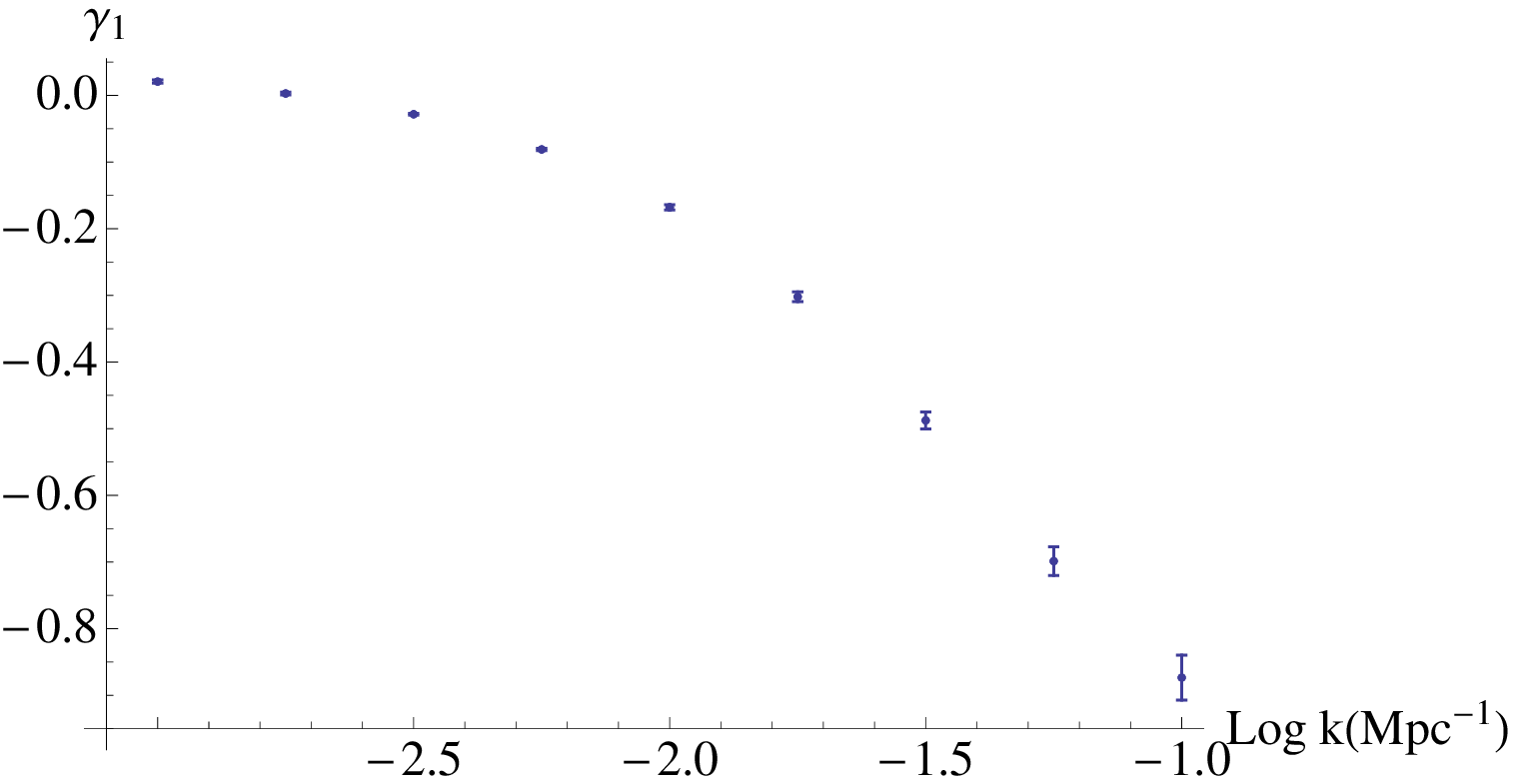}}
\quad
\subfigure[]{\includegraphics[width=0.45\textwidth]{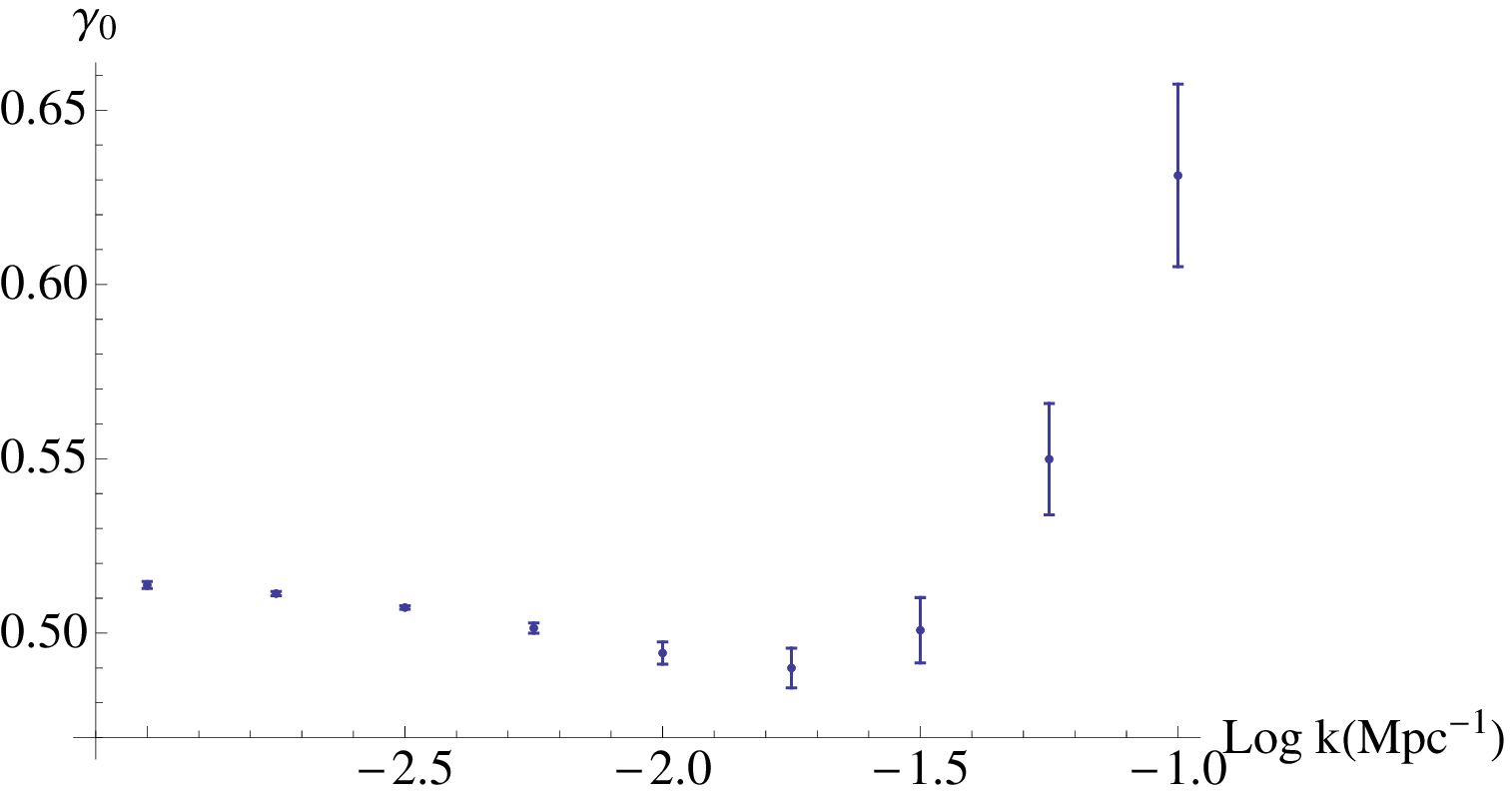}}
\quad
\subfigure[]{\includegraphics[width=0.45\textwidth]{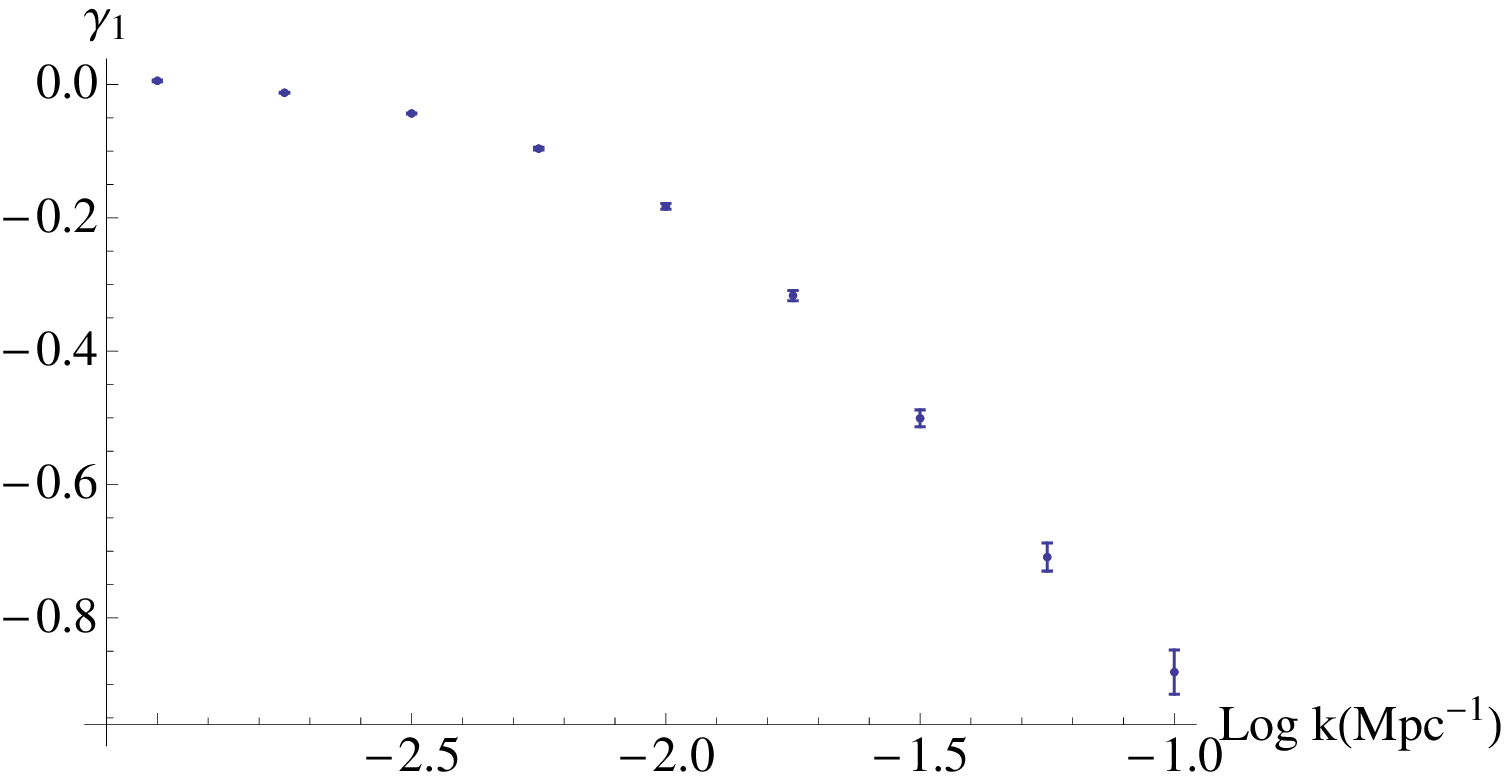}}
\caption{
Growth index fitting parameters in the case $\gamma = \gamma_0 + \gamma_1 z$ 
as a function of $\log k$ for the model $F_1(R)$ [(a) and (b)] 
and the model $F_2(R)$ [(c) and (d)]. 
Legend is the same as Fig.~\ref{constant_growth_index_vs_logk}. 
}
\label{HS_lineal_growth_index_vs_logk}
\end{figure}

In Fig.~\ref{HS_lineal_growth_index_vs_logk},  
we plot the parameters $\gamma_0$ and $\gamma_1$ for different values of 
$\log [k]$ in both of the models. 
Again,
the scale dependence of the parameters $\gamma_0$ and $\gamma_1$ is similar 
in these models. We can also find that $\gamma_0 \sim 0.46$ for the model $F_1(R)$ when $\log [k] \leq -2$, whereas $\gamma_0 \sim 0.51$ for the model $F_2(R)$ when $\log [k] \leq -2.5$. 
Moreover, 
the value of $\gamma_1$ has 
a strong dependence on $k$ in the range of $\log [k] > -2.25$, but in the range 
of $\log[ k] < -2.25$ this dependence becomes weaker in the both considered models. 

\begin{figure}[!h]
\subfigure[]{\includegraphics[width=0.3\textwidth]{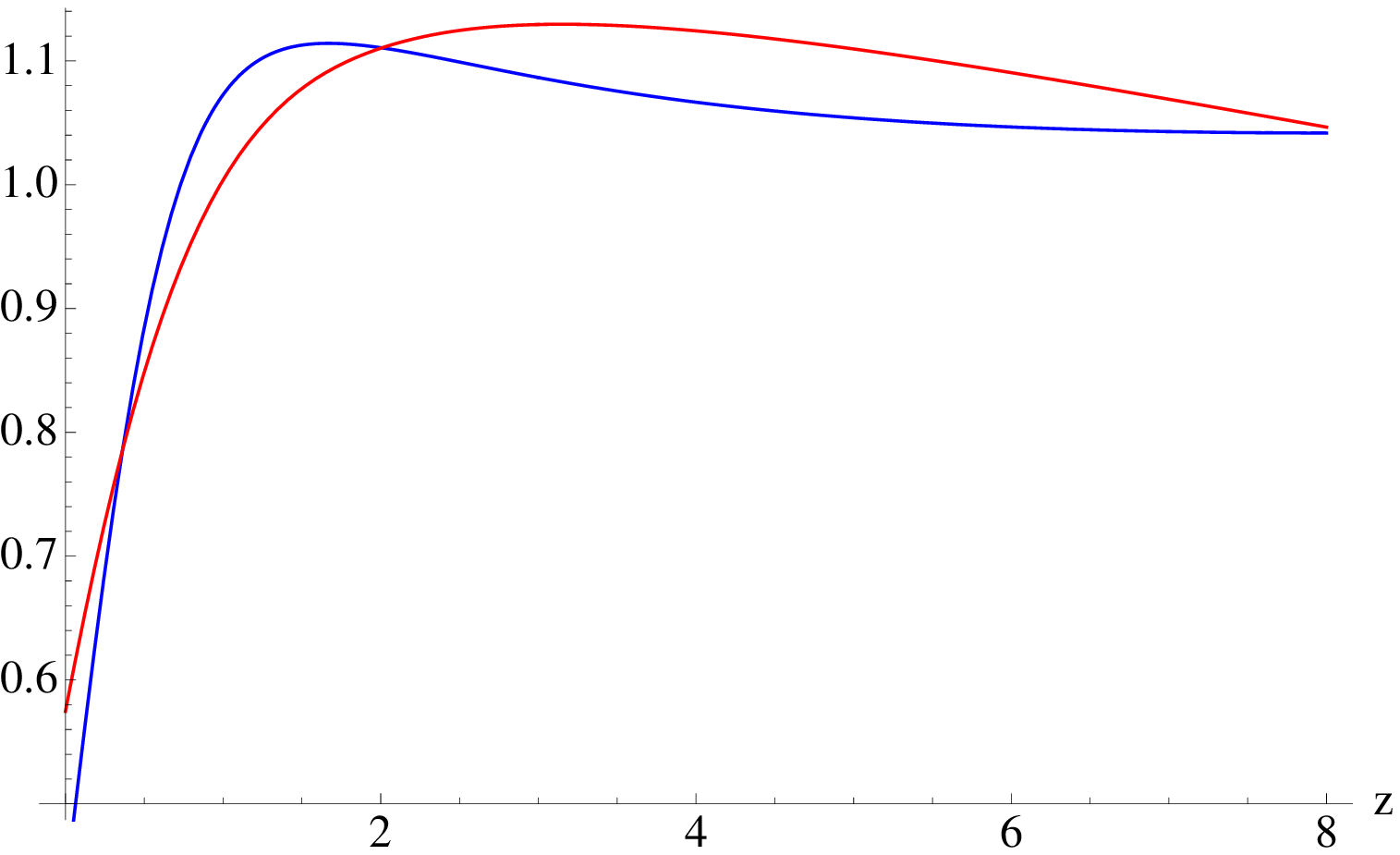}}
\quad
\subfigure[]{\includegraphics[width=0.3\textwidth]{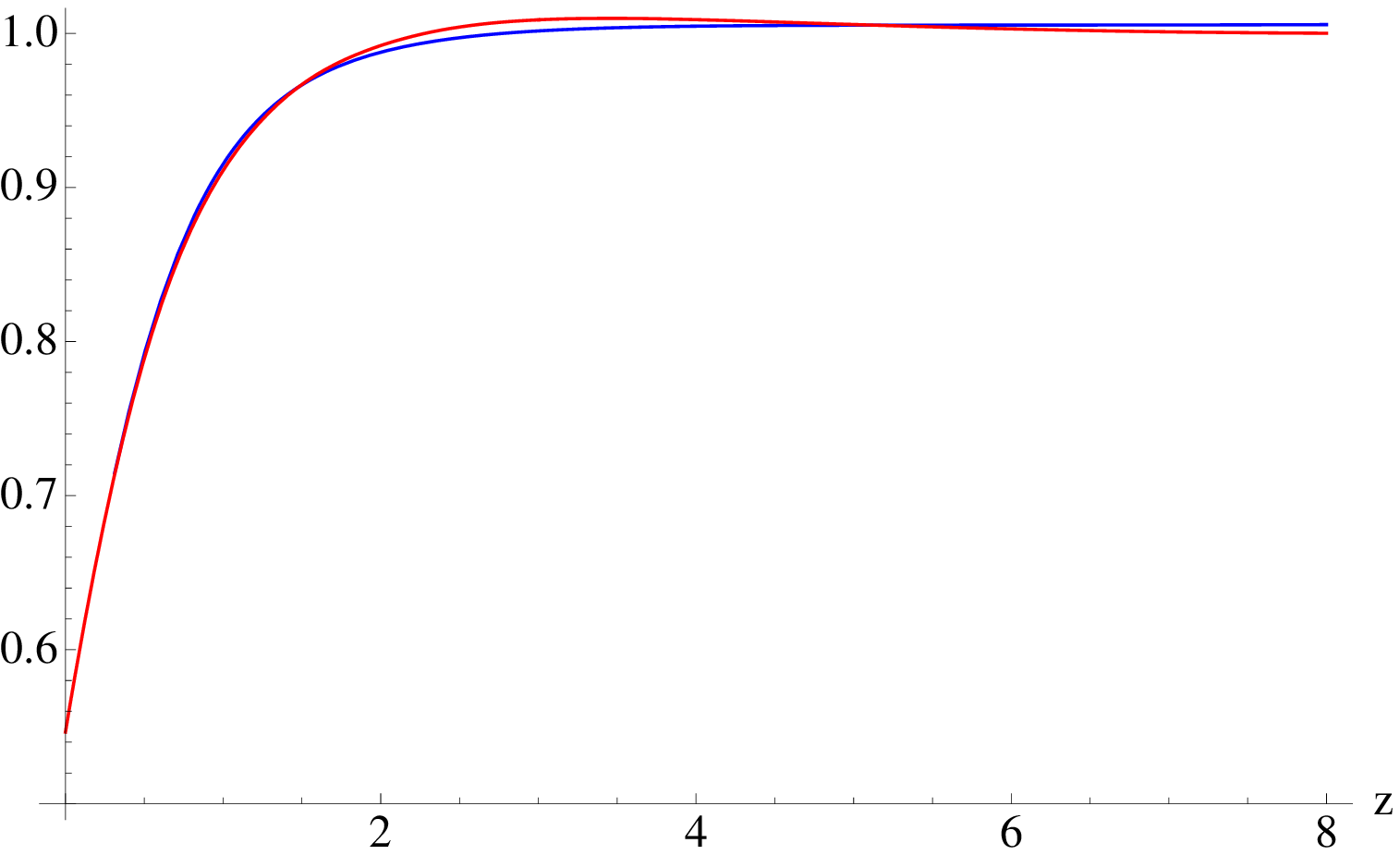}}
\quad
\subfigure[]{\includegraphics[width=0.3\textwidth]{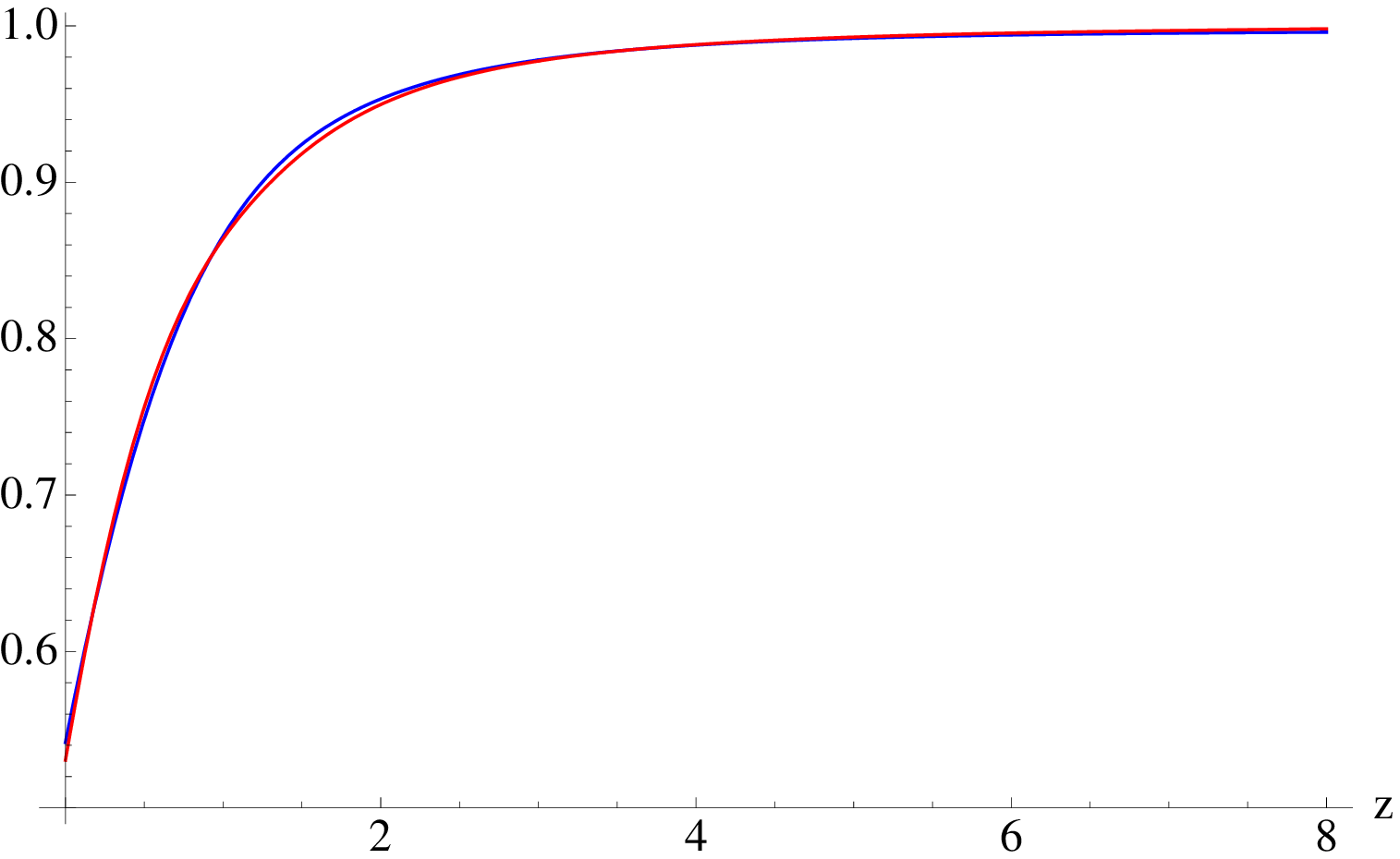}}
\quad
\subfigure[]{\includegraphics[width=0.3\textwidth]{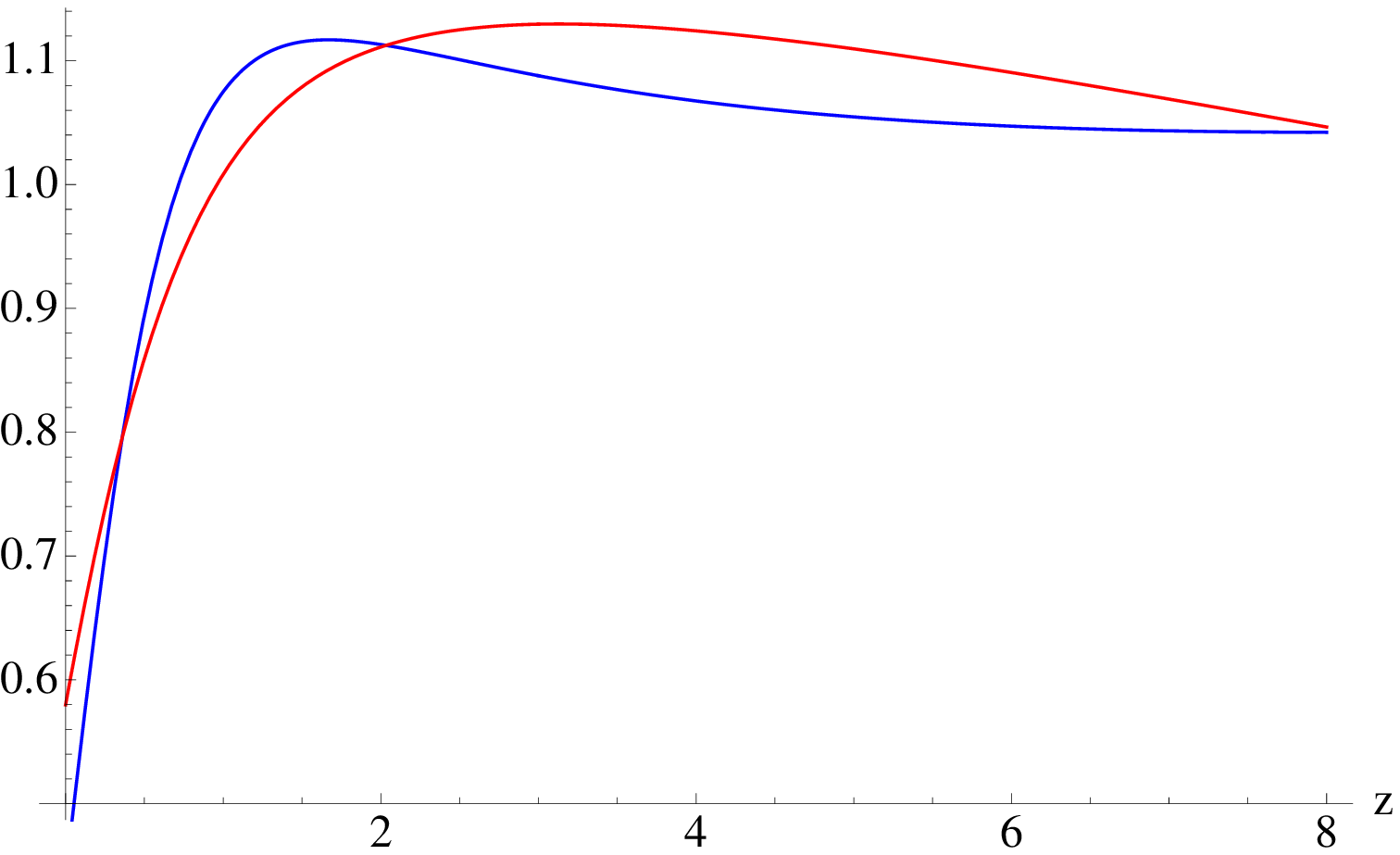}}
\quad
\subfigure[]{\includegraphics[width=0.3\textwidth]{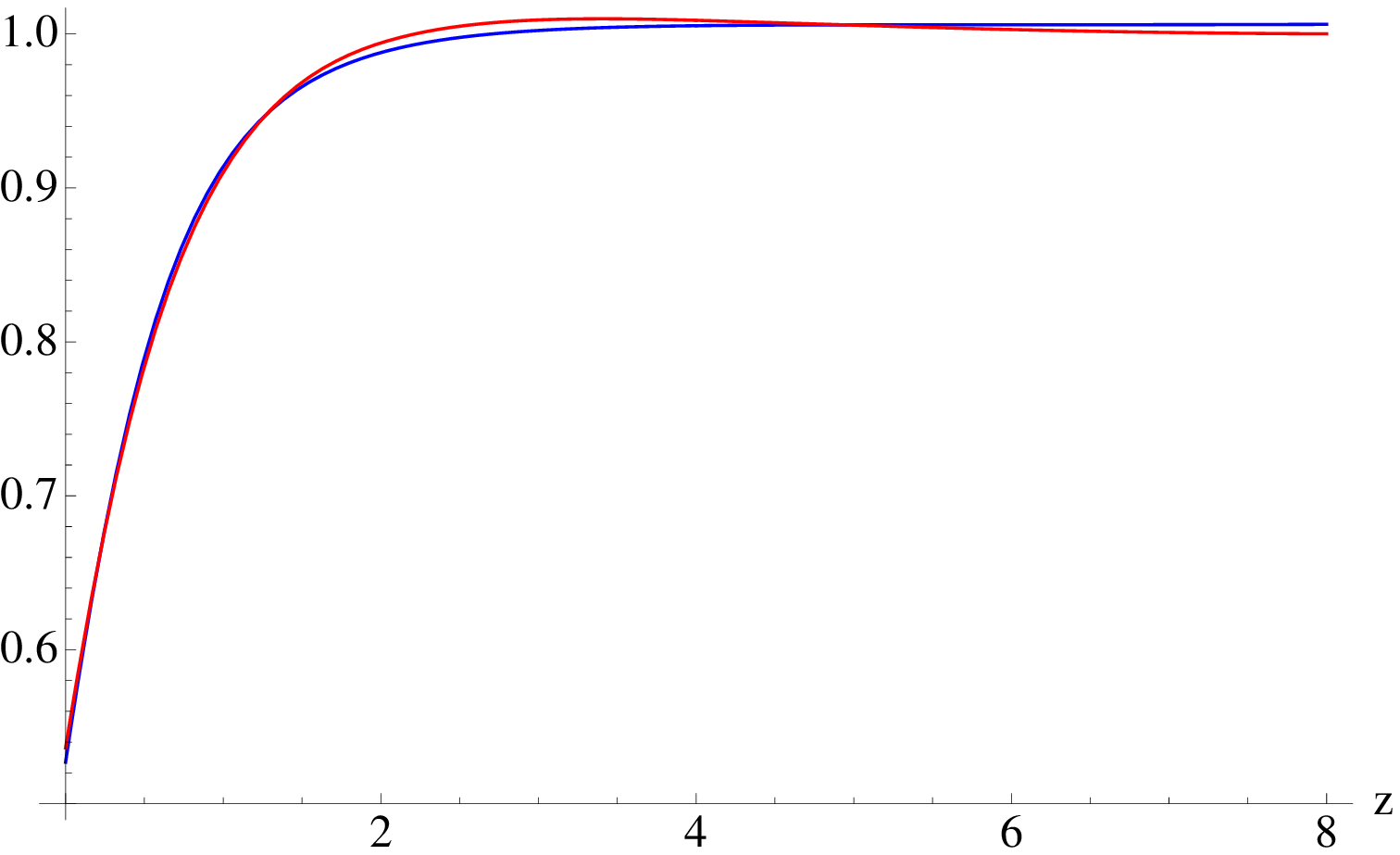}}
\quad
\subfigure[]{\includegraphics[width=0.3\textwidth]{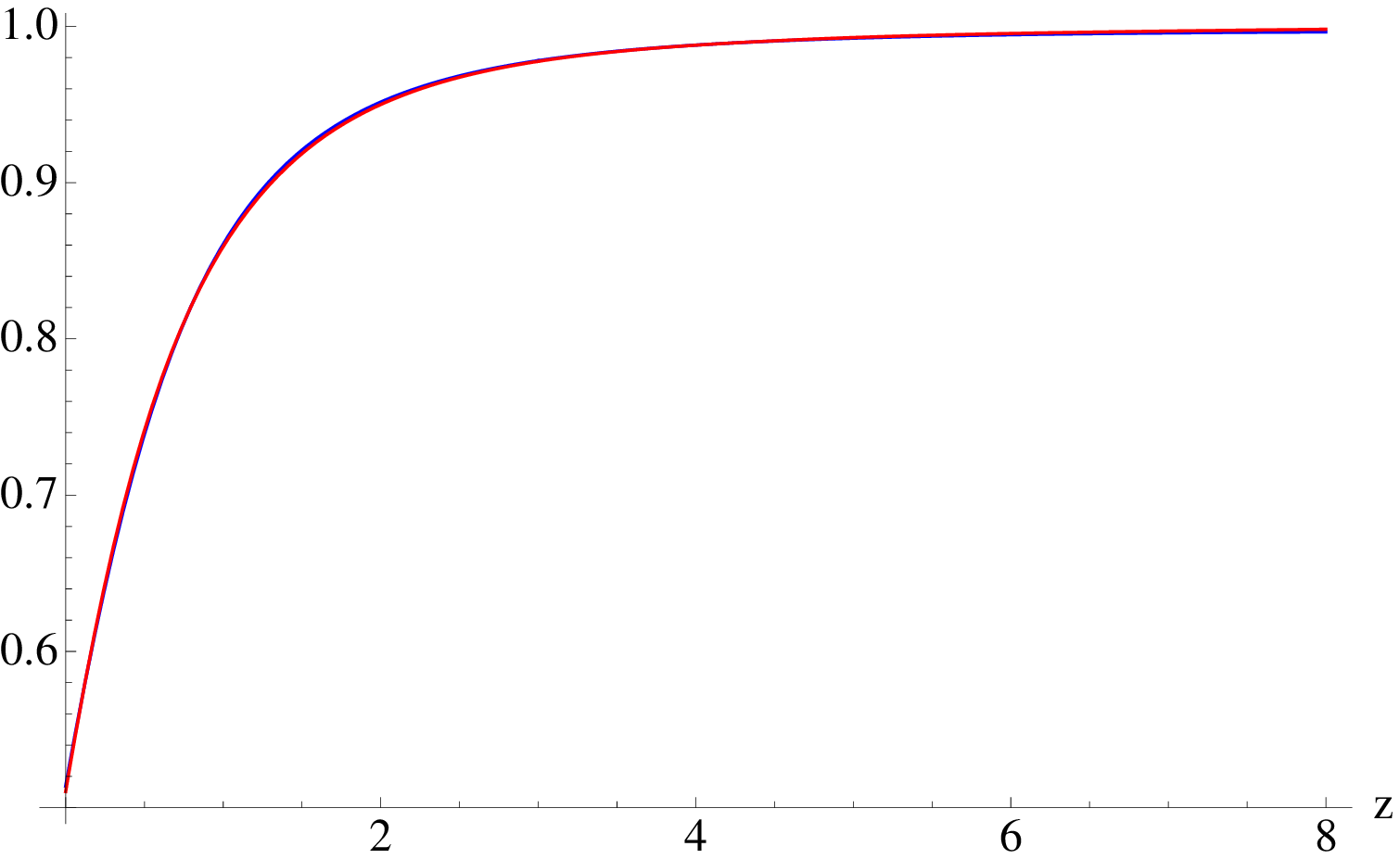}}
\caption{Cosmological evolutions of the 
growth rate $f_\mathrm{g}$ (red) and $\Omega_\mathrm{m}^\gamma$ (blue) with $\gamma = \gamma_0 + \gamma_1 z$ as functions of the redshift $z$ in the model $F_1(R)$ for $k = 0.1 \mathrm{Mpc}^{-1}$ (a), $k = 0.01 \mathrm{Mpc}^{-1}$ (b) and $k = 0.001 \mathrm{Mpc}^{-1}$ (c), and 
those in the model $F_2(R)$ for $k = 0.1 \mathrm{Mpc}^{-1}$ (d), $k = 0.01 \mathrm{Mpc}^{-1}$ (e) and $k = 0.001 \mathrm{Mpc}^{-1}$ (f).}
\label{HS_figure_lineal_growth_index}
\end{figure}

\begin{figure}[!h]
\subfigure[]{\includegraphics[width=0.45\textwidth]{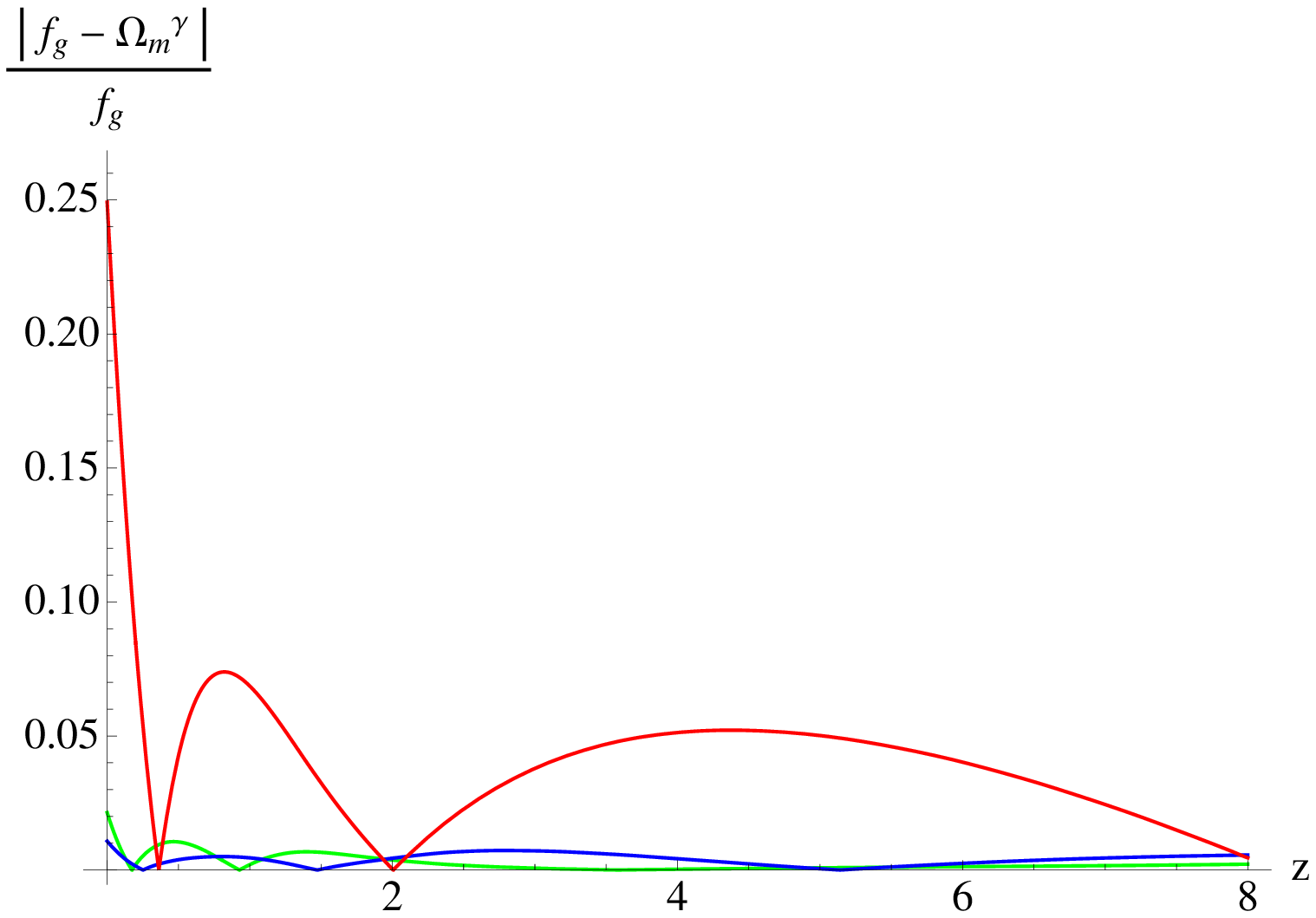}}
\quad
\subfigure[]{\includegraphics[width=0.45\textwidth]{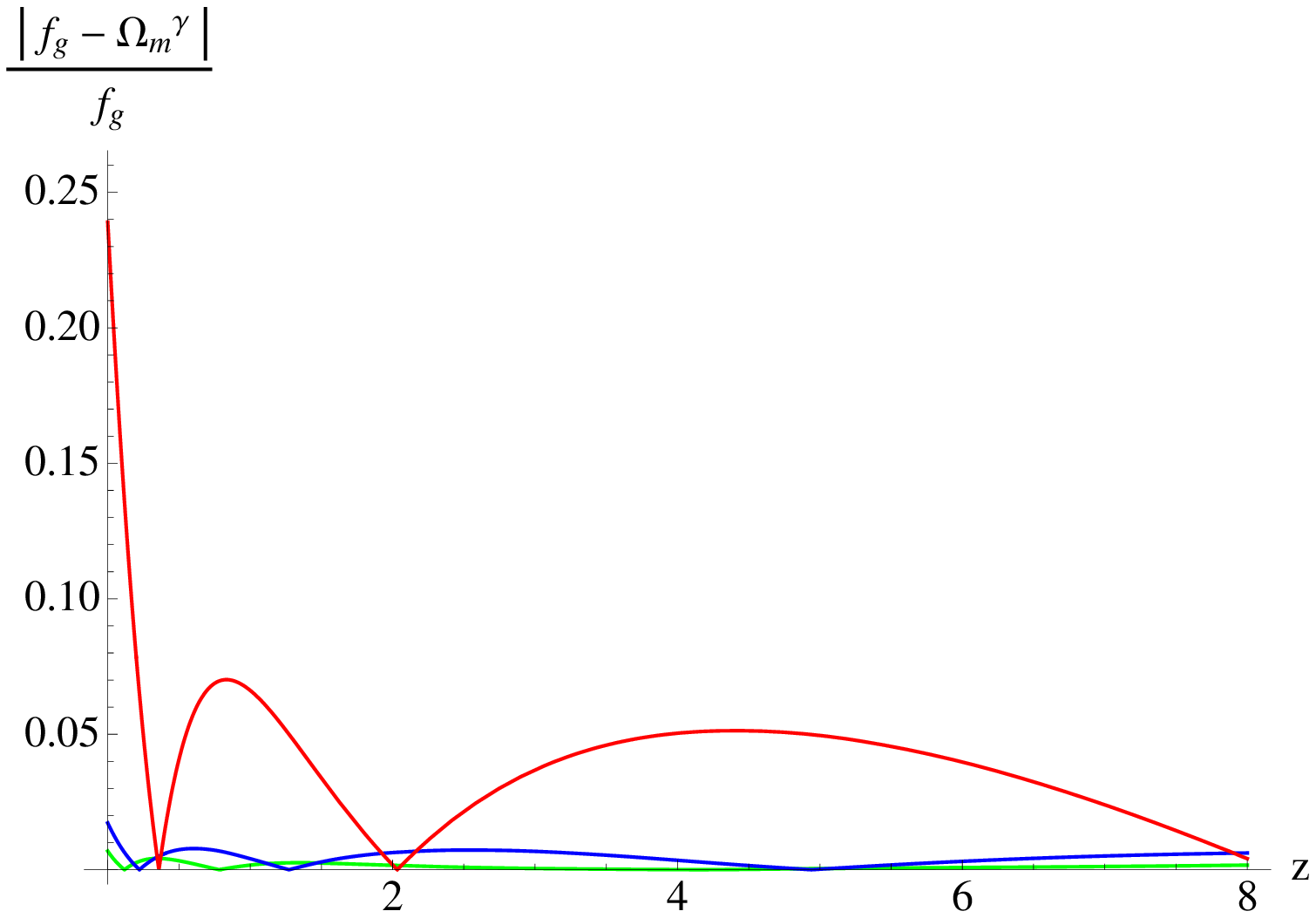}}
\caption{Cosmological evolution of the 
relative difference $\frac{\left| f_\mathrm{g} - \Omega_\mathrm{m}^\gamma \right|}{f_\mathrm{g}}$ with $\gamma = \gamma_0 + \gamma_1 z$ for $k = 0.1 \mathrm{Mpc}^{-1}$ (red), $k = 0.01 \mathrm{Mpc}^{-1}$ (blue) and $k = 0.001 \mathrm{Mpc}^{-1}$ (green) in the model $F_1(R)$ (a) and the model $F_2(R)$ (b).}
\label{lin_rel_dif}
\end{figure}

In Fig.~\ref{HS_figure_lineal_growth_index}, we depict cosmological evolutions of the growth rate $f_\mathrm{g}(z)$ and $\Omega_\mathrm{m}(z)^{\gamma(z)}$ as functions of the redshift $z$ together for 
the models $F_1(R)$ and $F_2(R)$. 
We can see that the fits for $\log [k] = 0.1$ is improved with respect to the same fits of the case with a constant growth index. 
Also, for $\log [k] < 0.1$, the fits continue to be quite good. 
In order to demonstrate these facts quantitatively, 
in Fig.~\ref{lin_rel_dif} we plot the cosmological evolution of 
the relative difference between $f_\mathrm{g}(z)$ and $\Omega_\mathrm{m}(z)^{\gamma(z)}$ as a function of $z$ 
for several values of $k$ in the models $F_1(R)$ and $F_2(R)$. 
In this case, for $\log [k] = -1$ the relative difference is smaller than $7.5\%$ in both the models if we do not consider lower values for $z$ (namely, $z < 0.2$). We also see that the linear growth index improves the fits in both the models for $\log [k] = -2$ in comparison with those for a constant growth index. 
In this case, the relative difference for the model $F_1(R)$ is always smaller than $1\%$, whereas that for model $F_2(R)$ is smaller than $2\%$. 
Finally, for $\log [k] = -3$, the results obtained for a constant growth index are similar to those for a linear dependence on $z$. 


\subsubsection{$\gamma = \left[\gamma_0 + \gamma_1 \frac{z}{1 + z}\right]$}

\paragraph{}As the last case, we examine the following Ansatz for the growth index,
\begin{equation} 
\gamma = \gamma_0 + \left[\gamma_1 \frac{z}{1 + z}\right]\,,
\end{equation}
where $\gamma_{0,1}$ are constants. 

\begin{figure}[!h]
\subfigure[]{\includegraphics[width=0.45\textwidth]{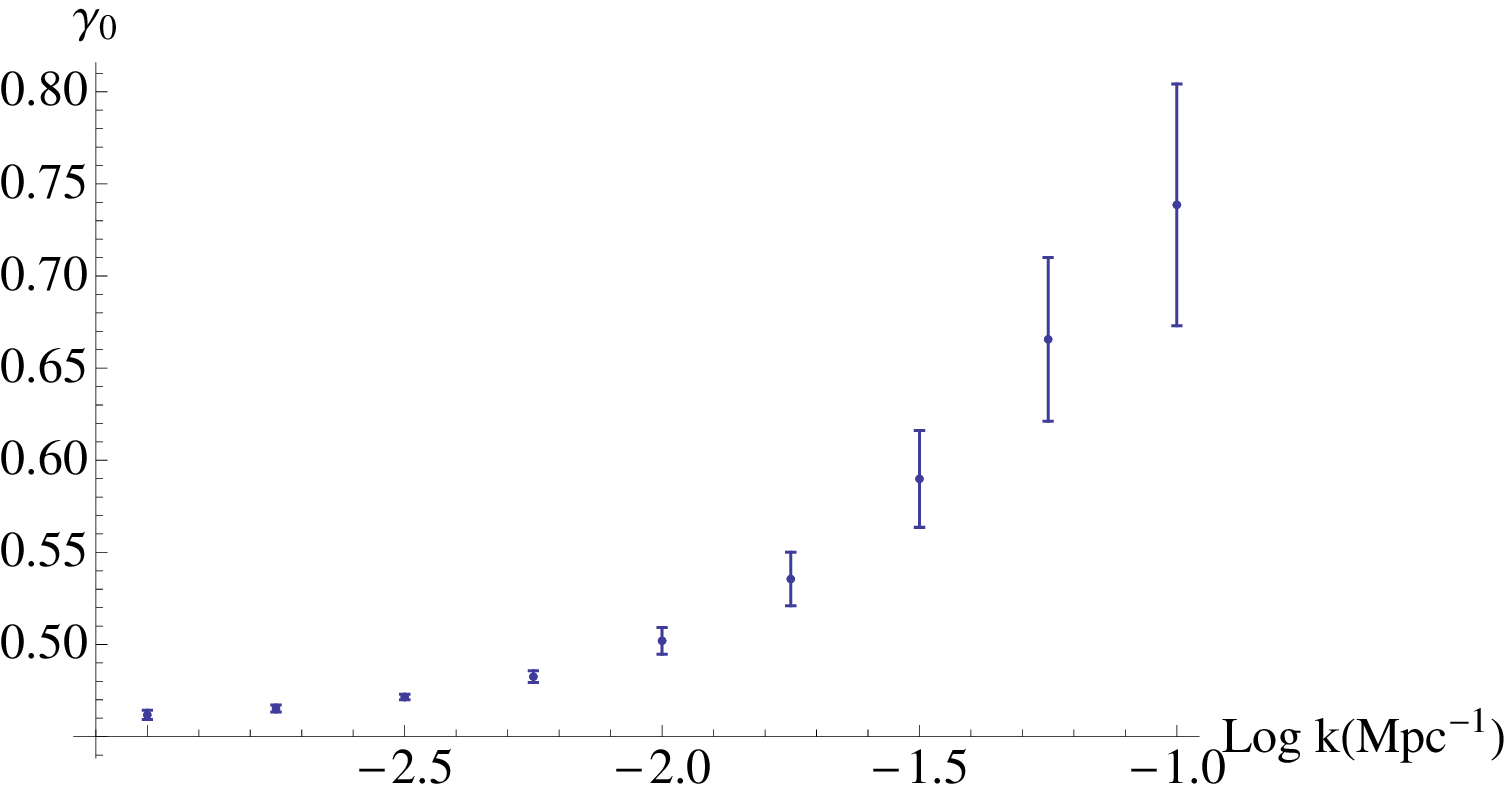}}
\quad
\subfigure[]{\includegraphics[width=0.45\textwidth]{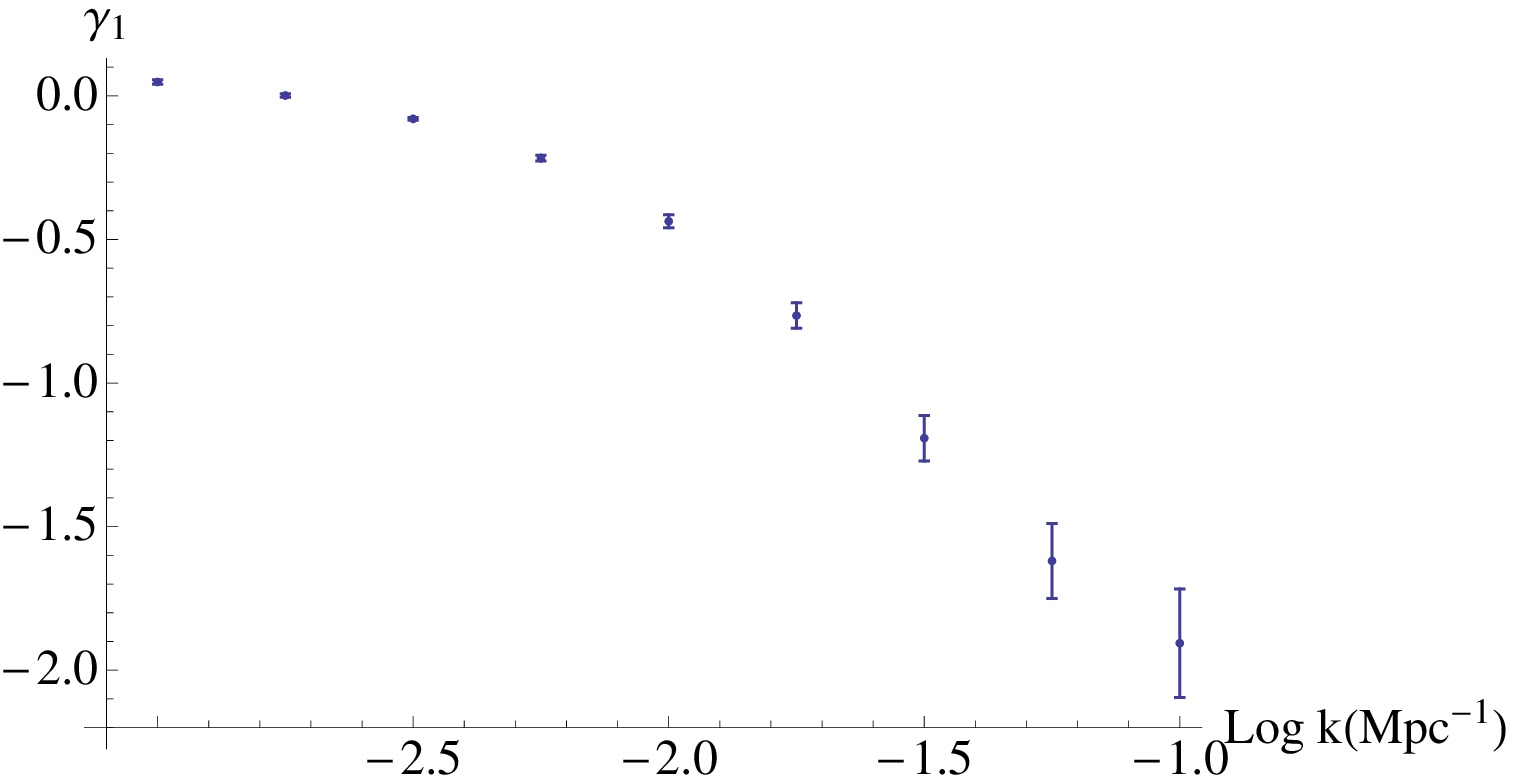}}
\subfigure[]{\includegraphics[width=0.45\textwidth]{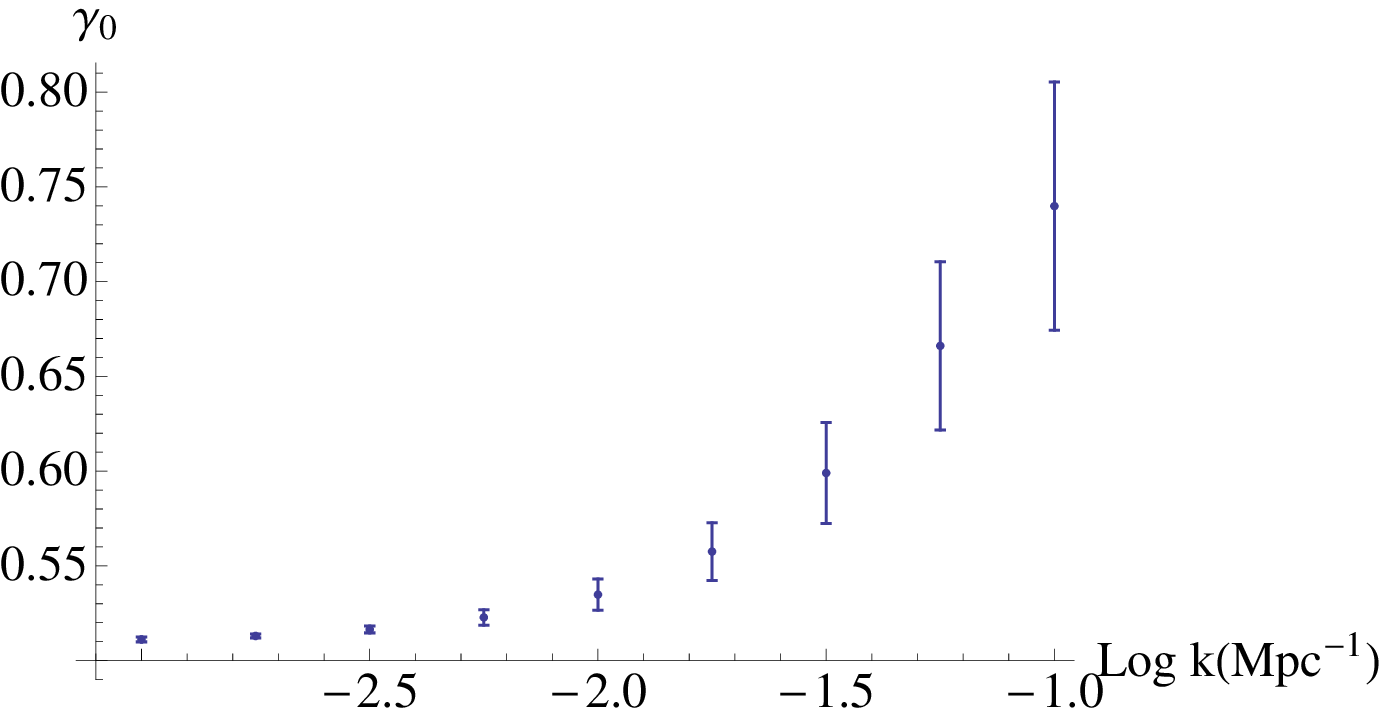}}
\quad
\subfigure[]{\includegraphics[width=0.45\textwidth]{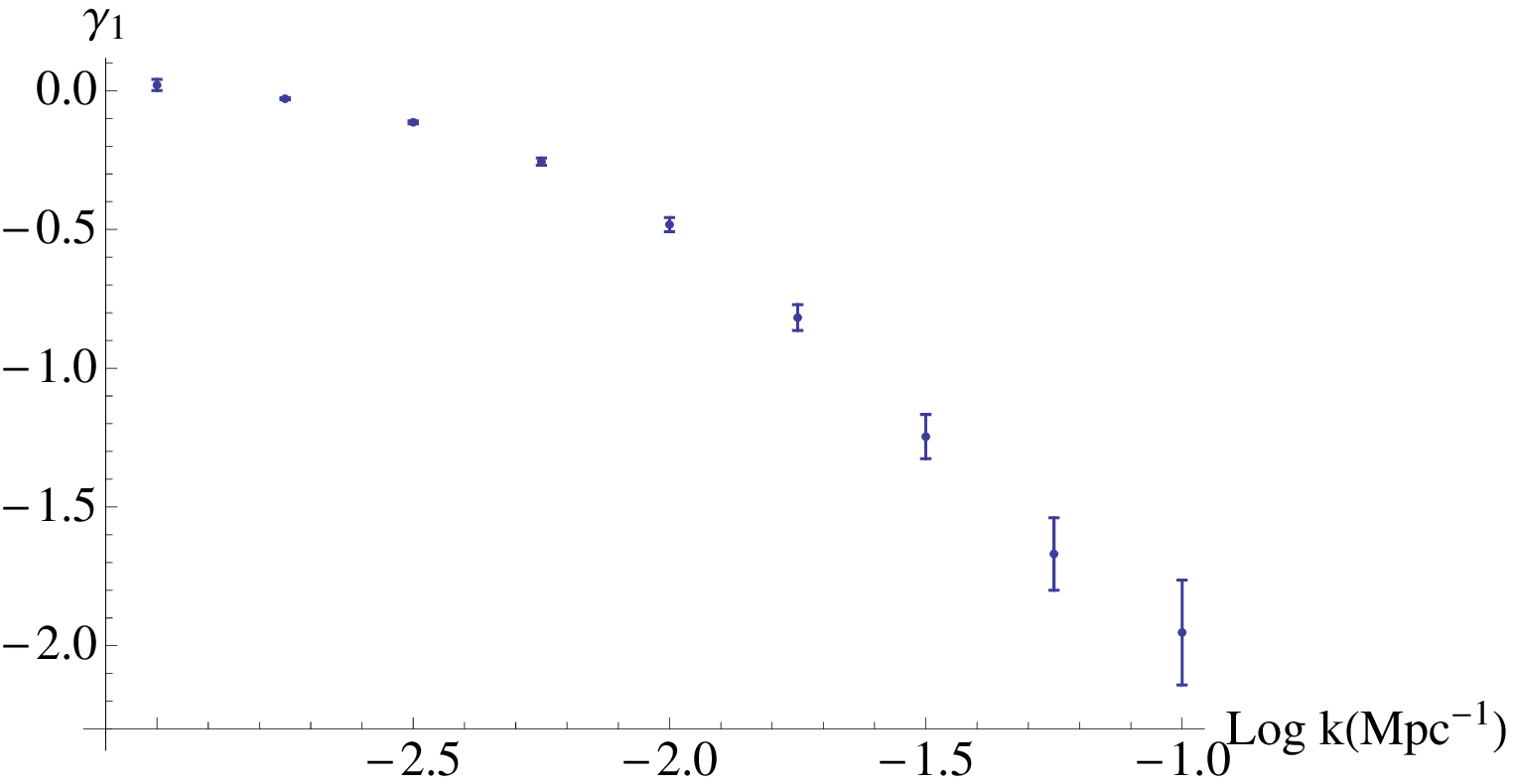}}
\caption{
Growth index fitting parameters in the case $\gamma = \gamma_0 + \gamma_1 \frac{z}{1 + z}$ as a function of $\log k$ for the model $F_1(R)$ [(a) and (b)] 
and the model $F_2(R)$ [(c) and (d)]. 
Legend is the same as Fig.~\ref{constant_growth_index_vs_logk}.
}
\label{HS_rational_growth_index_vs_logk}
\end{figure}

In Fig.~\ref{HS_rational_growth_index_vs_logk}, we plot 
the parameters $\gamma_0$ and $\gamma_1$ for several values of $\log [k]$ in both the models. 
The scale dependence of these parameters on $k$ is shown. 
The behavior of the parameter $\gamma_1$ is similar to 
that for the previous case, 
but we must take into account that the scale of the figures are different from each other, and that for the present Ansatz the scale dependence of $\gamma_1$ is stronger than that for the previous one. We also see that $\gamma_0 \sim 0.465$ for the model $F_1(R)$ and $\gamma_0 \sim 0.513$ for the model $F_2(R)$ in the scale 
$\log [k] < -2.5$. 

\begin{figure}[!h]
\subfigure[]{\includegraphics[width=0.3\textwidth]{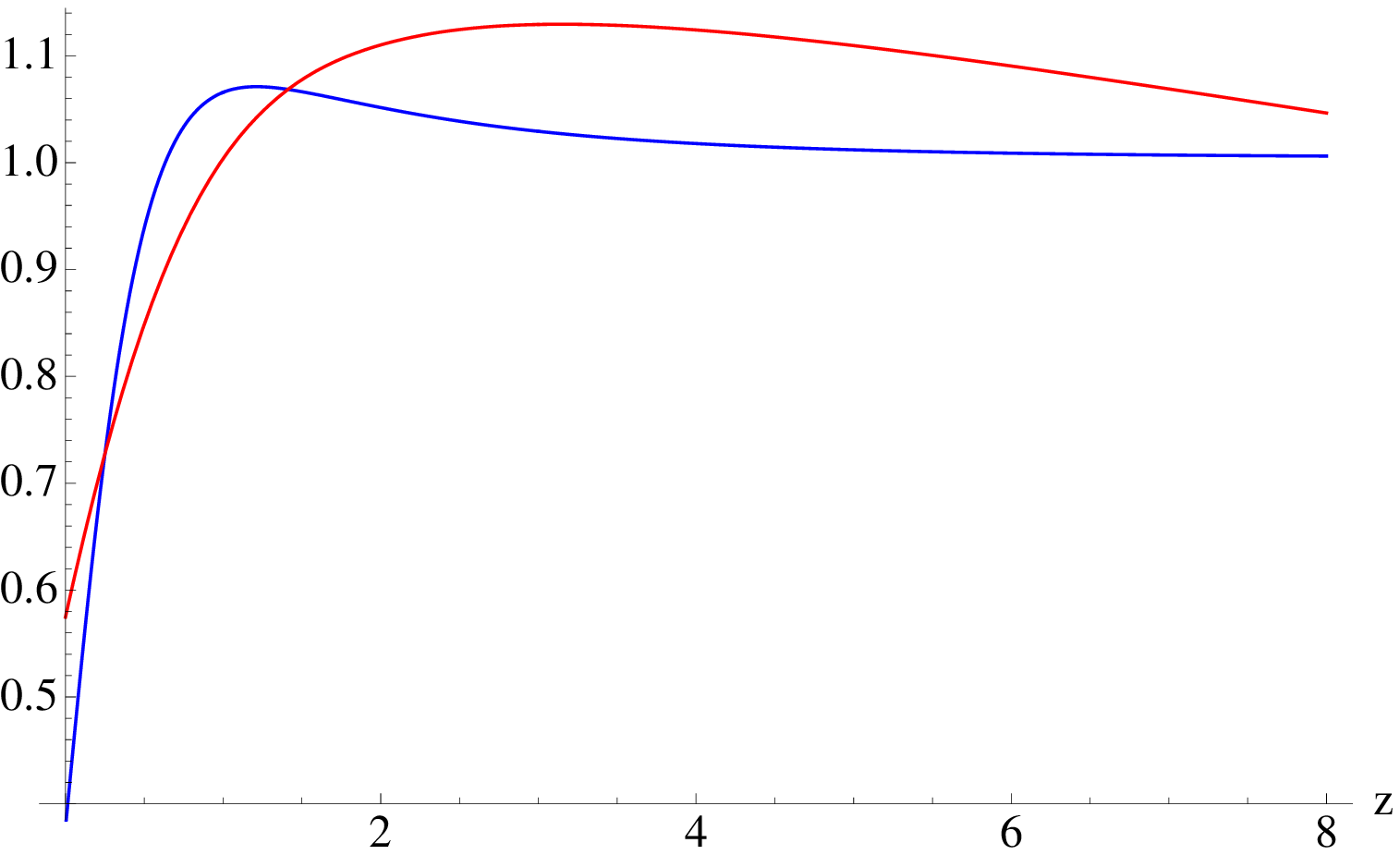}}
\quad
\subfigure[]{\includegraphics[width=0.3\textwidth]{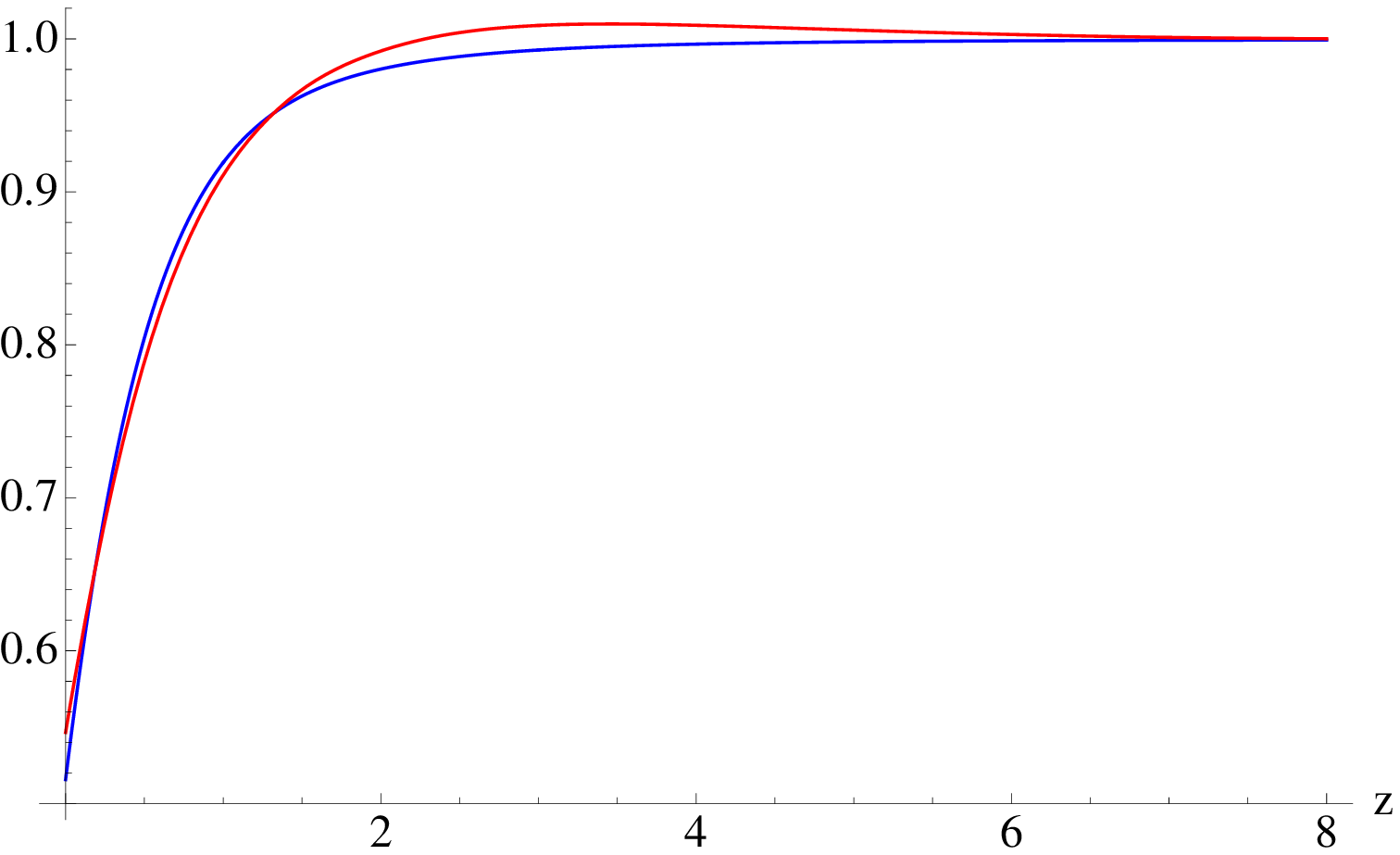}}
\quad
\subfigure[]{\includegraphics[width=0.3\textwidth]{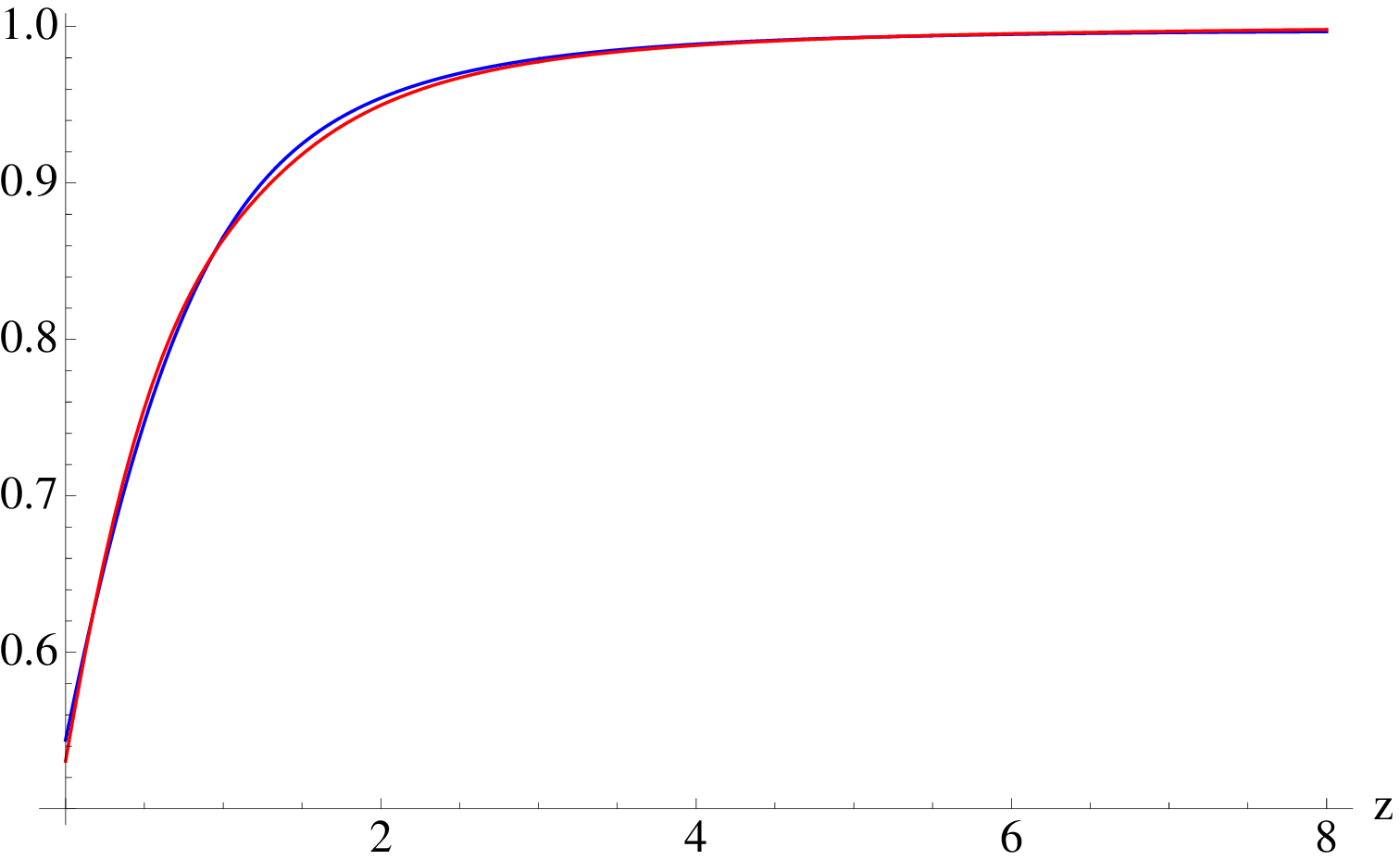}}
\quad
\subfigure[]{\includegraphics[width=0.3\textwidth]{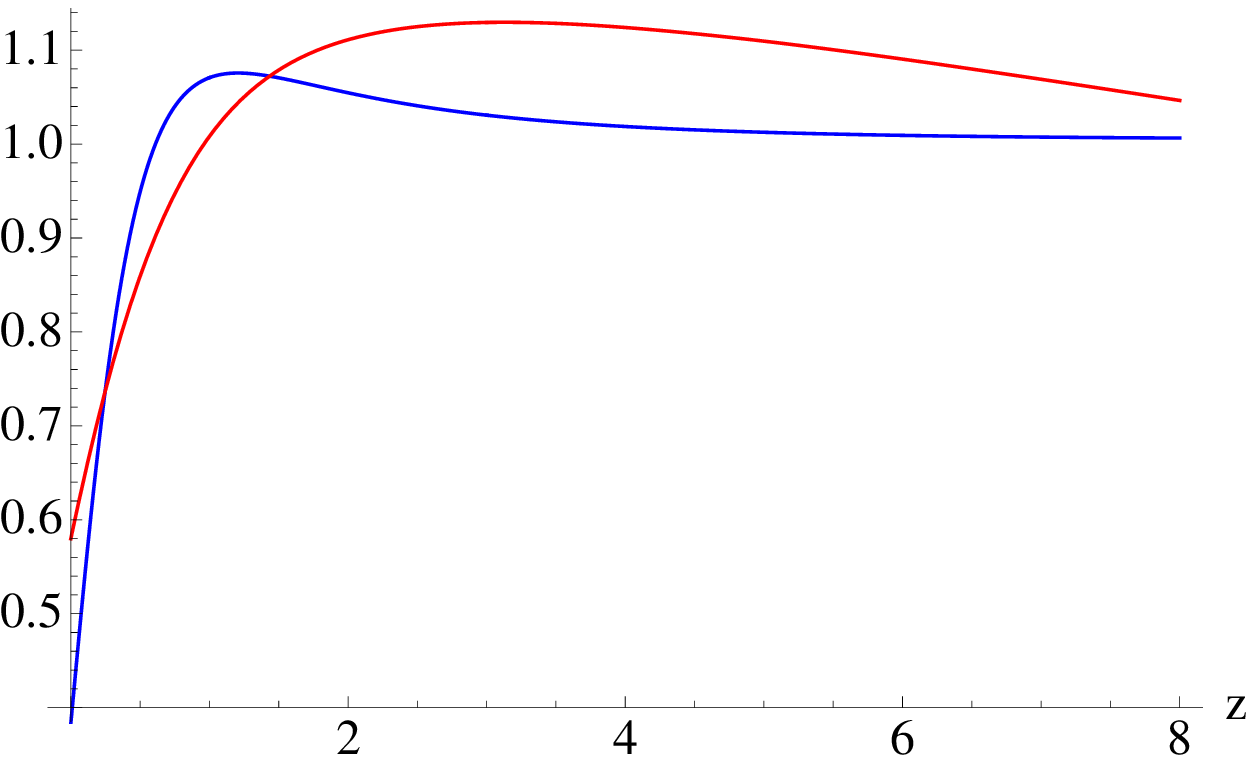}}
\quad
\subfigure[]{\includegraphics[width=0.3\textwidth]{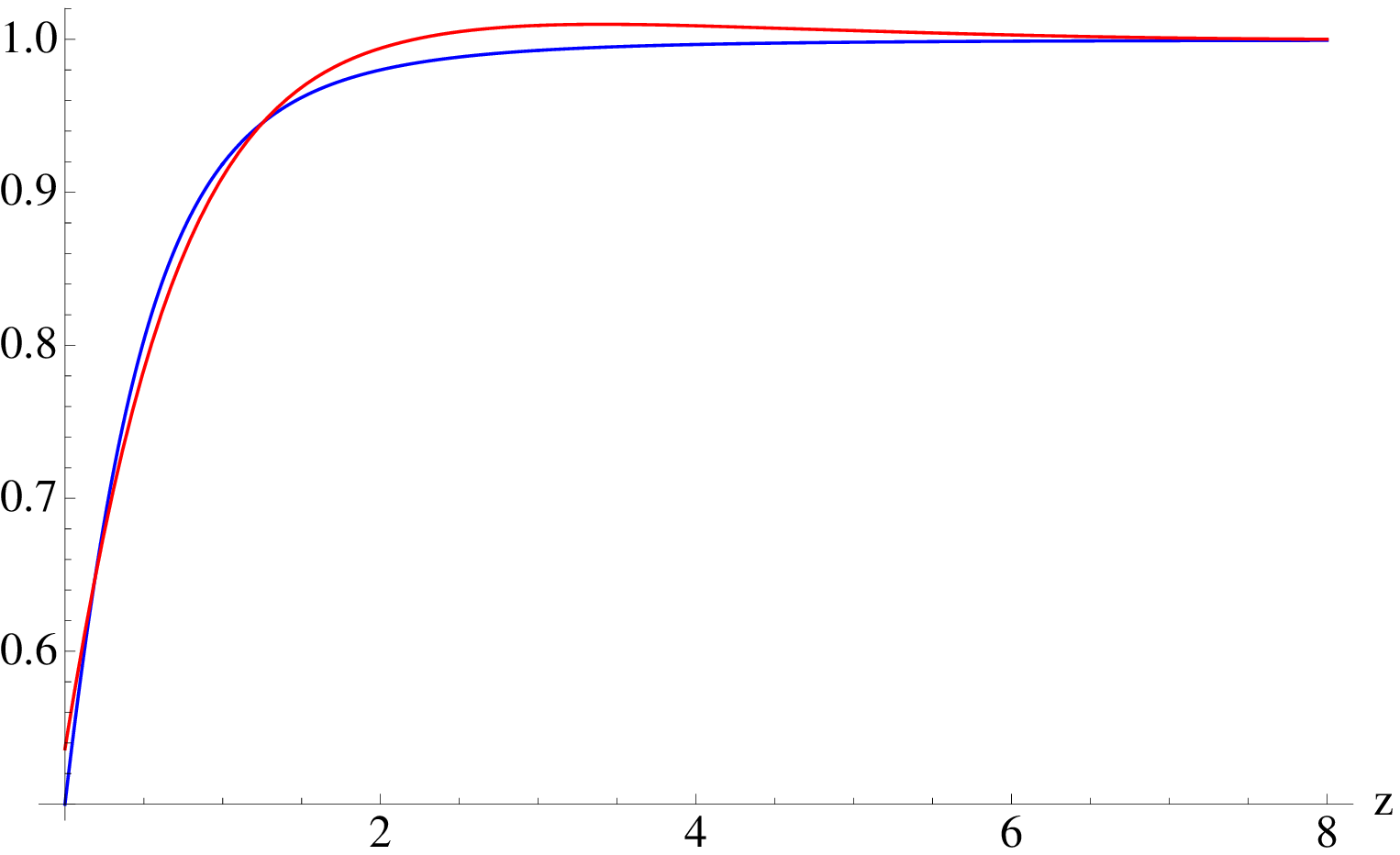}}
\quad
\subfigure[]{\includegraphics[width=0.3\textwidth]{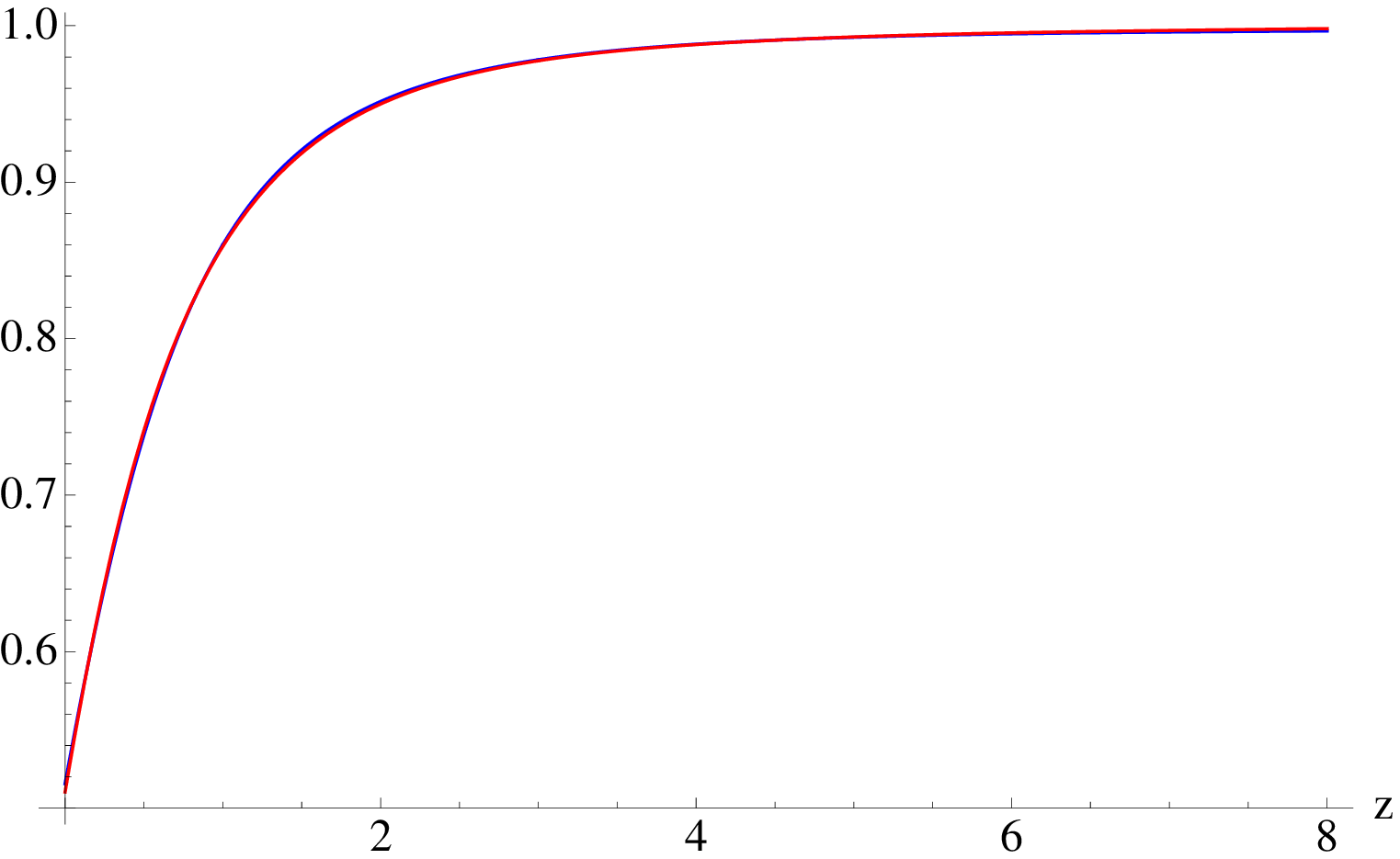}}
\caption{Cosmological evolutions of the 
growth rate $f_\mathrm{g}$ (red) and $\Omega_\mathrm{m}^\gamma$ (blue) with $\gamma = \gamma_0 + \gamma_1 \frac{z}{1 + z}$ as functions of the redshift $z$ in 
the model $F_1(R)$ for $k = 0.1 \mathrm{Mpc}^{-1}$ (a), $k = 0.01 \mathrm{Mpc}^{-1}$ (b) and $k = 0.001 \mathrm{Mpc}^{-1}$ (c), and 
those in the 
model $F_2(R)$ for $k = 0.1 \mathrm{Mpc}^{-1}$ (d), $k = 0.01 \mathrm{Mpc}^{-1}$ (e) and $k = 0.001 \mathrm{Mpc}^{-1}$ (f).}
\label{HS_figure_rational_growth_index}
\end{figure}

In Fig.~\ref{HS_figure_rational_growth_index}, we plot cosmological evolutions of the growth rate $f_\mathrm{g}(z)$ and $\Omega_\mathrm{m}(z)^{\gamma(z)}$ in the models $F_1(R)$ and $F_2(R)$ for different values of $k$. We observe that the fits for $\log [k] \leq -2$ are quite good, but, 
for higher values of $\log [k]$, it seems that the fits are similar to those for a constant growth rate 
and these fits do not reach the goodness of those for the case of $\gamma = \gamma_0 + \gamma_1 z$. 

\begin{figure}[!h]
\subfigure[]{\includegraphics[width=0.45\textwidth]{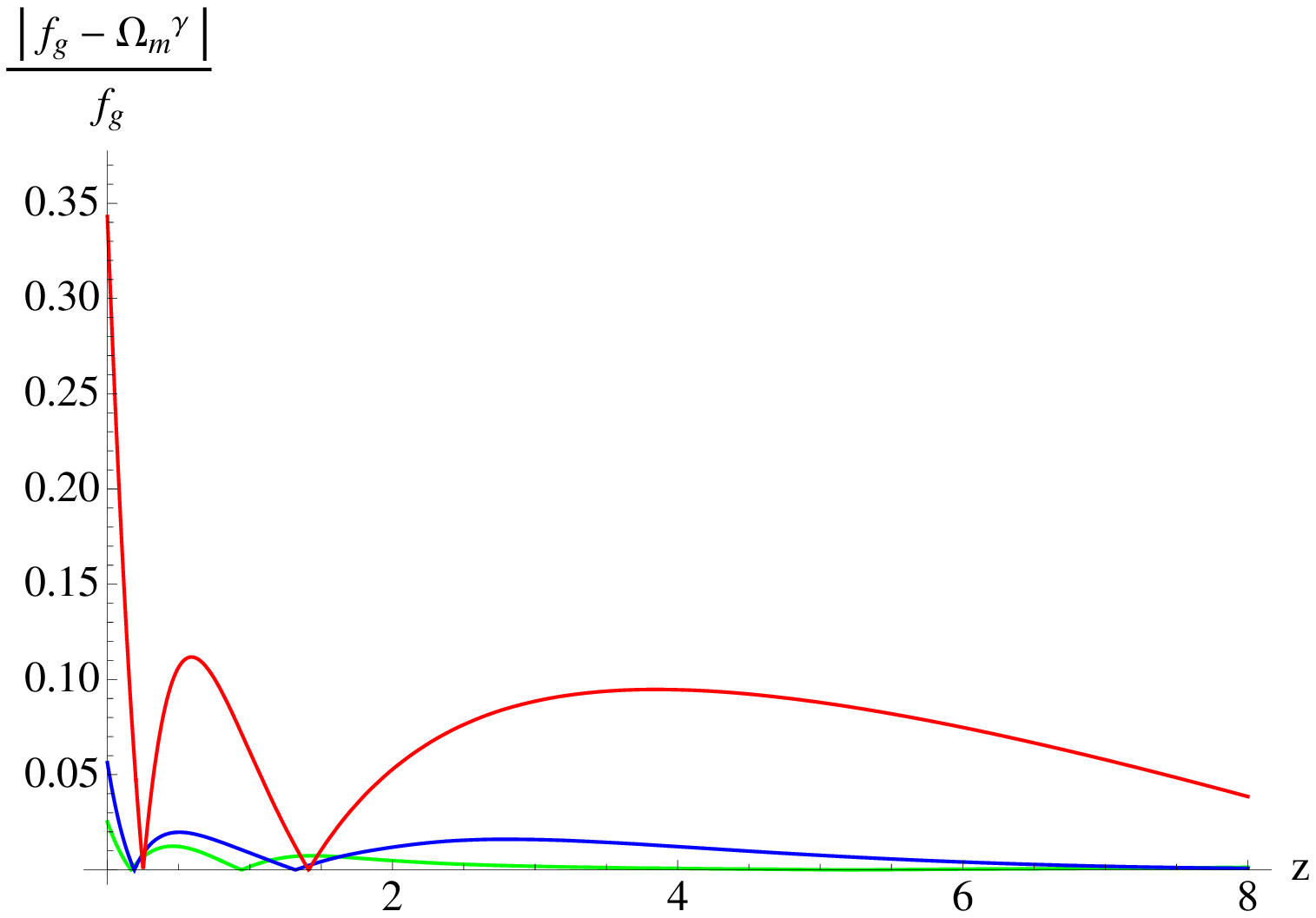}}
\quad
\subfigure[]{\includegraphics[width=0.45\textwidth]{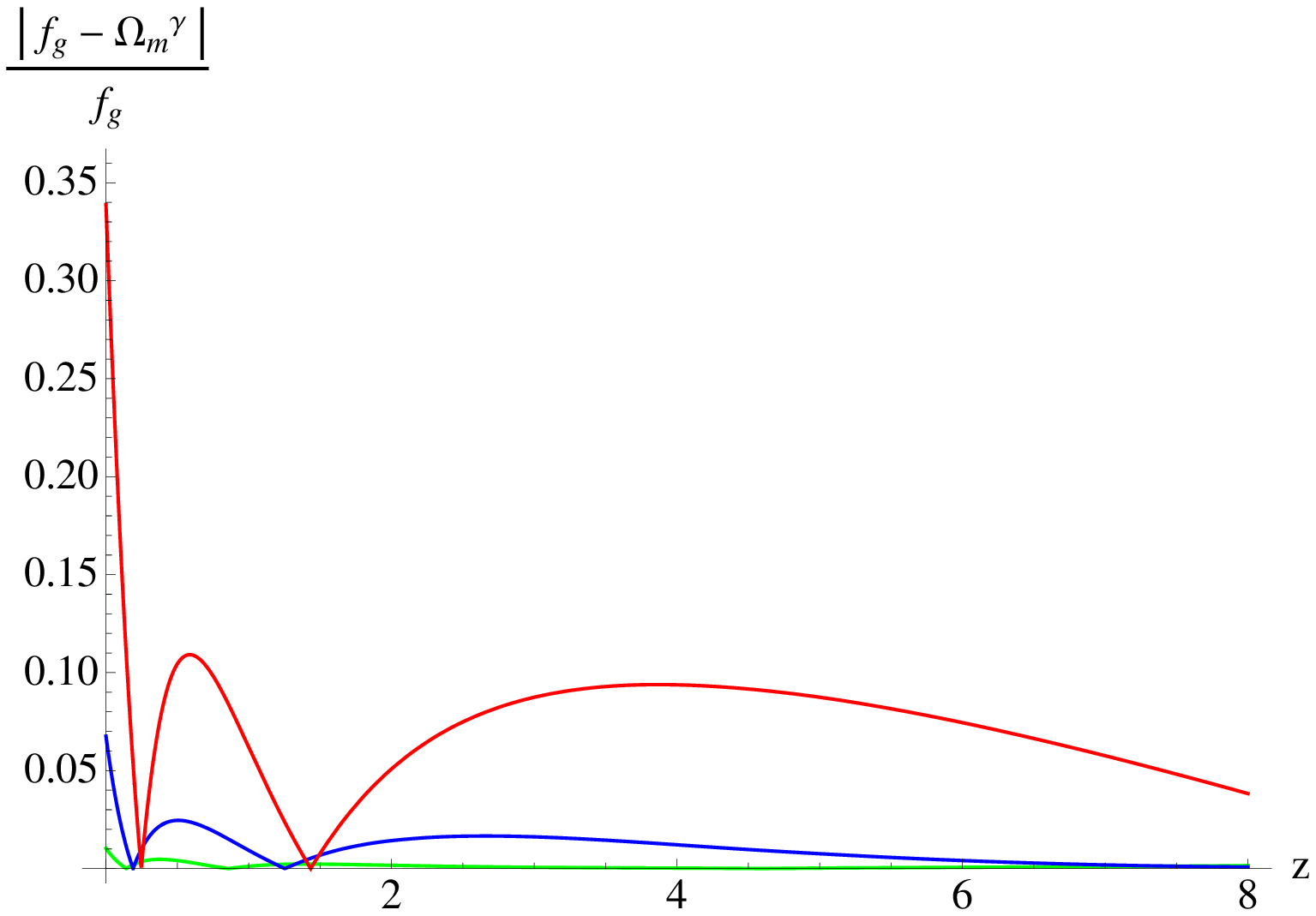}}
\caption{Cosmological evolution of the 
relative difference $\frac{\left| f_\mathrm{g} - \Omega_\mathrm{m}^\gamma \right|}{f_\mathrm{g}}$ with $\gamma = \gamma_0 + \gamma_1 \frac{z}{1 + z}$ for $k = 0.1 \mathrm{Mpc}^{-1}$ (red), $k = 0.01 \mathrm{Mpc}^{-1}$ (blue) and $k = 0.001 \mathrm{Mpc}^{-1}$ (green) in the model $F_1(R)$ (a) and the model $F_2(R)$ (b).}
\label{rat_rel_dif}
\end{figure}

In order to analyze the fits quantitatively, in Fig.~\ref{rat_rel_dif} 
we display the cosmological evolution of the relative difference between $f_\mathrm{g}(z)$ and $\Omega_\mathrm{m}(z)^{\gamma(z)}$ for several values of $k$ in the models $F_1(R)$ and $F_2(R)$. 
We see that, if we do not consider $z < 0.2$, the relative difference for $\log [k] = -1$ is smaller than $12\%$ 
in both of the models. 
Thus, it is confirmed that these fits are better than those for the constant growth rate, but these are worse than those for $\gamma = \gamma_0 + \gamma_1 z$. 
For lower values of $\log [k]$, the relative difference is smaller than $2\%$ 
for $z > 0.2$.\\ 

In conclusion, 
through the investigations of these different Ansatz for the growth index, 
we can say that $\gamma = \gamma_0 + \gamma_1 z$ is the 
parameterization that fits Eq.~(\ref{gi}) to the solution of 
Eq.~(\ref{gmp4}) better in a wide range of values for $k$. 
Moreover, despite the fact that the behavior of the parameters $\gamma_0$ and $\gamma_1$ in the 
models $F_1(R)$ and $F_2(R)$ is quite similar to each other, 
in order to distinguish between these models, 
in 
Fig.~\ref{HS_lineal_growth_index_vs_logk}
we can see that 
the more differences between these models come from 
the values of $\gamma_0$ for $\log [k] \leq -2$. 
In fact, as we observed before, for $\log [k] \leq -2.5$ we get
$\gamma_0 \sim 0.46$ for the model $F_1(R)$ and $\gamma_0 \sim 0.51$ 
for the model $F_2(R)$. 


\chapter{Inflation in $F(R)$-exponential gravity}

\paragraph*{}In this Chapter, we will ultimate the numerical analysis of viable $F(R)$-gravity by studying the early-time acceleration.
By applying the two viable (exponential) models described in Chapter {\bf 6} to inflationary cosmology 
and executing the numerical analysis of the inflation process, 
we illustrate that the exit from inflation can be realized. 
Concretely, we demonstrate that 
different numbers of $e$-folds during inflation can be obtained 
by taking different model parameters in the presence of ultrarelativistic 
matter, the existence of which makes inflation end and leads to 
the exit from it. 
Indeed, we observe that at the end of the inflation, the effective energy density as well as the curvature of the universe correctly decrease.

\section{Analysis of inflation}

\paragraph{}In order to study and understand the phenomenology of inflation, it is worth rewriting Eq.~(\ref{superEq}) by introducing a suitable scale factor $M^2$ at the inflation. For example, for  our inflation models (\ref{total})--(\ref{sin}), we can choose $M^2=\Lambda_\mathrm{i}$. The effective ``modified gravity'' energy density $y_H(z)$ is now defined as 
\begin{equation}
y_H (z)\equiv\frac{\rho_{\mathrm{MG}}}{M^2/\kappa^2}=\frac{3H^2}{M^2}
-\tilde\chi (z+1)^{4}\,,
\end{equation}
in analogy with (\ref{y}). In the above expression, we have replaced $\rho_{\mathrm{DE}}$ with $\rho_{\mathrm{MG}}$ and we have neglected the contribution of standard matter. We suppose the presence of ultrarelativistic matter/radiation in the hot universe scenario, whose energy density $\rho_{\mathrm{rad}}$ at the redshift equal to zero is related with the scale as 
\begin{equation}
\tilde\chi=\frac{\kappa^2\rho_{\mathrm{rad}}}{M^2}\,.
\end{equation}
Since the results are independent of the redshift scale, 
we set $z=0$ at some times around the end of inflation. 
Equation (\ref{superEq}) reads\\
\phantom{line}
\begin{eqnarray}
&&
y_H''(z)-\frac{y'(z)}{z+1}\left\{3+\frac{1-F'(R)}{2M^2F''(R)\left[y_H(z)+\tilde\chi(z+1)^4\right]}\right\} 
\nonumber\\ 
&&
{}+\frac{y_H(z)}{(z+1)^2}\left\{\frac{2-F'(R)}{M^2F''(R)\left[y_H(z)+\tilde\chi(z+1)^4\right]}\right\}
\nonumber\\ 
&&
{}
+\frac{(F'(R)-1)2\tilde\chi(z+1)^4+(F(R)-R)/M^2}{(z+1)^2\,2M^2F''(R)\left[y_H(z)+\tilde\chi(z+1)^4\right]}=0\,.
\label{inflsystem}
\end{eqnarray}
\phantom{line}\\
Moreover, the Ricci scalar is expressed as 
\begin{equation}
R=M^2\left[4y_H(z)-(z+1)\frac{d y_H(z)}{d z}\right]\,.
\end{equation}
Thus, 
by considering the perturbation $y_1(z)$ around the de Sitter solution $y_0$ describing inflation, as in Eq. (\ref{yexpansion}), from Eq. (\ref{inflsystem}), if we
assume the contribute of ultrarelativistic matter to be much smaller than $y_0$,
we easily find
\begin{eqnarray}
y_H(z)&\simeq&y_0+y_1(z)\,,\nonumber\\
y_1(z)&=&C_0(z+1)^x\,,\label{pertinfl}
\end{eqnarray}
whith
\begin{equation}
x=\frac{1}{2}\left(3-\sqrt{25-\frac{16F'(R_{\mathrm{dS}})}{R_{\mathrm{dS}}F''(\mathrm{R_{dS}})}}\right)\,,
\label{x}
\end{equation}
where $R_{\mathrm{dS}}=4 M^2 y_0$ and $x<0$ if the de Sitter point is unstable (in such a case, condition (\ref{dSstability}) is violated). Thus, the perturbation $y_1(z)$ in~(\ref{pertinfl}) grows up in expanding universe as~\cite{malloppone},
\begin{equation}
y_1(z)=y_1(z_\mathrm{i})\left[\frac{(z+1)}{(z_\mathrm{i}+1)}\right]^x\,.
\end{equation}
Here, we have considered $C_0=y_1(z_\mathrm{i})/(z_\mathrm{i}+1)^x$, $z_i$ being the redshift at the beginning of inflation where perturbation is bounded. 
When $y_1(z)$ is on the same order of the effective modified gravity energy density $y_0$ of 
the de Sitter solution, the model exits from 
inflation. 
Thus, we may evaluate the characteristic number of $e$-folds during inflation 
\begin{equation}
N=\log \left[\frac{z_\mathrm{i}+1}{z_\mathrm{e}+1}\right]\,,
\end{equation}
as
\begin{equation}
N\simeq \frac{1}{x}\log\left[\frac{y_1(z_\mathrm{i})}{y_0}\right]\,.
\label{Nfolding}
\end{equation}
The inflation ends at the redshift $z=z_e$. A value demanded in most inflationary scenarios is at least 
$N = 50$--$60$.

A classical perturbation on the (vacuum) de Sitter solution may be given by 
the presence of ultrarelativistic matter in the early universe.

\section{Numerical analysis of inflation in exponential gravity}

\paragraph{}In this Section we will analyze the Model I (\ref{total}) and the Model II (\ref{sin}) introduced in \S~\ref{inflation} in the context of exponential gravity.
This systems give rise to the de Sitter solution of the inflation where the universe expands in 
an accelerating way but, suddenly, it exits from inflation and tends towards 
the minimal attractor at $R = 0$ (the trivial de Sitter point). 
In this way, the small curvature regime arises and the physics of the $\Lambda$CDM Model is reproduced \cite{malloppone}. 

\subsection*{Model I}

\paragraph{}Let us consider the Model I given by (\ref{total}).
Following the analysis carried out in \S~\ref{inflation1}, we set $R_0=\Lambda\,, \Lambda=10^{-66}\text{eV}^2$ and $\Lambda_i=10^{100}\Lambda$ (however, since we choose the mass scale $M^2=\Lambda_{\mathrm{i}}$, all the results will be independent on the expected value of effective cosmological constant during inflation, while $\Lambda$ is negligible). Moreover, we take $\bar\gamma=1$, and choose $n=4$ and $R_{\mathrm{i}}=2\Lambda_{\mathrm{i}}$ to avoid the antigravity effects. We analyze the three different cases $\alpha=5/2\,,8/3\,,$ and $11/4$, so that $\tilde R_{\mathrm{i}}=4\Lambda_{\mathrm i}\,,6\Lambda_{\mathrm i}\,,$ and $8\Lambda_{\mathrm i}$, namely this paramter corresponds to the de Sitter solutions $R_{\mathrm{dS}}$ for which the inflation is realized in the three different cases under consideration.

Despite the fact that the values of $\alpha$ are very close 
each other, the values of $R_{\mathrm{dS}}$ and $x$ in Eq. (\ref{x}) significantly change and the reactions of the system to small perturbations are completely different.
By starting from Eq.~(\ref{Nfolding}), we may reconstruct the rate $y_1(z_\mathrm{i})/y_0$ between the abundances of ultrarelativistic matter/radiation and 
modified gravity energy at the beginning of inflation in order to obtain a determined 
number of $e$-folds during inflation in the three cases, by taking into account that $x=-0.086$, $-0.218$, and $-0.270$ for $\alpha=5/2$, $8/3$, and $11/4$, respectively. 
For example, in order to have $N=70$, 
for $\alpha=5/2$, 
a perturbation of $y_1(z_\mathrm{i})/y_0\sim 10^{-3}$ is necessary; 
for $\alpha=8/3$, 
a perturbation of $y_1(z_\mathrm{i})/y_0\sim 10^{-7}$ is sufficient; 
whereas for $\alpha=11/4$, 
$y_1(z_\mathrm{i})/y_0\sim 10^{-9}$. 
The system becomes more unstable, as $\left(3-\alpha\right)$ 
is closer to zero. 

In studying the behavior of the cosmic evolution in 
Model I, 
we put $M^2=\Lambda_{\mathrm i}$ in Eq.~(\ref{inflsystem})  and set $\tilde{\chi}=10^{-4}\,y_0/(z_\mathrm{i}+1)^4$ for the case $\alpha=5/2$ and $\tilde{\chi}=10^{-6}\,y_0/(z_\mathrm{i}+1)^4$ for the cases $\alpha=8/3, 11/4$. 
In these choices, 
the effective energy density originating from 
the modification of gravity is $10^4$ and $10^6$ times larger than 
that of ultrarelativistic matter/radiation during inflation. 
By using Eq.~(\ref{Nfolding}), we can predict the following numbers of 
$e$-folds: 
\begin{eqnarray}
N &\simeq& 107 \quad (\mathrm{for} \,\,\, \alpha=5/2)\,,\nonumber\\
N &\simeq& 64 \quad (\mathrm{for} \,\,\, \alpha=8/3)\,,\nonumber\\
N &\simeq& 51 \quad (\mathrm{for} \,\,\, \alpha=11/4)\,.\label{stime}
\end{eqnarray}
In order to solve Eq.~(\ref{inflsystem}) numerically, 
we use the initial conditions
\begin{eqnarray}
\left\{\begin{array}{l}
y_H(z)\Big\vert_{z_i} = \frac{R_{\mathrm{dS}}}{4\Lambda_\mathrm{i}}\,,\\ \\
\frac{d y_H(z)}{d (z)}\Big\vert_{z_i} = 0\,,
\end{array}\right.
\label{BC}
\end{eqnarray}
at the redshift $z_\mathrm{i}\gg 0$ when inflation starts. 
We put $z_\mathrm{i}=10^{46}$, $10^{27}$, and $10^{22}$ for $\alpha=5/2$, 
$8/3$, and $11/4$, respectively 
(just for a more comfortable reading of the graphics). 
We also remark that 
the initial conditions are subject to an artificial error that we can estimate to be in the order of $\exp\left[-\left(R_{\mathrm{dS}}/R_\mathrm{i}\right)^n\right]\sim 10^{-7}$. This is the reason for which we only consider $\tilde{\chi}>10^{-7}$. 

\begin{figure}[!h]
\subfigure[]{\includegraphics[width=0.3\textwidth]{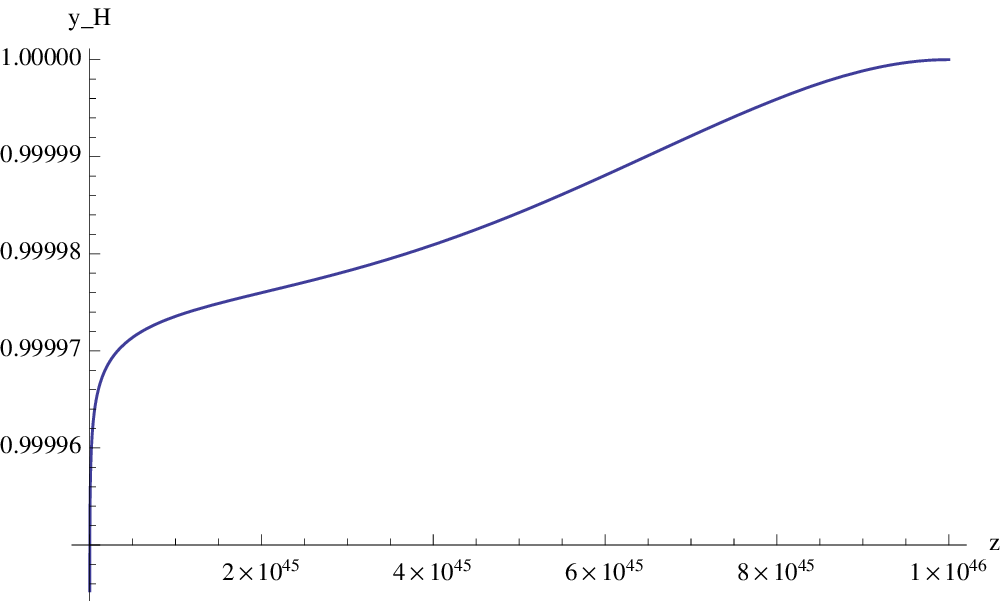}}
\centering
\subfigure[]{\includegraphics[width=0.3\textwidth]{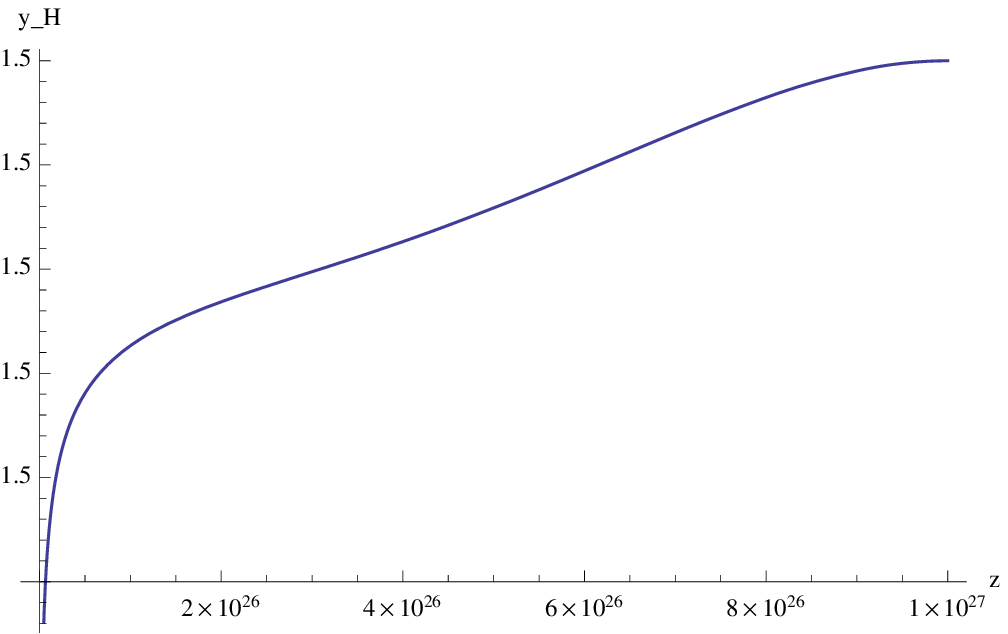}}
\centering
\subfigure[]{\includegraphics[width=0.3\textwidth]{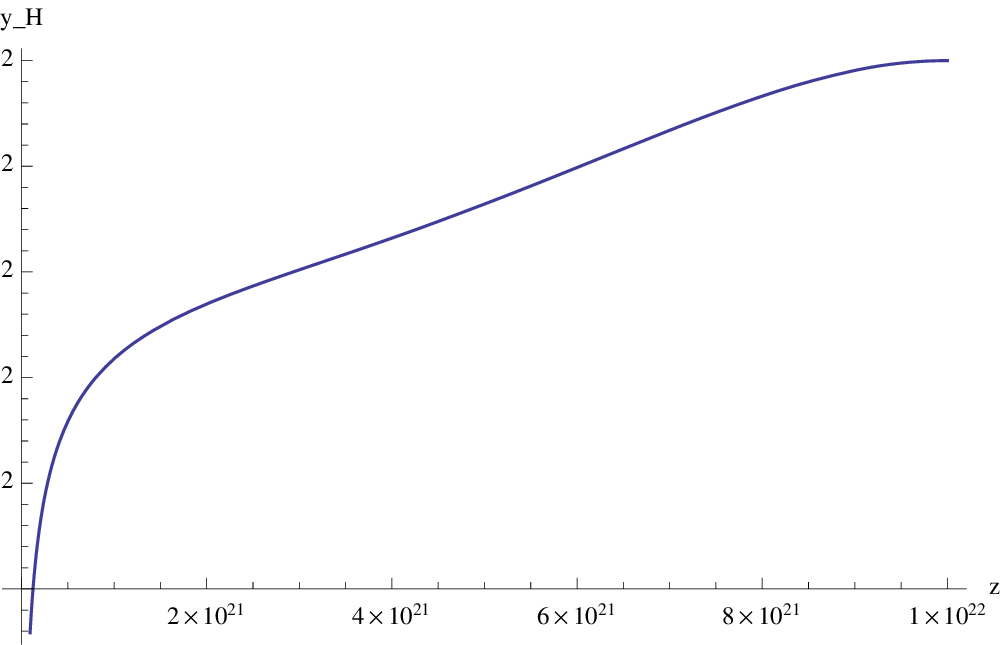}}
\quad
\subfigure[]{\includegraphics[width=0.3\textwidth]{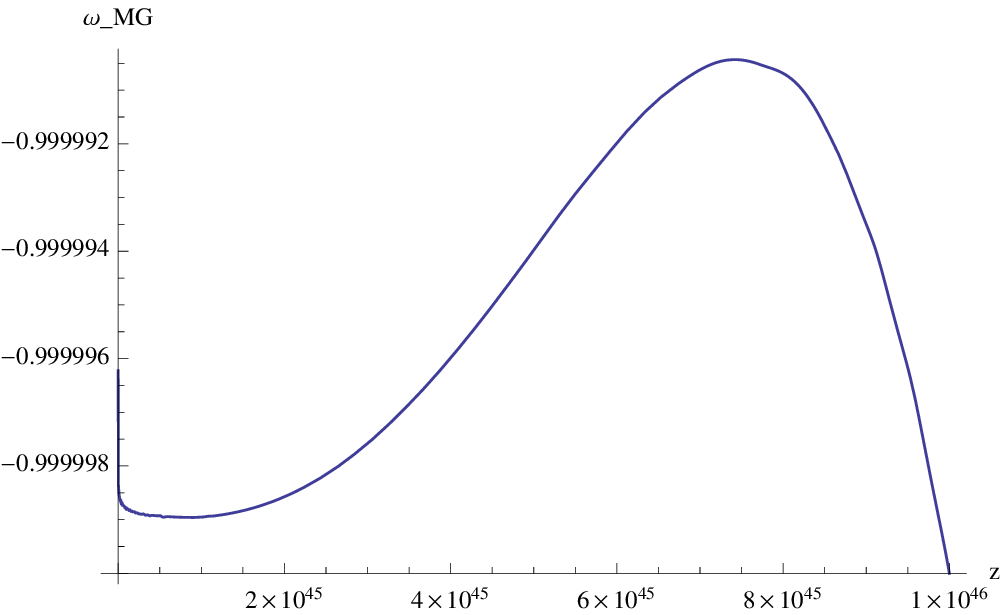}}
\centering
\subfigure[]{\includegraphics[width=0.3\textwidth]{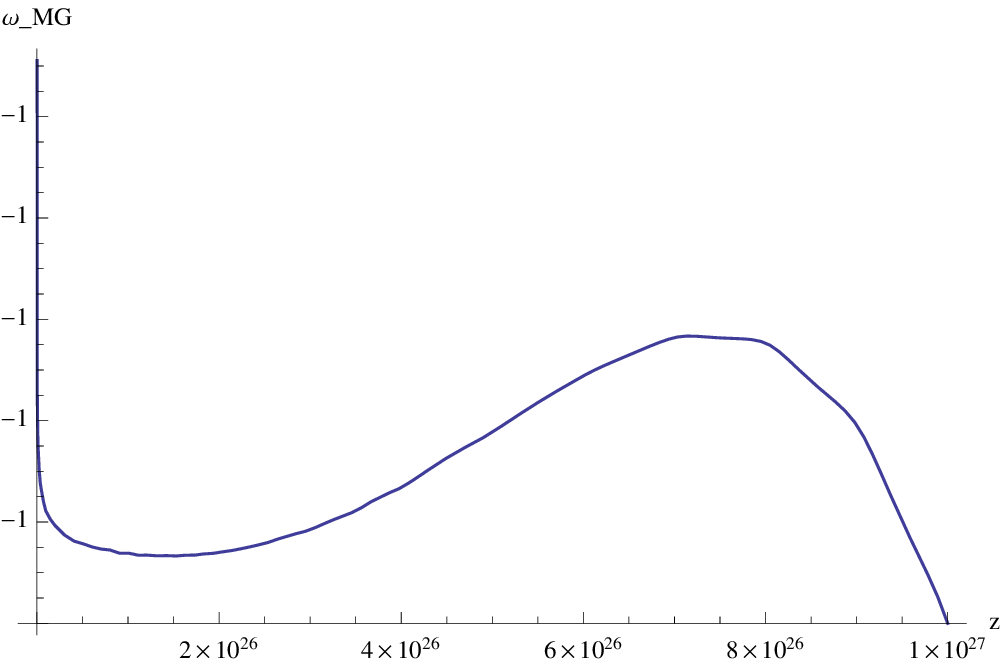}}
\centering
\subfigure[]{\includegraphics[width=0.3\textwidth]{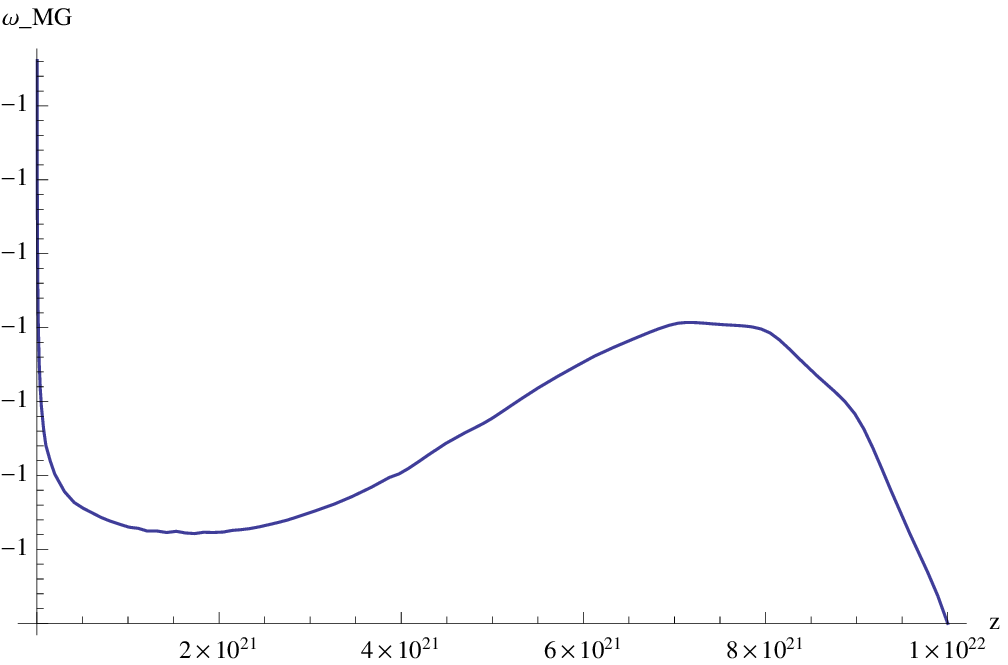}}
\caption{Plots of $y_H$ [a-c] and $\omega_{\mathrm{MG}}$ [d-f] as functions of the redshift $z$ for Model I with $\alpha=5/2$ [a-d], $\alpha=8/3$ [b-e] and $\alpha=11/4$ [c-f]. 
\label{001}}
\end{figure}

\begin{figure}[!h]
\subfigure[]{\includegraphics[width=0.3\textwidth]{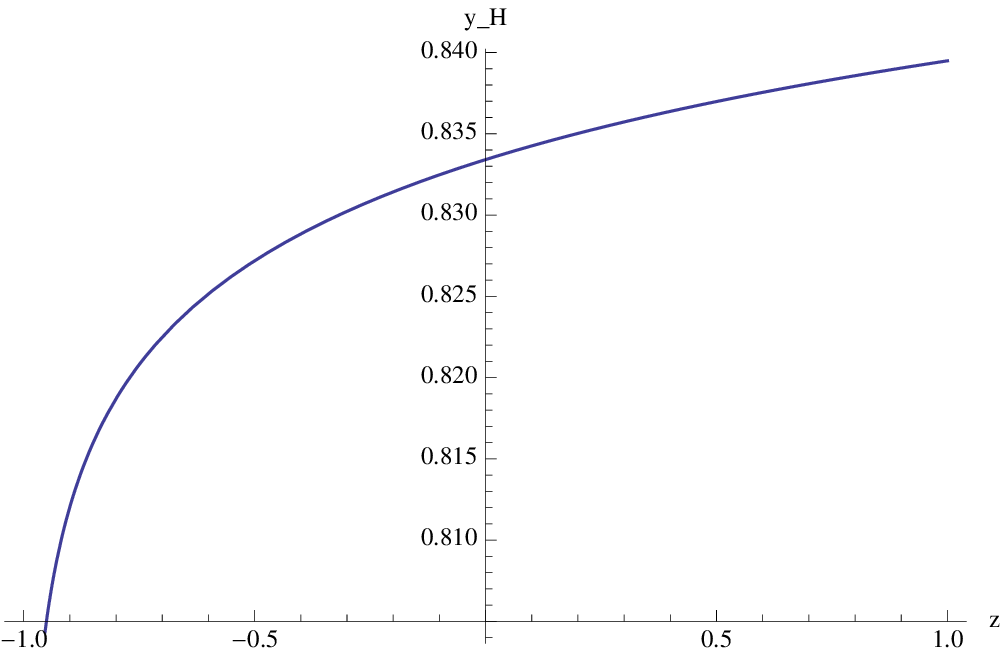}}
\centering
\subfigure[]{\includegraphics[width=0.3\textwidth]{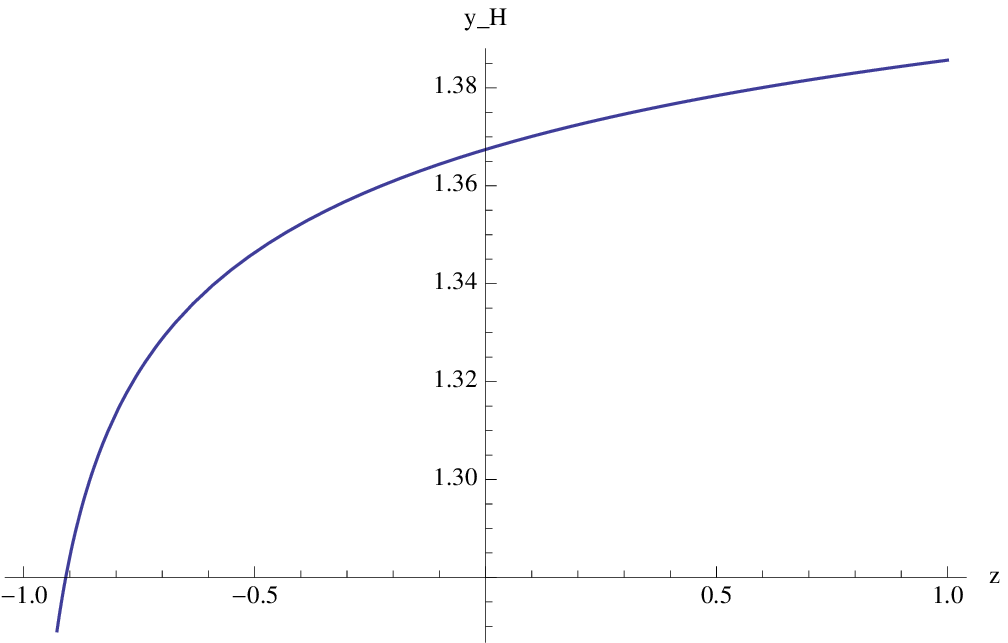}}
\centering
\subfigure[]{\includegraphics[width=0.3\textwidth]{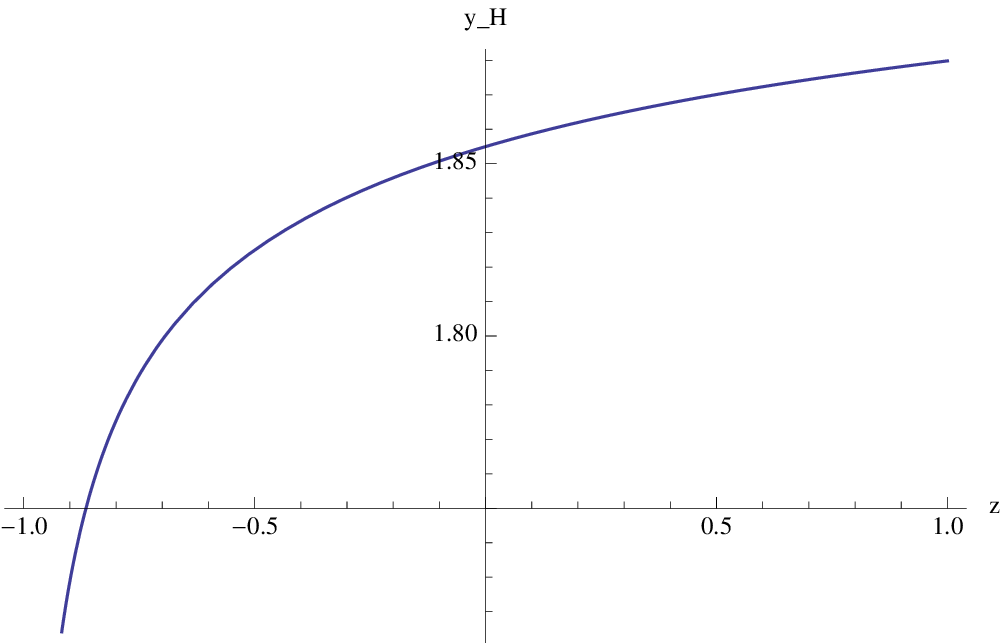}}
\quad
\subfigure[]{\includegraphics[width=0.3\textwidth]{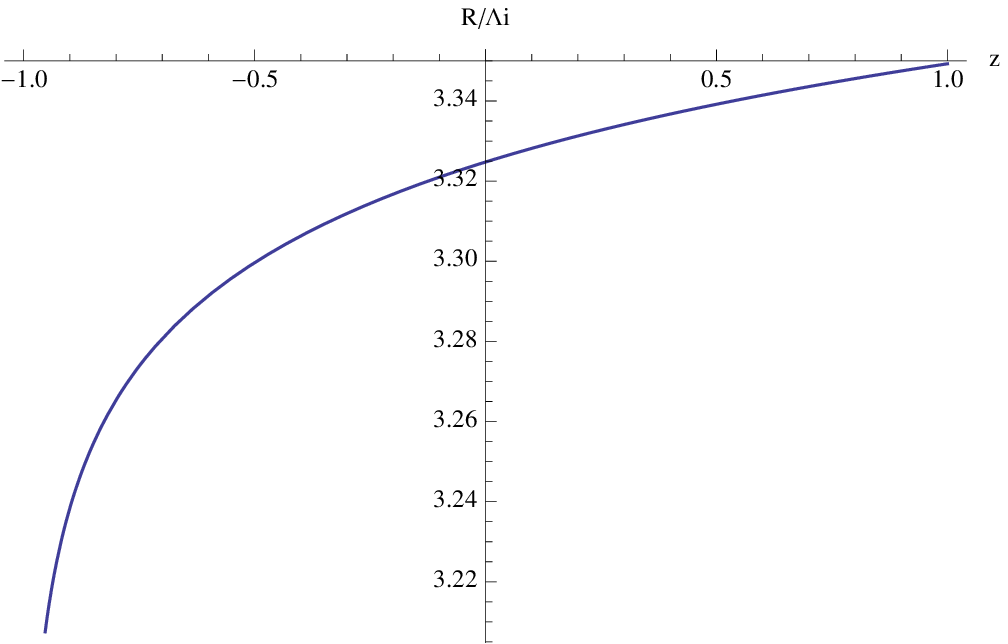}}
\centering
\subfigure[]{\includegraphics[width=0.3\textwidth]{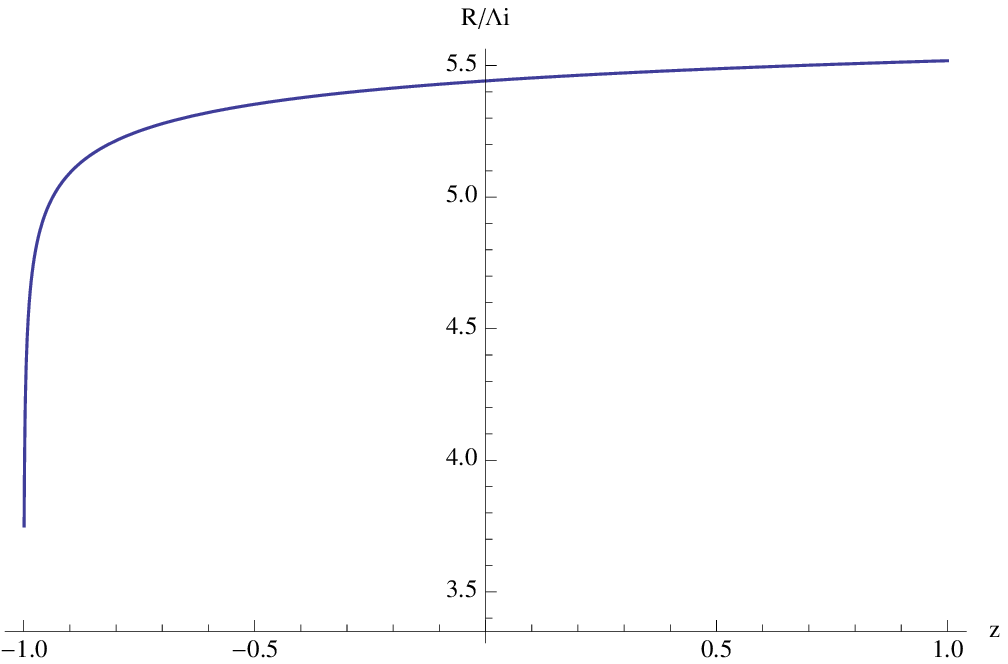}}
\centering
\subfigure[]{\includegraphics[width=0.3\textwidth]{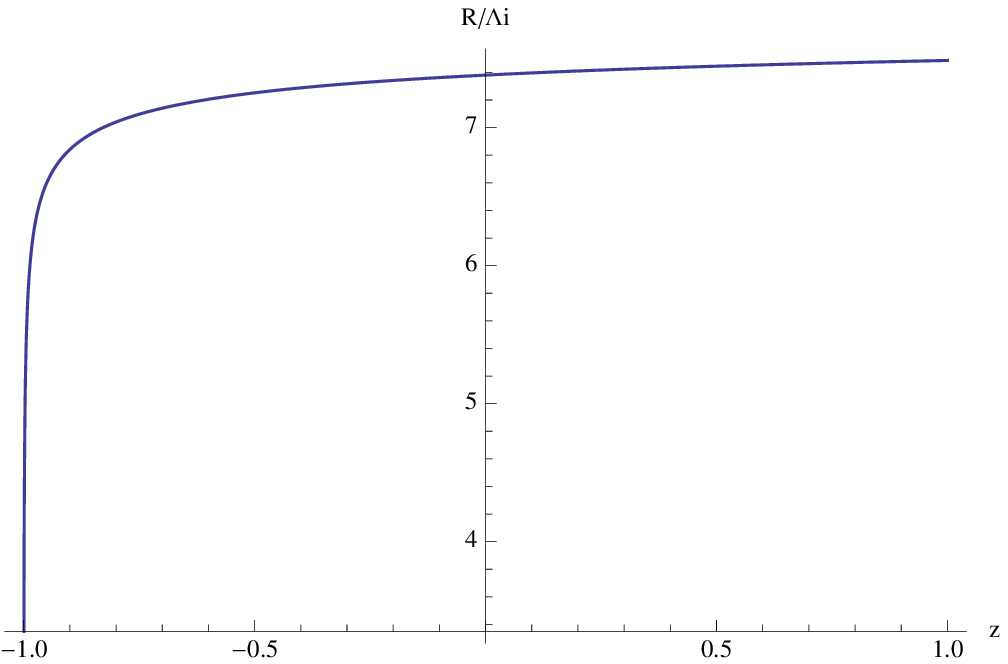}}
\caption{Cosmological evolution of $y_H$ [a-c] and $R/\Lambda_\mathrm{i}$ [d-f] as functions of the redshift 
$z$ in the region $-1<z<1$ for Model I with $\alpha=5/2$ [a-d], $\alpha=8/3$ [b-e] and $\alpha=11/4$ [c-f]. 
\label{0001}}
\end{figure}

In Fig.~\ref{001}, we illustrate the cosmological evolutions of $y_H$ and the corresponding modified gravity EoS parameter $\omega_{\mathrm{MG}}$ 
(defined in the same way of $\omega_{\mathrm{DE}}$ in Eq. (\ref{oo})) 
as functions of the redshift $z$ in the three cases. 
We can see, during inflation $\omega_{\mathrm{MG}}$ is indistinguishable from the value of -1 and $y_H$ tends to decrease very slowly with respect to 
$y_H=1\,, 3/2\,, 2$ for $\alpha=5/2\,, 8/3\,, 11/4$, so that the curvature can be the expected de Sitter one, $R_{\mathrm{dS}}(=4y_H)=4\Lambda_\mathrm{i}\,, 6\Lambda_\mathrm{i}\,, 8\Lambda_\mathrm{i}$. 
The expected values of $z_\mathrm{e}$ at the end of inflation may be derived from the number of $e$-folds in (\ref{stime}) during inflation 
and read $z_\mathrm{e} \simeq -0.47$ for $\alpha=5/2$; $z_\mathrm{e}\simeq -0.74$ for $\alpha=8/3$; $z_\mathrm{e}\simeq -0.39$ for $\alpha=11/4$. 
The numerical extrapolation yields 
\begin{eqnarray*}
y_H(z_\mathrm{e}) &=& 0.83y_H(z_\mathrm{i})\,,\quad R(z_\mathrm{e})=0.825 R_{\mathrm{dS}}\,,\quad (\mathrm{for} \,\,\, \alpha=5/2)\,;\nonumber\\
y_H(z_\mathrm{e}) &=& 0.88y_H(z_\mathrm{i})\,,\quad R(z_\mathrm{e})=0.853 R_{\mathrm{dS}}\,,\quad (\mathrm{for} \,\,\, \alpha=8/3)\,;\nonumber\\
y_H(z_\mathrm{e}) &=& 0.92y_H(z_\mathrm{i})\,,\quad R(z_\mathrm{e})=0.911 R_{\mathrm{dS}}\,,\quad (\mathrm{for} \,\,\, \alpha=11/4)\,.
\end{eqnarray*}
To confirm the exit from inflation,  in Fig.~\ref{0001}
we plot the cosmological evolutions of $y_H$ and $R/\Lambda_\mathrm{i}$ 
as functions of the redshift $z$ 
in the region $-1<z<1$, where $z_\mathrm{e}$ is included. 
The effective modified gravity energy density and the curvature decrease at the end of inflation and 
the physical processes described by the $\Lambda$CDM model can appear. 

\subsubsection*{Model II}

Next, we investigate Model II in Eq.~(\ref{sin}) analyzed in \S~\ref{inflation2}. 
Here, we put $n=3$ and $R_\mathrm{i}=2\Lambda_\mathrm{i}$, we take $\bar{\gamma}=1$ again and we execute the same numerical evaluation for $\alpha=5/2, 13/5, 21/8$ in this model as that in the previous case for Model I. 
The corresponding $\tilde R_{\mathrm i}$ parameters and de Sitter curvatures of inflation are 
$\tilde R_{\mathrm{i}}(=R_{\mathrm{dS}})=4\Lambda_\mathrm{i}, 5\Lambda_\mathrm{i}, 16\Lambda_\mathrm{i}/3$. Now, we obtain the factor in Eq.~(\ref{x}) for instability as 
$x=-0.086$, $-0.170$, and $-0.188$ for $\alpha=5/2$, $13/5$, and $21/8$, respectively. 
Hence, 
we set $\tilde{\chi}=10^{-3}\,y_0/(z_\mathrm{i}+1)^4$ for $\alpha=5/2$, 
$\tilde{\chi}=10^{-4}\,y_0/(z_\mathrm{i}+1)^4$ for $\alpha=13/5$, and 
$\tilde{\chi}=10^{-5}\,y_0/(z_\mathrm{i}+1)^4$ for $\alpha=21/8$. 
As a consequence, the numbers of $e$-folds during inflation result in 
$N=80$, $54$, and $61$. 
The initial conditions are the same as those in the previous case 
in (\ref{BC}). Furthermore, we put $z_\mathrm{i}=10^{34}$, $10^{22}$, and 
$10^{26}$ for $\alpha=5/2$, $13/5$, and $21/8$. 

\begin{figure}[!h]
\subfigure[]{\includegraphics[width=0.3\textwidth]{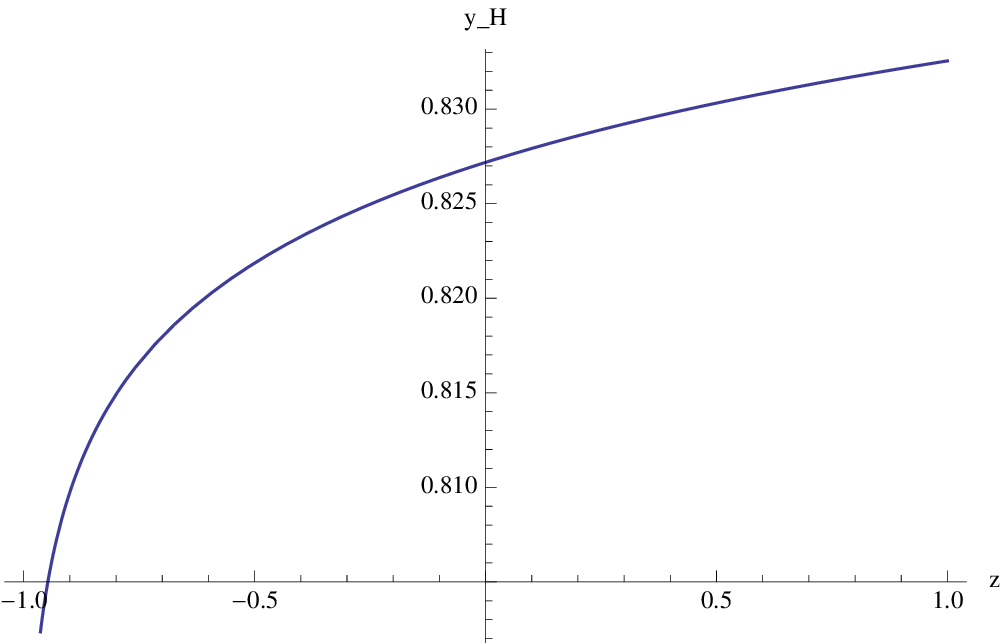}}
\centering
\subfigure[]{\includegraphics[width=0.3\textwidth]{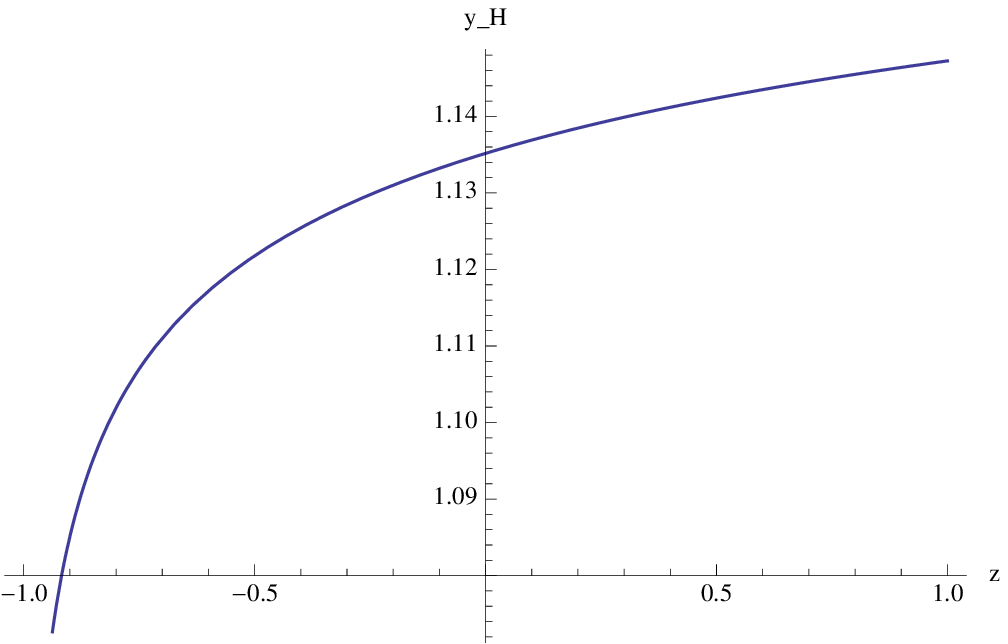}}
\centering
\subfigure[]{\includegraphics[width=0.3\textwidth]{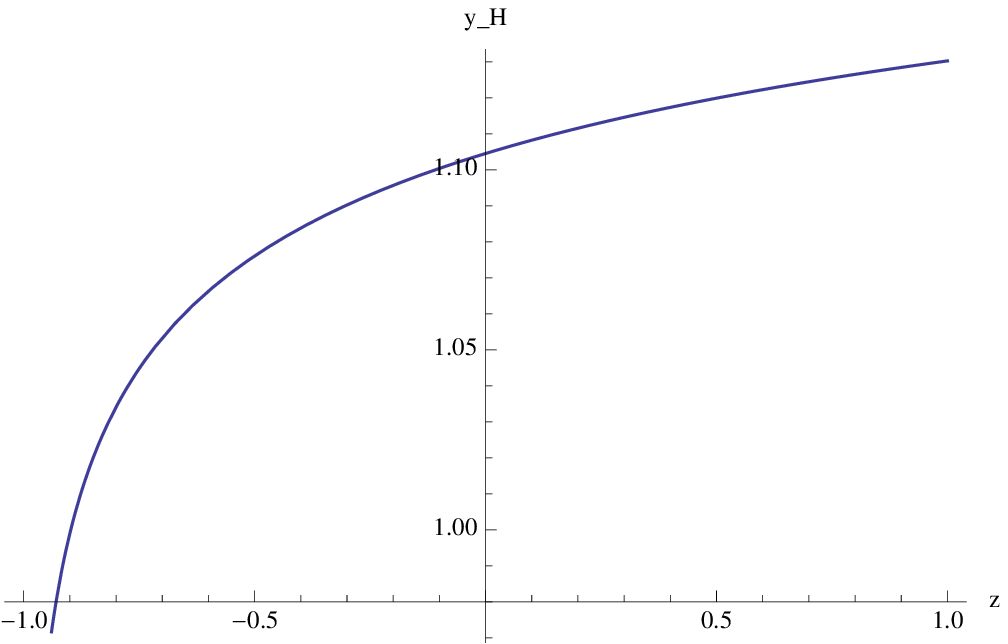}}
\quad
\subfigure[]{\includegraphics[width=0.3\textwidth]{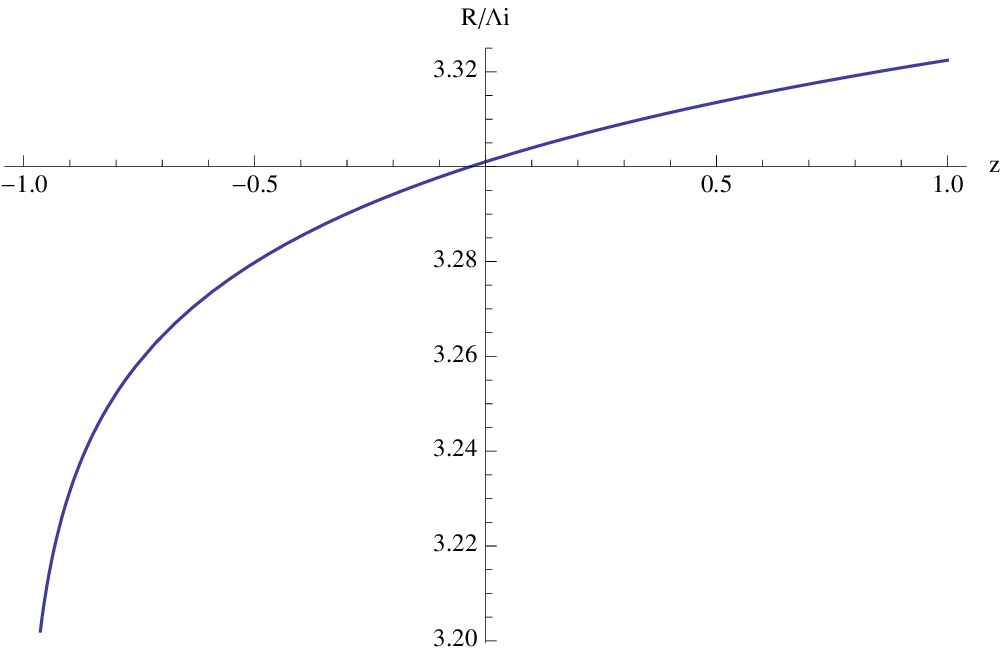}}
\centering
\subfigure[]{\includegraphics[width=0.3\textwidth]{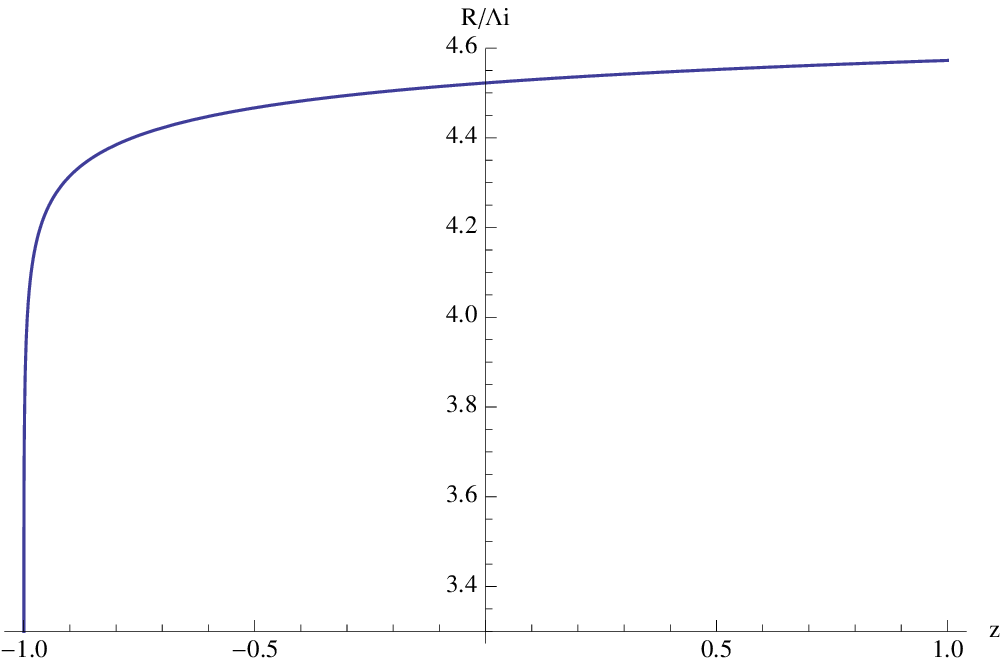}}
\centering
\subfigure[]{\includegraphics[width=0.3\textwidth]{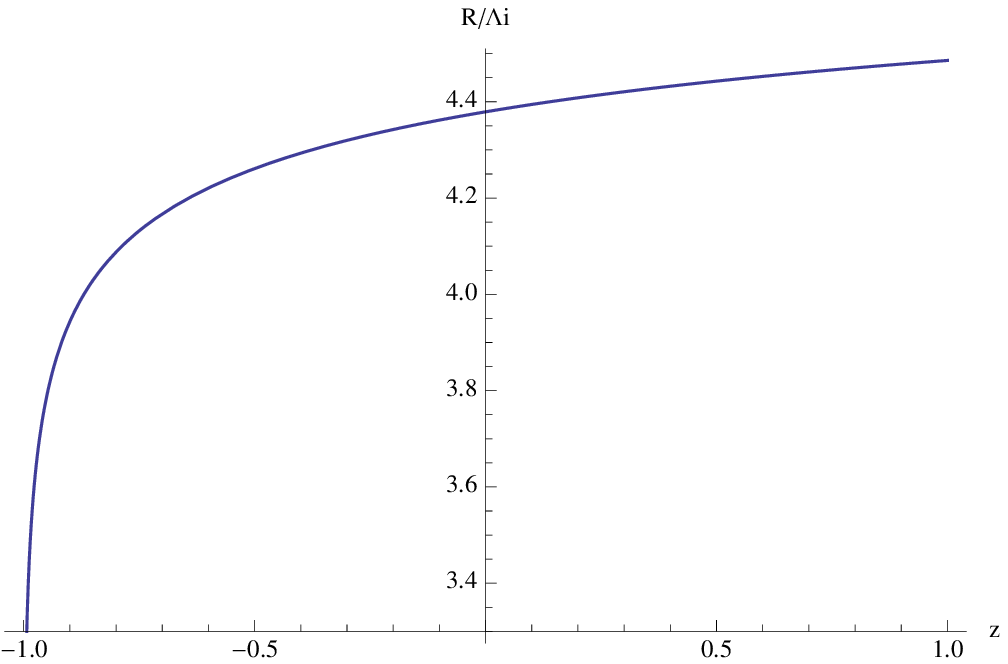}}
\caption{Cosmological evolution of $y_H$ [a-c] and $R/\Lambda_\mathrm{i}$ [d-f] as functions of the redshift 
$z$ in the region $-1<z<1$ for Model II with $\alpha=5/2$ [a-d], $\alpha=13/5$ [b-e] and $\alpha=21/8$ [c-f]. 
\label{0001bis}}
\end{figure}

Through the numerical extrapolation,
at the expected values of $z_\mathrm{e}$ at the end of inflation, 
$z_\mathrm{e}=-0.80$, $-0.97$, and $-0.71$, we aquire
the following values for the effective modified gravity energy density and the Ricci scalar: 
\begin{eqnarray*}
y_H(z_\mathrm{e}) &=& 0.82y_H(z_\mathrm{i})\,,\quad R(z_\mathrm{e})=0.813 R_{\mathrm{dS}}\,,\quad(\mathrm{for} \,\,\, \alpha=5/2)\,;\nonumber\\
y_H(z_\mathrm{e}) &=& 0.84y_H(z_\mathrm{i})\,,\quad R(z_\mathrm{e})=0.884 R_{\mathrm{dS}}\,,\quad(\mathrm{for} \,\,\, \alpha=13/5)\,;\nonumber\\
y_H(z_\mathrm{e}) &=& 0.79y_H(z_\mathrm{i})\,,\quad R(z_\mathrm{e})=0.780 R_{\mathrm{dS}}\,,\quad(\mathrm{for} \,\,\,\alpha=21/8)\,.
\end{eqnarray*}
For this model, in Fig.~\ref{0001bis}  we depict the cosmological evolutions of $y_H$ and $R/\Lambda_\mathrm{i}$ as functions of the redshift $z$ in the region $-1<z<1$ at the end of inflation. 
Again in this case, the effective modified gravity energy density and curvature decrease.

As a result, we have proved that it is possible to acquire 
a gravitational scenario for an unified description of inflation in 
the early universe with the late-time cosmic acceleration due to the 
the physical processes described by
$\Lambda$CDM-like dark energy domination. 

Here, we note that at the inflationary stage, radiation is negligible, 
as in the ordinary inflationary scenario. 
It causes the perturbations at the origin of instability. 
This point has been shown in a numerical way by using radiation, whose energy density is six order of magnitude smaller than that of dark energy.  

It should be emphasized that in \S~\ref{inflation} of Chapter {\bf 6}, 
as a first step, we have concentrated on the possibility 
of the realization of inflation, and here we have shown 
via numerical evaluation that realistic cosmological scenario may be realized in an unified therory by using exponential models.
However,  important issues in inflationary cosmology such as 
the graceful exit problem of 
inflation, the following reheating process, and the generation of the 
curvature perturbations, whose power spectrum has to be consistent with 
the anisotropies of the CMB radiation obtained from the 
Wilkinson Microwave Anisotropy Probe (WMAP) Observations, 
have not been analized and may be crucial future works for this kind of unified theories.\\

For the sake of completeness, in Appendix C, as an example, we also present the conformal transformation of Model I (\ref{total}) to study the slow-roll parameters of inflation.


\chapter{BH and dS-solutions in a covariant renormalizable field theory of gravity}

\paragraph*{} 
Some attempts to quantize gravity have been carried out in the past by considering the perturbations of a flat, Lorentz invariant background, and starting from the principles of General Relativity, but, unfortunately, unavoidable, non-renormalizable divergences
appear from the ultraviolet (UV) region in momentum space. To solve this problem, higher derivative theories have been invoked, but there emerges a new problem, the so called unitarity issue~\cite{Shapiro}. 

In 2009, Ho\v{r}ava proposed to directly modify the ultraviolet behavior of the graviton propagator in a Lorentz non-invariant way~\cite{Horava}, as $1/|{\vec k}|^{2z}$, with $|{\vec k}|$ the norm of the  the spatial momenta $\vec k$ and $z=2,3$ or higher. This exponent comes from the  (anisotropic) scaling properties of the space-time coordinates $\left(t, {\vec x}\right)$, namely  $t\to b^z t$, ${\vec x}\to b{\vec x}$, with $b$ the rescaling parameter. The advantage of this theory is that, for $z=3$, it is UV power-counting  renormalizable and the conjecture of renormalizability is aquired. In order to get the Lorentz non-invariance, some terms explicitly break the Lorentz invariance (or, more precisely, the full diffeomorphism invariance), by treating the temporal and the spatial coordinates differently. The Horawa model has diffeomorphism invariance with respect to the time coordinate $t$ only, while for the spatial coordinates one obtains for the variations $\delta x^i=\zeta^i(t,{\vec x})$, $\delta t=f(t)$, with $\zeta^i(t,{\vec x})$ and $f(t)$ arbitrary functions of $t$ and ${\vec x}$ and of $t$ only, respectively.

In Ref.~\cite{Nojiri:2009th}, a Ho\v{r}ava-like gravity model with full diffeomorphism invariance was proposed. Here, in considering perturbations from a flat, Lorentz
invariant background, the Lorentz invariance of the propagator is dynamically broken by a non-standard coupling with a perfect fluid. The propagator behaves as
$1/|{\vec k}|^{2z}$, with $z=2,3$ or higher in the ultraviolet region and it is possible to show that the model could be
perturbatively power counting (super-)renormalizable by requiring $z\geq 3$.
However, an
unknown fluid, which might have a stringy origin but cannot correspond to an usual fluid (i.e. radiation, baryons, dust...) emerges form the theory, since the model can be consistently constructed only if the EoS parameter of fluid $\omega_{\mathrm{F}}$ is $\omega_{\mathrm{F}}\neq -1\,,1/3$. For usual particles in the high energy region, the corresponding fluid must be ultrarelativistic matter/radiation, for which $\omega_{\mathrm{F}}= 1/3$, but here we need the non-relativistic fluid even in the high energy region.
Later, a dust fluid with $\omega_{\mathrm{F}}=0$ was constructed for the scalar theory by
introducing a Lagrange-multiplier field~\cite{Lim,Capozziello:2010uv}.
 
In Ref.~\cite{Nojiri:1004}, a fluid with arbitrary constant $\omega_{\mathrm{F}}$ from
a scalar field which satisfies a constraint has been investigated. It is shown that, due to the constraint, the scalar field is not dynamical and, even in the high energy
region, a non-relativistic fluid can be derived. Through coupling with
the fluid, a full diffeomorphism class of invariant Lagrangians results. 
It has been demonstrated that such kind of theory posseses
all the good properties of the Lorentz non-invariant gravities, like the conjecture of renormalizability \cite{Carloni}, and has the advantage of being at the same time a covariant theory. In addition, it has been conjectured that it may exhibit the spatially-flat FRW solution for accelerating universe.  

In the present Chapter, we will show that this is the case. We will consider the covariant renormalizable theory and we will demonstrate that Schwarzschild black hole and de Sitter solutions exist as exact solutions~\cite{covarianttheory}.

\section{Black hole solutions in covariant (power-counting) renormalizable gravity.}

\paragraph*{} To start, let us briefly review the covariant (power-counting) renormalizable gravity of Ref.~\cite{Nojiri:2009th}. It is described by the action\\
\phantom{line}
\begin{equation}
\label{Hrv1}
I =\frac{1}{2\kappa^2}\,\int_{\mathcal{M}} d^4x\,\sqrt{-g}\, \left\{R-2\Lambda
-\alpha\left[\left(R^{ij}-\frac\beta2\,Rg^{ij}\right)\nabla_i\phi\nabla_j\phi\right]^n
-\lambda\,\left(\frac12\,g^{ij}\nabla_i\phi\nabla_j\phi+U_0\right)\right\}\ ,
\end{equation}
\phantom{line}\\
where $\phi$ is a cosmological scalar field,
$\lambda$ a Lagrangian multiplier, $\alpha,\beta,\Lambda,U_0$ are arbitrary constants
and, finally, $n\geq1$ is an arbitrary number.
Variation of the action with respect to $\lambda$ gives the constraint
\begin{equation}\label{const}
g^{ij}\nabla_i\phi\nabla_j\phi=-2U_0\,,
\end{equation}
while the field equations for the scalar field read\\
\phantom{line}
\begin{eqnarray}\label{PHI}
0&=&\nabla_i\,\left\{\left[2n\alpha\,F^{n-1}\left(R^{ij}
       -\frac\beta2\,Rg^{ij}\right)+\lambda\,g^{ij}\right]\,\nabla_j\phi\right\}
\nonumber\\ &=&\frac1{\sqrt{-g}}\,\,
\partial_i\,\left\{\left[2n\alpha\,F^{n-1}\left(R^{ij}
       -\frac\beta2\,Rg^{ij}\right)+\lambda\,g^{ij}\right]\,\sqrt{-g}\,
                 \partial_j\phi\right\}\,,
\end{eqnarray}
\phantom{line}\\
where, for convenience, we have putted
\begin{equation}
F=T_{ij}R^{ij}-\frac\beta2\,RT\,,\qquad T_{ij}=\nabla_i\phi\nabla_j\phi\,,
\qquad T=g^{ij}T_{ij}=-2U_0\,.
\end{equation}
The field equations related to the gravitational field have the form\\
\phantom{line}
\begin{eqnarray}
\label{Gij}
G_{ij}+\Lambda g_{ij}+\frac{\alpha}{2}\,F^n\,g_{ij}&=&
n\alpha F^{n-1}\left[R^k_iT_{kj}+R^k_jT_{ki}-\frac\beta2\left(TR_{ij}+RT_{ij}\right)\right]
+\frac\lambda2\,T_{ij}\nonumber\\ \nonumber\\
&&+n\alpha\,\left[D_{rsij}(T^{rs}F^{n-1})-\frac\beta2\,D_{ij}(TF^{n-1})\right]
+\Omega^{rs}\,\frac{\delta T_{rs}}{\delta g^{ij}}\,,
\end{eqnarray}
\phantom{line}\\
where $G_{ij}$ is the usual Einstein's tensor, $\Omega_{rs}$ is a tensor which will play no role in the following, and we have introduced the differential operators\\
\phantom{line}
\begin{equation}
D_{ij}=g_{ij}\Box-\frac12\,(\nabla_i\nabla_j+\nabla_j\nabla_i)\,,
\end{equation}
\phantom{line}
\begin{eqnarray}
D_{rsij}&=&\frac{1}{4}\,[(g_{ir}g_{js}+g_{jr}g_{is})\Box
+g_{ij}(\nabla_r\nabla_s+\nabla_s\nabla_r)\nonumber\\ \nonumber\\ 
&&-(g_{ir}\nabla_s\nabla_j+g_{jr}\nabla_s\nabla_i
+g_{is}\nabla_r\nabla_j+g_{js}\nabla_r\nabla_i)]\,.
\end{eqnarray}
\phantom{line}\\
Note that the field equations (\ref{Gij}) are valid for an arbitrary, symmetric
`energy-momentum' tensor $T_{ij}$, but in our particular case such a tensor does not
depend on the metric and so the last term in Eq.~(\ref{Gij}), depending on $\Omega_{rs}$, drops out.
Now, we look for interesting physical solutions of the field equations above.

\subsubsection{Schwarzschild solution}
\paragraph*{} This is the simplest one and can be easily obtained for $\Lambda=0$ and $n>1$.
In fact, in all such cases $R_{ij}=0, \lambda=0$ satisfy all field equations and Schwarzshild solution (\ref{S2}) can be recovered.
The scalar field $\phi$ has to fulfill the constraint (\ref{const}) only.

\subsubsection{Einstein-space solutions} 
\paragraph*{} These are generalizations of the previous solution.
They have the form
\begin{equation}
R_{ij}=\frac14\,R_0\,g_{ij}\,.
\end{equation}
Here, $R=R_0$ is a constant Ricci scalar. In such a case,
\begin{equation}
F=\left(\beta-\frac12\right)\,R_0U_0\equiv F_0\,,
\end{equation}
where $F_0$ is a constant and, from Eq.~(\ref{PHI}) and Eq.~(\ref{Gij}), we get\\
\phantom{line}
\begin{equation}
\label{lamb}
g^{ij}\nabla_i\,\left[n\alpha\left(\frac12-\beta\right)\,R_0F_0^{n-1}
+\lambda\right]\nabla_j\phi=0\,,
\end{equation}
\phantom{line}
\begin{equation}
\left[\Lambda-\frac{R_0}{4}+\frac{\alpha}{2}\,
\left(1+\frac{n\beta}{1-2\beta}\right)\,F_0^n\right]\,g_{ij}=
\frac\lambda2\,T_{ij}
+n\alpha\,F^{n-1}\left(D_{rsij}\,T^{rs}+\frac{1-\beta}2\,R_0T_{ij}\right)\,.
\end{equation}
\phantom{line}\\
We see that non-trivial solutions effectively exist. For example,
if $\lambda$ and $\phi$ satisfy the equations
\begin{equation}
\label{ExCond1}
\lambda=n\alpha\left(\beta-\frac12\right)\,R_0F_0^{n-1}\,,
\end{equation}
\begin{equation}
\label{ExCond2}
D_{rsij}\,T^{rs}+\frac14\,R_0T_{ij}=\Sigma g_{ij}\,,
\end{equation}
$\Sigma$ being a constant, the curvature can be derived from the algebraic equation\\
\phantom{line}
\begin{equation}
\frac{R_0}4-\Lambda+\alpha\,\left\{n\Sigma+\frac{R_0U_0}4\,
\left[1-(n+2)\beta\right]\right\}\,
\left[\left(\beta-\frac12\right)R_0U_0\right]^{n-1}=0\,.
\end{equation}
\phantom{line}\\
Of course, this is a solution if the Eqs.~(\ref{ExCond1})--(\ref{ExCond2}) are compatible
with the constraint (\ref{const}).
In principle, more general solutions with non-constant $\Lambda$ may exist too.

\section{Cosmological applications}

\paragraph*{} We shall now look for cosmological solutions and thus we start with a FRW metric (\ref{metric}), and
a scalar field which depends on time only. Thus, $\phi=\phi(t)$ is
completely determined by the constraint (\ref{const}) and, as a consequence,
the tensor $T_{ij}$ has only one non-vanishing component, namely 
\begin{equation}
T_{00}\equiv\dot\phi^2=2U_0\,,\label{fluidsolution}
\end{equation}
where we have used the constraint (\ref{const}).

Since all quantities depend on time only, Eq.~(\ref{PHI}) gives
\begin{equation}
\label{lat}
\lambda
-n\alpha\left[6\left((\beta-1)\dot H+(2\beta-1)H^2\right)\right]^nU_0^{n-1}
=\frac{C_0}{a^3}\,,
\end{equation}
$H$ being, as usually, the Hubble parameter and $C_0$ an arbitrary integration constant.
Moreover, due to the symmetry of the metric in field equations (\ref{Gij}), only two equations are independent.
It is clear that, by choosing $\beta=1$, one has a simplification, namely
\begin{eqnarray}
0&=& \Lambda-3H^2+\frac12\,\alpha\,(1-4n)(6U_0H^2)^n+U_0\lambda\,,
\\
0&=& \Lambda-3H^2-2\dot H+\frac12\,\alpha(1-2n)(6U_0H^2)^n
+\frac13\,\alpha n(1-2n)\dot H(6U_0)^nH^{2n-1}\,.
\end{eqnarray}
Now, in the latter equations, $\lambda$ can be eliminated by means of Eq.~(\ref{lat}),
getting in this way the generalized Friedmann equations for the pure gravitational field. We have
\begin{eqnarray}
0&=& \Lambda-3H^2+\frac12\,\alpha\,(1-2n)(6U_0H^2)^n-\frac{C_0}{a^3}\,,
\\
0&=& \Lambda-3H^2-2\dot H+\frac12\,\alpha(1-2n)(6U_0H^2)^n
+\frac13\,\alpha n(1-2n)\dot H(6U_0)^nH^{2n-1}\,.
\end{eqnarray}
One easily sees that, in order to get de Sitter solutions, one has to choose
a vanishing integration constant, that is $C_0=0$. In this way the previous
equations become equivalent and one obtains a constant Hubble parameter $H_0$, namely $H=H_0$, by solving
\begin{equation}
\frac12\,\alpha\,(2n-1)(6U_0H_0^2)^n+3H_0^2-\Lambda=0\,.\label{dSC=0}
\end{equation}
On the contrary, choosing $C_0\neq0$, one gets a second-order
differential equation in the scale factor $a(t)$. A simple way to get such equation is to make use of the well known minisuperspace approach, which we have briefly described in \S~\ref{1.4} referring to SSS solutions of $\mathcal{F}(R,G)$-gravity.\\

Recall we  are dealing with the FRW space-time (\ref{metric}) with non constant $N(t)$ function,
which describes the reparametrization invariance of the model. As a result, for $\beta$ generic, one has
\begin{equation}
F=K^{ij}\partial_i \phi \partial_j \phi =\left(R^{ij}-\frac{\beta}{2}Rg^{ij}\right)\partial_i \phi \partial_j \phi=
\frac{3 (\dot \phi)^{2}}{N^4}\left[(\dot a)^{2}a^{-2}+(\beta-1)\left(\frac{ \ddot a}{a}-\frac{\dot a \dot N}{aN}\right)\right]\,.
\end{equation}
Here, $N=N(t)$, $a=a(t)$ and $\phi=\phi(t)$ are functions of time $t$ only.
One can see the particular role played by the dimensionless parameter $\beta$. If one makes the choice $\beta=1$,
namely $K_{ij}=G_{ij}$, where $G_{ij}$ is the Einstein's tensor, the dependence on the acceleration $\ddot{a}$ and $\dot N$ drops out. In fact, due precisely to the diffeomorphism invariance of the model, $ G_{00}$ is the Hamiltoniam constraint of GR and the modified
gravitational fluid model becomes very  simple, so that one has  the following simplified (effective) minisuperspace action\\
\phantom{line}
\begin{equation}
\hat I=\frac{1}{2\kappa^2}\int_{\mathcal{M}} dt \left[ -6 a (\dot a)^2 N^{-1}-2\Lambda a^3 N -\alpha 3^n N^{(1-4n)}(\dot a)^{2n}a^{-2n+3}
(\dot \phi)^{2n}-\lambda a^3 N\left(U_0-\frac{(\dot \phi)^2}{2N^2}\right)\right]\,.
\end{equation}
\phantom{line}\\
In this case,  one has two Lagrangian multipliers $\lambda$ and $N$,
the first one implements the constraint
\begin{equation}
U_0=\frac{(\dot \phi)^2}{2N^2} \,,
\label{c1}
\end{equation}
while the second gives the Hamilonian constraint of our covariant model. After the variation,  one has to take the gauge $N=1$.
The other two Lagrangian coordinates are $\phi$ and $a$, and one has the corresponding equations of motion.
Let us continue with the equation of motion associated with $N$.
On shell, one gets
\begin{equation}
6H^2-\alpha(1-4n)(6U_0)^nH^{2n}-2\Lambda=2\lambda U_0\,.
\label{bbb}
\end{equation}
On the other hand, since the Lagrangian does not depend on $\phi$, the associated equation of motion reads
\begin{equation}
C_0=\frac{\partial \mathcal{L}}{\partial \dot \phi}\,,
\end{equation}
where $\mathcal{L}$ is the Lagrangian and $C_0$ is a constant of integration. On shell,
\begin{equation}
-2n\alpha(6U_0)^nH^{2n}+2\lambda U_0=\frac{C_0 \sqrt{2 U_0}}{a^3}\,.
\label{p}
\end{equation}
Making use of the two last equations, we arrive at
\begin{equation}
6H^2-\alpha(1-2n)(6U_0)^nH^{2n}-2\Lambda=\frac{C_0 \sqrt{2 U_0}}{a^3}\,.
\label{nol}
\end{equation}
Finally, the last equation of motion is the one associated with $a$. It reads
\begin{equation}
(6H^2-\alpha(1-2n)(6U_0)^nH^{2n}-2\Lambda)=-\left(4+\alpha\frac{2n}{3}(2n-1)(6U_0)^nH^{2n-2}\right)\dot H\,.
\label{aaa}
\end{equation}
Making use of above equations, we also have
\begin{equation}
\frac{C_0 \sqrt{2 U_0}}{a^3}=-\left(\alpha\frac{2n}{3}(2n-1)(6U_0)^nH^{2n-2}+4\right)\dot H\,.
\label{a1}
\end{equation}
Some remarks are here in order. The equations we have obtained are identical to the ones coming directly from the equations of motion. In particular, as in General Relativity, the equation of motion associated with $a$ is not an independent one, since it can be obtained by taking the derivative with respect to $t$ of the other equations and de Sitter solution, for which $\dot{H}=0$ and $H=H_0$, where $H_0$ is a constant, corresponds to the choice $C_0=0$. In this case, Eq.~(\ref{a1})
is satisfied, and we find Eq.~(\ref{dSC=0}).\\

With regard to the dS-solution of Eq.~(\ref{dSC=0}), one needs to look for positive $H_0^2$ solutions with $\alpha >0$, a necessary condition in order to have a correct non linear
graviton dispersion relation~\cite{Nojiri:2009th}. With regard to this issue, let us consider the simplest non trivial case,
namely $n=2$. One has as a solution
\begin{equation}
H_0^2=\frac{-1+ \sqrt{1+24\alpha U_0^2\Lambda}}{36\alpha U_0^2}\,.\label{dSn=2}
\end{equation}
Note that, for $\Lambda=0$, the de Sitter solution  exists only for $\alpha<0$, which would correspond to an unusual dispersion relation for the graviton.

The stability of all de Sitter solutions is not difficult to study. In fact taking the first variation of Eq.~(\ref{aaa}) around $H=H_0$, one obtains
\begin{equation}
\delta\dot{H}=-3 H_0 \delta H\,.
\end{equation}
As a consequence, all the de Sitter solutions are stable.\\

Let us investigate the case when $C_0$ is non-vanishing. In this case a de Sitter solution does not exist. Then, we may take $\Lambda=0$. First, let us study the
model with  $n=2$. In this case, with $\alpha >0$, one has the differential equation from Eq.~(\ref{a1}),
\begin{equation}
\frac{d H}{d t}=-\frac{3}{2}\frac{ H^2+18\alpha U_0^2 H^4}{1+36\alpha U_0^2 H^2}\,.
\end{equation}
Separating variables, one gets
\begin{equation}
\frac{1}{H}-6U_0\sqrt{\frac{\alpha}{2}}\mbox{arctan}\left(6U_0\sqrt{\frac{\alpha}{2}}H\right)
=\frac{3}{2}t+B_0\,,
\label{generalsolution}
\end{equation}
where $B_0$ is an integration constant. The solution is given in an implicit way only.  However, even then it is easy to show that the model is protected against future-time singularities. In fact, let us look for Big Rip solutions in which the Hubble parameter is as in Eq.~(\ref{Hsingular}), namely $H(t)=h_0/(t_{0}-t)^{\beta}$.
When $\beta>0$, $1/H$ tends to zero and the arctangent tends to a constant and the sign of the first leading term on the left hand side of Eq.~(\ref{generalsolution}) is inconsistent with the sign of the right hand side. Moreover, when $ \beta<0$, the left side of Eq.~(\ref{generalsolution}) diverges. As a consequence, no singular future solution can exist. 

In the general case, we can investigate the possible presence of acceleration. In fact, with  $\Lambda=0$, one has
\begin{equation}
\frac{\dot H}{H^2}+1=\frac{1}{(2+\alpha\frac{n}{3}(2n-1)(6U_0)^nH^{2n-2})}\left[ -1+\alpha(2n-1)(2n-3)6^{n-1}U_0^nH^{2n-2}\right]\,.
\end{equation}
As a result, one may have acceleration as long as
\begin{equation}
H^{2n-2}>\frac{1}{\alpha(2n-1)(2n-3)6^{n-1}U_0^n }\,.
\end{equation}
In particular, for $n=2$ this condition becomes
\begin{equation}
H^2>\frac{1}{18 \alpha U_0^2}\,.
\end{equation}
Coming back to the general model, it turns out that for $\beta \neq 1$ calculations are much more involved, since $ \ddot a$ is present in the Lagrangian, and the model
becomes a higher-derivative system in the sense of Ostrogradsky. However, we may carry out a direct calculation, which shows that a dS-solution is not possible there.

\section{Entropy calculation}

\paragraph*{} It is of interest to evaluate the black hole entropy associated with the different solutions we have discussed. Since we are dealing with a covariant theory, we can make use of the Noether charge Wald methods, as in Chapter {\bf 3}. A direct evaluation of formula (\ref{Wald}) yields
(cf.~with Ref.~\cite{cogno}),\\
\phantom{line}
\begin{equation}
S_W=-2\pi\int_{\mathcal{M}}\frac{\partial \mathcal{L}}{\partial R_{ijrs}}\Big\vert_H \varepsilon_{ij}\varepsilon_{rs}d\Sigma=
-\frac{1}{8 G_N}\int_{\mathcal{M}}\,\left[
\varepsilon_{ij}\varepsilon^{ij}
-n\alpha F^{n-1} \frac{\partial F}{\partial R_{ijrs}}\varepsilon_{ij}\varepsilon_{rs}\right]\Big\vert_H d\Sigma\,.
\end{equation}
\phantom{line}\\
The first term is the GR contribution, while the other one is due to the modification of GR in the considered model. However, in the case of the Schwarzschild solution one has $F=0$ and Eq.~(\ref{AreaLaw}) is found. As a consequence, in this modified gravity model, the entropy of  the Schwarzschild black hole satisfies the usual Area Law, $S_W=\mathcal{A}_H/(4G_N)$.

Let us now consider the dS-solution we have found for $\beta=1$ and $n=2$ in Eq.~(\ref{dSn=2}). The simplest way to perform the calculation is to make use of the static gauge, namely
\begin{equation}
ds^2=-V(\rho) dt^2_s+\frac{d \rho^2}{V(\rho)}+\rho^2 d\Omega^2\,,
\end{equation}
being  $V(\rho)=1-H_0^2\rho^2 $ and $d\Omega^2=(d\theta^2+\sin^2\theta d\phi^2)$. This static form of the dS-metric can be obtained from the coordinate transformation of FRW metric
\begin{equation}
\rho=r e^{H_0 t}\,,\quad t_s=t-\frac{1}{2 H_0}\ln V(\rho)\,.
\end{equation}
The solution of Eq.~(\ref{fluidsolution}) corresponding to the scalar fluid reads
\begin{equation}
\phi(t_s,\rho)=\sqrt{2 U_0}\left[t_s+\frac{1}{2 H_0}\ln V(\rho)\right]\,.
\end{equation}
The relevant scalar quantity to be evaluated is
\begin{equation}
\frac{\partial F}{\partial R_{ijrs}}\varepsilon_{ij}\varepsilon_{rs}=-
2U_0+\varepsilon_{ij}\varepsilon_{rs}\partial^i \phi \partial^r \phi g^{js}\,,
\end{equation}
In general, the binormal tensor is given by  $\varepsilon_{ij}=v_i u_j-v_j u_i$ and, in a static gauge, it is easy to show that one may choose  $v_i=(\sqrt{V},0,0,0)$ and $u_i=\left(0,\frac{1}{\sqrt{V}},0,0\right)$. A direct calculation yields
\begin{equation}
\varepsilon_{ij}\varepsilon_{rs}\partial^i \phi \partial^r \phi g^{js}=2 U_0 \,.
\end{equation}
Thus, the Area Law is also satisfied for the de Sitter solution we have found, confirming that, for $\beta=1$, we are dealing with a minimal modification of General Relativity.


\chapter{An introduction to $F(T)$-gravity: dark matter in teleparallel gravity}

\paragraph*{} 
Recently, a new type of modified gravity theories, namely the $F(T)$-theories of gravity, has been
proposed. 
These models are based on the ``teleparallel'' equivalent of General
Relativity (TEGR) \cite{F(T)1,F(T)2}, where, instead of using
the curvature defined via the Levi-Civita connection \cite{LeviCivita1}-\cite{LeviCivita4}, one uses the
Weitzenb$\ddot{o}$ck connection, which has no curvature $R$ but only
torsion $T$, in order to explain inflation and the late time accelerated expansion of the universe.
The  field equations of $F(T)$-gravity are  2nd order differential equations
and this fact makes these theories simpler than the ones where
modification is via curvature invariants (see Refs. \cite{Myrzakulovonly1}-\cite{altriF(T)} for recent developments).
The proprieties of $F(T)$-gravity 
have been diversely explored, for example the local Lorentz invariance~\cite{Local-Lorentz-invarianceT1}-\cite{Local-Lorentz-invarianceT4}, non-trivial conformal frames and thermodynamics~\cite{Thermodynamicsf(T)1}-\cite{Thermodynamicsf(T)4}.
In this  Chapter we will furnish a brief review of $F(T)$-gravity, we will look for the field equations of the theory in FRW universe and we will show a nice application of teleparallel gravity in describing the galaxies 
and the dark matter fenomenology.

\section{General aspects of $F(T)$ gravity}

\paragraph{}The action of $F(T)$-gravity theory reads
\begin {equation}
I=\int_{\mathcal M} d^4x\left(e\right)\,\left[\frac{F(T)}{2\kappa^{2}}+\mathcal{L}^{\mathrm{(matter)}}\right]\,,
\end{equation}
where, as usually, $\mathcal{L}^{\mathrm{(matter)}}$ is the matter Lagrangian. 
Moreover, $T$ is the torsion scalar (see below)
and $e$ (we will find, $e=\sqrt{-g}$, where $g$ is the usual determinant of metric tensor) is defined as
$e=\det{[e^{i}_{\mu}]}$, such that $e^{i}_{\mu}=e^{i}_{\mu}(x^\mu)$ are the components of the vierbein vector field $e_{A}(x^\mu)$ in the coordinate basis $e_{A}\equiv e^{\mu}_{A}(x^\mu)\partial_{\mu}$. Here, the index $i,\mu$ and therefore the index `A' run over $0, ..., 3$. Note that in the teleparallel gravity, the dynamical variable is given by the vierbein field $e_{A}(x^{\mu})$, since the metric element reads 
\begin{equation}
ds^2\equiv g_{\mu\nu}dx^{\mu}dx^{\nu}=\eta_{ab}\theta^{a}\theta^{b}\label{ele},
\end{equation}
where
\begin{equation}
\theta^{a}=e^{a}_{\;\;\mu}dx^{\mu}\;,\quad dx^{\mu}=e_{a}^{\;\;\mu}\theta^{a}\label{the},
\end{equation}
$g_{\mu\nu}$ being the metric of the space-time, $\eta_{ab}$ the Minkowski's metric, $\theta^{a}$ the tetrads and $e^{a}_{\;\;\mu}$ and their inverses $e_{a}^{\;\;\mu}$ the tetrads basis. The tetrad basis satisfy the folllowing relations:
 \begin{equation}
 e^{a}_{\;\;\mu}e_{a}^{\;\;\nu}=\delta^{\nu}_{\mu}, \quad e^{a}_{\;\;\mu}e_{b}^{\;\;\mu}=\delta^{a}_{b}.
\end{equation} 
A a consequence, the root of the metric determinant which effectively appears in the Lagrangian  is given by
 \begin{equation}
 \sqrt{-g}=\det[e^{a}_{\;\;\mu}]=\left(e\right).
\end{equation}  The standard Weitzenbok's connection reads
\begin{equation}
\Gamma^{\alpha}_{\mu\nu}=e_{i}^{\;\;\alpha}\partial_{\nu}e^{i}_{\;\;\mu}=-e^{i}_{\;\;\mu}\partial_{\nu}e_{i}^{\;\;\alpha}\label{co}.
\end{equation}
As a result, the covariant derivative, denoted by $D_\mu$, satisfies the equation
\begin{equation}
D_\mu e^i_\nu:=\partial_{\mu}e^{i}_{\nu}-\Gamma^{\lambda}_{\nu\mu}e^{i}_{\lambda}=0.
\end{equation}
Thus,  the components of the torsion and the contorsion are given by\\
\phantom{line}
\begin{eqnarray}
\left\{\begin{array}{l}
T^{\alpha}_{\;\;\mu\nu}=\Gamma^{\alpha}_{\nu\mu}-\Gamma^{\alpha}_{\mu\nu}=e_{i}^{\;\;\alpha}\left(\partial_{\mu} e^{i}_{\;\;\nu}-\partial_{\nu} e^{i}_{\;\;\mu}\right)\label{tor}\;,\\ \\
K^{\mu\nu}_{\;\;\;\;\alpha}=-\frac{1}{2}\left(T^{\mu\nu}_{\;\;\;\;\alpha}-T^{\nu\mu}_{\;\;\;\;\alpha}-T_{\alpha}^{\;\;\mu\nu}\right)\label{cont}\; .
\end{array}\right.
\end{eqnarray}
\phantom{line}\\
Now, if we define the following tensor from the components of torsion and contorsion 
\begin{equation}
S_{\alpha}^{\;\;\mu\nu}=\frac{1}{2}\left( K_{\;\;\;\;\alpha}^{\mu\nu}+\delta^{\mu}_{\alpha}T^{\beta\nu}_{\;\;\;\;\beta}-\delta^{\nu}_{\alpha}T^{\beta\mu}_{\;\;\;\;\beta}\right)\label{s}\;,
\end{equation}
we finally  obtain the torsion scalar $T$,
\begin{equation}
T=T^{\alpha}_{\;\;\mu\nu}S_{\alpha}^{\;\;\mu\nu}\label{t1}\,.
\end{equation}
In order to derive the EOMs, we need the following quantities,\\
\phantom{line}
\begin{equation}
\frac{\partial \mathcal L}{\partial e^{a}_{\;\;\mu}}=F(T)\,e\,e_{a}^{\;\;\mu}+4e\,F_{T}(T)e_{a}^{\;\;\alpha}T^{\sigma}_{\;\;\nu\alpha}S_{\sigma}^{\;\;\mu\nu}+\frac{\partial{\mathcal L^{\mathrm{(matter)}}}}{\partial e^{a}_{\;\;\mu}},\label{1}
\end{equation}
\phantom{line}\\
and\\
\phantom{line}
\begin{equation}
\partial_{\alpha}\left[\frac{\partial {\mathcal L}}{\partial (\partial_{\alpha}e^{a}_{\;\;\mu})}\right]=-4F_{T}(T)\partial_{\alpha}\left(e\,e_{a}^{\;\;\sigma}S_{\sigma}^{\;\;\mu\nu}\right)-4ee_{a}^{\;\;\sigma}S_{\sigma}^{\;\;\mu\alpha}\partial_{\alpha}T\,F_{TT}(T)+\partial_{\alpha}\left[\frac{\partial\mathcal L^{\mathrm{(matter)}}}{\partial (\partial_{\alpha}e^{a}_{\;\;\mu})}\right]\label{2},
\end{equation}
\phantom{line}\\
where $\mathcal L$ is the total Lagrangian density of the theory and we have putted\\
\phantom{line}
\begin{equation}
F_T(T)=\frac{d F(T)}{d T}\,,\quad F_{TT}(T)=\frac{d^2 F(T)}{d T^2}\,.\label{conventionsT}
\end{equation}
\phantom{line}\\
In what follows, where is not necessary, we will drop out the argument of such functions.
By making use of the  Euler-Lagrange equations with \eqref{1}--\eqref{2}, we get
\begin{eqnarray}
\hspace{-0.5cm}S_{\beta}^{\;\;\mu\alpha}\partial_{\alpha}T\,F_{TT}(T)+\left[e^{-1}e^{a}_{\;\;\beta}\partial_{\alpha}\left(ee_{a}^{\;\;\sigma}S_{\sigma}^{\;\;\mu\alpha}\right)+T^{\sigma}_{\;\;\nu\beta}S_{\sigma}^{\;\;\mu\nu}\right]F_{T}(T)+\frac{1}{4}\delta^{\mu}_{\beta}F(T)=\frac{\kappa^2}{2} T^{\mathrm{(matter)}\mu}_{\beta}\label{em}\;,
\end{eqnarray}
where\\
\phantom{line}
\begin{eqnarray}
T^{\mathrm{(matter)}\mu}_{\beta}=-\frac{e^{-1}e^{a}_{\;\;\beta}}{2\kappa^2}\left\{ \frac{\partial \mathcal{L}^{\mathrm{(matter)}}}{\partial e^{a}_{\;\;\mu}}-\partial_{\alpha}\left[\frac{\partial \mathcal{L}^{\mathrm{(matter)}}}{\partial (\partial_{\alpha}e^{a}_{\;\;\mu})}\right]\right\}\,,
\end{eqnarray}
\phantom{line}\\
turns out to be the matter stress energy tensor. If we consider the case of teleparallel gravity (TG), namely  $F(T)=T$, then the trace of gravitional field equations reduce to
\begin{equation}
T-2e^{-1}\partial_{\sigma}(eT_{\rho}{}^{\rho\sigma})=\kappa^2 T^{\mathrm{(matter)}\mu}_{\mu},\label{TGGR}
\end{equation}
which shows an equivalence between GR and TG~\cite{equivalenceTGR}, due to the fact that
\begin{equation}
-R=T-2e^{-1}\partial_{\sigma}(eT_{\rho}{}^{\rho\sigma})\,,\label{TGGR2}
\end{equation}
and we see that (\ref{TGGR}) is the trace of Einstein's equations. 

\section{The FRW universe in generalized teleparallel gravity}

\paragraph{}In generalized teleparallel gravity, the field equations for homogeneous and isotropic FRW metric (\ref{metric}) are derived as
\begin{eqnarray}
    -2TF_{T}+F&=&2\kappa^2 \rho_\mathrm{m},\label{fieldT1} \\
    -8\dot{H}TF_{TT}+(2T-4\dot{H})F_{T}-F&=&2\kappa^2p_{\mathrm{m}},\label{fieldT2} 
\end{eqnarray}
where $\rho_{\mathrm{m}}$ and $p_{\mathrm{m}}$ are, as usually, the energy density and pressure of matter which satisfy the continuity equation (\ref{conservationlaw}.)
Moreover, the torsion scalar reads
\begin{equation}
T=-6H^2\,.\label{eqT}
\end{equation}
The field equations (\ref{fieldT1})--(\ref{fieldT2}) with conservation law (\ref{conservationlaw}) are equivalent to
\begin{eqnarray}
\hat{M}_1F&=&2\kappa^2 \rho_m\,, \label{M1}\\
\hat{M}_2F&=&-\hat{M}_{3}\hat{M}_{1}F=2\kappa^2p_m\,,\label{M2} \\
\hat{M}_3\rho_{m}&=&-p_{m}\,,\label{M3}
\end{eqnarray}
where\\
\phantom{line}
\begin{eqnarray}
\left\{\begin{array}{l}
\hat{M}_1=    -2T\partial_{T}+1\,, \\ \\
\hat{M}_2=    -8\dot{H}T\partial^2_{TT}+(2T-4\dot{H})\partial_{T}-1=(4\dot{H}\partial_{T}-1)\hat{M}_{1}=
-(\frac{1}{3H}\partial_{t}+1)\hat{M}_{1}
=-\hat{M}_{3}\hat{M}_{1}\,,\\ \\
\hat{M}_3=    \frac{1}{3H}\partial_{t}+1\,.
\end{array}\right.
\end{eqnarray}
By using these equations we may construct high hierarchy of $F(T)$-gravity. For the vacuum case ($\rho_\mathrm{m}=p_\mathrm{m}=0$), such hierarchy reads
\begin{equation}\left(
\hat{M}_1\right)^nF_n=0\,,\label{M1bis}
\end{equation}
where $F_n$ indicates the solution of the corresponding equation. 
For $n=1$, we have  $F_1=F$ and we recover Eq. (\ref{M1}). Then, we have
\begin{eqnarray}
-2TF_{1T}+F_1&=&0\,,\quad n=1\,; \nonumber\\
4T^2F_{2TT}+F_2&=&0\,,\quad n=2\,; \nonumber\\
-8T^3F_{3TTT}-12T^2F_{3TT}-2TF_{3T}+F_3&=&0\,,\quad n=3\,;\nonumber\\
&&...
\nonumber
\end{eqnarray}
and so on. Of course, the solutions of Eq. (\ref{M1}) are also solution of (\ref{M1bis}).

\subsection{$F(T)$-models in FRW universe}

\paragraph{}Some explicit models of $F(T)$-gravity appeared in the recent literature \cite{Myrzakulovonly1}--\cite{altriF(T)}. Here, we would like to present some applications of the formalism of modified teleparallel gravity applied to FRW universe \cite{Myrzakulovonly2}-\cite{Myrzakulovonly5}.

\subsubsection*{Example 1: M$_{13}$-model}

\paragraph{}The Lagrangian of M$_{13}$-model reads\\
\phantom{line}
\begin {equation}
F(T)=\sum_{j=-m}^{n}\nu_j T^{j}=\nu_{-m} T^{-m}+ ... +\nu_{-1} T^{-1}+\nu_0+\nu_{1}T + ... + \nu_nT^{n}.
\end{equation}
\phantom{line}\\
Here, $\nu_{j}$ are generic constants.
As an example, let us consider the particular case $m=n=1$. Thus, we have\\
\phantom{line}
\begin {equation}
F=\nu_{-1}T^{-1}+\nu_{0}+\nu_{1}T, \quad
F_{T}=-\nu_{-1}T^{-2}+\nu_{1}, \quad F_{TT}=2\nu_{-1}T^{-3}.
\end{equation}
\phantom{line}\\
By substituting these expressions into (\ref{fieldT1})--(\ref{fieldT2}) we obtain
\begin{align}
\frac{3}{\kappa^2}H^2=\rho_{\mathrm{MGT}}+ \rho_m\,,
\quad
-\frac{1}{\kappa^2}(2\dot{H}+3H^2)=p_{\mathrm{MGT}}+p_m,\label{effectiveeqsT}
\end{align}
where $\rho_{\mathrm{eff}}$ and $p_{\mathrm{eff}}$ are the effective energy density and pressure given by the modification via $F(T)$-gravity, namely, for the specific case,
\begin{align}
\rho_{\mathrm{MGT}}=\frac{1}{\kappa^2}[3H^2-1.5\nu_{-1}T^{-1}+0.5\nu_1 T-0.5\nu_0]\,,\nonumber
\end{align}
\begin{align}
p_{\mathrm{MGT}}=\frac{1}{\kappa^2}[6\nu_{-1}\dot{H}T^{-2}+1.5\nu_{-1} T^{-1}-0.5\nu_1T+0.5\nu_0+2(\nu_1-1)\dot{H}-3H^2]\,.
\end{align}
The scalar torsion $T$ is given by (\ref{eqT}). 

\subsubsection*{Example 2: M$_{21}$-model}
 \paragraph{}Our next example is the M$_{21}$ - model\,,
  \begin{align}
F(T)=T+\alpha T^{\delta}\ln[T]\,,
\end{align}
where $\alpha$ and $\delta$ are constant paramters. We get
  \begin {equation}
F_{T}=1+\alpha\,\delta T^{\delta-1}\ln{T}+\alpha T^{\delta-1}, \quad
F_{TT}=\alpha\,\delta(\delta-1) T^{\delta-2}\ln{T}+\alpha(2\delta-1) T^{\delta-2}.
\end{equation}
As a consequence, Eqs. (\ref{fieldT1})--(\ref{fieldT2}) take the form (\ref{effectiveeqsT}) with
\begin{align}
\rho_{\mathrm{MGT}}=\frac{1}{2\kappa^2}[2\alpha T^{\delta}+\alpha(2\delta-1) T^{\delta}\ln{T}]\,,\nonumber
\end{align}
\begin{align}
p_{\mathrm{MGT}}=-\frac{1}{2\kappa^2}\alpha T^{\delta-1}[(2\delta-1)(T-4\delta\dot{H})\ln{T}+2T-4(4\delta-1)\dot{H}]\,.
\end{align}
The special case $\delta=1/2$ deserves a separate consideration. In this case the above expressions assume a simple form,
\begin{align}
\rho_{\mathrm{MGT}}=\frac{1}{\kappa^2}\alpha\, T^{1/\delta}, \quad p_{\mathrm{MGT}}=-\frac{1}{\kappa^2}\alpha\, T^{-1/\delta}(T-2\dot{H}).
\end{align}

\subsubsection*{Example 3: M$_{22}$-model}
\paragraph{}Now we consider  the M$_{22}$-model,
  \begin{align}
F(T)=T+f(y), \quad y=\tanh[T].
\end{align}
Thus,
 \begin {equation}
F_{T}=1+f_y(1-y^2),\quad
F_{TT}=f_{yy}(1-y^2)^2-2y(1-y^2)f_y\,,
\end{equation}
where the index `$y$' indicates the derivative with respect to $y$. 
Equations (\ref{fieldT1})--(\ref{fieldT2}) take the form (\ref{effectiveeqsT}) with
\begin{align}
\rho_{\mathrm{MGT}}=\frac{1}{2\kappa^2}\left[2(1-y^2)Tf_y-f\right]\,,
\end{align}
\begin{align}
p_{\mathrm{MGT}}=\frac{1}{2\kappa^2}\left[8(1-y^2)^2T\dot{H}f_{yy}-(16y\dot{H}T+2T-4\dot{H})(1-y^2)f_{y}+f\right]\,.
\end{align}

\subsection*{Example 4: M$_{25}$ - model}
\paragraph{}To consclude, we will consider the M$_{25}$-model,
\begin{equation}
F(T)=\sum_{j=-m}^{n}\nu_{j}\xi^{j}\,,\quad\xi=\ln{T}\,,
\end{equation}
where $\nu_j$ are generic constants. We take the case $m=n=1$, namely
  \begin {equation}
F=\nu_{-1}\xi^{-1}+\nu_0+\nu_1\xi\,,\quad
F_{T}=(-\nu_{-1}\xi^{-2}+\nu_1)e^{-\xi},\quad
F_{TT}=(2\nu_{-1}\xi^{-3}+\nu_{-1}\xi^{-2}-\nu_1)e^{-2\xi}.
\end{equation}
In this case, the EOMs (\ref{fieldT1})--(\ref{fieldT2}) lead to
 \begin{equation}
    2\nu_{-1}\xi^{-2}+\nu_{-1}\xi^{-1}+\nu_0-2\nu_1+\nu_1\xi=2\kappa^2 \rho_m,
\end{equation}
\begin{equation}
    -4\dot{H}(4\nu_{-1}\xi^{-3}+\nu_{-1}\xi^{-2}-\nu_1)e^{-\xi}-2\nu_{-1}\xi^{-2}-\nu_{-1}\xi^{-1}+2\nu_1-\nu_0-\nu_1\xi=2\kappa^2p_m.
\end{equation}

\section{Dark matter in $F(T)$-teleparallel gravity}

\paragraph{}As it is well known, dark (non-luminous and non-absorbing) matter is an
old idea even before the dark energy problem.
The most accepted observational evidence
for the existence of dark matter comes from the astrophysical
measurements of several  galactic rotation curves. From the point of view of the
classical mechanics, we expect that the rotational velocity
$v_\varphi$ of any astrophysical  object moving in  a (quasi)
stable Newtonian circular orbit with radius $r$ must be in the form
\begin{equation}
v_{\varphi}(r) \propto \sqrt{
  \frac{M(r)}{r}}\,,
\end{equation}
where $M(r)$ is identified with the (effective) mass profile
thoroughly located inside the orbit and depending on the radius. 
However, for many spiral and elliptical galaxies,
the velocity $v_{\varphi}$ is approximately constant 
when the radius is very large, and this fact suggests the presence of an undetected form of exotic matter, the dark matter, which actually composes about the $20\%$ of our universe. 

In order to
solve the DM problem \cite{DM} several proposals have been introduced. 
In this Chapter, we will focus our attention on $f(T)$-gravity and
we will show that in the context of this non-Riemannian
extension of the General Relativity, it is possible to explain
the rotation curves of the galaxies without introducing dark matter.
In this Section, we will furnish a very simple example of (teleparallel) $F(T)$-model able to describe the galaxies and the dark matter \cite{davoodDMmodel}.

\subsection{Geometry of the galaxies}

\paragraph{} A typical spiral galaxy contains two
forms of matter: luminous matter from stars and stellar
clusters; dark matter
in the galactic halo which
encapsulates the galaxy disk. 
Since the precise form of
distribution of dark matter in the halos is unknown, we assume
that the spatial geometry of galactic halo is spherically symmetric.
Moreover, the dark matter halo is isotropic: the spherical DM halo
expands (hypothetically) only radially, without having tangential or
orthogonal motions relative to the radial one. 
Basing on the above assumptions, in the (quasi)-static case,
the metric of a static spherically symmetric space-time can be described, without loss of generality, as
\begin{equation}
ds^{2}=\mathrm{e}^{a(r)}dt^{2}-\mathrm{e}^{b(r)}dr^{2}-r^{2}\left(d\theta^{2}+\sin^{2}\theta d\phi^{2}\right)\label{g}\;,
\end{equation}
where $a(r)$ and $b(r)$ are functions of the radius $r$.
Four our attempt, this metric results to be in a more useful form than the one in (\ref{SSS}).
The  torsion scalar reads
\begin{eqnarray}
T(r) &=& \frac{2\mathrm{e}^{-b(r)}}{r}\left(\frac{d a(r)}{d r}+\frac{1}{r}\right)\label{te}\;.
\end{eqnarray}
The components of field equations for an
anisotropic fluid are \cite{h}:\\
\phantom{line}
\begin{eqnarray}
\frac{\kappa^2}{2}\rho &=& \frac{F}{4}-\left[T-\frac{1}{r^2}-\frac{\mathrm{e}^{-b(r)}}{r}\left(\frac{d a(r)}{d r}+\frac{d b(r)}{d r}\right)\right]\frac{F_T}{2}\,,\label{dens} \\ \nonumber\\
\frac{\kappa^2}{2} p_{\mathrm{rad}} &=& \left(T-\frac{1}{r^2}\right)\frac{F_T}{2}-\frac{F}{4}\label{presr}\,, \\ \nonumber \\
\frac{\kappa^2}{2} p_{\mathrm{tg}} &=& \frac{F_T}{2}\left\{\frac{T}{2}+\mathrm{e}^{-b(r)}\left[\frac{1}{2}\frac{d^2 a(r)}{d r^2}+\left(\frac{1}{4}\frac{d a(r)}{d r}+\frac{1}{2r}\right) \left(\frac{d a(r)}{d r}-\frac{d b(r)}{d r}\right)\right]\right\}-\frac{F}{4}\label{prest}\,,\\ \nonumber \\
0 &=&\left(\frac{\cot\theta}{2r^2}\right)\frac{d T}{d r}F_{TT}\,.\label{impos}
\end{eqnarray}
\phantom{line}\\
In the above expressions, $T=T(r)$ is given by (\ref{te}), $p_{\mathrm{rad}}$ and $p_{\mathrm{tg}}$  are the radial and tangential pressures, respectively, and $\rho$ is the density profile. This last quantity is very important in astrophysical predictions. 
We focus on
the following possible form of the model, which arises  from Eq.~(\ref{impos}), namely
\begin{eqnarray}
F_{TT}=0\,,\quad F(T)=a_0+b_0 T\;.\label{case2T}
\end{eqnarray}
Here, $a_0$ and $b_0$ are constant parameters. In the next Subsection, by starting from this Ansatz, we will solve the 
equations (\ref{dens}), (\ref{presr}) and (\ref{prest}) for the
metric function $a(r)$ (and $b(r)$). 

\subsection{Toy model of teleparallel gravity}

\paragraph{} By assuming $a(r)=-b(r)$
and by imposing the isotropicity for the pressure
components, namely $p_\mathrm{rad}=p_\mathrm{tg}$, 
the quasi global solution for the metric (\ref{g}) when the torsion is constant (and Eq.~(\ref{impos}) automatically is satisfied) is the Schwarzschild-AdS/dS solution 
presented in Ref. \cite{h}.
Obviously, this metric cannot generate the
rotation curve of the spiral galaxies. Indeed, we expect
that the DM effects come from a non constant torsion, $T=T(r)\neq 0$. 
In order to check other solutions, we assume the isotropic Ansatz for the matter distribution,
\begin{equation}
p_\mathrm{rad}=p_\mathrm{tg}\,,\label{p=p}
\end{equation}
and we put $F(T)=T$, which corresponds to (\ref{case2T}) with
$a_0=0$ and $b_0=1$. In fact, we are considering teleparallel gravity, which is related to GR by some equivalence (see Eqs. (\ref{TGGR})--(\ref{TGGR2})): however, we must not forget that we are dealing 
with a non-Riemannian manifold different to the usual Einstein's one, and in our attempt the effects of torsion 
may be used to explain the phenomenology of DM.

Moreover, we introduce the following form for $b(r)$,
\begin{equation}
b(r)=\mathrm{c_0}\label{bT}\,,
\end{equation}
where $c_0$ is a generic constant.
This choice is motivated by physical reasons which
can be found in the corresponding renormalizable action for the theory and are well explained in Ref. \cite{davoodDMmodel} and references therein. Thus, due to the assumption (\ref{p=p}), by equaling the right sides of Eq. (\ref{presr}) and Eq. (\ref{prest}), we obtain a differential equation for $a(r)$, which finally reads
\begin{equation}
a(r)=\log\left[ \frac{r^{3-\sqrt{13-4\mathrm{e}^{c_0}}}}{16(-13+4\mathrm{e}^{c_0})^2}\left(c_1 r^{\frac{\sqrt{13-4\mathrm{e}^{c}}}{2}}
-c_2\right)^4\right]\,,\label{ea}
\end{equation}
where $c_{1,2}$ are integration constants of the solution. As a consequence, from Eq. (\ref{dens}), we get the density profile of the galaxy as 
\begin{eqnarray}
\rho(r)=\frac{1}{2\kappa^2}\Big[c_{2}+\frac{2c_1(1-e^{-c_0})}{r^2}\Big]\,.\label{ourmodel}
\end{eqnarray}
In order to set the values of $c_{0,1,2}$ 
we may use the large set of cosmological data arisen from the local tests of
$F(T)$-gravity and based on the cosmographic description
\cite{cosmography}. Here, we avoid to enter in the details of the fitting of the parameters of our toy model for teleparlallel gravity and we limit to 
plot the results in the figures. 

\begin{figure}[!h]
\includegraphics[scale=0.4]{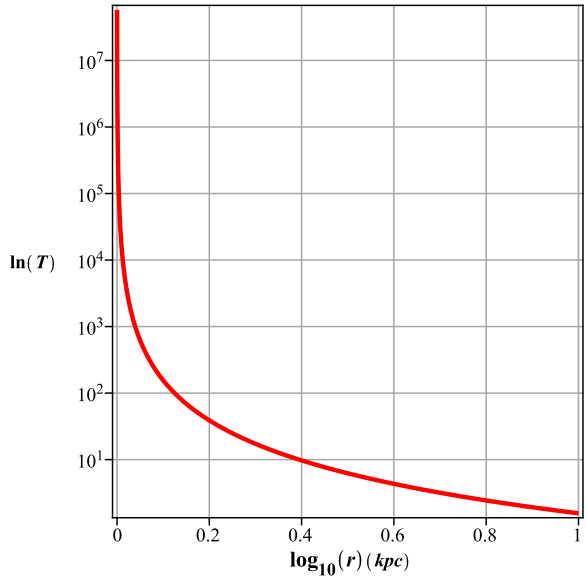}
\caption{Plot of torsion scalar $T$ as a function of $r$ in toy model for teleparallel gravity. \label{FigTorsion}}
\end{figure}

The standard plot of torsion $T(r)$ for the toy model under invastigation is depicted in Fig. \ref{FigTorsion}. Its form is derived from  Eq. (\ref{te}) and it depends on the radial coordinate.

Let us return now to the dark matter problem. 
The expression for  rotation curves of galaxies is given by \cite{v1,v2}
\begin{equation}
v_\varphi=\sqrt{\frac{r}{2}\frac{d a(r)}{d r}}\,,\label{v}
\end{equation}
as in the Einstein gravity. 

\begin{figure}[h!]
\includegraphics[scale=0.4]{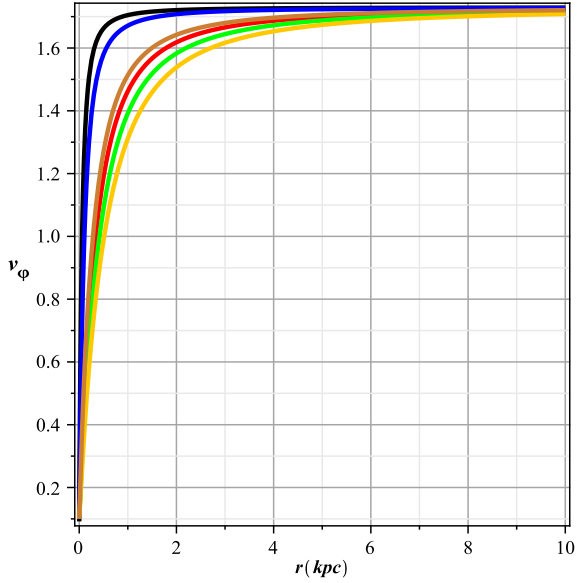}
\caption{Rotation curves for large spiral galaxy in toy model for teleparallel gravity. The units of the vertical axis must be multiplied by a factor of $\times 10 \frac{Km}{s}$. \label{rotationgal}}
\end{figure}

By substituting (\ref{ea}) in (\ref{v}), we can plot
the rotation (tangential) velocity of galaxy 
for our toy model (we assume different viable parameters). 
Fig. \ref{rotationgal} resembles the rotation curves for large
(spiral) galaxies \cite{rotv1,rotv2,rotv3}. The velocity tends to a constant value 
for $r\approx 3 \mathrm{kpc}$, in a reasonable agreement with the
data. 
It is important to remark that the velocity profile shown in
this figure is constructed by starting from phenomenological toy models, and it is interesting to see how the theory
predicts the asymptotic constant bahaviour of the velocity without invoking the dark matter.

As the last point, we mention that from observational  data we know that there exists a core with
(roughly) constant mass density inside the galaxy. Many models
have been proposed to reconstruct this mass profile density. These
models are used in the numerical simulations, for example
\begin{eqnarray}
\rho_{\mathrm{NFW}}(r)&=&\frac{\rho_\mathrm{i}}{\left(\frac{r}{r_\mathrm{s}}\right)\left(1+\frac{r}{r_\mathrm{s}}\right)^2}\label{nfw}\,,\quad \mathrm{(NFW)}\,,\label{MM11}\\
\rho_{\mathrm{\Lambda CDM}}(r)&=&\frac{\rho_0}{1+\left(\frac{r}{r_\mathrm{c}}\right)^2}\,,\,\quad\quad \mathrm{(\Lambda CDM)}\,.\label{MM22}\label{model2}
\end{eqnarray}
Here, $\rho_\mathrm{i}$ represents the density of Universe at the collapse
time, $ \rho_0$ is the central density of the halo, $r_\mathrm{s}$ is a
characteristic radius for the halo and $r_\mathrm{c}$ is the radius of the
core. 
\begin{figure}[h!]
\includegraphics[scale=0.4]{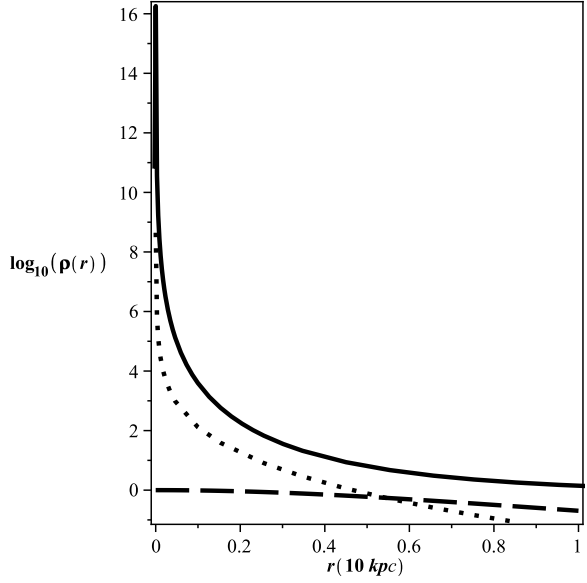}
\caption{The astrophysical halo density profiles of NFW (dot line) and $\Lambda$CDM Model 
(dash line) compared with toy model for teleparallel gravity (solid line).\label{comparison}}
\end{figure}

In Fig. (\ref{comparison}), we compare the density profile coming from our toy model 
with the ones of the models in (\ref{MM11})--(\ref{MM22}), whose fittings can be derived from astrophisical data as in Ref. \cite{rho1,rho2,rho3}. 
We can appreciate the fact that the density profile of our toy model 
is very close to NFW prediction and for large radius is also comparable with the pseudo-isothermal sphere approximation introduced by $\Lambda$CDM Model.


\fancyhead[EC, OC]{Appendices}

\setcounter{equation}{0}

\section*{Appendix A}
\addcontentsline{toc}{chapter}{Appendix A: The Tunneling method}

\section*{The Tunneling method}

\paragraph*{} In this Appendix, we present  a short review of the tunneling method in its Hamilton-Jacobi
variant. The method is based on the computation of the classical action $I$ along a trajectory starting slightly behind the trapping horizon but ending in the bulk of a dynamical black hole, and the associated WKB approximation ($c=1$, $\hbar=6.582\,\mathrm{x}\,10^{-16}\mathrm{eVsec}$),
\begin{equation*}
\mbox{Amplitude} \propto \mathrm{e}^{i \frac{I}{\hbar}}\, .
\end{equation*} 
The related semi-classical emission rate $\Gamma$ reads 
\begin{equation*}
\Gamma \propto |\mbox{Amplitude}|^2 \propto 
\mathrm{e}^{-2\frac{\Im\,I}{\hbar}} \,. 
\end{equation*}
The imaginary part of the classical action is due to deformation of the integration path according to the Feynman 
prescription, in order to avoid the divergence present on the horizon. As a result, one asymptotically gets a 
Boltzmann factor $\beta$, and an energy $\omega_K$ appears, namely
\begin{equation*}
\Gamma \propto \mathrm{e}^{- \frac{\beta}{\hbar} \omega_K}\,.
\end{equation*}
Thus, the Killing/Hawking temperature $T_K$ is identified as 
\begin{equation*}
T_K=\frac{1}{\beta}\,.
\end{equation*}
To evaluate the action $I$, let us start with a generic 
static, spherically symmetric  solution 
in $D$-dimension, written in Eddington-Finkelstein 
gauge, which, as it is well known,  is regular gauge on the  horizon
\begin{eqnarray*}
ds^2=\gamma_{ij}(x^i)dx^idx^j+r^2d\Omega^2_{D-2}=-B(r)e^{2\alpha(r)}dv^2+2e^{\alpha(r)} dr\,dv+r^2d\Omega^2_{D-2}\,.
\end{eqnarray*}
Here, $x^i=(v,r)$, where $v$ is the advanced time. Since we are dealing with 
static, spherically symmetric  solution 
space-times, one may restrict to  radial trajectories, and only the two-dimensional normal metric is relevant, and the Hamilton-Jacobi equation for a (massless) particle reads,
 \begin{eqnarray*}
\gamma^{ij}\partial_i I \partial_j I=+2e^{\alpha(r)}\partial_v I \partial_r I+e^{2\alpha(r)}B(r)(\partial_r I)^2 =0\,.
\end{eqnarray*}
As a consequence,
\begin{eqnarray*}
 \partial_r I=\frac{2\omega_K}{e^{\alpha(r)}B(r)}\,,
\end{eqnarray*}
in which $\omega_K=-\partial_v I$ is the Killing energy of the emitted particle. In the near horizon approximation,
$B(r)\simeq B'(r_H)(r-r_H)$ and, by making use of Feynman prescription for the simple pole in $(r-r_H)$, one has
\begin{eqnarray*}
I=\int dr \partial_r I=\int dr \frac{2\omega_K}{e^{\alpha(r)}B'(r_H)(r-r_H-i\varepsilon)}\,,
\end{eqnarray*}
where the range of integration over $r$ contains the location of the horizon $r_H$. Thus,
\begin{eqnarray*}
\Im\,I= \frac{2\pi \omega_K}{e^{\alpha(r_H)}B'(r_H)}\,,
\end{eqnarray*}
and the Killing/Hawking temperature finally is
\begin{eqnarray*}
T_K=\frac{e^{\alpha(r_H)}B'(r_H)}{4 \pi}\,.
\end{eqnarray*}
If one had introduced the Kodama energy $\omega_H=e^{-\alpha_H}\omega_K$, one would have obtained the Kodama/Hayward temperature
\begin{equation*} 
T_H=\frac{B'_H}{4 \pi}\,. 
\end{equation*}


\setcounter{equation}{0}

\section*{Appendix B}
\addcontentsline{toc}{chapter}{Appendix B: Energy conditions near the finite-time future singularities}

\section*{Energy conditions near the finite-time future singularities}

\paragraph*{} We briefly discuss the energy conditions related with occurrence of singularities. We have four types of energy conditions:
\begin{enumerate}
\item Weak energy condition (WEC): $\rho_{\mathrm{eff}}\geqslant 0$ and $\rho_{\mathrm{eff}}+p_{\mathrm{eff}}\geqslant 0$; 
\item Strong energy condition (SEC): $\rho_{\mathrm{eff}}+p_{\mathrm{eff}}\geqslant 0$ and $\rho_{\mathrm{eff}}+3p_{\mathrm{eff}}\geqslant 0$;
\item Null energy condition (NEC): $\rho_{\mathrm{eff}}+p_{\mathrm{eff}}\geqslant 0$;
\item Dominant energy condition (DEC): $\rho_{\mathrm{eff}}\geqslant |p_{\mathrm{eff}}|$.
\end{enumerate}
On the singular solution $H(t)=h_0/(t_0-t)^{\beta}+H_{0}$, we get\\
\phantom{line}
\begin{equation*}
\rho_{\mathrm{eff}}+p_{\mathrm{eff}}=-\frac{2}{\kappa^2}\frac{h_0\beta}{(t_{0}-t)^{\beta+1}}\,,
\end{equation*}
where $\rho_{\mathrm{eff}}$ and $p_{\mathrm{eff}}$ are the effective energy density and pressure of the universe (deriving from modified gravity, fluids, scalar fields...).

The effective dark energy related with Type I and III singularities ($\beta>0\,,\beta\neq 1$) violate the SEC and the NEC also, whereas DE related with Types II and III satisfy the NEC.

Note that\\
\phantom{line}
\begin{equation*}
 \rho_{\mathrm{eff}}+3p_{\mathrm{eff}}=-\frac{6}{\kappa^2}\left(H_{0}^{2}+2\frac{h_0 H_{0}}{(t_{0}-t)^{\beta}}+\frac{h_0^{2}}{(t_{0}-t)^{2\beta}}+\frac{h_0\beta}{(t_{0}-t)^{\beta+1}}\right)\,.\label{tatata}
\end{equation*}
\phantom{line}\\
The effective DE related with Type II singularities ($-1<\beta<0$) violate the SEC for small value of $t$. Only when $t$ is close to $t_{0}$, the last term of this equation is dominant and the SEC is satisfied on the singular solution.

In the case of Type IV singularities ($\beta<-1$), when $t$ is really close to $t_{0}$, the term $H_{0}^2$ could be dominant and the SEC is violated, expecially if $|\beta|\ll 1$. 

At last, it is easy to see that, on the singular solutions, when $t$ is near to $t_{0}$, the DEC is always violated except for large value of $H_{0}$ in the case of Type IV singularities, but also in this case the behaviour of universe approaching the singular solution violate the DEC. As a consequence, since $\rho_{\mathrm{eff}}$ has to be positive, the WEC always is satisfied on singular solutions.


\setcounter{equation}{0}

\section*{Appendix C}
\addcontentsline{toc}{chapter}{Appendix C: Conformal transformation of exponential model for inflation}

\section*{Conformal transformation of exponential model for inflation}

\paragraph{} In several cases, a suitable conformal frame to study inflation 
may be the so-called ``Einstein frame''. 
A $F(R)$-gravity theory can be rewritten in the scalar field 
form via the conformal transformation (see \S~\ref{conformal}). 

By starting from the ``Jordan frame'' of $F(R)$-gravity in vacuum, 
the ``Einstein frame'' action of the scalar field 
$\sigma$ (here, we use a suitable renormalization),
\begin{equation*}
\sigma = -\frac{\sqrt{3}}{\sqrt{2\kappa^2}}\ln [F'(R)]\,,
\end{equation*}
is given by
\begin{eqnarray*}
I_\mathrm{EF}  
&=& 
\int_{\mathcal{M}} d^4 x \sqrt{-\tilde{g}} \left( \frac{\tilde{R}}{2\kappa^2} -
\frac{1}{2}\tilde{g}^{\mu\nu}
\partial_\mu \sigma \partial_\nu \sigma + V(\sigma)\right)\,,
\end{eqnarray*}
where
\begin{equation*}
V(\sigma)=-\frac{1}{2\kappa^2}\left\{\mathrm{e}^{\sigma}R(\mathrm{e}^{-\sigma})
-\mathrm{e}^{2\sigma}F[R(\mathrm{e}^{-\sigma})]\right\}\,.
\end{equation*}
We remember that the form of conformal metric is $\tilde g_{\mu\nu}=\mathrm{e}^{-\sigma}g_{\mu\nu}$ and $\tilde R$ is the Ricci scalar defined by $\tilde g_{\mu\nu}$.

As an example, 
we explore our unified model (\ref{total}) with $\bar\gamma=1$. 
Since we are interested in the de Sitter solution at inflation, 
we take $\exp[-(R/R_\mathrm{i})^n]\rightarrow 0$ and neglect 
the Cosmological Constant $\Lambda$ of the first part. 
In this case, 
the potential $V(\sigma)$ reads 
\begin{equation*}
V(\sigma)=-\frac{1}{2\kappa^2}\left[\tilde R_\mathrm i\left(\frac{\e^{-\tilde \sigma}-1}{\alpha}\right)^{\frac{1}{\alpha-1}}\left(\e^{\tilde \sigma}-2\e^{2\tilde \sigma}\right)+\Lambda_\mathrm{i}\,\e^{2\tilde \sigma}\right]\,,
\end{equation*}
where $\tilde\sigma=\left(\sqrt{2\kappa^2/3}\right)\sigma$.
According with \S~\ref{inflation1}, 
we put $\tilde R_\mathrm{i}=R_{\mathrm{dS}}$. 
It is clearly seen that, on the de Sitter solution describing inflation, the corresponding value of the field $\sigma$ reads
\begin{equation*}
\sigma_{\mathrm{dS}}=-\sqrt{\frac{3}{2\kappa^2}}\log[1+\alpha]\,,\quad R=R_{\mathrm{dS}}\,,
\end{equation*}
and $V_{\sigma}(\sigma_{\mathrm{dS}})=0$, 
where the index `$\sigma$' denotes the derivative with respect to the inflation field $\sigma$. 
Since $V_{\sigma\sigma}(\sigma_{\mathrm{dS}})>0$, 
the scalar potential has a minimum, that is a necessary condition for 
a slow-roll inflation. 
For slow-roll parameters, we have to require 
\begin{eqnarray*}
\epsilon(\sigma) &=& \frac{1}{2\kappa^2}\left[\frac{V_{\sigma}(\sigma)}{V(\sigma)}\right]^2\ll 1\,,\nonumber\\
|\eta(\sigma)| &=& \frac{1}{\kappa^2}\left|\frac{V_{\sigma\sigma}(\sigma)}{V(\sigma)}\right|\ll 1\,.
\end{eqnarray*}
By defining the energy density and pressure of $\sigma$ as 
$\rho_{\sigma}=\dot{\sigma}^2/2-V(\sigma)$ and 
$p_{\sigma}=\dot{\sigma}^2/2+V(\sigma)$, 
these conditions imply that the gravitational field equations in the flat 
FRW space-time are given by $3H^2/\kappa^2=-V(\sigma)$, 
$3H\dot{\sigma}\simeq-V_{\sigma}(\sigma)$, and that $\ddot{a}(t)>0$, 
and hence guarantee a sufficiently long time inflation. 
In our case, since $V(\sigma_{\mathrm{dS}})\neq 0$, these two conditions are 
well satisfied around the de Sitter solution. 
Thus, since $\dot{\sigma}\simeq 0$, we find 
$\tilde H_{\mathrm{dS}}=R_{\mathrm{dS}}/\left[12(1+\alpha)\right]=\tilde R_{\mathrm{dS}}/12$. 


\fancyhead[EC, OC]{Acknowledgements}
\newpage
\section*{Acknowledgements}
\addcontentsline{toc}{chapter}{Acknowledgements}

\paragraph*{}
We would like to thank all our friends 
K.~Bamba,
E.~Bellini,
R.~Di Criscienzo,
E.~Elizalde, 
O.~Gorbunova, 
M.~Jamil,
A.~Lopez-Revelles, 
D.~Momeni,
S.~Nojiri
and M.~R.~Setare,
who collaborated with us in the papers based on which the present work has been realized.
In particular, we would like to thank 
G.~Cognola and   S.~D.~Odintsov for precious advice and valuable suggestions.
 

\renewcommand{\headrulewidth}{0.6pt}
\renewcommand{\footrulewidth}{0.pt}
\fancyhead[EC, OC]{Bibliography}
\fancyfoot[C]{\thepage} 

\addcontentsline{toc}{chapter}{Bibliography}

\bibliographystyle{plain}
\bibliography{thesis}

\renewcommand{\headrulewidth}{0.pt}
\renewcommand{\footrulewidth}{0.pt}
\fancyhead[EC, OC]{\null}
\fancyfoot[EC, OC]{\null}
\cleardoublepage

\end{document}